\documentstyle[graphicx,psfig]{oldelsart}

\newcommand{\prL}{Phys.\ Rev.\ Lett.\ }
\newcommand{\pr}{Phys.\ Rev.\ }
\newcommand{\jpb}{J.\ Phys.\ B: Atom.\ Mol.\ Opt.\ Phys.}
\newcommand{\epl}{Europhys.\ Lett.\ }

\newcommand{\ch}{CHAOS\ }

\newcommand{\phr}{Phys.\ Rep.\ }
\newcommand{\amop}{Adv.\ At.\ Mol.\ Opt.\ Phys.\ }
\newcommand{\natw}{Die Naturwissenschaften\ }
\newcommand{\pla}{Physics Letters A\ }
\newcommand{\anp}{Ann.\ Phys.\ }

\newcommand{\josab}{J.\ Opt.\ Soc.\ Am.\ B\ }
\newcommand{\zpd}{Z.\ Phys.\ D\ }

\newcommand{\csf}{Chaos, Solitons and Fractals\ }
\newcommand{\psc}{Physica Scripta\ }

\newcommand{\jpa}{J.\ Phys.\ A\ }

\newcommand{\phd}{Physica D\ }

\newcommand{\rmp}{Rev.\ Mod.\ Phys.\ }

\newcommand{\zp}{Z.\ Phys.\ }
\def\bea{\begin{eqnarray}}
\def\eea{\end{eqnarray}}
\def\be{\begin{equation}}
\def\ee{\end{equation}}
\def\c#1{\setbox0=\hbox{#1}\ifdim\ht0=1ex\accent24 #1%
  \else{\ooalign{\hidewidth\char24\hidewidth\crcr\unhbox0}}\fi}
\begin{document}
\date{}

\begin{frontmatter}

\title{Non-dispersive wave packets in periodically driven quantum systems
\vskip -1truecm}

\author{Andreas Buchleitner$^1$, Dominique Delande $^{1,2}$,
 and Jakub Zakrzewski$^3$}

\address{$^1$Max-Planck-Institut f\"ur Physik komplexer Systeme,
        Dresden, Germany }
\address{$^2$Laboratoire Kastler-Brossel, Tour 12, \'Etage 1, Universit\'e Pierre et
Marie Curie,\\
4 Place Jussieu, 75005 Paris, France
}
 \address{$^3$Instytut Fizyki imienia Mariana Smoluchowskiego, Uniwersytet
Jagiello\'nski,
 Reymonta 4, PL-30-059 Krak\'ow, Poland
}

\tableofcontents

\begin{abstract}
With the exception of the harmonic oscillator, quantum wave-packets
usually spread as time evolves. This is due to the
non-linear character of the classical equations of motion which
makes the various components of the wave-packet evolve at various
frequencies. We show here that, using the nonlinear 
resonance between an internal frequency of a system and an external 
periodic driving, it is possible to overcome this spreading
and build non-dispersive (or non-spreading)
wave-packets which are well localized and follow a classical
periodic orbit without spreading. From the quantum mechanical
point of view, the non-dispersive wave-packets are time periodic
eigenstates of the Floquet Hamiltonian, localized in the nonlinear
resonance island.

We discuss the general mechanism which produces the non-dispersive wave-packets,
with emphasis on simple realization in the electronic motion
of a Rydberg electron driven by a microwave field.
We show the robustness of such wavepackets  for a model 
one-dimensional as well as for realistic three dimensional atoms.
We consider their  essential properties such as
the stability versus ionization, the characteristic
energy spectrum and long lifetimes. The requirements
for experiments aimed at observing such non-dispersive wave-packets are
also considered.

The analysis is extended to situations in which the driving frequency is
a multiple of the internal atomic frequency. Such a case allows us to discuss
non-dispersive states composed of several, macroscopically separated
wave-packets communicating among themselves by tunneling. Similarly we
briefly discuss other closely related phenomena in atomic and molecular
physics as well as possible further extensions of the theory.    
\end{abstract}

\begin{keyword}
wave-packet, dispersion, spreading, coherent states, Rydberg atoms,
non-linear resonance, atom-field interaction
\PACS{
05.45.Mt, 03.65.Sq, 32.80.Qk, 32.80.Rm, 42.50.Hz}
\end{keyword}

\end{frontmatter}

\section{Introduction}
\subsection{What is a wave packet?}
\label{INTRO1}
It is commonly accepted that, for macroscopic systems like comets, cars, 
cats, and dogs \cite{englert95}, quantum objects
behave like classical ones. 
Throughout this report, we will understand by
``quantum objects'' 
physical systems governed by the Schr\"odinger
equation: the system can then be entirely described by its state 
$|\psi\rangle$, which, mathematically speaking,
is just a vector in Hilbert space.
We will furthermore restrict ourselves to the dynamics 
of a single, spin-less particle, 
such as to have an immediate representation of $|\psi\rangle$ in configuration 
and momentum space by the wave functions 
$\langle\vec{r}|\psi\rangle =\psi(\vec{r})$ 
and $\langle\vec{p}|\psi\rangle =\psi(\vec{p})$, respectively. 
The same object is in classical mechanics described by its phase space coordinates 
$\vec{r}$ and $\vec{p}$ (or variants thereof), 
and our central concern will be to understand
how faithfully we can mimic the classical time evolution of $\vec{r}$ and $\vec{p}$ by a 
{\em single} quantum state $|\psi\rangle$, in the {\em microscopic} realm.

Whereas classical dynamics are described by Hamilton's equations of motion, which determine
the values 
of $\vec{r}$ and $\vec{p}$ at any time, given some initial condition $(\vec{r_0},\vec{p_0})$, 
the quantum evolution is described by the Schr\"odinger equation,
which propagates the wave function. Hence, it is suggestive to
associate a classical particle with a quantum state $|\psi\rangle$
which is optimally localized around the classical particle's phase
space position, at any time $t$.              
However, quantum mechanics imposes a fundamental limit
on 
localization, expressed by 
Heisenberg's uncertainty relation
\begin{equation}
\Delta z\ .\ \Delta p \ge \frac{\hbar}{2},
\label{heisenberg}
\end{equation}
where $\Delta z$ and $\Delta p$ are the uncertainties 
(i.e., square roots of the variances) of the 
probability distributions of $z$ and its conjugate momentum $p$ in
state $|\psi\rangle$, 
respectively 
(similar relations hold for 
other choices of canonically conjugate coordinates). Consequently, the
best we can hope for is a quantum state localized with a finite width
$(\Delta z,\Delta p)$ around the particle's classical position
$(z,p)$, with $\Delta z$ and $\Delta p$ much smaller than the typical
scales of the classical trajectory. This, however, would satisfy our
aim of constructing a quantum state that mimics the classical motion,
provided $|\psi\rangle$ keeps track of the classical time evolution
of $z$ and $p$, and $\Delta z$ and $\Delta p$ remain small as time
proceeds. After all, also classical bodies follow their center of
mass trajectory even if they have a finite volume. Quantum states
which exhibit these properties at least on a finite time scale are
called ``wave-packets'', simply due to their localization properties
in phase space \cite{schroe26}.

More formally, a localized solution of a wave equation like the
Schr\"odinger equation can be conceived as a linear superposition of
plane waves (eigenstates of the momentum operator) or of any other
suitable basis states. From a purely technical point of view, such a
superposition may be seen as a {\em packet of waves}, hence, a
wave-packet. Note, however, that any strongly localized object is a
wave-packet in this formal sense, though not all superpositions of
plane waves qualify as localized objects. In addition, this formal
definition quite obviously depends on the basis used for the
decomposition. Therefore, the only sensible definition of a
wave-packet can be through its localization properties in phase space,
as outlined above. 

What can we say about the localization properties of a quantum state
$|\psi\rangle$ as time evolves? For simplicity, let us assume that
the Hamiltonian describing the dynamics has the time-independent form 
\begin{equation}
H = \frac{{\vec p}^2}{2m} + V({\vec r}),
\end{equation} 
with $V({\vec r})$ 
some potential.
The 
time evolution of $|\psi\rangle$ is then described by the 
Schr\"odinger equation 
\begin{equation}
\left[ -\frac{{\hbar}^2}{2m} \Delta + V({\vec r})  \right]\psi({\vec r},t) 
= i \hbar \frac{\partial \psi({\vec r},t)}{\partial t}.
\end{equation}
The expectation values of position and momentum in this state are
given by 
\begin{eqnarray}
<{\vec r}(t)> = \langle \psi(t)|\ {\vec r}\ |\psi(t) \rangle , \\ 
<{\vec p}(t)> = \langle \psi(t)|\ {\vec p}\ |\psi(t) \rangle ,
\end{eqnarray}
with time evolution 
\begin{eqnarray}
&&\frac{d<{\vec r}>}{dt} = \frac{1}{i\hbar}\ <[{\vec r},H]> = 
\frac{<{\vec p}>}{m},\\
&&\frac{d<{\vec p}>}{dt} = \frac{1}{i\hbar}\ <[{\vec p},H]> = 
- <\nabla V({\vec r})>,
\label{ehrenfest1}
\end{eqnarray}
and $[.,.]$ 
the commutator. 
These 
are almost the classical equations of motion generated by $H$, apart
from the right hand side of eq.~(\ref{ehrenfest1}), and apply for {\em
any} $|\psi\rangle$, irrespective of its localization
properties. If we additionally assume $|\psi\rangle$ to be
localized within a spatial region where $\nabla V(\vec{r})$ is
essentially constant, we have $<\nabla V({\vec r})> \simeq \nabla
V(<{\vec r}>)$, and therefore
\begin{eqnarray}
\frac{d<{\vec r}>}{dt} & = & \frac{<{\vec p}>}{m},\\
\frac{d<{\vec p}>}{dt} & \simeq & - \nabla V(<{\vec r}>),
\label{ehrenfest2}
\end{eqnarray}
precisely identical to the classical equations of motion. This is
nothing but 
Ehrenfest's theorem and tells us that the quantum expectation values
of $\vec{r}$ and $\vec{p}$ of an initially localized wave-packet
evolve according to the classical dynamics, as long as
$|\psi\rangle$ remains localized within a range where $\nabla
V(\vec{r})$ is approximately constant. However, these equations do not
yet give us any clue on the time evolution of the uncertainties
$(\Delta z,\Delta p)$ (and of those in the remaining degrees of
freedom), and, consequently, neither on the time scales on which they
are reliable. 

On the other hand, given a localized wave-packet at time $t=0$, a
decomposition 
\begin{equation}
|\psi (t=0)\rangle = \sum_n c_n \ |\phi_n\rangle
\label{decompo}
\end{equation}
with coefficients 
\begin{equation}
c_n = \langle \phi_n | \psi(t=0) \rangle
\label{coeffs}
\end{equation}
in unperturbed energy eigenstates 
\begin{equation}
H |\phi_n\rangle = E_n |\phi_n\rangle,\ n=1,2,\ldots,
\end{equation}
tells us immediately that 
\begin{equation}
|\psi (t)\rangle =  
\sum_n c_n \exp \left( -i \frac{E_n t}{\hbar}\right) \ |\phi_n\rangle 
\label{psit}
\end{equation}
cannot be stationary, {\em except} for 
\begin{equation}
|\psi (t=0)\rangle = |\phi_j\rangle ,
\label{single}
\end{equation}
for some suitable $j$. 

The eigenstates $|\phi_j\rangle$ are typically delocalized over a 
large part of phase space (for example, over
a classical trajectory, see section~\ref{SQ}), and thus
are not wave-packets. There is however, an exception:
in the vicinity of a (stable) fixed point of the classical
dynamics (defined \cite{lichtenberg83} as a point in phase space 
where the 
time derivatives 
of positions and momenta vanish simultaneously), there exist localized 
eigenstates, see section~\ref{GM}.

For a particle moving in a one-dimensional, binding potential bounded
from below, there is a stable fixed
point at any potential minimum. 
The quantum mechanical ground state of this system is 
localized near
the fixed point at the global minimum of the potential and is a
wave-packet, though a very special one: 
it does
not evolve in time. Note that there is no need for the potential to be 
harmonic, any potential minimum will do.
The same argument can be used for a one-dimensional binding potential 
whose origin 
moves with uniform velocity.
The problem can be 
reduced to the previous one  by transforming 
to the moving frame where 
the potential is
stationary. 
Back in the laboratory frame, 
the ground state of the particle in the 
moving frame will appear
as a wave-packet which moves 
at uniform velocity. Obviously, expanding the wave-packet in a
stationary 
basis
in the laboratory 
frame will result in an awfully complicated decomposition, with
time-dependent coefficients, and this example clearly 
illustrates
the importance of the proper choice of the referential.\footnote{In passing, 
note
that such a situation is actually realized in particle accelerators: 
electromagnetic
fields are applied to the particles, such 
that these are trapped at some fixed 
point 
(preferably stable) in
an accelerated frame \cite{lichtenberg83}.}

If eq.~(\ref{single}) is not fulfilled, the initial localization of
$|\psi(t=0)\rangle$ (which is equivalent to an appropriate choice
of the $c_n$ in eq.~(\ref{coeffs})) will progressively deteriorate as
time evolves, the wave-packet will {\em spread}, 
due to the accumulation of relative phases of the
different contributions to the sum in eq.~(\ref{psit}). Whereas the
classical dynamics in a one-dimensional binding potential $V(\vec{r})$ are described
by periodic orbits (at any energy), the quantum dynamics are in
general {\em not} periodic. A return to the initial state is {\em only}
possible if {\em all} the phases $\exp(-i E_n t/\hbar)$ simultaneously take
the same value. This implies that all the energy levels $E_n$ 
(with $c_n\neq 0$) are
equally spaced, or that all the 
level spacings are integer multiples of some quantity.
In practice, this is realized only for the harmonic oscillator 
(in any dimension), and for tops or rotors where the Hamiltonian
is proportional to some component of an angular 
momentum variable. Another possibility is to use the linear Stark effect
in the hydrogen atom which produces manifolds of equally spaced energy levels.
However, experimental imperfections (higher order Stark effect and
effect of the ionic core on non-hydrogenic Rydberg atoms) break the
equality of the spacings and consequently lead to dispersion \cite{raman97}.
 
The equality of consecutive spacings has a simple classical interpretation: 
since all classical trajectories are
periodic with the {\em same} period, the system is exactly back in its initial state
after an integer number of periods. In other words, in those special cases, 
there is no wave-packet
spreading at long times.

However, for more generic systems, the energy levels are not equally spaced,
neither are the spacings simply related, and a wave-packet {\em will} spread. 
For a 
one-dimensional,
time-independent system, it is even possible to estimate the time after which 
the wave-packet
has significantly spread (this phenomenon is also known as the ``collapse'' 
of the
wave-packet \cite{yeazell90,alber91}). 
This is done by expanding the various energies $E_n$ around
the ``central'' energy $E_{n_0}$ of the wave-packet:
\begin{equation}
E_n \simeq E_{n_0} + (n-n_0) \frac{dE_n}{dn}(n_0) + \frac{(n-n_0)^2}{2} 
\frac{d^2E_n}{dn^2}(n_0).
\label{esec}
\end{equation}

The wave-packet being initially localized, its energy is more or less
well defined and only a relatively small number $\Delta n \ll n_0$
of the coefficients $c_n$ have significant values.
At short times, the contribution of the second order term in eq.~(\ref{esec})
to the evolution can be neglected. Within this approximation, the important energy levels
can be considered as equally spaced, and one obtains a periodic motion of the
wave-packet, with period:
\begin{equation}
T_{\rm recurrence} = \frac{2\pi \hbar}{\frac{dE_n}{dn}(n_0)}.
\label{trec}
\end{equation}
In the standard semiclassical 
WKB approximation (discussed in section \ref{wkb}) \cite{landau2}, this is
nothing but the classical period of the motion at energy $E_{n_0}$, 
and one recovers the similarity
between the quantum motion of the wave-packet and the classical motion of a
particle.

At longer times, the contributions of the various eigenstates to the dynamics 
of the
wave-packet will come out of phase because of the second order term in 
eq.~(\ref{esec}),
resulting in spreading and collapse of the wave-packet. A rough estimate of 
the collapse time
is thus when the relevant phases have changed by $2\pi.$
One obtains:
\begin{equation}
T_{\rm collapse} \simeq \frac{1}{(\Delta n)^2}\ \frac{2 \hbar}{\frac{d^2E_n}{dn^2}(n_0)}.
\label{tcol}
\end{equation}
Using the standard WKB approximation, one can show that this expression 
actually
corresponds to the time needed
for the corresponding classical phase space density to 
significantly
spread under the influence of the {\em classical} evolution. 

At still longer times, a pure quantum phenomenon appears, due to the 
discrete
nature of the energy spectrum. Since the $(n-n_0)^2$ factors in 
eq.~(\ref{esec}) are all
integers, the second order contributions to the
phase are all integer multiples of the phase of the $n-n_0=1$ term. If the 
latter is an
integer multiple of $2\pi$, {\em all} the second order contributions 
will rephase,
inducing a revival of the wave-packet in its original shape. A refined 
estimation of the revival time actually shows that this analysis 
overestimates the
revival time by a factor two.\footnote{It must also be noted that, at simple rational multiples 
(such as 1/3, 1/2, 2/3) of the revival time, one observes ``fractional revivals'' 
\cite{parker86,alber86,averbukh89,yeazell91},
where only part of the
various amplitudes which contribute to eq.~(\ref{psit}) rephase. This
generates a wave-function split
into several individual wave-packets, localized at different positions along 
the
classical orbit.}
 The correct result is
\cite{parker86,alber86,averbukh89,yeazell91}:
\begin{equation}
T_{\rm revival} = \frac{2\pi \hbar}{\frac{d^2E_n}{dn^2}(n_0)}.
\label{trev}
\end{equation}

Based on the very elementary considerations above, we can so far
draw the following conclusions:
\begin{itemize}
\item An initially localized wave-packet will follow the classical
equations of motion for a finite time $t\sim T_{\rm recurrence}$;
\item its localization properties {\em cannot} be stationary as time
evolves;
\item in general, the initial quasi-classical motion is followed by collapse
and revival, with the corresponding time scales $T_{\rm
recurrence}<T_{\rm collapse}<T_{\rm revival}$.
\end{itemize}

In the sequel of this report, we will show how under very general
conditions it is indeed possible to create wave-packets as single
eigenstates of quantum systems, i.e., as localized ground states in an
appropriately defined reference frame.
The most suitable framework is to consider
quantum evolution in classical phase space, that provides a picture
which is {\em independent} of the choice of the basis and allows for
an immediate comparison with  the classical Hamiltonian flow. In
addition, such a picture motivates a semiclassical interpretation,
which we will expand upon in sec.~\ref{SQ}. The appropriate technical
tool for a phase space description are quasiprobability distributions
\cite{hillery84} 
as the Wigner representation
of the state $|\psi(t)\rangle$,
\begin{equation}
W({\vec r},{\vec p}) = \frac{1}{(2\pi \hbar)^f}
\ \int \psi^*\left({\vec r}+ \frac{\vec x}{2}\right)  
\psi \left({\vec r}- \frac{\vec x}{2}\right)
\exp{\left( i \frac{{\vec x}.{\vec p}}{\hbar}\right)}
\ {\rm d}^f{\vec x},
\end{equation}
where $f$ is the number of degrees of freedom. 

The Wigner density $W({\vec r},{\vec p})$ is real, but not necessarily
positive \cite{hillery84,moyal49}.
Its  
time-evolution follows from the Schr\"odinger equation \cite{hillery84}:
\begin{equation}
\hbar \frac{\partial W({\vec r},{\vec p},t)}{\partial t} =
-2 H({\vec r},{\vec p},t) \sin \left(\frac{ \hbar \Lambda}{2}\right )\ 
W({\vec r},{\vec p},t),
\label{dwsdt}
\end{equation}
where 
\begin{equation}
\Lambda = \overleftarrow{\nabla_{\vec p}} \overrightarrow{\nabla_{\vec r}} 
- \overleftarrow{\nabla_{\vec r}} \overrightarrow{\nabla_{\vec p}}, 
\end{equation}
and the arrows indicate in which direction the derivatives act.
Eq.~(\protect\ref{dwsdt}) can serve to
motivate the semiclassical approach. Indeed,
the $\sin$ function can be expanded in a Taylor series, i.e. a power 
expansion in $\hbar.$ 
At lowest non--vanishing order, only terms linear in $\Lambda$ 
contribute and
one obtains:
\begin{equation}
\frac{\partial W({\vec r},{\vec p},t)}{\partial t} = \{H, W({\vec r},{\vec p},t)\},
\label{liouville}
\end{equation}
where $\{.,.\}$ denotes the classical Poisson bracket \cite{lichtenberg83}:
\footnote{We choose here the most common definition of the Poisson bracket.
Note, however, that 
some
authors \protect{\cite{landau1,haake90}} use the opposite sign!}
\begin{equation}
\{f,g\} = \sum_{i=1...f}{\frac{\partial f}{\partial r_i} 
\frac{\partial g}{\partial p_i} - \frac{\partial f}{\partial p_i} 
\frac{\partial g}{\partial r_i} }.
\label{poisson_brackets}
\end{equation}
Eq.~(\ref{liouville}) is nothing but 
the classical Liouville equation \cite{lichtenberg83}
which describes the  classical evolution of a phase 
space density.
Hence, in the ``semiclassical limit'' $\hbar \rightarrow 0,$ the
Wigner density evolves classically.
Corrections of higher power in $\hbar$ can be calculated systematically.
For example, the next order is $\hbar^3 H\Lambda^3W/24$ in eq.~(\ref{dwsdt}), 
and
generates terms which contain third order 
derivatives (in either position and/or
momentum) of the
Hamiltonian. Therefore, for a Hamiltonian of maximal degree two
in 
position and/or momentum, all higher order terms in eq.~(\ref{dwsdt}) vanish 
and the
Wigner distribution follows the classical evolution for an arbitrary 
initial
phase space density, 
and for arbitrarily long times. The harmonic oscillator 
is an example
of such a 
system \cite{schroe26,glauber63}, in agreement with our discussion of
eq.~(\ref{esec}) above.

Now, once again, why does a wave-packet spread? At first sight, it could be thought that
this is due to the higher order terms in eq.~(\ref{dwsdt}), and thus of
quantum origin. This is not true and spreading of a wave-packet
has a purely classical origin, as illustrated 
by the following example. Let us consider
a one-dimensional, free particle (i.e. no potential), initially described by a
Gaussian wave-function with average position $z_0$,
average momentum $p_0>0$, and spatial width $\sigma$:
\begin{equation}
\psi(z,t=0) = \frac{1}{\pi^{1/4} \sqrt{\sigma}}
\exp \left( i \frac {p_0 z}{\hbar}
- \frac{(z-z_0)^2}{2\sigma^2}\right ).
\label{wp_gaussian}
\end{equation}
The corresponding Wigner distribution is a Gaussian in phase space:
\begin{equation}
W(z,p,t=0)=\frac{1}{\pi \hbar} \ \exp \left( - \frac {(z-z_0)^2}{\sigma^2} 
-\frac {\sigma^2 (p-p_0)^2}{\hbar^2} \right).
\label{wigner_gaussian}
\end{equation}
As the Hamiltonian is quadratic in the momentum, without potential, 
this distribution
evolves precisely alike the equivalent 
classical phase space density. Hence, the part of the 
wave-packet
with $p>p_0$ will evolve 
with a 
larger velocity than the part with $p<p_0.$ 
Even if both parts are initially localized close to $z_0$, 
the  
contribution of different velocity classes 
implies that their distance will increase
without bound at long times. The wave-packet will therefore 
{\em spread}, because
the various {\em classical} trajectories have different velocities. Spreading is thus
a completely classical phenomenon. 

\begin{figure}
\centerline{\psfig{figure=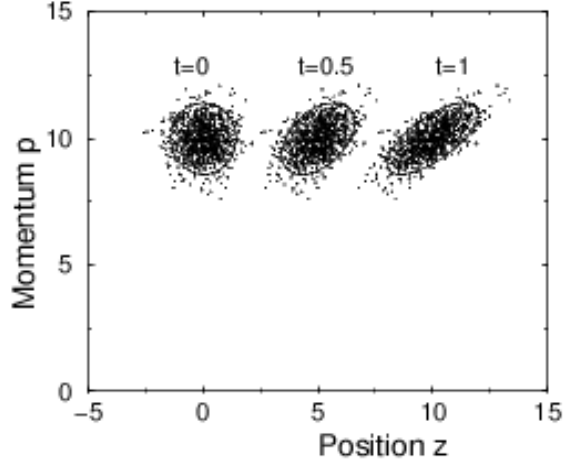,width=8cm}}
\caption{Evolution of the Wigner density of a wave-packet for a free
particle moving in a one-dimensional configuration space, compared 
to the classical 
evolution of a swarm
of classical particles with the same initial probability density. 
While the uncertainty in momentum does not vary
with time, the Wigner density stretches along the position coordinate
and loses its initial minimum uncertainty character. This implies
spreading of the wave-packet. This has a purely classical origin, as shown
by the classical evolution of the swarm of particles, which closely follow the
quantum evolution. The contour of the Wigner density is chosen to contain
86\% of the probability.}
\label{classical_spreading}
\end{figure}

This can be seen quantitatively by calculating the exact quantum evolution.
One obtains
\begin{equation}
W(z,p,t)=\frac{1}{\pi \hbar} \exp \left( - \frac {(z-z_0-pt/m)^2}{\sigma^2} 
-\frac {\sigma^2 (p-p_0)^2}{\hbar^2} \right)
\end{equation}
for the Wigner distribution, and 
\begin{equation}
\psi(z,t) = \frac{1}{\pi^{1/4}}\ 
\frac{{\rm e}^{i\phi}}{\left(\sigma^2+\frac{\hbar^2t^2}{m^2\sigma^2}\right)^{1/4}}
\ \exp \left\{ i \frac{p_0 z}{\hbar}- \frac{\left(z-z_0-\frac{p_0t}{m}\right)^2}{2\sigma^2+\frac{2i\hbar t}{m}}\right\}
\end{equation}
for the wave-function 
(${\rm e}^{i\phi}$ is an irrelevant, complicated
phase factor).
The former is 
represented in fig.~\ref{classical_spreading}, together with the
evolution of a swarm of classical particles with an initial
phase space density identical to the one of the initial quantum
wave-packet. Since the quantum evolution follows exactly the classical one,
the phase space volume of the wave-packet is preserved. However, the
Wigner distribution is progressively stretched along the $z$ axis.
This results in a less and less localized wave-packet, with
\begin{equation}
\left\{
\begin{array}{l}
\displaystyle
\Delta z (t) = \frac{\sigma}{\sqrt{2}} \sqrt{1+\frac{\hbar^2t^2}{m^2\sigma^4}},\\
\displaystyle
\Delta p (t) = \frac{\hbar}{\sigma \sqrt{2}}.
\end{array}
\right.
\end{equation}
The product $\Delta z\Delta p,$ initially minimum ($\hbar/2)$, continuously increases
and localization is eventually lost.

\psfull

\subsection{Gaussian wave-packets -- Coherent states}
\label{coherent_states}

\begin{figure}
\centerline{\psfig{figure=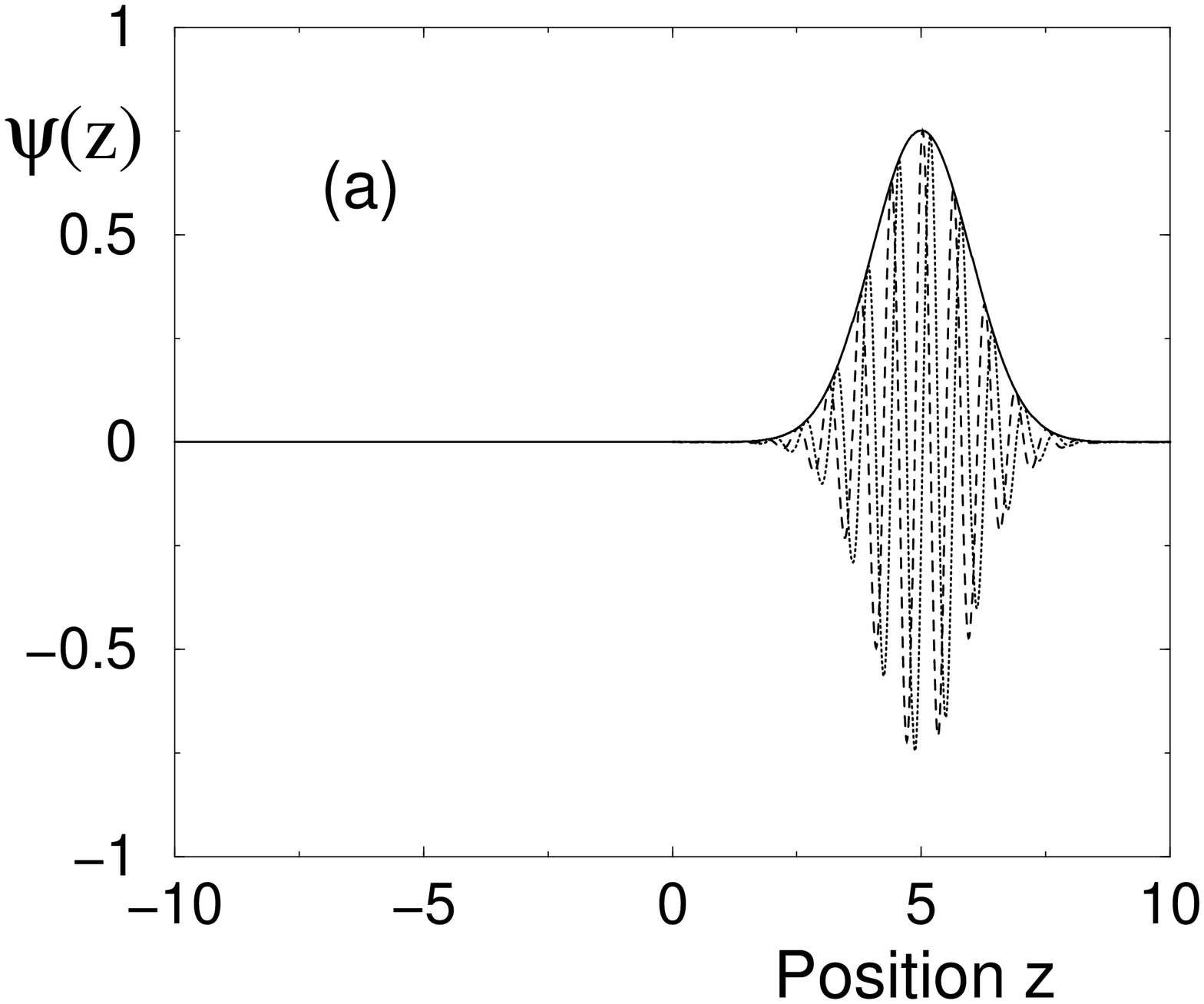,width=7cm,angle=-90}
\psfig{figure=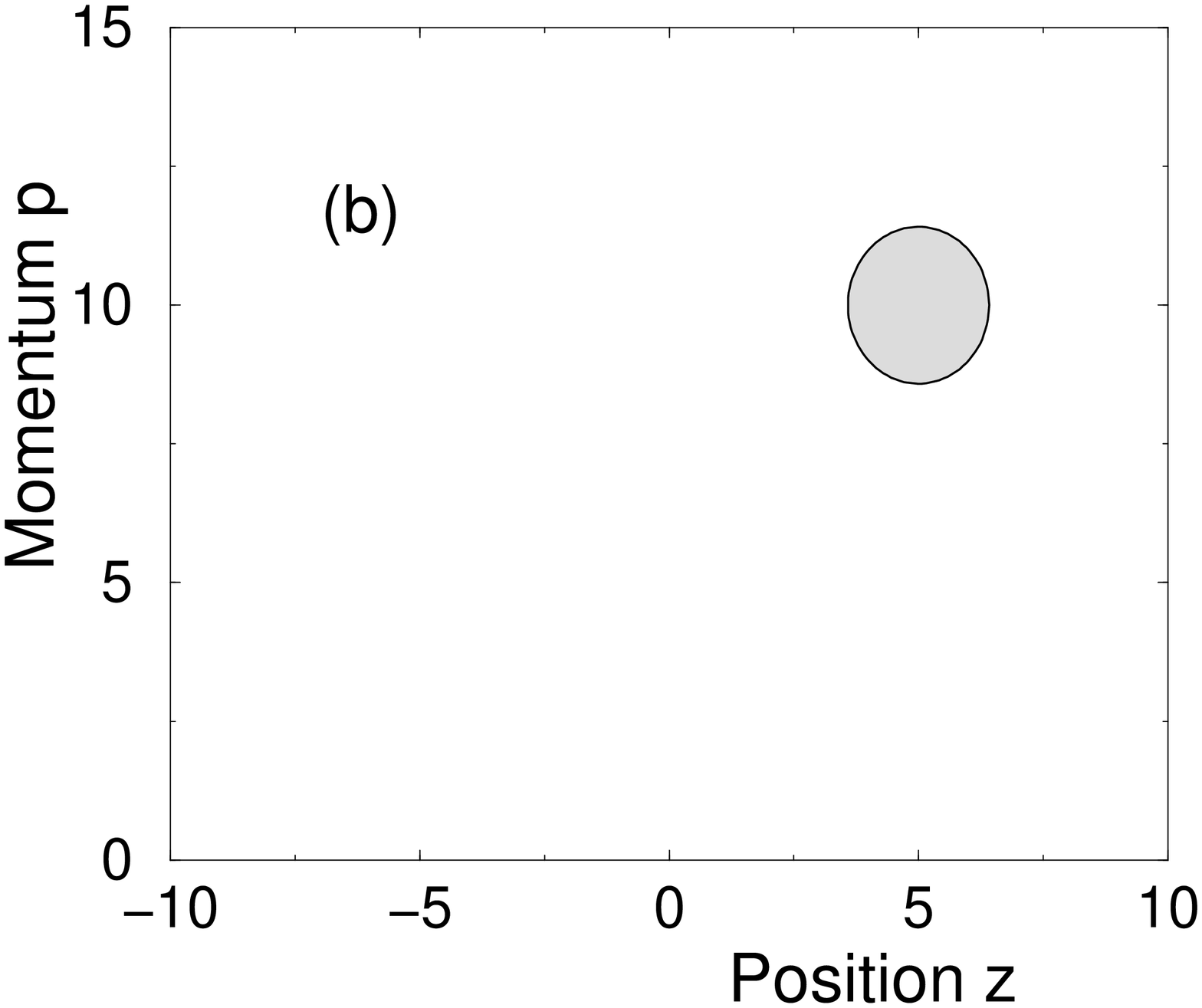,width=7cm,angle=-90}}
\caption{(a): Coherent state of the harmonic oscillator 
\protect\cite{schroe26}. The
probability density and the modulus of the wave-function
(solid line) have a Gaussian distribution. The real (dashed line)
and the imaginary (dotted line) part of the wave-function show, in addition, 
oscillations which reflect
the non-vanishing momentum. Whereas the envelope of the wave-function
preserves its shape under time evolution, the frequency of the oscillations
exhibits the same 
time dependence as the momentum of the corresponding classical particle. 
(b): Contour of the corresponding Wigner distribution which shows
localization in both position and momentum. The isovalue for the contour
is chosen to enclose 86\% of the total probability.} 
\label{coherent_state} 
\end{figure}

We have already realized above that,
for the harmonic oscillator, the second derivative $d^2E_n/dn^2$ in 
eq.~(\ref{esec}) 
vanishes
identically, and
a wave-packet does not spread, undergoing periodic motion. 
For this specific 
system,
one can define a restricted class of wave-packets, which are minimum
uncertainty states
(i.e., $\Delta z\Delta p = \hbar/2)$, and remain minimal under
time-evolution \cite{schroe26}. 
Nowadays, these states are known as ``coherent'' states of the harmonic oscillator
\cite{glauber63}, and are frequently employed in the analysis of simple quantum
systems such as the quantized electromagnetic field \cite{mandel90,cct92}.
They have Gaussian wave-functions, see fig.~\ref{coherent_state}(a), 
given by eq.~(\ref{wp_gaussian}), and characterized by an average position
$z_0$, an average momentum $p_0$, and a spatial width
\begin{equation}
\sigma = \sqrt{\frac{\hbar}{m\omega}},
\label{cohsigma}
\end{equation}
where $\omega$ is the classical eigenfrequency of the harmonic oscillator. 
The corresponding
Wigner distribution, 
eq.~(\ref{wigner_gaussian}), also
has 
Gaussian shape. The properties
of coherent states are widely discussed in the litterature, see
\cite{cct73,liboff80}.

In the ``naturally scaled'', dimensionless coordinates 
$z\sqrt{m\omega/\hbar}$ and $p/\sqrt{m\omega \hbar}$,
the classical trajectories of the harmonic oscillator 
are circles, and the Wigner distribution is an
isotropic Gaussian centered at $(z_0,p_0),$ see 
fig.~\ref{coherent_state}(b). 
Under time evolution, which follows precisely the classical
dynamics, its isotropic Gaussian shape is preserved.

\psfull

An important point when discussing wave-packets is to avoid the
confusion between localized wave-packets (as defined above) and minimum 
uncertainty
(coherent or squeezed \cite{cct92}) states. The latter are
just a very restricted class of localized states. They are the best ones in 
the sense
that they have optimum localization. On the other hand, as soon as dynamics 
is considered,
they have nice properties only for harmonic oscillators. In generic systems, 
they
spread exactly like other wave-packets.
Considering only coherent states as good semiclassical analogs of classical
particles is in our opinion 
a too formal point of view. Whether
the product $\Delta z\Delta p$ is exactly $\hbar/2$ or slightly
larger is certainly of secondary relevance for the semiclassical 
character of the wave-packet. What counts is that, in the semiclassical limit
$\hbar \rightarrow 0,$ the wave-packet is asymptotically perfectly
localized in all directions of phase space. This was Schr\"odinger's original 
concern, without reference to the actual value of $\Delta z\Delta p$ 
\cite{schroe26}.

Finally, for future applications, let us define the so called Husimi representation
of the quantum wave-function~\cite{husimi40}. It is the squared projection of a given
quantum state over a set of coherent states. 
Let us denote the gaussian wavefunction of eq.~(\ref{wp_gaussian}) 
(with $\sigma$ given by eq.~(\ref{cohsigma})) as 
$|{\mathrm Coh}(z_0,p_0)\rangle$. Then the
Husimi representation of $\psi(z)$ is defined as
\begin{equation}
{\mathrm Hus}(z,p)=\frac{1}{\pi}|\langle {\mathrm Coh}(z,p)|\psi\rangle|^2,
\label{husimi_def}
\end{equation} 
where the factor $1/\pi$ is due to the resolution of unity in the coherent states basis
\cite{cct73} and is often omitted (to confuse the reader). Alternatively,
the Husimi function may be looked upon as a Wigner function convoluted with
a Gaussian~\cite{hillery84}.

\subsection{A simple example: the one-dimensional hydrogen atom}
\label{INTRO2}

We now illustrate the ideas discussed in the preceding sections, using the
specific example of a one-dimensional hydrogen atom.
This object is both, representative of generic systems, and useful for
atomic systems to be discussed later in this paper.
We choose the simplest hydrogen atom: we neglect all relativistic, spin and QED
effects, and assume that the nucleus is infinitely massive. The Hamiltonian 
reads:
\begin{equation}
H = \frac{p^2}{2m} - \frac{e^2}{z},
\label{ham_h_1d}
\end{equation}
where $m$ is the mass of the electron, $e^2=q^2/4\pi\epsilon_0$, with $q$
the elementary charge, and
$z$ is restricted to the positive real axis. The validity of this
model as compared to the real 3D atom will be discussed in sec.~\ref{LIN}.

Here and in the rest of this paper, we will use atomic units, defined by 
$m$, $e^2$
and $\hbar$. The unit of length is
the Bohr radius $a_0=\hbar^2/me^2=5.2917\times
 10^{-11}\ {\rm m}$, the unit of time
is 
$\hbar^3/me^4 = 2.4189\times 10^{-17}\ {\rm s}$, the unit of 
energy is the 
Hartree
$me^4/\hbar^2 = 27.2\ {\rm eV},$ {\em twice} the ionization energy of the
hydrogen atom, and the unit of frequency is 
$me^4/2\pi\hbar^3 = 6.5796\times 10^{16}\ {\rm Hz}$ \cite{bethe77}.

With these premises, the energy levels are:
\footnote{The present analysis
is restricted to bound states of the atom. Continuum (i.e., scattering) 
states also
exist  but usually do not significantly contribute to the
wave-packet dynamics. If needed, they can be incorporated without any
fundamental difficulty~\cite{goldberger50,faisal87}.}
\begin{equation}
E_n= - \frac{1}{2n^2},\ \ \ \ \ \ {\rm for}\ n\geq 1.
\label{energy_levels}
\end{equation}
Clearly, the levels are not equally spaced, and therefore (see eq.~(\ref{esec})) 
any wave-packet
will spread. 

\begin{figure}
\centerline{\psfig{figure=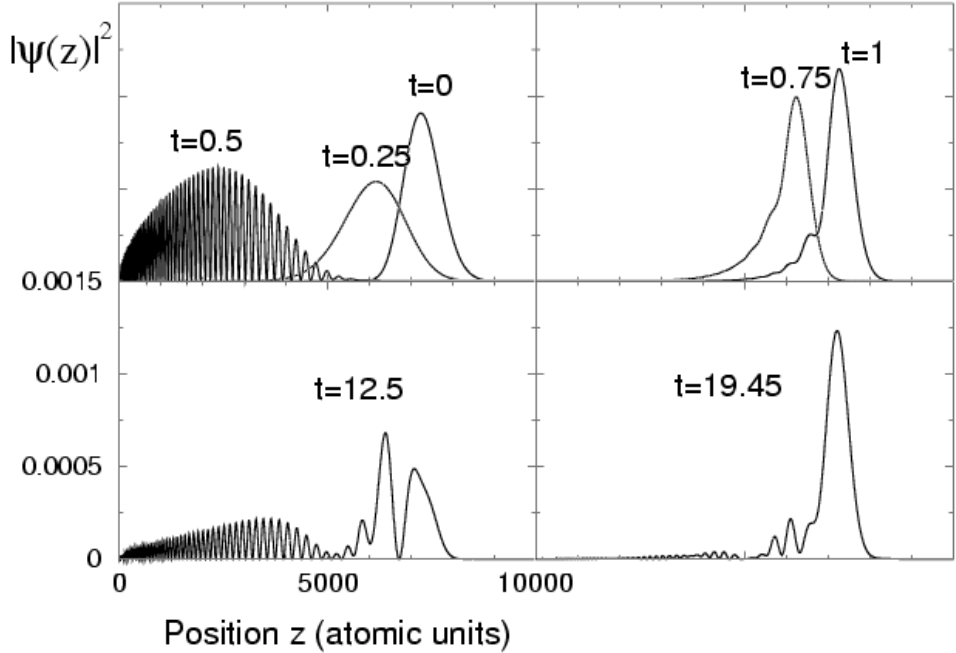,width=14cm,angle=-90}}
\caption{Time evolution of an initially localized
wave-packet in the one-dimensional hydrogen atom,
eq.~(\protect\ref{ham_h_1d}). The wave-packet is constructed 
as a linear superposition of energy eigenstates of $H$, with a Gaussian 
distribution (centered at $n_0=60$, with a 
width $\Delta n=1.8$ of the 
$|c_n|^2$) of the coefficients $c_n$ in eq.~(\protect\ref{decompo}). Time $t$ is
measured in units of the classical Kepler period $T_{\rm recurrence}$,
eq.~(\protect\ref{hrecur}). Note the quasiclassical approach of the wave-packet 
to the nucleus, during the first half period (top left), with the
appearance of interference fringes as the particle is accelerated towards the
Coulomb center. After one period (top right) the wave-packet almost resumes
its initial shape at the outer turning point of the classical motion, but
exhibits considerable dispersion (collapse) 
after few  Kepler cycles (bottom
left). Leaving a little more time to the quantum evolution, we observe a
non-classical revival after approx. $20$ Kepler cycles (bottom right).
Recurrence, collapse and revival times are very well predicted by
eqs.~(\protect\ref{hrecur}), (\protect\ref{hcol}) and (\protect\ref{hreviv}).}
\label{wp_h1d} 
\end{figure}

Fig.~\ref{wp_h1d} shows the evolution of a wave-packet
built from a linear combination of eigenstates of $H$, using a Gaussian
distribution of the coefficients $c_n$ in eq.~(\ref{psit}). The distribution
is  
centered at $n_0=60$, with
a width $\Delta n = 1.8$ for the $|c_n|^2.$ 
The calculation is done numerically, but is simple 
in the hydrogen atom since
all ingredients -- energy levels and eigenstates -- are known analytically. 
At time $t=0$, the wave-packet is localized at the outer turning point
(roughly at a distance $2n_0^2$ from the origin), and has 
zero initial momentum; its shape is roughly Gaussian.
After a quarter of a classical Kepler 
period $T_{\rm recurrence}$, it is significantly closer to the nucleus,
with negative velocity, following the classical trajectory. 
After half a period, it has reached the nucleus (it is essentially 
localized near
the origin). However, interference fringes are clearly visible: they 
originate from
the interference between the head of the wave-packet, which has already 
been reflected
off the nucleus, and its tail, which has not yet reached the nucleus.
After 3/4 of a period, the interference 
fringes have disappeared, and the wave-packet propagates
to the right. It has already spread significantly. After one period,
it is close to its initial position, but no more as well localized as 
initially.
This recurrence time is given by
eqs.~(\ref{trec}) and (\ref{energy_levels}):
\begin{equation}
T_{\rm recurrence} =2\pi n_0^3.
\label{hrecur}
\end{equation}

\psfull

\begin{figure}
\centerline{\psfig{figure=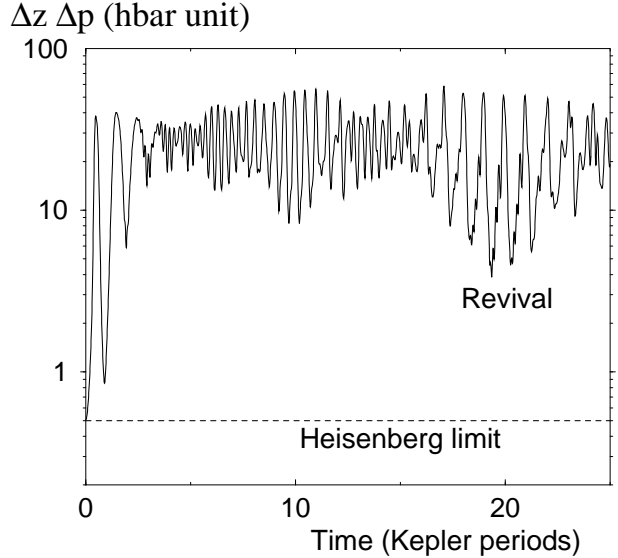,width=8cm,angle=-90}}
\caption{Time evolution of the uncertainty product $\Delta z\Delta p$ 
(in units of $\hbar$) of the wave-packet shown in fig.~\protect\ref{wp_h1d}. 
Starting out from minimum uncertainty, $\Delta z\Delta p\simeq
\hbar/2$ (the Heisenberg limit, eq.~(\protect\ref{heisenberg})), the
wave-packet exhibits some transient spreading on the time scale of a 
Kepler period $T_{\rm recurrence}$, thus reflecting the classical motion
(compare top left of fig.~\protect\ref{wp_h1d}), collapses on a time scale of
few Kepler cycles (manifest in the damping of the oscillations of $\Delta
z\Delta p$ during the first five classical periods), shows a fractional
revival around $t\simeq 10\times T_{\rm recurrence}$, and a full revival at 
$t\simeq 20\times T_{\rm recurrence}$. Note that, nontheless, even at the full
revival the contrast of the oscillations of the uncertainty product is reduced
as compared to the initial stage of the evolution, as a consequence of 
higher-order corrections which are neglected in eq.~(\protect\ref{esec}).}
\label{dpdx} 
\end{figure}

After few periods, the wave-packet has considerably spread and is now
completely delocalized along
the classical trajectory. The time for the collapse of the
wave-packet is well predicted by eq.~(\ref{tcol}):
\begin{eqnarray}
T_{\rm collapse} & \simeq & \frac{2 n_0^4}{3(\Delta n)^2} = 
\frac{n_0}{3 \pi (\Delta n)^2}\times T_{\rm recurrence}\nonumber \\
& = & 1.96\times T_{\rm recurrence},\ 
{\rm for}\
n_0=60\ {\rm and}\ \Delta n=1.8.
\label{hcol}
\end{eqnarray}
Finally, after 20 periods, the
wave-packet revives with a shape similar to its initial state. Again, this 
revival
time is in good agreement with the theoretical prediction, 
eq.~(\ref{trev}):
\begin{eqnarray}
T_{\rm revival} & = & \frac{2 \pi n_0^4}{3} = \frac{n_0}{3}\times T_{\rm
recurrence} \nonumber \\ 
& = & 20\times T_{\rm recurrence},\ {\rm for}\
n_0=60\ {\rm and}\ \Delta n=1.8.
\label{hreviv}
\end{eqnarray}
At longer times, the wave-packet continues to alternate between collapses 
and revivals.
In fig.~\ref{dpdx}, we show the temporal evolution 
of the product $\Delta z \Delta p.$
It is initially close to the Heisenberg limit (minimum value) $\hbar/2$, 
and oscillates at the
frequency of the classical motion with a global increase. 
When the wave-packet has completely
spread, the uncertainty product is roughly constant, with apparently 
erratic fluctuations. 
At the revival time, the uncertainty undergoes again rather orderly
oscillations of a relatively large magnitute reaching, at minima,
values close to $\hbar$. That is a manifestation of
 its partial relocalization.

\psfull
 
\begin{figure}
\centerline{\psfig{figure=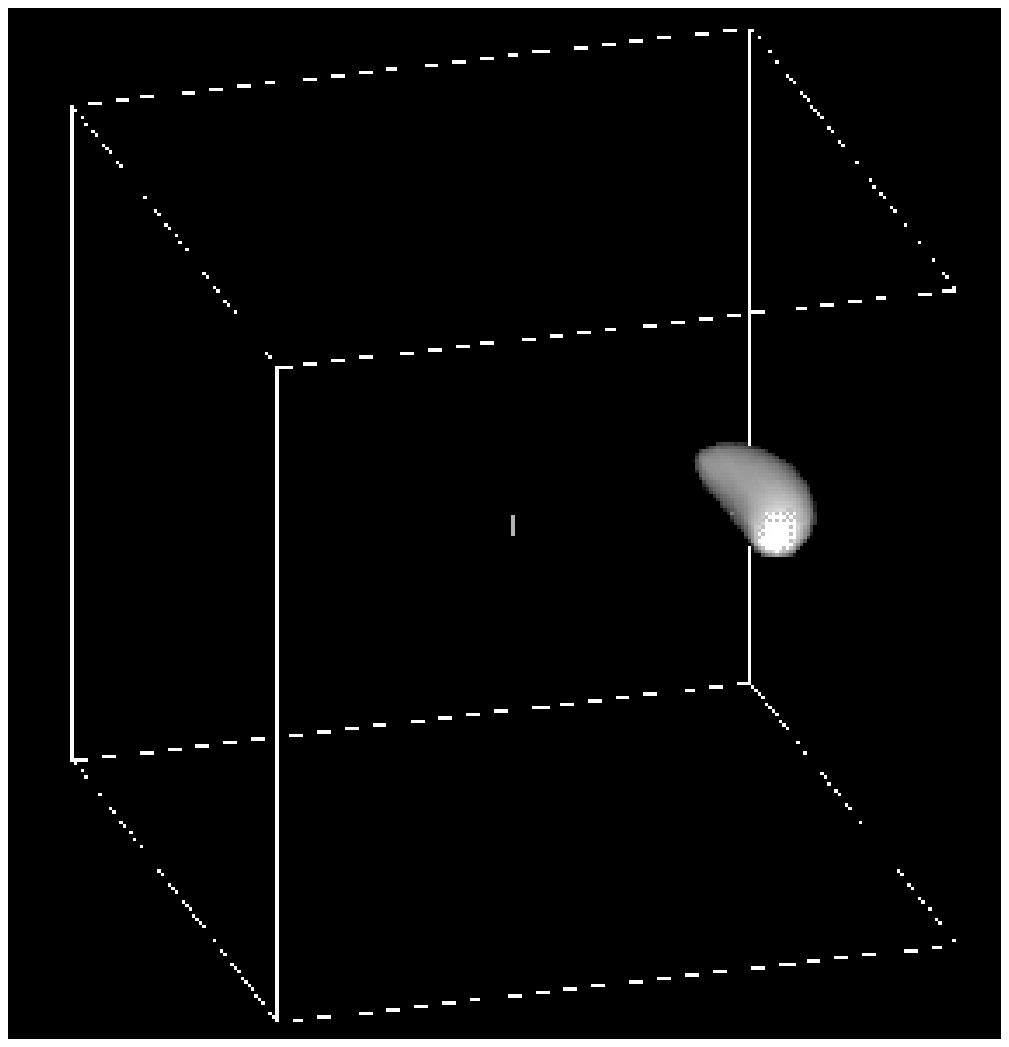,width=6cm,angle=0}
\psfig{figure=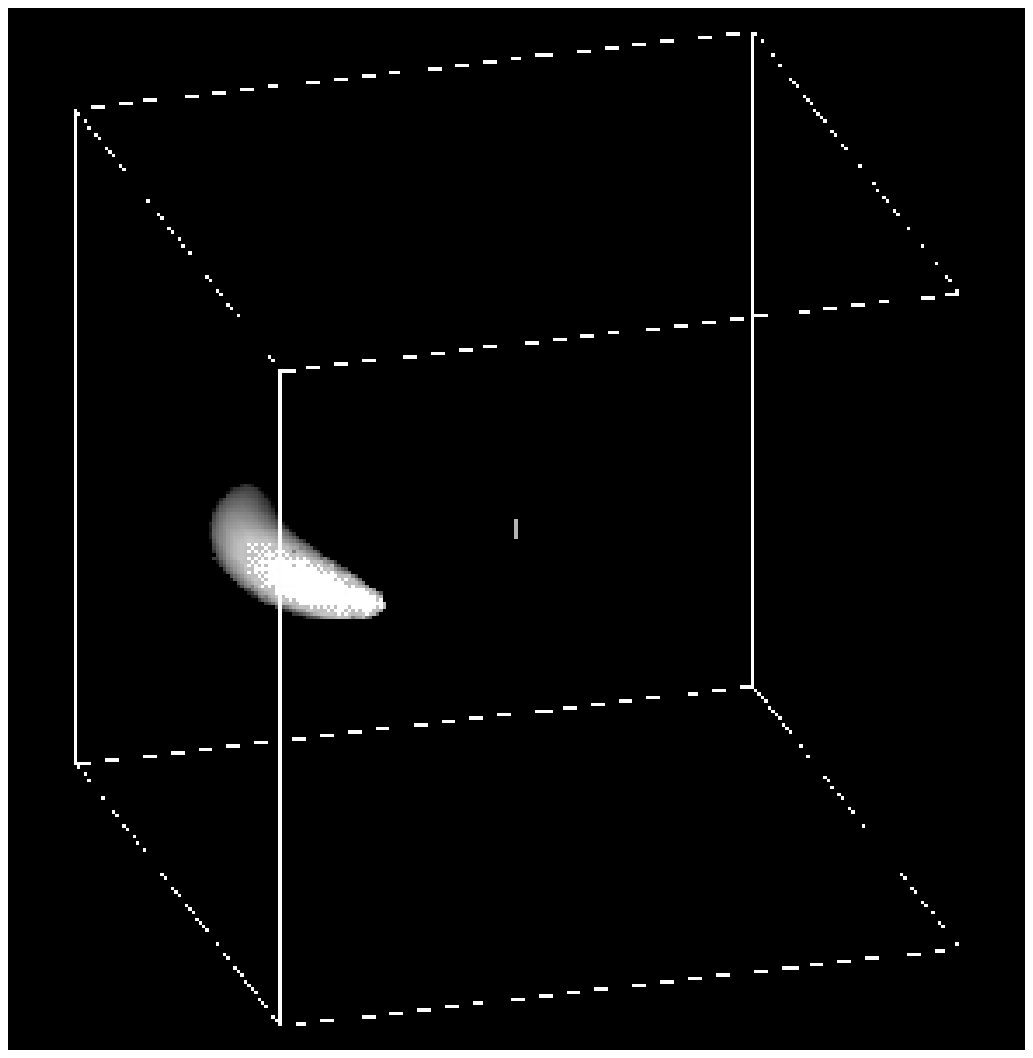,width=6cm,angle=0}}

\smallskip

\centerline{\psfig{figure=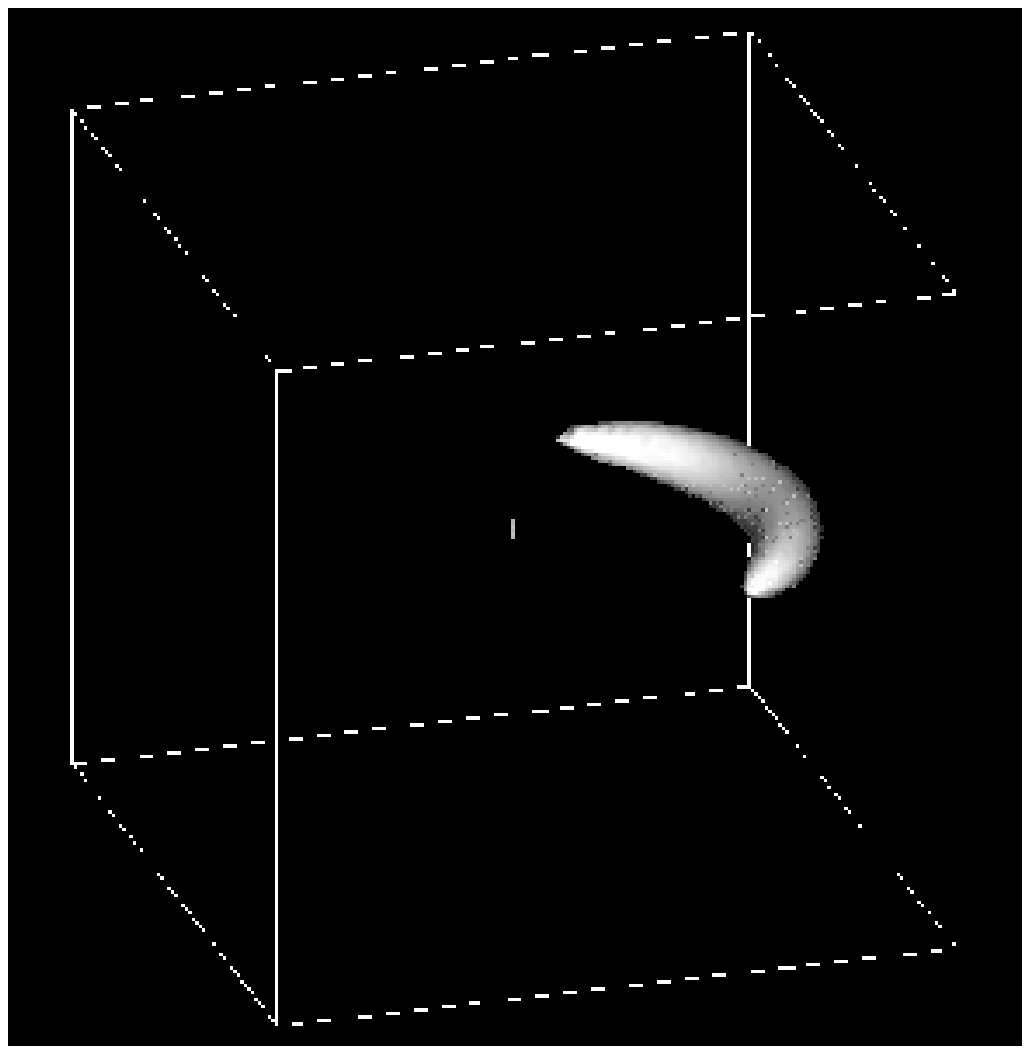,width=6cm,angle=0}
\psfig{figure=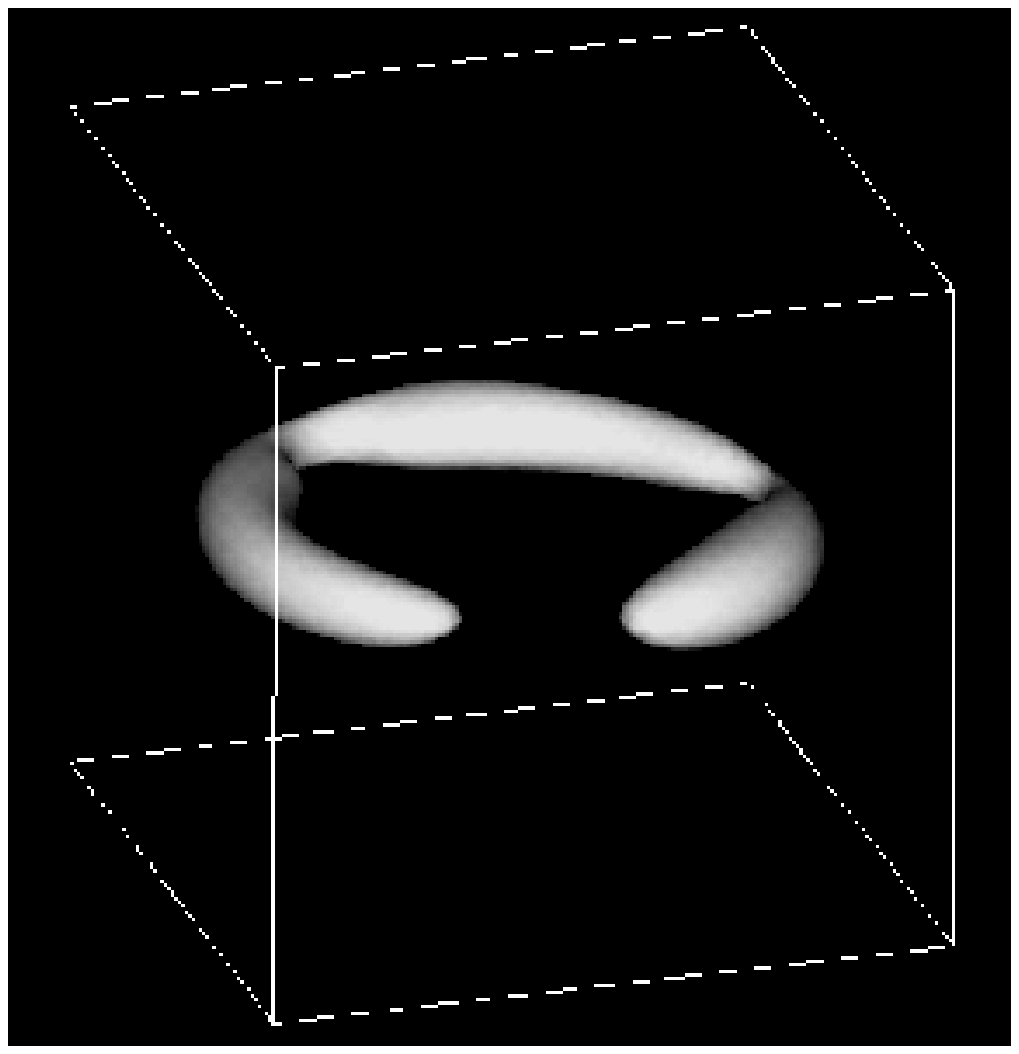,width=6cm,angle=0}}

\smallskip

\centerline{\psfig{figure=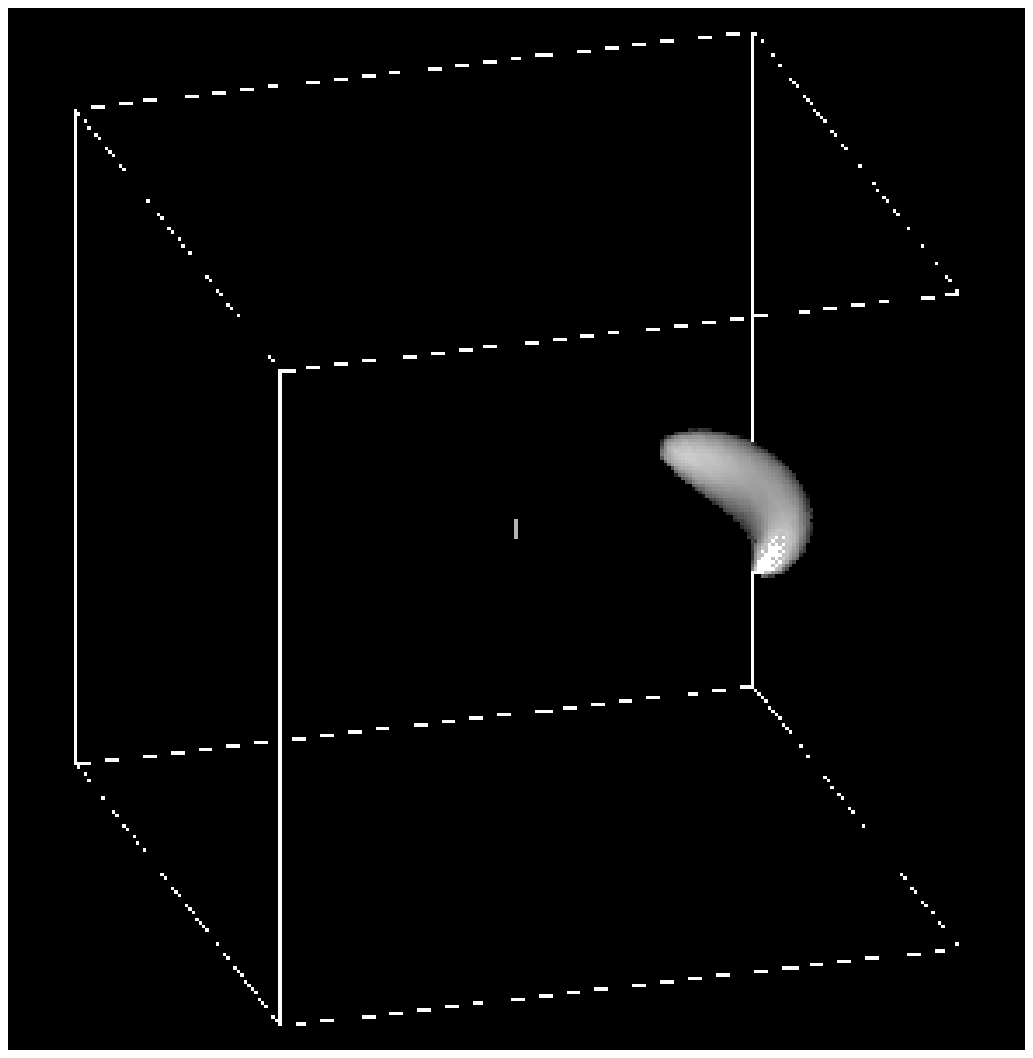,width=5cm,angle=0}}
\caption{Time evolution of a wave-packet launched along a
circular Kepler trajectory, with the same Gaussian weights $c_n$,
eq.~(\protect\ref{coeffs}), as employed for the one-dimensional 
example displayed
in fig.~\protect\ref{wp_h1d}, i.e., centered around the principal
quantum number $n_0=60$.
Since the relative
phases accumulated during the time evolution only depend on $n_0$ -- see
eq.~(\protect\ref{psit}) -- we observe precisely the same behaviour as in the
one-dimensional case: classical propagation 
at short times (top), followed by spreading and collapse
(middle), and revival (bottom). The snapshots of the
wave function are taken at times (in units of $T_{\rm recurrence}$) 
$t=0$ (top left), $t=0.5$ (top right), 
$t=1$ (middle left), $t=12.5$ (middle right), and $t=19.45$ (bottom).
The cube size is 10000 Bohr radii, centered on the nucleus (marked with 
a cross). The radius of the circular wave-packet trajectory 
equals approx.
3600 Bohr radii.}
\label{wp_3d} 
\end{figure}

\begin{figure}
\centerline{\psfig{figure=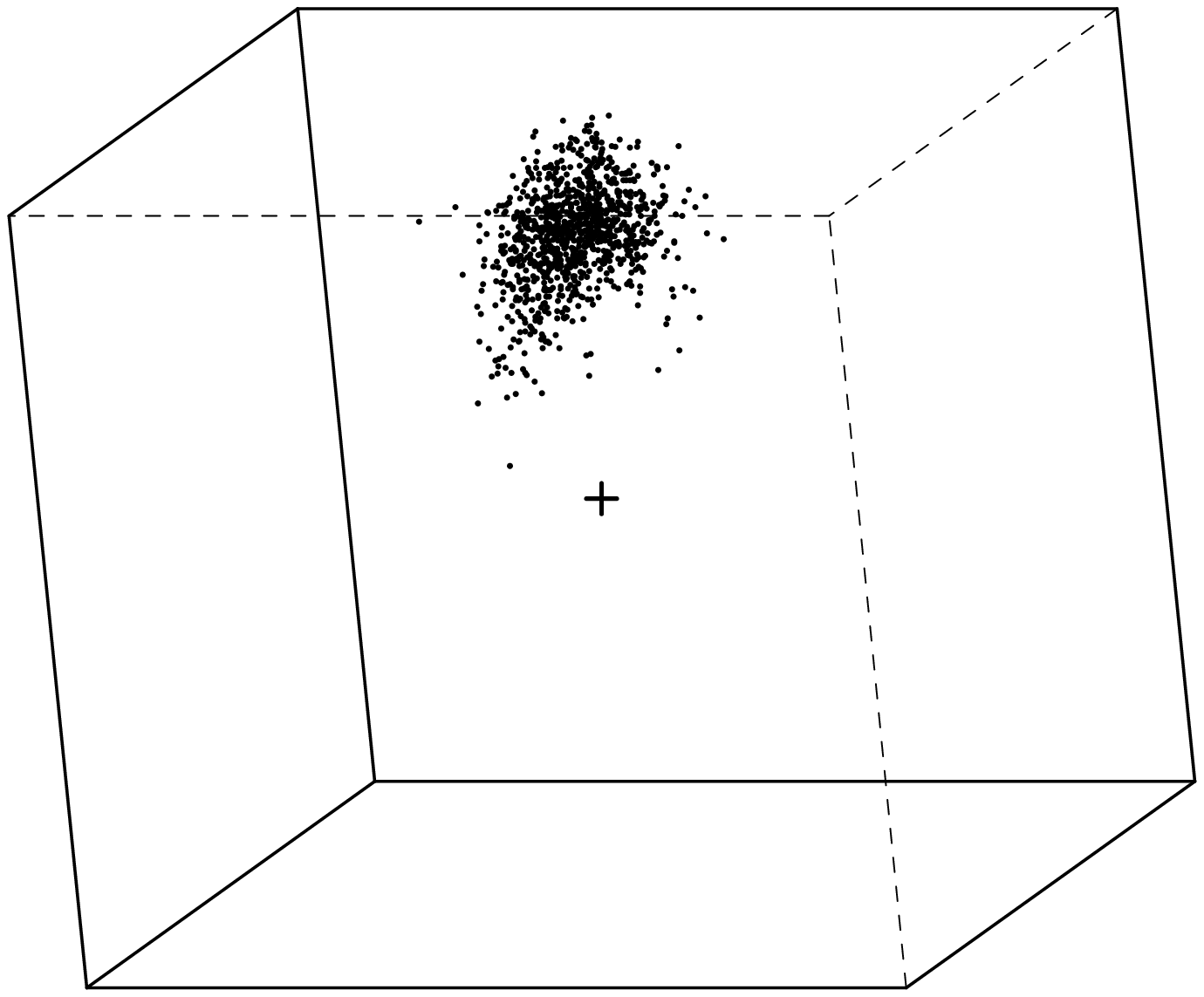,width=6cm,angle=-90}
\psfig{figure=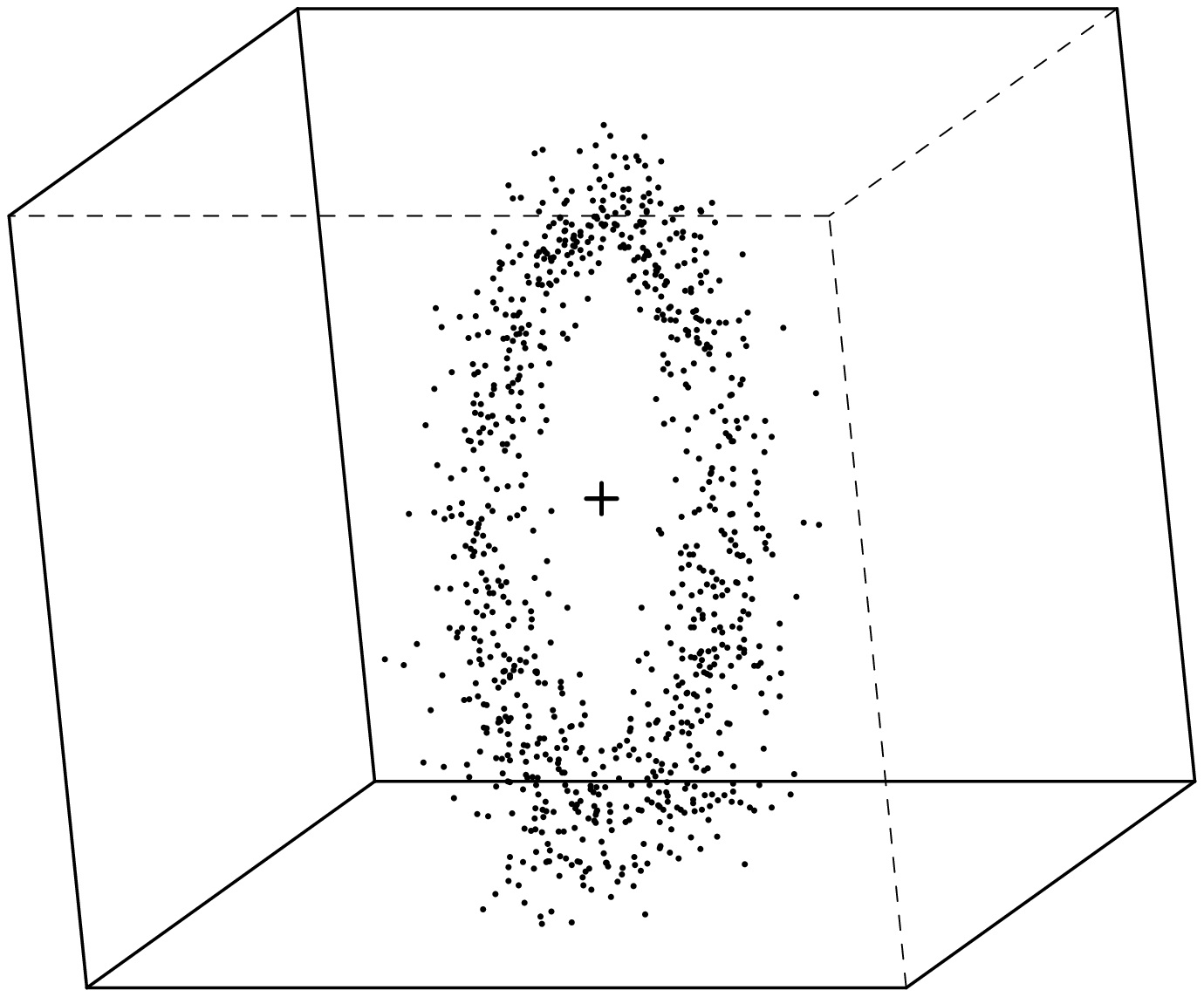,width=6cm,angle=-90}}
\caption{Classical time evolution of a Gaussian (in spherical coordinates) 
phase space density fitted to the minimum uncertainty wave-packet of
fig.~\protect\ref{wp_3d} at time $t=0$ (left). As time evolves, the classical 
phase space density spreads along the circular Kepler orbit ($t=12.5$, right),
but exhibits no revival. Hence, wave-packet spreading is of purely classical
origin, 
only the revival is 
a quantum feature.  
The cube size is 10000 Bohr radii, centered on the nucleus (marked with 
a cross).  The radius of the circular wave-packet trajectory 
equals approx.
3600 Bohr radii.
}
\label{swarm_3d_disp} 
\end{figure}

For the three-dimensional hydrogen atom, the energy spectrum is exactly the same as
in one dimension. This implies that the 
temporal dynamics is built from exactly the same
frequencies; thus, the 3D dynamics is essentially the same 
as the 1D dynamics.\footnote{In a generic,
multidimensional, integrable
system, there are several different classical 
frequencies along the various degrees
of freedom. Hence, only partial revivals of the 
wave-packet at various times
are observed. The 3D hydrogen atom is {\em not} generic, 
because the three frequencies
are degenerate, which opens the possibility of 
a {\em complete} revival, {\em simultaneously}
along all three coordinates.}
Indeed, collapses and revivals of the wave-packet were also observed, under
various experimental conditions, in the laboratory
\cite{yeazell90,yeazell91,yeazell88,marmet94}.
Fig.~\ref{wp_3d} shows the evolution of a 
minimum uncertainty 
wave-packet of the 3D atom, initially localized
on a circular Kepler orbit of the electron. It is built as a linear
combination of circular hydrogenic states (i.e., states with
maximum angular and magnetic quantum numbers $L=M=n-1$), using the
same Gaussian distribution of the coefficients as in fig.~\ref{wp_h1d}.
As expected, the wave-packet 
spreads along the
circular trajectory (but not transversally to it) and eventually
re-establishes its
initial shape after $T_{\rm revival}.$ Figure~\ref{swarm_3d_disp} 
shows the corresponding 
evolution of a swarm of classical particles, 
for the same initial phase space density.
The spreading 
of the classical distribution and of the quantum wave-packet
proceeds very similarly, 
whereas the
revival is completely absent in the classical evolution, which once more
illustrates 
its purely quantum origin.

\psfull

\psfull

Finally, let us notice that collapse and revival of a 3D wave-packet 
depend
on the principal quantum number $n_0$ only -- see eqs.~(\ref{hcol}) and 
(\ref{hreviv}) -- and
are independent
of other parameters which characterize 
the classical motion, such as the eccentricity and the orientation
of the classical elliptical trajectory. This establishes that a 3D 
wave-packet 
with low average
angular momentum (and, a fortiori, a 1D wave-packet as shown in
fig.~\ref{wp_h1d}) -- which deeply explores the non-linearity of the 
Coulomb force -- does not
disperse faster than a circular wave-packet which essentially feels
a constant force. Hence, arguments on the non-linear character
of the interaction should be used 
with some caution.

There have been several experimental realizations of electronic wave-packets
in atoms \cite{raman97,yeazell90,yeazell91,yeazell88,mallalieu94,weinacht99},
either along the pure radial coordinate or even along angular coordinates too.
However, all these wave-packets dispersed rather quickly. 

\subsection{How to overcome dispersion}
\label{INTRO3}

Soon after the discovery of quantum mechanics, the spreading of 
wave-packets was realized and attempts were made to overcome 
it~\cite{schroe26}.
From eq.~(\ref{psit}), it is however clear that this is only possible
if the populated energy levels are equally spaced. In practice, this condition
is 
only met for the harmonic oscillator (or simple tops and rotors). In 
any other system, the anharmonicity of the
energy ladder will 
induce dispersion. Hence, the situation
seems hopeless.
 
Surprisingly, it is
classical 
mechanics which provides us with a possible solution. Indeed, as discussed
above, a quantum wave-packet spreads exactly as the corresponding
swarm of classical particles. Hence, dispersion can be overcome if all 
classical trajectories
behave similarly in the long time limit. In other words, if an initial 
volume of phase
space remains well localized under time evolution, it is reasonable to expect
that a wave-packet built on this initial volume will not spread either. 
The simplest
example is to consider a stable fixed point: 
by definition \cite{lichtenberg83}, 
every initial condition in its vicinity
will forever remain close to it. The corresponding wave-packet indeed does 
not spread
at long times \ldots though
this is of limited interest, as it is 
simply at rest!

Another possibility is to use a set of classical trajectories which all
exhibit the
same periodic motion, {\it with the same period for all trajectories}. This
condition, however, is too restrictive, since it leads us back 
to the harmonic oscillator.
Though, we 
can slightly relax this constraint by allowing classical trajectories
which are not strictly periodic but quasi-periodic and staying forever
in the vicinity of a well defined periodic orbit: A wave-packet built
on such orbits should evolve along the classical periodic orbit
while keeping a finite dispersion around it.

It happens that there is a simple possibility 
to generate such 
classical trajectories {\em locked} on a periodic orbit, which is to
drive the system by an external periodic driving. The general 
theory of nonlinear dynamical systems (described in section \ref{CD}) 
\cite{lichtenberg83,ottb} shows 
that when a nonlinear system (the internal frequency of which depends on 
the initial conditions) is subject to an external
periodic driving, a {\em phase locking} phenomenon -- known as
a nonlinear resonance -- takes place. For initial conditions where the
internal frequency is close to the driving frequency (quasi-resonant
trajectories), the effect of the coupling is to force the motion towards the
external frequency. In other words, trajectories which, in the absence
of the coupling, would oscillate at a frequency slightly lower than
the driving are pushed forward by the nonlinear coupling, and 
trajectories
with slightly larger frequency are pulled backward. In a certain region
of phase space -- 
termed 
``nonlinear resonance island'' -- all trajectories
are trapped, and locked on the external driving. At the center of the resonance
island, there is a stable periodic orbit 
which precisely evolves with the driving frequency.
If the driving is a small perturbation, this periodic orbit
is just the periodic orbit which, in the absence of driving, has exactly
the driving frequency. All the trajectories in the resonance island are
winding around the central orbit with their phases locked on the
external driving. The crucial point for our purposes 
is that the resonance island occupies
a finite volume of phase space, i.e., it traps all trajectories in a window
of internal frequencies centered around the driving frequency. The size of this
frequency window {\em increases} 
with the amplitude of the system-driving coupling, and, as we shall see in
section~\ref{GM}, can be made large enough to support wave-packet
eigenstates of the corresponding quantum system.

\begin{figure}
\centerline{\psfig{figure=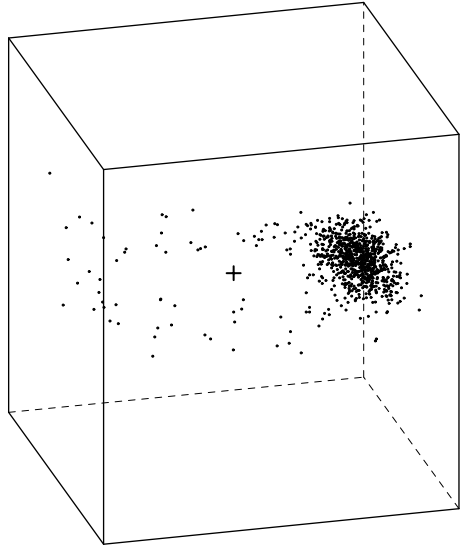,width=6cm,angle=-90}}
\caption{The initially Gaussian distributed swarm of classical particles,
shown in fig.~\protect\ref{swarm_3d_disp}, after evolution
in the presence of the Coulomb field of the nucleus, with a
resonant, circularly polarized microwave
field in the plane of the circular Kepler trajectory added. 
The nonlinear resonance between the unperturbed Kepler motion and
the driving field locks the phase of the particles on the phase of the driving
field. As opposed to the free (classical) evolution depicted in 
fig.~\protect\ref{swarm_3d_disp}, 
the classical distribution does {\em not} exhibit
dispersion along the orbit, except for few particles launched from the
tail of the initial Gaussian distribution, 
which are not trapped by the principal resonance island.}
\label{wp_clas} 
\end{figure}

The classical trapping mechanism is illustrated
in fig.~\ref{wp_clas} which shows a swarm of classical particles launched
along a circular Kepler orbit of a
three-dimensional hydrogen atom exposed to a resonant, circularly polarized
microwave field: the effect of the microwave field is to lock the particles
in the vicinity of a circular trajectory. Note that also the phase along
the classical circular trajectory is locked: the particles
are grouped in the direction of the microwave field and follow
its circular motion without any drift. There is a striking difference
with the situation shown previously in fig.~\ref{swarm_3d_disp}, where the
cloud of particles rapidly spreads in the absence of the microwave field
(the same swarm of initial conditions is used in the two figures). In
figure \ref{wp_clas}, there
are few particles (about 10\%) in the swarm which are not phase 
locked with the microwave field. This is due to the finite 
subvolume of phase space which is effectively phase locked. 
Particles in the tail
of the initial Gaussian distribution may not be trapped 
\cite{lee97}.

\psfull

Although
the microwave field applied in fig.~\ref{wp_clas} 
amounts to less than 5\% of the Coulomb field
along the classical trajectory, it is sufficient to
synchronize the classical motion. The same phenomenon 
carries over to quantum mechanics, and 
allows the creation of non-dispersive wave-packets, as will
be explained in detail in section \ref{QD}.

\subsection{The interest of non-dispersive wave-packets}
\label{INTRO4}

Schr\"odinger dreamt of the possibility of building quantum wave-packets 
following classical trajectories \cite{schroe26}. He succeded for the harmonic 
oscillator, but failed for other systems \cite{schroe68}. 
It was then believed that
wave-packets must spread if the system is nonlinear, and this is
correct for time-independent systems.
However, this is not true {\em in general}, and we have seen in the
previous section that clever use of the non-linearity may, 
on the contrary, {\em stabilize} a wave-packet and preserve it
from spreading. Such non-dispersive wave-packets are thus
a realization of Schr\"odinger's dream. 

One has to emphasize strongly that they are {\em not} some variant
of the coherent states of the harmonic oscillator. They are
of intrinsically completely different origin. Paradoxically,
they exist only if there is some non-linearity, i.e. some
unharmonicity,  in the classical system. They have
some resemblance with classical solitons which are localized
solutions of a non-linear equation that propagate without
spreading. However, they are {\em not} solitons, as they are
solutions of the {\em linear} Schr\"odinger equation. 
They are simply new objects. 

Non-dispersive wave-packets in atomic systems were identified
for the hydrogen atom exposed
to a linearly polarized \cite{delande94,abu95d} and 
circularly polarized \cite{ibb94}  microwave fields quite independently
and using different physical pictures. The former approach associated
the wave-packets with single Floquet states localized in the vicinity of
the periodic orbit corresponding to atom-microwave nonlinear resonance.
The latter treatment relied on the fact  that a transformation
to a frame corotating with the microwave field removes
the explicit time-dependence of the Hamiltonian for the
circular polarization (see section~\ref{CP}). The states localized 
near the equilibria of the rotating system
were baptized ``Trojan wave-packets" to stress the analogy
of the stability mechanism with Trojan asteroids. Such an approach
is, however, 
restricted to a narrow class of systems where the time-dependence
can be removed and lacks the identification of the non-linear
resonance as the relevant mechanism. We thus prefer to use
in this review the more general term ``non-dispersive wave-packets"
noting also that in several other papers ``non-spreading wave-packets''
appear equally often.

Apart from their possible practical applications (for example,
for the purpose of quantum control of atomic or molecular fragmentation 
processes
\cite{weinacht99}, or for information 
storage \cite{ahn00,meyer00,kwiat00,bucksbaum00} in a confined volume 
of (phase) space
for long times), they show the fruitful character of classical
nonlinear dynamics. Indeed, here the nonlinearity is not 
a nuisance to be minimized, but rather 
the essential ingredient. From complex nonlinear dynamics,
a simple object is born. The existence of such non-dispersive wave-packets
it extremely difficult to understand (let alone to predict) 
from quantum mechanics
and the Schr\"odinger equation alone. The classical nonlinear dynamics
point of view is by far more illuminating and predictive. It is 
the classical mechanics inside which led Berman and Zaslavsky \cite{berman77}
to the pioneering discussion of states associated with the classical
resonance island, the subsequent studies \cite{henkel92,holthaus94,holthaus95}
further identified such states for driven one-dimensional systems using the
Mathieu approach without, however, discussing the wave-packet aspects of
the states. 
The best proof of the importance of the classical mechanics inside
is that the non-dispersive wave-packets  
could have been discovered for a very long
time (immediately after the formulation of the Schr\"odinger equation),
but were actually identified only during the last ten years 
\cite{abuth,delande94,abu95d,ibb94,farrelly95a,ibb95,farrelly95,kalinski95a,kalinski95b,delande95,kuba95a,ibb96,eberly96,kalinski96a,brunello96,kalinski96b,kalinski97,kalinski98,ibb97a,ibb97b,delande97,kuba97a,cerjan97,kuba97b,abu96,kuba98,kuba98a,abu98a,delande98,sachath,hornbergerda,hornberger98,sacha98a,sacha98b,sacha99a,yeazell00}, after
the recent major developments of nonlinear dynamics.

\section{Semiclassical quantization}
\label{SQ}

In this section, we briefly recall the basic results
on the semiclassical quantization of Hamiltonian systems, 
which we will 
need
for
the construction of non-spreading wave-packets in classical phase space, as well as
to 
understand
their properties. This section does not contain
any original material.

\subsection{WKB quantization}

\label{wkb}

For a one-dimensional, bounded, time-independent system, the Hamilton function
(equivalent to the total energy)
is a classical constant of motion, 
and the  
dynamics are periodic.
It is possible to define canonically conjugate action-angle variables
$(I,\theta)$, such that the Hamilton function depends 
on the action 
alone. The usual definition of the action along a periodic orbit (p.o.)
writes:
\begin{equation}
I=\frac{1}{2\pi}\ \oint_{\mathrm p.o.}{p\ dz},
\label{action_def}
\end{equation}
where $p$ is the momentum along the trajectory. 

The WKB (for Wentzel, Kramers, and Brillouin) method \cite{Berry_WKB}
allows to construct 
an approximate solution of the Schr\"odinger
equation, in terms of the classical action-angle variables and of
Planck's constant $\hbar:$
\begin{equation}
\psi(z) = \frac{1}{\sqrt{p}}\ \exp{\left(\frac{i}{\hbar}
\int{p\ dz}\right)}
\label{psi_wkb}
\end{equation}
as an integral along the classical trajectory.
This construction is possible
if and only if the phase accumulated along a period of the orbit is an integer
multiple of $2\pi$. This means that  the quantized states are those where
the action variable $I$ is an integer multiple of $\hbar.$ This simple
picture has to be slightly amended because the semiclassical
WKB approximation for the wave-function breaks down at the turning points
of the classical motion,
where the velocity of the classical particle 
vanishes and, consequently, the
expression~(\ref{psi_wkb}) diverges. This failure can be repaired \cite{Berry_WKB}
by adding an additional phase $\pi/2$ for each turning point. This
leads to the final quantization condition
\begin{equation}
I=\frac{1}{2\pi}\ \oint_{\mathrm p.o.}{p\ dz} = 
\left(n+\frac{\mu}{4}\right) \hbar,
\label{WKB}
\end{equation}
where $n$ is a non-negative integer and $\mu$ -- the ``Maslov index'' --
counts the number of turning points along the periodic orbit
($\mu =2$ for a simple 1D periodic orbit).

Thus, the WKB recipe is extremely simple: when the classical Hamilton
function $H(I)$
is expressed in terms of the action $I$, the semiclassical energy
levels are obtained by calculating $H(I)$ for the quantized values of 
$I$:
\begin{equation}
E_n = H\left(I=\left(n+\frac{\mu}{4}\right)\hbar\right).
\label{semen}
\end{equation}

Finally, as a consequence of eqs.~(\ref{WKB}),(\ref{semen}), 
the spacing between two
consecutive semiclassical energy 
levels is simply related to the classical frequency $\Omega$ of the motion,
\begin{equation}
E_{n+1}-E_n \simeq \frac{d E_n}{d n} = \hbar \frac{d H}{d I} = \hbar
\Omega(I),
\label{cprinc}
\end{equation}
a result which 
establishes the immediate correspondence between a resonant 
transition between two quantum mechanical eigenstates in the semiclassical
regime, and resonant driving of the associated classical trajectory.

In the vicinity of a fixed (equilibrium) point, the Hamilton function 
can be expanded
at second order (the first order terms
are zero, by definition of the fixed point), 
leading to the ``harmonic approximation''. If the fixed point is
stable, the semiclassical WKB quantization of the harmonic approximation gives 
exactly the quantum result, although the semiclassical wave-function,
eq.~(\ref{psi_wkb}), is incorrect. This remarkable feature is not true
for an unstable fixed point (where classical trajectories escape far
from the fixed point), and the WKB approximation fails in this case.    

\subsection{EBK quantization}
\label{EBKsec}

For a multi-dimensional system, 
it is a much more complicated task to extract the quantum mechanical 
eigenenergies 
from the classical dynamics of a Hamiltonian system.
The problem can be solved for {\em integrable\/} systems,
where there are as many constants of motion as degrees of 
freedom~\cite{ozorio88}.
This is known as EBK (for Einstein, Brillouin, and Keller) quantization
\cite{einstein17}, and is a simple extension
of the WKB quantization 
scheme.
Let us choose two degrees of freedom for simplicity, the extension
to higher dimensions being straightforward. If the system
is integrable, the Liouville-Arnold theorem~\cite{lichtenberg83}
assures the existence of two pairs of canonically conjugate
action-angle variables, $(I_1,\theta_1)$ and $(I_2,\theta_2)$, 
such that the classical Hamilton function depends only on the actions:
\begin{equation}
H = H(I_1,I_2).
\end{equation}
The classical motion is periodic along each angle (the actions being constants
of the motion)
with frequencies
\begin{eqnarray}
\Omega_1 & = & \frac{\partial H}{\partial I_1},\\
\Omega_2 & = & \frac{\partial H}{\partial I_2}.
\end{eqnarray}
In the generic case, these two frequencies are incommensurate, such that the
full motion in the four-dimensional phase space 
is quasi-periodic, and densely fills the
so-called ``invariant torus" defined by the constant values $I_1$ and $I_2$.
The semiclassical wave-function is constructed similarly to the WKB 
wave-function.
Turning points are now replaced by caustics \cite{ottb,einstein17,bornwolf} 
of the classical
motion (where the projection of the invariant torus on configuration
space is singular), but the conclusions are essentially
identical. The single-valued character of the wave-function requires the
following quantization of the actions:
\begin{eqnarray}
I_1 & = & \left(n_1 + \frac{\mu_1}{4}\right) \hbar,\\
I_2 & = & \left(n_2 + \frac{\mu_2}{4}\right) \hbar,
\label{EBK}
\end{eqnarray}
where $n_1,n_2$ are two non-negative integers, and
$\mu_1,\mu_2$ the Maslov indices (counting the number of caustics encountered
on the torus)
along the $\theta_1,\theta_2$ directions.
Once again, 
the semiclassical energy levels (which now depend on two quantum
numbers) are obtained by substitution of these quantized values into the classical Hamilton
function.

An alternative formulation of the EBK criterium 
is possible using the original position-momentum 
coordinates. Indeed, eq.~(\ref{EBK}) just expresses that, along any closed
loop on the invariant torus, the phase accumulated 
by the wave-function is an integer
multiple of $2\pi$ (modulo the Maslov contribution). 
Using the canonical invariance of the total action
\cite{ozorio88}, 
the EBK quantization conditions can be written as:
\begin{equation}
\frac{1}{2\pi} \oint_{{\mathrm closed\ path}\ \gamma_i}{\vec{p}. \d \vec{r}} = 
\left(n_i + \frac{\mu_i}{4}\right) \hbar
\label{ebkgen}
\end{equation}
where the integral has to be taken along two topologically 
independent closed paths 
$(\gamma_1,\gamma_2)$ on the
invariant torus.   

Note that, as opposed to the WKB procedure in a 1D situation, 
the EBK quantization uses the invariant tori of the classical
dynamics, {\em not\/} the trajectories themselves.
When there is a stable periodic orbit, it is 
surrounded by invariant tori. 
The smallest quantized torus around the stable orbit is associated with
a quantum number equal to zero for the motion transverse to the orbit: 
it defines 
a narrow tube around the orbit, whose projection on configuration
space will be 
localized in the immediate vicinity
of the orbit. Thus, the corresponding wave-function
will also be localized close to this narrow tube, i.e.,
along the stable periodic orbit in configuration space.
Transversely to the orbit, the wave-function (or the Wigner function)
will essentially look like the ground state of an harmonic oscillator,
i.e. like a Gaussian. 

Finally, let us note that it is also possible to develop an analogous EBK
scheme for periodically time dependent Hamiltonians \cite{breuer91}
using the notion of an extended phase space \cite{lichtenberg83}.
Such an approach will be extensively used in the next section, so it is 
discussed in detail there.

\subsection{Scars}

\label{scars}

When a periodic orbit is unstable, there is no torus closely surrounding it.
However, it often happens that quantum eigenstates exhibit an increased
probability density in the vicinity of unstable periodic orbits. This scarring
phenomenon  is nowadays relatively well understood, and the interested
reader may 
consult references \cite{heller89,stoeckmann99}.

Similarly, some quantum states have
an increased probability density 
in the vicinity of an unstable equilibrium point
\cite{abu95d,jensen89b,leopold94}.
This localization is only partial. Indeed, since a quantum eigenstate
is a stationary structure, some probability density
{\em must} localize along the unstable directions of the classical Hamiltonian
flow \cite{ozorio88}, and the localization cannot be perfect.
This is in sharp contrast with stable equilibrium points and stable
periodic orbits which -- see above -- optimally support
localized eigenstates.

Note that there is, however, a big difference between scarring and localization
in the vicinity of an unstable fixed point. The latter phenomenon
is of purely classical origin. Indeed, close to an 
equilibrium point, the velocity goes to zero and the particle consequently 
spends
more time close to the equilibrium point than further away from it. The quantum
eigenfunctions have the same property: the probability density is large
near the equilibrium point. This trivial enhancement of the
probability density is already well known 
for a one-dimensional
system where the WKB wave-function, eq.~(\ref{psi_wkb}),
diverges when the momentum
tends to zero. 
The localization effect
near an unstable point is just the quantum manifestation of the
critical slowing down of the classical particle~\cite{delande97}.

\section{Non-dispersive wave-packets and their realization in various 
atomic systems}

\subsection{General model -- nonlinear resonances}
\label{GM}

In this section, we present the general theory of non-dispersive wave-packets.
As explained in section \ref{INTRO3}, the basic ingredients
for building a non-dispersive wave-packet are a non-linear
dynamical system and an external periodic driving which is
resonant with an internal frequency of the dynamical system.
We present here a very general theory starting out from classical
mechanics which provides us with the most suggestive 
approach to non-linear resonances. In a second step, we choose a pure
quantum approach 
giving essentially the same physics. 

We use a one-dimensional model, which displays all the interesting
features of non-linear resonances. 
While the direct link between classical nonlinear resonances, the corresponding
Floquet states, and non-dispersive wave-packets has been identified only 
recently \cite{abuth,delande94,abu95d,delande95,abu96} some 
aspects of the developments presented below
may be found in earlier studies  
\cite{berman77,zaslavsky81,henkel92,holthaus95}.

Several complications not included in the simple one-dimensional
model are important features of ``real systems". 
They are discussed at a later stage in this paper:
\begin{itemize}
\item the effect of additional degrees
of freedom, in sections \ref{LIN3D}-\ref{EP};
\item higher 
nonlinear resonances (where the driving frequency is a multiple
of the internal frequency), in section \ref{HOR};
\item an unbounded phase space, leading to the decay of 
non-dispersive wave-packets (as ``open quantum systems''), 
in section \ref{ION};
\item sources of  ``decoherence'', such as
spontaneous emission of atomic wave-packets, in section \ref{SPO};
\item deviations from 
temporal periodicity, in section \ref{PTP}.
\end{itemize}

In 
particular cases, an apparently simpler approach is also possible
(such as the use of the rotating frame for a Rydberg atom exposed to a
circularly polarized electromagnetic field, see section \ref{CP}).
Despite all its advantages, it may be quite specific and 
too restricted to reveal nonlinear resonances 
as the actual cause of the
phenomenon. Here, we seek the most general description.

\subsubsection{Classical dynamics}
\label{CD}

Let us start from a time-independent, bounded, one-dimensional system
described by the Hamilton function
$H_0(p,z)$.
Since energy is conserved, the motion is confined to a
one-dimensional manifold in two-dimensional phase space.
Except for energies which define a fixed point of the
Hamiltonian dynamics (such that $\partial H_0/\partial z = 0$
and $\partial H_0/\partial p = 0;$ these fixed points generically
only exist at some 
isolated values of energy, for example at $E=0$
for the harmonic oscillator), the motion is periodic
in time, and the phase space trajectory is a simple closed loop.

It is always possible to find a set of canonically
conjugate phase space coordinates adapted to the dynamics of the system.
These are the action-angle coordinates $(I,\theta)$,
whose existence is guaranteed
by the Liouville-Arnold theorem \cite{lichtenberg83}, with:
\begin{eqnarray}
&&0 \leq I, \\
&&0 \leq \theta \leq 2\pi, \label{thetaint} \\
&&\left\{ \theta , I \right\} = 1,
\end{eqnarray}
and $\{ .,.\}$ the usual Poisson brackets, eq.~(\ref{poisson_brackets}).

A fundamental property is that the Hamilton function in these coordinates
depends on $I$ alone, not on $\theta:$
\begin{equation}
H_0 = H_0(I).
\end{equation}
As a consequence of 
Hamilton's
equations of motion, $I$ is a
constant of motion, and 
\begin{equation}
\theta =\Omega t +\theta_0
\label{thetatime}
\end{equation}
evolves linearly in time,
with the angular velocity
\begin{equation}
\Omega(I) = \frac{\partial H_0}{\partial I}(I),
\label{internal_omega}
\end{equation}
which depends on the 
action $I$.
The period of the motion at a given value of $I$ reads
\begin{equation}
T = \frac{2\pi}{\Omega(I)}.
\end{equation}

In simple words, the action $I$ is nothing but the properly
``rescaled" total energy, and the angle $\theta$ just measures how time
evolves
along the (periodic) orbits. 
In a one-dimensional system, the action variable
can be expressed as an
integral along the orbit, see eq.~(\ref{action_def}).

Suppose now that the system is exposed to a periodic driving force, such
that the Hamilton function, in the original coordinates, writes
\begin{equation}
\label{ham}
H = H_0(p,z) + \lambda V(p,z) \cos \omega t,
\label{h_gen}
\end{equation}
with $\omega$ the frequency of the periodic drive and
$\lambda$ some small parameter which determines
the strength of the perturbation.
For simplicity, we choose a single cosine function to define the
periodic driving.  For a more
complicated dependence on time \cite{ringot00}, 
it is enough to expand it in a Fourier
series, see section \ref{EP}.
The equations become slightly more complicated, but the physics
is essentially identical.

We now express the perturbation $V(p,z)$ in action-angle coordinates.
Since $\theta$ is $2\pi-$periodic, eq.~(\ref{thetaint}), we obtain a Fourier series:
\begin{equation}
V(I,\theta) = \sum_{m=-\infty}^{+\infty}
{V_m(I)\ \exp (im\theta)}.
\label{fourier_components}
\end{equation}
Note that, as $\theta$ evolves linearly with time $t$ for the unperturbed
motion (and therefore parametrizes an unperturbed periodic orbit),
the $V_m$ can also be seen as the Fourier
components of $V(t)$ evaluated along the classical, unperturbed trajectory.
Furthermore, since the Hamilton function is real, $V_{-m}=V_m^*.$
Again for the sake of simplicity, we will assume that both are
real and thus equal.
The general case can be studied as well, at the price of slightly
more complicated formulas.

Plugging eq.~(\ref{fourier_components}) in eq.~(\ref{h_gen}) results in the
following Hamilton function in action-angle coordinates,
\begin{equation}
H = H_0(I) + \lambda  \sum_{m=-\infty}^{+\infty}
{V_m(I)\ \exp (im\theta)} \cos \omega t,
\end{equation}
which (assuming $V_{-m}=V_m$ -- see above) can be rewritten as
\begin{equation}
H = H_0(I) + \lambda \sum_{m=-\infty}^{+\infty}
{V_m(I)\ \cos (m\theta-\omega t)}.
\label{hamfou}
\end{equation}
For $\lambda$ sufficiently small, the phase space trajectories of the perturbed dynamics 
will remain close to the unperturbed ones (for short times).
This means
that $m\theta -\omega t$ evolves approximately linearly in time
as 
$(m\Omega - \omega)t$ (see eq.~(\ref{thetatime})),
while $I$ is slowly varying.
It is therefore reasonable to expect that all the terms
$ V_m(I)\ \cos (m\theta-\omega t)$ will oscillate rapidly and average out
to zero,
leading to an effective approximate Hamiltonian identical to the unperturbed
one.
Of course, this approach is too simple. Indeed, close to a  
``resonance'', 
where $(s\Omega - \omega)$ is small, the various terms  
$V_m\ \cos (m\theta-\omega t)$ oscillate, except for the $m=s$ term which
may evolve very {\em slowly} and affect the dynamics considerably.
For simplicity, we restrict the present analysis to the principal
resonance such that $\Omega \simeq \omega$.  The extension
to higher  resonances (with $s\Omega\simeq\omega$) is discussed
in section \ref{HOR}.

Our preceding remark is the basis of the ``secular approximation''
\cite{lichtenberg83,cct92}.
 The
guiding idea
is to perform a canonical change of coordinates involving
the slowly varying variable $\theta -\omega t.$ Because of the explicit
time dependence, this requires first 
the passage to an extended phase space,
which comprises time as an additional coordinate. The Hamilton function
in extended phase space is defined by
\begin{equation}
{\cal H} = P_t + H,
\label{hamext}
\end{equation}
with $P_t$ the momentum canonically conjugate to the new coordinate
 - time $t$. The physical time $t$ is now parametrized by
some
fictitious time, say $\xi$. However,
\begin{equation}
\frac{\partial{\cal H}}{\partial P_t}=\frac{dt}{d\xi}=1,
\end{equation}
 i.e., $t$ and
$\xi$ are essentially identical.
${\cal H}$, being independent of $\xi$,
 is conserved as  $\xi$ evolves.
The requested transformation to slowly varying variables $\theta -\omega t$ 
reads:\footnote{This
{\em canonical}
change of coordinates is often refered to as
``passing to the rotating frame". It should however be emphasized that
this
suggests the correct picture {\em only}
in phase space
spanned by the action-angle coordinates $(I,\theta)$. In the original
coordinates $(p,z)$, the transformation is usually very complicated, and
only rarely
a standard rotation in configuration space (see also section \ref{CP}).}
\begin{eqnarray}
&&\hat{\theta} = \theta - \omega t, \label{rotframe_a} \\
&&\hat{I} = I,\label{rotframe_b} \\
&&\hat{P}_t = P_t + \omega I,
\label{rotframe_c}
\end{eqnarray}
 which transforms ${\cal H}$ into
\begin{equation}
 \hat{\cal H} = \hat{P}_t + H_0(\hat{I}) -\omega \hat{I}+ \lambda
\sum_{m=-\infty}^{+\infty}
{V_m(\hat{I})\ \cos (m\hat{\theta}+(m-1)\omega t)}.
\label{floq_sc_gen}
\end{equation}
This Hamilton function does not
involve any approximation yet. Only in the next step
we average $\hat{\cal H}$ over the fast variable $t,$ i.e., over one period
of the external driving. This has the effect of canceling all oscillating
terms in the sum, except
the resonant one, defined by $m=1$.
Consequently, we are left with the approximate, 
``secular'' Hamilton
function:
\begin{equation}
{\cal H}_{\mathrm sec} = \hat{P}_t + H_0(\hat{I}) -\omega \hat{I}+
\lambda
V_1(\hat{I}) \cos \hat{\theta}.
\label{hsec_ap}
\end{equation}

The secular Hamilton function no longer depends on time. Hence,
$\hat{P}_t$ is a constant
of motion and we are left with an integrable Hamiltonian system living in a
two-dimensional phase space, spanned by $(\hat{I},\hat{\theta})$.
The above averaging procedure is valid at first order in $\lambda$.
Higher order expansions, using, e.g., the Lie algebraic
transformation method~\cite{lichtenberg83}, 
are possible.\footnote{An example is given
in \cite{abu97}, in a slightly different situation,
where the perturbation is not resonant with the
internal frequency.} Basically, the interesting physical phenomena
are already present at lowest non-vanishing order, to which we will
restrain in the following.

The dynamics
generated by the secular Hamilton function
is rather simple. At order zero in $\lambda,$
$\hat{I}$ is constant and $\hat{\theta}$ evolves linearly with time.
As we can read from eq.~(\ref{hsec_ap}),  a
continuous family (parametrized by the value of
$0\leq \hat{\theta}<2\pi$)
of fixed points exists
if $d\hat{\theta}/dt=\partial{\cal H}_{\mathrm
sec}/\partial\hat{I}$ vanishes,
i.e., at actions $\hat{I}_1$ such that
\begin{equation}
\Omega(\hat{I}_1) = \frac{\partial H_0}{\partial I}(\hat{I}_1) = \omega.
\label{resonance_condition}
\end{equation}
Thus, unperturbed trajectories that are {\em resonant} with the
external drive are fixed points
of the unperturbed secular
dynamics. This is precisely why slowly
varying variables
are introduced.

Typically, eq.~(\ref{resonance_condition}) has only isolated solutions --
we will
assume that in the following. Such is the case when
$\partial^2 H_0/\partial I^2$ does not vanish -- excluding
the
pathological situation of the harmonic oscillator, where all trajectories
are simultaneously resonant.
Hence,
if $\partial^2 H_0/\partial I^2$ is positive, the line 
($\hat{I}=\hat{I}_1,0\leq \hat{\theta}<2\pi$, parametrized by $\hat\theta$)
is a minimum of the unperturbed secular Hamilton function ${\cal H}_{\mathrm
sec}$;
if $\partial^2 H_0/\partial I^2$
is negative, it is a maximum.

At first order in $\lambda$, the fixed points of the secular
Hamiltonian should have an action close to $\hat{I_1}.$ Hence, it is
reasonable to perform
a power expansion of the unperturbed Hamiltonian in the vicinity of
$\hat{I}=\hat{I_1}.$
We obtain the following approximate  Hamiltonian:
\begin{equation}
{\cal H}_{\mathrm pend} = \hat{P}_t + H_0(\hat{I_1}) -\omega\hat{I}_1 + \frac{1}{2}
H^{''}_0(\hat{I_1})\ (\hat{I}-\hat{I_1})^2 + \lambda
V_1(\hat{I_1}) \cos \hat{\theta},
\label{eqpend}
\end{equation}
with:
\begin{equation}
H^{''}_0 = \frac{\partial^2 H_0}{\partial I^2}.
\label{second_derivative}
\end{equation}
Consistently at lowest order in $\lambda,$ it is not necessary 
to take into
account the dependence of $V_1$ on $\hat{I}.$

As already anticipated by the label, ${\cal H}_{\mathrm pend}$ defined
in eq.~(\ref{eqpend}) describes a usual, one-dimensional
pendulum: $\hat\theta$ represents the angle of the pendulum 
with the vertical axis,
$\hat{I}-\hat{I_1}$ its angular velocity, $1/H^{''}_0(\hat{I_1})$
its momentum of inertia
and $\lambda V_1(\hat{I_1})$ the gravitational field.
This equivalence of the secular Hamilton function with that of a pendulum,
 in the
vicinity of the resonant action $\hat{I}_1$, is extremely useful to gain
some physical insight
in the dynamics of any Hamiltonian system close to a resonance. In particular,
it will render our
analysis of non-dispersive wave-packets rather simple.

\begin{figure}
\centerline{\psfig{figure=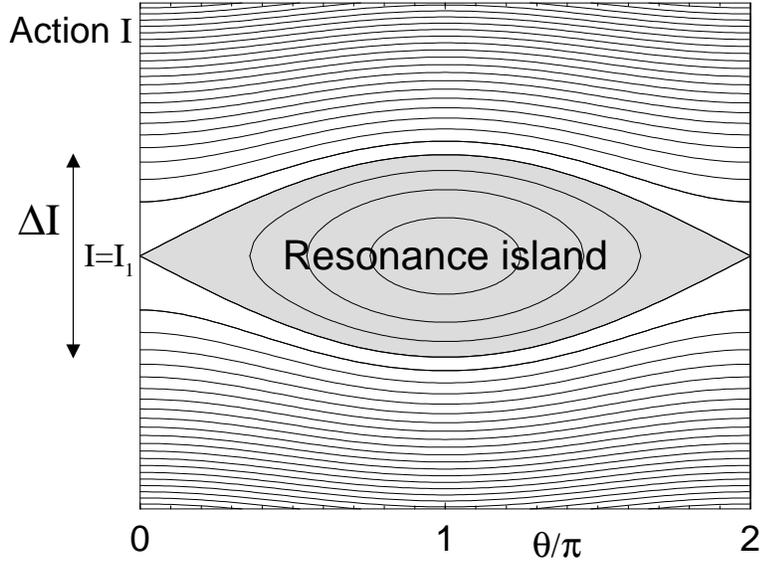,width=10cm,angle=-90}}
\caption{Isovalues of the Hamiltonian ${\cal H}_{\mathrm pend}$ of a
pendulum in a gravitational field. This Hamiltonian is a good
approximation for the motion of a periodically driven system when
the driving frequency is resonant with the internal frequency
of the system. The stable equilibrium point of the pendulum is surrounded
by an island of librational motion (shaded region). 
This defines the resonance island of the periodically driven system, where
the internal motion is {\em locked} on the external driving. This 
nonlinear phase-locking
phenomenon is essential for the existence of non-dispersive wave-packets.}
\label{pendulum}
\end{figure}

Figure~\ref{pendulum} shows the isovalue lines of
${\cal H}_{\mathrm pend}$ in the $(\hat{I},\hat{\theta})$ plane, i.e. the
classical phase space trajectories in the presence of the
resonant perturbation. In
the absence of the resonant perturbation, these should be horizontal 
straight lines at
constant  $\hat{I}.$
We observe that the effect of the resonant perturbation is mainly
to create a new structure, called the ``resonance island", located around
the resonant action $\hat{I}_1.$ 

\psfull

To characterize this structure, let us examine the
fixed points of the Hamiltonian~(\ref{eqpend}). They are easily calculated
(imposing $\partial{\cal H}_{\mathrm pend}/\partial\hat{I}=\partial{\cal
H}_{\mathrm pend}/\partial\hat\theta=0$), and
located at
\begin{equation}
\hat{I}=\hat{I_1},\ \ \ \hat{\theta}=0,\ \ {\mathrm with\ energy}\
H_0(\hat{I_1}) -\omega\hat{I}_1 + \lambda V_1(\hat{I_1}),
\label{stablep}
\end{equation}
and
\begin{equation}
\hat{I}=\hat{I_1},\ \ \ \hat{\theta}=\pi,\ \ {\mathrm with\ energy}\
H_0(\hat{I_1}) -\omega\hat{I}_1- \lambda V_1(\hat{I_1}),
\label{resoact}
\end{equation}
respectively.

If $H^{''}_0(\hat{I_1})$ (and thus the ``kinetic energy'' part
in ${\cal H}_{\mathrm pend}$)
is positive, the minimum of the potential
 $\lambda V_1(\hat{I_1})\cos\hat{\theta}$ corresponds to a global
 minimum of ${\cal H}_{\mathrm pend}$, and thus to a stable
equilibrium point. The maximum of $\lambda V_1(\hat{I_1})\cos\hat{\theta}$
is a saddle point of ${\cal H}_{\mathrm pend}$, 
and thus 
represents an unstable equilibrium point,
as the standard intuition suggests. For $H^{''}_0(\hat{I}_1)<0$
the situation is reversed -- and less intuitive 
since the
``kinetic energy'' is negative -- and the {\em maximum}
of ${\cal H}_{\mathrm pend}$ is now a stable equilibrium point,
as the reader may easily check by
standard linear stability analysis in the vicinity of the
fixed point. 
Thus, in compact form, if
$\lambda V_1(\hat{I_1}) H^{''}_0(\hat{I_1})$ is positive,
$\hat{\theta}=\pi$ is
a stable equilibrium point, while $\hat{\theta}=0$ is
unstable. 
If $\lambda V_1(\hat{I_1}) H^{''}_0(\hat{I_1})$ is negative, the stable and unstable
points are interchanged.

There are two qualitatively different types of motion:
\begin{itemize}
\item Close to the stable equilibrium point of the pendulum, $\hat{\theta}$
oscillates
periodically, with an amplitude smaller than $\pi$.
This is the ``librational motion'' of the pendulum inside the
resonance
island. Any trajectory started within 
this region of phase space (the shaded area in
fig.~\ref{pendulum})
exhibits 
librational motion. It should be realized
that the resonance island confines the motion to finite intervals in $\hat{I}$
and $\hat{\theta}$, and thereby strongly affects all trajectories with action
close to the resonant action. According to eqs.~(\ref{stablep},\ref{resoact}),
the resonance island is associated with the energy range
$[H_0(\hat{I_1})-\omega\hat{I}_1- |\lambda V_1(\hat{I_1})|,
H_0(\hat{I_1})-\omega\hat{I}_1+ |\lambda
V_1(\hat{I_1})|].$
\item For any initial energy outside that energy range
 the pendulum has sufficient
kinetic energy to rotate. This is the ``rotational
motion'' of the pendulum outside the resonance island, where $\hat{\theta}$ is
an unbounded and
monotonous function of time. Far from the center of the island, the motion
occurs at almost constant unperturbed action $\hat{I}$, with an almost
constant
angular velocity in $\hat{\theta},$ 
tending to the unperturbed
motion. This illustrates that the effect of the perturbation is 
important
for initial conditions close to the resonance island, but negligible for 
non-resonant trajectories.
\end{itemize}

The size of the resonance island can be simply estimated from
eq.~(\ref{eqpend}) and
fig.~\ref{pendulum}. The extension in $\hat{\theta}$ is $2\pi$, its width in
$\hat{I}$ (which depends on $\hat{\theta})$ is:
\begin{equation}
\Delta \hat{I} = 4 \sqrt{\left|
\frac{\lambda V_1(\hat{I_1})}{ H^{''}_0(\hat{I_1})}\right|},
\label{width-island}
\end{equation}
and the total area \cite{lichtenberg83}:
\begin{equation}
A(\lambda) = 16 \sqrt{\left|
\frac{\lambda V_1(\hat{I_1})}{ H^{''}_0(\hat{I_1})}\right|}.
\label{area}
\end{equation}

The dependence of $A(\lambda)$ on $\sqrt{|\lambda|}$ implies that even
a small perturbation may
induce
significant changes in the phase space structure, provided the pertubation
is resonant.

The above picture is valid in the rotating frame defined by
eqs.~(\ref{rotframe_a}-\ref{rotframe_c}). If we go back to the original
action-angle coordinates $(I,\theta),$ the stable (resp. unstable) fixed
point of the secular Hamiltonian is mapped on a
stable (resp. unstable) periodic orbit
whose period is {\em exactly} equal to the period of the driving perturbation,
as a consequence of eq.~(\ref{resonance_condition}).
Any trajectory started in the vicinity of the stable periodic orbit will
correspond to an initial point close to the fixed
point in the rotating frame,
and thus will remain trapped within the resonance island. In the original
coordinate frame, it will appear
as a trajectory evolving close to the stable periodic orbit forever.
In particular, 
the difference in $\theta$ between the stable periodic orbit and any orbit 
trapped in the resonance island remains bounded within $(-\pi,+\pi)$, for arbitrarily 
long times. This means that the phase of any trapped trajectory cannot drift with 
respect to the phase of the periodic orbit.
 As the latter evolves at the driving frequency, we
reach the conclusion that the phase of any trajectory started in the
resonance island
will be {\em locked} on the phase of the driving field. This is the very origin
of the phase locking phenomenon discussed in section~\ref{INTRO3}
above. A crucial point herein
is that
the resonance island
covers a significant part of phase space with finite volume:
it is the
whole structure, not few trajectories, which is phase locked. This is why, 
further down, we will be able to build quantum wave-packets on this structure, 
which will be phase locked to the classical orbit 
and will not spread. The classical version of a non-spreading
wave-packet
thus consists of a family of trajectories, trapped within the resonance island, such
that
this family is invariant under the evolution generated by the pendulum
Hamiltonian. The simplest
possibility
is to sample all trajectories within the ``energy" range
$[H_0(\hat{I_1}) -\omega \hat{I}_1 - |\lambda V_1(\hat{I_1})|,
H_0(\hat{I_1})-\omega \hat{I}_1+ |\lambda
V_1(\hat{I_1})|]$ of the pendulum.
In real space, this will appear as a localized probability density
following the classical stable periodic orbit, reproducing 
its shape exactly after each period of the drive.

So far, to derive the characteristics
of the resonance island, we have consistently used
first order
perturbation theory, which is valid for small $\lambda.$ At higher values of
$\lambda,$ higher order terms come into play and modify the shape and the
precise location
of the resonance island. However,
it is crucial to note that the island itself considered as a structure is
robust,
and will survive up to rather high values of $\lambda$ (as a consequence of
the KAM theorem \cite{lichtenberg83}).
Since the size of the resonance island grows with
$|\lambda|$, 
eq.~(\ref{area}), the island may occupy
 a significant
area in phase space and eventually interact with
islands associated
with other resonances, for sufficiently large $|\lambda|$. The mechanism of this ``resonance
overlap'' is rather well understood \cite{chirikov59,chirikov79}: in general,
the motion close to the separatrix (where the period of the classical motion
of the pendulum tends to infinity) is 
most sensitive to higher order 
corrections.
The general scenario
is thus the 
nonintegrable perturbation of the separatrix and the 
emergence of a
``stochastic" layer of
chaotic motion in phase space, as 
$|\lambda|$ is increased. At
still larger
values of $|\lambda|$, 
chaos may invade large parts of phase space, 
and the resonance island may shrink
and finally
disappear.
While considering  
realistic examples later-on, we shall 
enter the
non-perturbative regime. Let us, however, consider first the quantum
perturbative picture.

\subsubsection{Quantum dynamics}
\label{QD}

As shown in the previous section, the dynamics of
a one-dimensional system exposed to a weak, resonant, periodic driving
is essentially regular and analogous to the one of a pendulum,
eq.~(\ref{second_derivative})
(in the rotating frame, eqs.~(\ref{rotframe_a}-\ref{rotframe_c})).
In the present section, we will
show that the same
physical picture can be employed in quantum mechanics, to construct
non-dispersive wave-packets. They will follow the stable
classical trajectory locked on the external drive,
and 
exactly reproduce their initial shape 
after 
each period.

Our starting point is the time-dependent Schr\"odinger equation
associated with the Hamiltonian (\ref{ham}):\footnote{For simplicity,
we use the same notation for
classical and quantum quantities, the distinction between them
 will
become clear from the context.}
\begin{equation}
i\hbar \frac{d|\psi(t)\rangle}{dt} = (H_0 + \lambda V  \cos \omega t)
|\psi(t)\rangle.
\label{se}
\end{equation}
Since the Hamiltonian (\ref{ham}) is periodic in time, the Floquet
theorem\footnote{The Floquet theorem \protect\cite{floquet1883} in the time
domain is strictly equivalent to the Bloch theorem for potentials
periodic in space \protect\cite{mermin76}.}
guarantees that the general solution of eq.~(\ref{se}) is given by
a linear combination
of elementary, time-periodic states -- the so-called ``Floquet eigenstates'' of
the system -- multiplied by oscillatory functions:
\begin{equation}
|\psi(t)\rangle = \sum_j c_j \exp\left( - i \frac{{\cal E}_jt}
{\hbar}\right) |{\cal E}_j(t)\rangle,
\label{solse}
\end{equation}
with
\begin{equation}
|{\cal E}_j(t+T)\rangle =|{\cal E}_j(t)\rangle .
\label{flstate}
\end{equation}
The ${\cal E}_j$ are the 
``quasi-energies" of the system.
Floquet states  and quasi-energies are eigenstates
and eigenenergies
of the Floquet Hamiltonian
\begin{equation}
 {\cal H} = H_0 + \lambda V  \cos \omega t - i \hbar \frac{\partial}{\partial
 t}.
\label{calhq}
\end{equation}
Note that, because
of the time-periodicity with period $T=2\pi/\omega$, the quasi-energies are
defined modulo $\hbar \omega$ \cite{shirley65}.

The Floquet Hamiltonian (\ref{calhq}) is nothing but the quantum analog of
the classical Hamiltonian (\ref{hamext})
in 
extended phase space.
Indeed, $- i \hbar \partial /\partial t$ is the quantum version of the
canonical momentum $P_t$ conjugate to time $t.$
In 
strict analogy with the classical discussion of the previous section,
it is the Floquet Hamiltonian in 
extended phase space which will be the
central object of our discussion. It contains all the
relevant information on the system, encoded in its eigenstates.

In a quantum optics or atomic physics context -- 
with the external perturbation 
given by quantized modes of the electromagnetic field -- the
concept of ``dressed atom" is widely used \cite{cct92}. There, 
a given field mode 
and the atom are treated on an equal footing, as a composite quantum system, 
leading to a time-independent Hamiltonian (energy is conserved for the entire system
comprising atom {\em and} field).
This picture
is 
indeed very close to the Floquet picture. If the field mode is in a 
coherent state \cite{cct92} with a large average number of
photons, 
the electromagnetic field can be treated (semi)classically -- i.e., replaced 
by a $\cos$ time dependence and a fixed amplitude $F$ -- and the
energy spectrum of the dressed atom exactly coincides with the
spectrum of the Floquet Hamiltonian \cite{shirley65}.

By its mere definition, eq.~(\ref{flstate}),
each Floquet eigenstate is
associated with a strictly time-periodic probability density in configuration
space.
Due to this periodicity with the period of the driving field,
the probability density of a Floquet eigenstate in general changes its shape
as time evolves, but recovers its initial shape after 
each period.
Hence, the Floquet picture provides clearly the simplest approach to
non-dispersive wave-packets. Given the ability to build a Floquet
state which is well localized at a given phase of the driving field, it will
{\em automatically} represent a non-dispersive wave-packet.
In our opinion, this is a much simpler approach than the attempt to
build an {\it a priori} localized wave-packet and try to minimize
its spreading during the subsequent evolution
\cite{farrelly95a,ibb95,ibb94,kalinski95a}.

Note that also
the reverse property holds true. Any state with $T$-periodic probability
density (and, in particular, any localized wave-packet propagating along a
$T$-periodic classical orbit)
has to be a single Floquet eigenstate:
Such a state can be expanded into the Floquet eigenbasis, and during one
period, the various components of the expansion accumulate phase factors
$\exp(-i{\mathcal E}_iT/\hbar).$ Hence, 
the only solution which allows for a $T$-periodic density is
a one-component expansion, i.e., a single Floquet eigenstate
\footnote{One
might argue that Floquet states differing in energy by an
integer multiple of $\hbar\omega$ could be used. However, 
the Floquet spectrum is $\hbar\omega$-periodic by construction
\cite{shirley65}, and two
such states
represent the same physical state.}.

To summarize, the construction of non-dispersive
wave-packets in a time-periodic system is equivalent to finding
localized Floquet eigenstates. The existence of such states is far
from obvious, as the Floquet spectrum is usually very complex, composed of
quasi-bound states, resonances and continua. This is why
a semiclassical analysis can be very helpful in finding these objects.

\subsubsection{Semiclassical approximation}
\label{sapp}

Dealing with highly excited states, a semiclassical
approximation can be used to determine quasi-energies and
Floquet eigenstates~\cite{breuer91}. If the driving perturbation is sufficiently weak,
we have shown in section~\ref{CD} that the classical dynamics close to a
nonlinear resonance is
essentially regular and accurately described by the pendulum
Hamiltonian~(\ref{eqpend}). It describes a system with two
degrees of freedom (along $t$ and $\hat\theta$, with their
conjugate momenta $\hat{P}_t$ and $\hat{I}$, respectively)
which is essentially regular. For semiclassical quantization, 
we may then
use the standard EBK rules, eq.~(\ref{EBK}), introduced in section~\ref{SQ}.

The momentum $\hat{P}_t$ is a constant of the motion,
and the isovalue curves of $\hat{P}_t,\hat{I},\hat{\theta}$
lying on the
invariant tori
can be used for the EBK quantization scheme. 
Along such a curve, $t$
evolves from $0$ to $2\pi/\omega$, with $\hat{\theta}=\theta-\omega t$
kept constant. Thus, $\theta$ itself is changed by $2\pi$, what
implies that  the Maslov index $\mu$ of the unperturbed $(I,\theta)$ motion
has to be included,
leading to the following quantization condition for $\hat{P}_t$,
\begin{equation}
\frac{1}{2\pi}\int_0^{T}\hat{P}_tdt=\frac{\hat{P}_tT}{2\pi}=
\left(k+\frac{\mu}{4}\right)\hbar ,
\label{ebkpt}
\end{equation}
with integer $k$.
Since $T=2\pi/\omega$ is just the period of the resonant driving,
we get the quantized values of $\hat{P}_t:$ 
\begin{equation}
\hat{P}_t = \left(k+\frac{\mu}{4}\right)\ \hbar\omega
\label{ebkpt2}
\end{equation}
which are equally spaced by
$\hbar \omega.$ Thus, we recover semiclassically 
the $\omega$-periodicity of the Floquet spectrum.

For the motion in the $(\hat{I},\hat{\theta})$ plane, we can use
the isocontour lines
of the pendulum Hamiltonian $H_{\mathrm pend}$, eq.~(\ref{eqpend}), 
as closed paths,
keeping $\hat{P}_t$ and $t$ constant.
Depending on the nature of the pendulum motion (librational or
rotational), the topology of the closed paths is different,
leading to distinct expressions:
\begin{itemize}
\item For trapped librational motion, inside the resonance island,
the path is isomorphic to a circle in the $(\hat{I},\hat{\theta})$ plane,
with a Maslov index equal to two. The quantization condition
is:
\begin{equation}
\frac{1}{2\pi}\oint \hat{I}d\hat{\theta}=\left(N+\frac{1}{2}\right)\hbar,\
({\rm librational\ motion})
\label{ebkires}
\end{equation}
with $N$ a non-negative integer.
Of special interest is the ``fundamental" state, $N=0$, which exhibits
maximum localization within the resonance island and
is therefore
expected to represent the optimal non-dispersive wave-packet.
\item For unbounded rotational motion, outside the
resonance island, the path includes a $2\pi$ phase change for $\theta$
and acquires the Maslov index of the unperturbed motion:
\begin{equation}
\frac{1}{2\pi}\int_{0}^{2\pi} \hat{I}d\hat{\theta}=\left(
N+\frac{\mu}{4}\right)\hbar,\ ({\rm rotational \ motion}).
\label{ebkinres}
\end{equation}
\end{itemize}

This semiclassical quantization scheme is expected to work provided
the classical phase space velocity is sufficiently large, see section~\ref{SQ}.
This may fail close to the stable and unstable fixed points where
the velocity vanishes. Near the stable equilibrium
point, the expansion of the Hamiltonian at second order leads to an
approximate harmonic Hamiltonian with frequency:
\begin{equation}
\omega_{\mathrm harm} = \sqrt{|\lambda V_1(\hat{I_1})H^{''}_0(\hat{I_1})|}.
\label{omega_harmonic}
\end{equation}

In this harmonic approximation, the semiclassical quantization
is known to be exact~\cite{landau2}.
 Thus, close to the stable
equilibrium, the quasi-energy
levels, labeled by the non-negative integer $N$, are
given by the harmonic approximation. There are various cases which
depend on the signs
of $H^{''}_0(\hat{I_1})$, $V_1(\hat{I_1})$, and $\lambda$,
 with the general
result
given by
\begin{equation}
{\mathcal E}_{N,k} = H_0(\hat{I_1}) - \omega \hat{I_1} +
\left(k+\frac{\mu}{4}\right)\hbar\omega
-\ {\mathrm sign}(H^{''}_0(\hat{I_1}))
\left[ |\lambda V_1(\hat{I_1})| -
\left(N+\frac{1}{2}\right) \hbar \omega_{\mathrm harm} \right].
\label{spectrum_harmonic}
\end{equation}

For $N=0,k=0,$ this gives
 a fairly accurate estimate of the
energy of the non-dispersive wave-packet with optimum localization.
The EBK semiclassical scheme provides us also with some
interesting information on the eigenstate. Indeed, the invariant
tori considered here are tubes surrounding the resonant
stable periodic orbit. They cover the $[0,2\pi]$ range
of the $t$ variable but are well localized in the
transverse $(\hat{I}=I,\hat{\theta}=\theta-\omega t)$ plane,
with an approximately Gaussian phase space distribution.
Hence, at any fixed time $t$, the Floquet
eigenstate will appear as a Gaussian distribution
localized around the point $(I=\hat{I_1},\theta=\omega t).$ 
As this point precisely defines the resonant, stable
periodic orbit, one expects the $N=0$ state to be
a Gaussian wave-packet following the classical orbit.
In the original $(p,z)$ coordinates, the width of the
wave-packet will depend on the system under consideration
through the change of variables $(p,z)\to (I,\theta),$
but the Gaussian character is expected to be approximately valid
for both the phase space density and the configuration space
wave-function, as long as the change of variables is smooth.

Let us note that low-$N$ states may be considered as excitations
of the $N=0$ ``ground'' state. Such states has been termed
``flotons'' in \cite{holthaus95} where
their wave-packet character was, however, not considered.

The number of eigenstates trapped within the resonance island -- i.e.
the number of non-dispersive wave-packets -- is easily evaluated
in the semiclassical limit, as it is the maximum $N$ with librational
motion.  It is roughly the area of the
resonance island, eq.~(\ref{area}), divided by $2\pi\hbar$:
\begin{equation}
{\mathrm Number\ of\ trapped\ states} \simeq \frac{8}{\pi \hbar}
\sqrt{\left|\frac{\lambda V_1(\hat{I_1})}{ H^{''}_0(\hat{I_1})}\right|}.
\label{number_of_trapped_states}
\end{equation}

Near the unstable fixed point -- that is at the energy which separates
librational and rotational motion -- the semiclassical quantization fails 
because of the critical slowing down in its  vicinity~\cite{marion}.
The corresponding
quantum states -- known as separatrix states~\cite{leopold94} -- are
expected to be dominantly
localized near the unstable fixed point, simply because the classical
motion there slows down, and the pendulum spends more time close to its
upright position. This localization is once again of purely classical origin,
but not perfect: some part of the wave-function
must be also localized 
along the separatrix, which autointersects
at the hyperbolic fixed point. Hence, the Floquet eigenstates associated with
the unstable fixed points
are not expected to form non-dispersive wave-packets with optimum
localization (see also sec.~\ref{scars}).

\subsubsection{The Mathieu approach}
\label{section_mathieu}
The pendulum approximation, eq.~(\ref{eqpend}), for a resonantly driven system can also be found
by a pure quantum description~\cite{berman77,holthaus95}. Let us consider
a Floquet state of the system.
Its spatial part can
be expanded in the eigenbasis of the unperturbed Hamiltonian $H_0$,
\begin{equation}
H_0 |\phi_n\rangle = E_n|\phi_n\rangle ,
\end{equation}
while the time-periodic wave-function can be
expanded in a Fourier series. One obtains:
\begin{equation}
|\psi(t)\rangle =\ \sum_{n,k} c_{n,k}
\ \exp (-ik\omega t)\ \ |\phi_n\rangle,
\label{wfmat}
\end{equation}
where the coefficients $c_{n,k}$ are to be determined.
The Schr\"odinger equation for the Floquet states (eigenstates of ${\cal H}$),
eq.~(\ref{calhq}), with quasi-energies ${\cal E}$ and 
time dependence (\ref{solse}),
reads:
\begin{equation}
c_{n,k}({\cal E}+k\hbar \omega - E_n) = \frac{\lambda}{2} \ \sum_{p}
{\langle \phi_n|V|\phi_{n+p}\rangle\ (c_{n+p,k+1} + c_{n+p,k-1})}.
\label{coupled_equations}
\end{equation}

For $\lambda=0,$ the solutions of eq.~(\ref{coupled_equations})
are trivial: ${\cal E}=E_n-k\hbar\omega$, which is nothing
but the unperturbed energy spectrum modulo $\hbar\omega.$
\footnote{Note that, as a consequence of the negative sign of the argument 
of the exponential factor in eq.~(\ref{wfmat}), the energy shift $k\hbar\omega$ 
appears here with a negative sign in the expression for $\cal E$ -- in contrast 
to semiclassical expressions alike eq.~(\ref{spectrum_harmonic}), where we chose
the more suggestive positive sign. Since $k=-\infty\ldots +\infty$, both conventions 
are strictly equivalent.}
In the presence of a small perturbation, only quasi-degenerate states
with
values close to
$E_n-k\hbar\omega$ will be efficiently coupled. In the semiclassical limit,
see section \ref{SQ}, eq.~(\ref{cprinc}),
the unperturbed eigenenergies $E_n,$ labeled by a non-negative
integer,
are locally approximately spaced by $\hbar \Omega,$
where $\Omega$ is the frequency of the unperturbed classical motion.
Close to 
resonance, $\Omega \simeq \omega$ (eq.~(\ref{resonance_condition})), and 
thus: 
\begin{equation}
E_{n}-k \hbar \omega \simeq E_{n+1} - (k+1) \hbar \omega \simeq  E_{n+2} -
(k+2)\hbar \omega\simeq  ...,
\label{selrule}
\end{equation}
so that only states with the same value of $n-k$ will be efficiently coupled.

The first approximation is thus to neglect the couplings which do not
preserve $n-k.$ This is just the quantum version of the
secular approximation for the classical dynamics.
Then, the set of equations~(\ref{coupled_equations})
can be rearranged in independent blocks, each subset being characterized by $n-k$.
The various subsets are in fact identical, except for a shift in energy
by an integer multiple of $\hbar\omega.$ This is nothing but the
$\hbar\omega$-periodicity  
of the Floquet spectrum already encountered in secs.~\ref{QD} and \ref{sapp}.
As a consequence, we can
consider
the
$n-k=0$ block alone.

Consistently, since eq.~(\ref{selrule}) is valid close to the center
of the resonance only, one can expand the quantities of interest in the
vicinity of the center of the resonance, and use semiclassical 
approximations for 
matrix elements of $V.$ Let $n_0$ denote the effective, resonant quantum number
such that, with eqs.~(\ref{cprinc}) and (\ref{resonance_condition}),
\begin{equation}
\left. \frac{dE_n}{dn} \right|_{n=n_0}= \hbar \omega.
\label{space}
\end{equation}
Note that, by this definition, $n_0$ is {\em not} necessarily an integer.
 In the semiclassical
limit, where $n_0$ is large, the WKB approximation connects $n_0$ to the center
of the classical resonance island, see eq.~(\ref{semen}),
\begin{eqnarray}
&&n_0 + \frac{\mu}{4} = \frac{\hat{I_1}}{\hbar},\\
&&E_{n_0} = H_0(\hat{I_1}),
\end{eqnarray}
with $\mu$ being the Maslov index along the resonant trajectory.
Furthermore, for $n$ close to $n_0,$ we can expand
the unperturbed energy at second order in $(n-n_0)$,
\begin{equation}
E_n \simeq E_{n_0} + (n-n_0) \hbar \omega + \frac{1}{2}
\ \left.\frac{d^2E_n}{dn^2}\right|_
{n_0} \ (n-n_0)^2,
\end{equation}
where
the second derivative $d^2E_n/dn^2$
is directly related to the classical quantity $H^{''}_0$, see
eq.~(\ref{second_derivative}), within the semiclassical WKB
approximation, eq.~(\ref{semen}). Similarly, the matrix elements of $V$
are related to the classical Fourier components
of the potential~\cite{landau2}, eq.~(\ref{fourier_components}),
\begin{equation}
\langle \phi_n|V|\phi_{n+1} \rangle \simeq 
\langle \phi_{n+1}|V|\phi_{n+2} \rangle \simeq V_1(\hat{I_1}),
\label{matel_mat}
\end{equation}
evaluated at the center $\hat{I}=\hat{I}_1$ -- 
see eq.~(\ref{resonance_condition}) --
of the resonance zone.

With these ingredients and $r=n-n_0$, eq.~(\ref{coupled_equations}) 
is transformed
in the following set of approximate equations:
\begin{equation}
\left[ {\cal E} - H_0(\hat{I_1}) + 
\omega\left(\hat{I}_1 -\frac{\mu\hbar}{4}\right) - 
\frac{\hbar^2}{2} H^{''}_0(\hat{I_1}) r^2
\right] d_r =
\lambda V_1 (d_{r+1} + d_{r-1}),
\label{tridiag}
\end{equation}
where
\begin{equation}
d_r \equiv c_{n_0+r,n_0+r}.
\label{diag_mat}
\end{equation}

Note that, because of eq.~(\ref{space}),
the $r$ values are not necessarily integers, but all have the same
fractional part.

The tridiagonal set of coupled equations (\ref{tridiag})
can be rewritten as a differential
equation. Indeed, if one introduces the following function associated with
the Fourier
components $d_r$,
\begin{equation}
f(\phi) = \sum_r{\exp (ir\phi) d_r},
\label{dual}
\end{equation}
eq.~(\ref{tridiag}) can be written as
\begin{equation}
\left[- \frac{\hbar^2}{2} H^{''}_0(\hat{I_1}) \frac{d^2 }{d\phi^2} +
H_0(\hat{I_1}) -
\omega\left(\hat{I}_1 -\frac{\mu\hbar}{4}\right)
+ \lambda V_1 \cos \phi \right] f(\phi) = {\cal E} f(\phi),
\label{mathieu}
\end{equation}
which is nothing but the quantum version of the pendulum
Hamiltonian,
eq.~(\ref{eqpend}).
Thus, the present calculation is just the purely quantum 
description of the non-linear resonance phenomenon.
The dummy variable
$\phi$
introduced for convenience coincides with the classical angle
variable  $\hat{\theta}$. 
In general, $r$ is not an integer, so that the various
$\exp (ir\phi)$ in eq.~(\ref{dual}) are not periodic functions
of $\phi.$ However, as all $r$ values have the same fractional part,
if follows that $f(\phi)$ must satisfy ``modified'' 
periodic boundary conditions of the form
\begin{equation}
f(\phi + 2\pi) = \exp (-2i\pi n_0) f(\phi).
\label{boundary}
\end{equation}
The reason for this surprising boundary condition is clear:
$r=n-n_0$ is the quantum analog of $\hat{I}-\hat{I}_1.$
In general, the resonant action $\hat{I}_1$ 
is {\em not} an integer or half-integer multiple
of $\hbar$ -- exactly as $n_0$ is not an integer. The semiclassical
quantization, eq.~(\ref{ebkinres}), which expresses
the $\hat{\theta}$ periodicity of the eigenstate,
applies for the $\hat{I}$ variable. When expressed 
in terms of the variable $\hat{I}-\hat{I}_1$,
it contains the additional phase shift 
present in eq.~(\ref{boundary}).
  
Few words of caution are in order: the equivalence
of the semiclassical quantization with the pure
quantum approach holds in the semiclassical limit only,
when the quantum problem
can be mapped on a pendulum problem. In the general case, it is not possible to
define a quantum angle variable \cite{loudon}.
Hence, the quantum treatment presented here
is no
more general or more powerful than the semiclassical treatment. They both rely
on the same approximations and have the same limitations: perturbative
regime (no overlap of resonances) and semiclassical approximation.

Finally, eq.~(\ref{mathieu}) can be written in its standard form, known as
the ``Mathieu equation" \cite{abramowitz72}:
\begin{equation}
\frac{d^2y}{dv^2} + (a-2q\cos 2v) y = 0.
\label{mat_eq}
\end{equation}
The correspondence with eq.~(\ref{mathieu}) is established via:
\begin{eqnarray}
&&\phi = 2v,
\label{map_mathieu1}\\
&&a=\frac{8\left[{\cal E}-H_0(\hat{I_1})+
\omega\left(\hat{I}_1 -\frac{\mu\hbar}{4}\right)
\right]}{\hbar^2 H^{''}_0(\hat{I_1})},
\label{map_mathieu2}\\
&&q=\frac{4\lambda
V_1}{\hbar^2 H^{''}_0(\hat{I_1})}.
\label{map_mathieu3}
\end{eqnarray}

The boundary condition, eq.~(\ref{boundary}),
is fixed by the so-called ``characteristic exponent"
in the Mathieu equation,
\begin{equation}
\nu = -2 n_0\ \ \ ({\mathrm mod}\ 2).
\label{cexp}
\end{equation}

\begin{figure}
\centerline{\psfig{figure=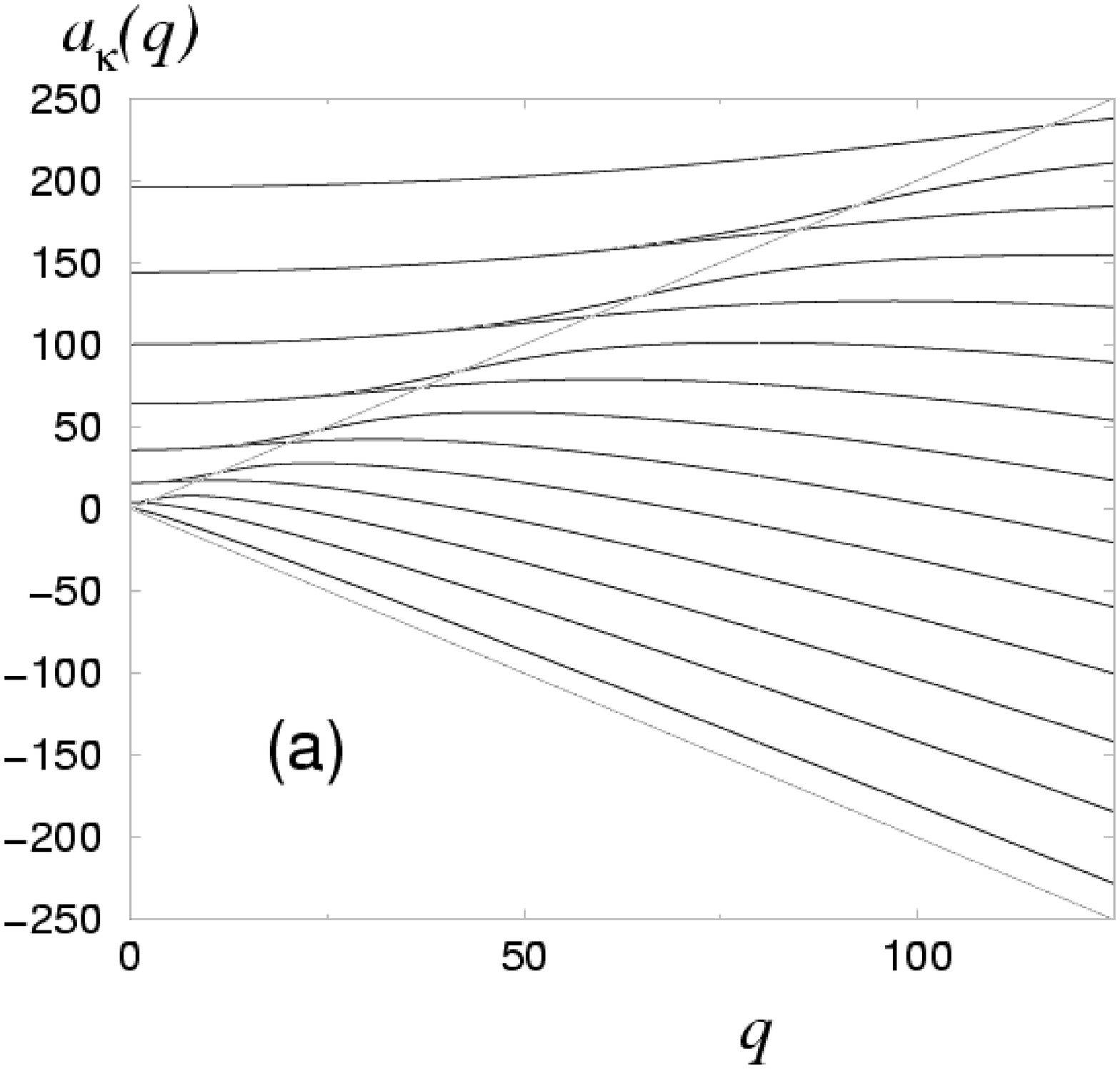,width=7cm,angle=-90}
\psfig{figure=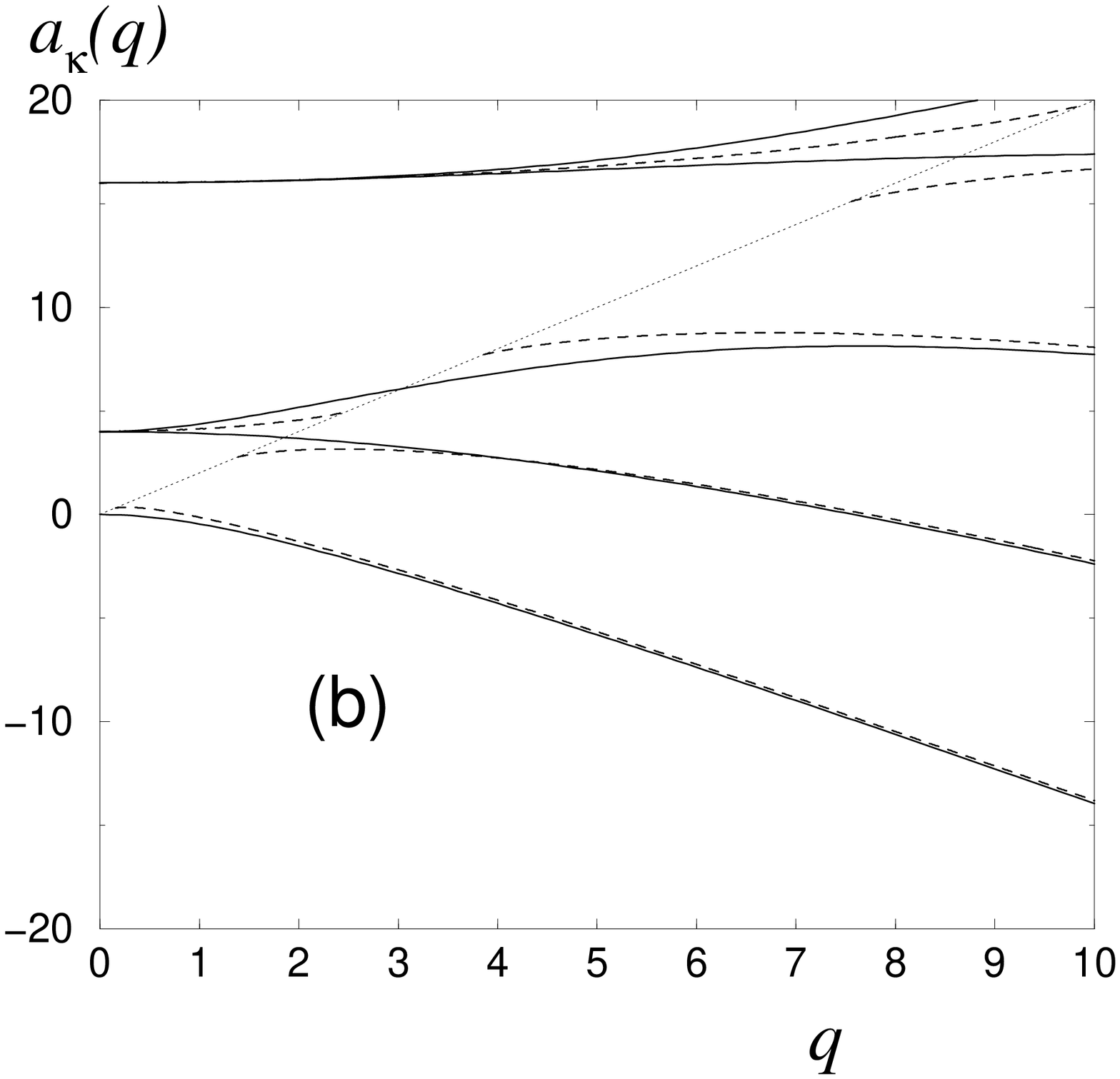,width=7cm,angle=-90}}
\caption{Eigenvalues $a_{\kappa}(q)$ of the Mathieu equation, 
for a characteristic
exponent $\nu=0.$ These represent the energy levels of a pendulum as a function
of the gravitational field, see
eqs.~(\protect\ref{eqpend},\protect\ref{mathieu}). 
(a): Eigenvalues in the range $[-2q,2q]$ are
associated with the librational bounded motion of the pendulum, while
eigenvalues above $2q$ are associated with rotational modes.
The dotted lines represent the energies of the stable equilibrium
point (lower line) and of the unstable equilibrium point (separatrix, upper
line). Near the separatrix, the classical motion slows down and the
quantal energy levels get closer. (b): Details for the first excited states
together with the semiclassical WKB prediction for the energy levels
(dashed lines, eqs.~(\protect\ref{semen})). 
The semiclassical prediction is very accurate, except
in the vicinity of the separatrix.}
\label{fig_mathieu}
\end{figure}

The Mathieu equation has solutions (for a given characteristic exponent)
for a discrete set of values of $a$ only. That implies quantization of
the quasi-energy levels, according to eq.~(\ref{map_mathieu2}).
The quantized values $a_{\kappa}(\nu,q)$
depend on $q$ and $\nu$, and are labeled\footnote{In 
the standard text books as \protect\cite{abramowitz72},
the various solutions of the Mathieu equation are divided in 
odd and even solutions, and furthermore in $\pi$- and $2\pi$-periodic 
functions. In our case, only the
``$a_{2p}$" and ``$b_{2p}$" (in the language of \protect\cite{abramowitz72})
are to be considered.} 
by a non-negative
integer $\kappa.$
They are well known
-- especially asymptotic expansions are available both in the
small and in the large $q$ regime -- and
can be found in standard handbooks
\cite{abramowitz72}. For example, fig.~\ref{fig_mathieu}(a) shows the
first $a_{\kappa}(q)\equiv a_{\kappa}(\nu=0,q)$ curves for the case of 
``optimal" resonance (see below), where
$n_0$ is an integer and thus
the characteristic exponent $\nu$ vanishes.
Equivalently, the figure can be interpreted as the evolution of the
energy levels of a pendulum
with the gravitational field.

\psfull

The quasi-energy levels of the driven
system can now be expressed as a function of $a_{\kappa}(\nu,q):$
\begin{equation}
{\cal E}_{\kappa} = H_0(\hat{I_1}) - 
\omega\left(\hat{I}_1 -\frac{\mu\hbar}{4}\right)
+ \frac{\hbar^2}{8} H^{''}_0(\hat{I_1})\ a_{\kappa}(\nu,q).
\label{prediction_mathieu}
\end{equation}
Together with eqs.~(\ref{map_mathieu1})-(\ref{map_mathieu3}),
this equation gives the quasienergy
levels of a periodically driven system in the vicinity of the resonance zone.
The full, quasi-resonant part of the Floquet spectrum of the driven system is built
from these quantized values through shifts $k\hbar\omega,$ with
arbitrary integer values of $k$.

A visual inspection of fig.~\ref{fig_mathieu} immediately shows the existence
of two regions in the energy diagram: within the ``inner region",
$|a_{\kappa}(\nu,q)| \leq 2q,$ the energy levels
form a regular fan of curves 
and tend to decrease with $q.$ On the contrary,
for $a_{\kappa}(\nu,q) > 2q,$ the energy levels increase with $q$. 
Around $a_{\kappa}(\nu,q)=2q,$
a transition region is visible with a series of apparent avoided crossings
between the levels. This has a simple semiclassical explanation.
The stable fixed point of the pendulum described by the Mathieu
equation (\ref{mat_eq}) lies at $v=0$ with energy $-2q$,
what explains why
the $a_{\kappa}(\nu,q)$ values are always larger. The unstable
fixed point has an energy $+2q.$ Thus, in the range 
$a_{\kappa}(\nu,q) \in [-2q,2q],$
the pendulum
is trapped in a region of
librational motion. The energy levels can be approximated
using the standard WKB quantization in the $(\hat{I},\hat{\theta)}$
plane, as described in section~\ref{CD}. 
The number of such states is given by eq.~(\ref{number_of_trapped_states})
which can be rewritten, using eq.~(\ref{map_mathieu3}), as :
\begin{equation}
{\mathrm Number\ of\ trapped\ states} \simeq \frac{4\sqrt{|q|}}{\pi}.
\label{number_of_trapped_states_2}
\end{equation}
At the center of the island (states with small $\kappa$
and/or large $|q|$), an
approximate expression for $a_{\kappa}(\nu,q)$ reads \cite{abramowitz72}:
\begin{equation}
a_{\kappa}(\nu,q) \approx -2|q|+4\left(\kappa+\frac{1}{2}\right) \sqrt{|q|}.
\label{math_asy}
\end{equation}
It does not depend on $\nu$ (what physically means in our case 
that the states deeply inside the resonance island are insensitive
to the boundary condition).
When inserted in eq.~(\ref{prediction_mathieu}), it
yields exactly the energy levels of eq.~(\ref{ebkires}), with
$\kappa=N.$ Hence, the
Mathieu approach agrees with the harmonic approximation within the
resonance island, for sufficiently large islands.
 
For $a_{\kappa}(\nu,q) > 2q,$ the
pendulum undergoes
a rotational, unbounded motion, which again can be quantized using WKB.
For small $|q|$, 
$a_{\kappa}(\nu,q)\approx 4([(\kappa+1)/2]-\nu/2)^2$ (where
$[\ ]$ stands for the integer part), see \cite{abramowitz72}. 
With this  in
eq.~(\ref{prediction_mathieu}), 
one recovers the known Floquet spectrum, in the limit of a
vanishingly  weak perturbation. 
Note, however, that in this rotational mode, the eigenstates
{\em are} sensitive to the boundary conditions (and the $a_{\kappa}(\nu,q)$
values depend on $\nu$). This is essential for
the correct $\lambda \rightarrow 0$ limit.

Around $a_{\kappa}(\nu,q) = 2q,$ 
the pendulum is close to the separatrix between
librational and rotational motion: the period of the classical motion
tends to infinity (critical slowing down). That explains the locally enhanced
density of states apparent in fig.~\ref{fig_mathieu}(a).

In fig.~\ref{fig_mathieu}(b), we also plot the semiclassical WKB
prediction for the
quantized $a_{\kappa}(q)$ values. 
Obviously, the  agreement with the exact ``quantum"
Mathieu result is very good, even for weakly excited states, except
in the vicinity of the separatrix. This is not unexpected because the
semiclassical
approximation is known
to break down near the unstable fixed point, see
sections~\ref{SQ} and \ref{sapp} above.

The Mathieu equation
yields accurate predictions for
properties of
non-dispersive wave-packets in  periodically driven
quantum systems. Indeed, in the range  $a_{\kappa}(\nu,q) \in [-2q,2q],$ 
the classical
motion is trapped inside the resonance island, and the corresponding
quantum eigenstates are expected to be non-dispersive wave-packets.
In particular,
the lowest state in the resonance island, associated
with the ground state of the pendulum $\kappa=0$, 
corresponds to the semiclassical eigenstate $N=0$, see eq.~(\ref{ebkires}), 
and represents 
the non-dispersive wave-packet
with the best localization properties.
As can be seen in fig.~\ref{fig_mathieu}(b), the
semiclassical quantization for this state is in excellent agreement with the
exact Mathieu result. This signifies that eq.~(\ref{spectrum_harmonic})
can be used for {\em quantitative} predictions of the quasi-energy of this
eigenstate.

As already mentioned,
the characteristic exponent $\nu$ does not play a major role inside the
resonance island, as the eigenvalues  $a_{\kappa}(\nu,q)$ there depends
very little on $\nu.$ However, it is an important parameter outside the
resonance, close to the separatrix especially at small $q.$ Indeed,
at $q=0$, the minimum eigenvalue is obtained
for $\nu=0: a_{0}(\nu=0,q=0)=0.$ 
This implies that, even for very small $q,$ the ground state
enters most rapidly the resonance island. On the opposite, the
worst case is $\nu=1$ where the lowest eigenvalue is doubly degenerate:
$a_{0}(\nu=1,q=0)=a_{1}(\nu=1,q=0)=1.$ 
As we are interested in the ground state of the
pendulum (the one with maximum localization), 
the situation for $\nu=0$ is preferable: not only the state enters rapidly the
resonance island, but it is also separated from the other states by an
energy gap and is thus more robust versus any perturbation.
We will call this situation ``optimal resonance". From eq.~(\ref{cexp}),
it is associated with an integer value of $n_0.$ On the contrary, a
 half-integer value of $n_0$ corresponds to $\nu=1$ and the
 least optimal case.

\subsection{Rydberg states in external fields}
\label{RSEF}

\subsubsection{Rydberg atoms}

In order to construct non-dispersive wave-packets, a quantum
system subject to periodic driving with classically non-linear dynamics is
needed. The latter requirement rules out the harmonic oscillator,
and all its variants. The simplest periodic driving is certainly
provided by
an externally applied, monochromatic electromagnetic field.
Extremely stable, tunable and well controlled sources
exist over a wide range of frequencies.

Furthermore, incoherent
processes which destroy the phase coherence of the quantum wave-function have
to be minimized. Otherwise, they will spoil
the localization properties
of the non-dispersive wave-packets and -- in the worst case -- destroy
them completely. Therefore, the characteristic time scales of the
incoherent processes should be at least much longer than the period of the
driving field.
In this respect, atomic electrons appear as
very good candidates,
since -- given suitable experimental conditions -- atoms can be
considered as practically isolated from the external world, with
spontaneous emission of photons as the only incoherent process.
Spontaneous emission is usually a very slow 
mechanism, especially for highly
excited states: the spontaneous life-time of typical atomic states
is at least four or five orders
of magnitude longer than the classical Kepler period (typically
nanoseconds vs. femtoseconds for weakly excited states~\cite{bethe77}).

The Coulomb interaction between  
the nucleus and the
electrons is highly
non-linear, which is very favourable. The efficiency
of the coupling with an external electromagnetic
field is known to increase rapidly with the degree of excitation of the atom
\cite{bethe77}.
As we have seen in the preceeding sections~\ref{CD} and \ref{QD},
non-dispersive wave-packets are the quantum mechanical counterparts
of nonlinear resonances in periodically driven Hamiltonian systems, where
the period of the drive matches some intrinsic time scale of the unperturbed
Hamiltonian dynamics.
Due to the immediate correspondence between the classical Kepler problem
and the hydrogen atom, the relevant time scale in this simplest atomic system
is the
unperturbed classical Kepler period, which 
-- compare eq.~(\ref{cprinc}) --
coincides with the inverse level spacing between neighbouring
eigenstates of the unperturbed atom, for large quantum numbers.
Hence, the driving field frequency has to be chosen resonant
with an atomic transition
in the Rydberg regime, typically around the
principal quantum number $n_0=60$. This is the microwave domain,
 where excellent sources exist.
Thus, we believe that atomic Rydberg states are very well suited
for the experimental preparation of non-dispersive wave-packets\footnote{Note,
however, that this is a specific choice. Any driven quantum system 
with a sufficiently high denisty of states and mixed regular-chaotic 
classical dynamics will exhibit nondispersive wave-packets. Since a mixed 
phase space structure is the generic scenario for dynamical systems,
nondispersive wave-packets are expected to be a completely general and 
ubiquitous phenomenon.}.

In most cases, the energy scale involved in the dynamics of Rydberg
electrons is so small that the inner electrons of the ionic core
can be considered as frozen and ignored. Thus, we will consider
mainly the hydrogen atom as the simplest prototype. Multi-electron
effects are discussed in section~\ref{HE}. Alternative systems
for observing non-dispersive wave-packets are considered in section
\ref{OGB}.

Compared to our simple one-dimensional model introduced in section~\ref{GM} 
above,
an {\em atom}
displays a couple of additional features:
\begin{itemize}
\item A real hydrogen atom is a three-dimensional (3D) system. 
However, it is a degenerate system, 
because the energy depends on the principal quantum number
$n$ only, but not on the angular or magnetic quantum numbers $L$ and $M$,
respectively.
Thus, the structure of the energy levels, which is crucial for the
properties of the non-dispersive wave-packet, see section \ref{QD},
is identical in the 1D and 3D cases. Before discussing the properties
of 3D wave-packets in sections~\ref{LIN3D} to \ref{MA}, we will
consider a simplified 1D model of the hydrogen atom in section~\ref{LIN1D}.
\item Although spontaneous emission is a weak incoherent process,
it 
nontheless limits the life time of the non-dispersive wave-packets which may
decay to lower lying states, losing the phase coherence
of the electronic wave-function. Non-dispersive wave-packets
exhibit specific spontaneous decay properties which are studied in
section~\ref{SPO};
\item The electron in a hydrogen atom is not necessarily bound.
It may ionize, especially when the atom is exposed to
a microwave field. This is a {\em coherent} decay process 
where the ionized electron keeps its phase coherence.
There are no exact bound states in the system, but rather resonances.
From a quantum point of view, the Floquet spectrum is no longer
discrete but continuous and we actually deal with an {\em open system}. 
In a more elementary language,
the atom can successively absorb several photons so that its energy
exceeds the ionization threshold. 
If initially prepared in a wave-packet eigenstate, this is a pure quantum
phenomenon, since the classical dynamics remain trapped within the
resonance island forever. The multi-photon ionization 
may then be 
considered as a tunneling process from 
inside the resonance island
to the non-resonant part of phase space, 
where the Rydberg electron eventually escapes
to infinity. 
This picture is elaborated in  section~\ref{ION}.
\end{itemize}

\subsubsection{Hamiltonian, basis sets and selection rules}

In the presence of a microwave field, the dipole 
approximation~\cite{cct92,loudon}
can be used to describe  the atom-field interaction. Different gauges
can be used, the physics being of course independent of the choice
of gauge. The most common choices are the length and the velocity gauges.
For simplicity, in our discussion,
we shall use the length gauge, although actual
quantum calculations are usually a bit easier in the velocity gauge
\cite{shakeshaft88,abu95c,cormier96}.
The Hamiltonian reads:
\begin{equation}
H = \frac{\vec{p}^2}{2m} - \frac{q^2}{4\pi \epsilon_0 r}
-q \vec{r}\cdot\vec{F}(t)
\label{ham_3d_si}
\end{equation}
where $q$ is the (negative) charge of the electron, $m$ its mass, and
$\vec{F}(t)$ the microwave electric field acting on the atom.
We neglect here all relativistic, spin, QED effects, etc, and assume
an infinitely massive nucleus.
Later, unless specified otherwise, we shall use atomic units, where
$|q|\equiv 4\pi\epsilon_0\equiv m\equiv \hbar\equiv 1$, and the
Hamiltonian reduces to
\begin{equation}
H = \frac{\vec{p}^2}{2} - \frac{1}{r}
+\vec{r}\cdot\vec{F}(t).
\label{h}
\end{equation}
Here, 
\begin{equation}
H_0= \frac{\vec{p}^2}{2} - \frac{1}{r}
\label{h0}
\end{equation}
describes the unperturbed atomic part, with 
bound energy spectrum: 
\begin{equation}
E_n= - \frac{1}{2n^2},\ n\geq 1.
\label{spectrum}
\end{equation}
The external driving
\begin{equation}
V = \vec{r}\cdot \vec{F}(t)
\label{v}
\end{equation}
is characterized by
the amplitude $F$ of the microwave field, and by its  
frequency
$\omega.$ 
We shall consider in detail the case of linear polarization, 
\begin{equation}
V=F z \cos \omega t,
\label{vlinear} \end{equation}
in section \ref{LIN}, and that of circular polarization
\begin{equation}
V=F( x \cos \omega t +  y \sin \omega t),
\label{vcirc} 
\end{equation}
in section~\ref{CP}. The general case of elliptic polarization
will be 
studied in section~\ref{EP}.

In the treatment of the perturbed Coulomb problem, eq.~(\ref{h}),
there is an apparent difficulty with the singularity of the Coulomb
potential at the origin, especially for the restricted one-dimensional
model of the atom.
This led several authors,
see, e.g.,
\cite{javanainen88,su91,reed91,burnett92,grobe93,grobe94}, to consider
unphysical potentials without singularity, for example
of the type $(r^2+a^2)^{-1/2}.$ This is completely unnecessary
and potentially dangerous. Indeed, such a potential breaks
the Coulomb degeneracy which is responsible for the closed character
of elliptical Kepler orbits (in the classical world), and
for the degeneracy of the energy levels (in the quantum world).
Such an unphysical symmetry breaking strongly modifies the structure
of the non-linear resonance island, and affects the existence
and properties of non-dispersive wave-packets outlined in section~\ref{LIN3D}.

The Coulomb singularity can be rigorously regularized (in any
dimension), both in classical and in quantum mechanics. In classical
mechanics, this is made possible through the well-known
Kustaanheimo-Stiefel transformation \cite{kusta65}, used by
various authors for perturbed Coulomb problems, see, e.g.,
\cite{delande84,rath88,griffiths92,gebarowski95,schlagheck99}.
 In quantum mechanics, one may  use a basis set
of non-orthogonal functions, known as the Sturmian functions.
Ultimately, the whole analysis relies on the dynamical symmetry
properties of the Coulomb interaction and the associated SO(4,2) group
\cite{barut79,chen80,chen81,delandeth}.
It not only allows to treat the Coulomb singularity properly, but also
to define a basis set of Sturmian functions 
 extremely efficient for
numerical calculations. The most common set 
are ``spherical" Sturmian functions characterized by three quantum numbers,
$L$ and $M$ for the angular structure (the associated wave-functions are
the usual spherical harmonics), and the positive integer $n$, for the radial
part. 

As discussed in section~\ref{QD}, the quantum properties
of the non-dispersive wave-packets are 
encoded in the
spectrum of the Floquet Hamiltonian (\ref{calhq}), which acts
in configuration space extended by the time axis. The temporal
properties are completely independent of the spatial dimension,
and any Floquet eigenstate can be expanded in a Fourier series
indexed by the integer $k$, 
as in eq.~(\ref{wfmat}). The Schr\"odinger equation
for the Floquet Hamiltonian is tridiagonal in $k,$ as in
eq.~(\ref{coupled_equations}).
When the spatial part of the wave-function is 
expressed in a Sturmian basis,
one finally obtains
a generalized eigenvalue problem $(A-{\cal E} B)|\psi> = 0$, where both,
$A$ and $B$, are sparse matrices, the elements of which are known analytically
and obey the following selection rules:
\begin{equation}
|\Delta M|\leq 1,\ \ |\Delta L| \leq 1,\ \ |\Delta n| \leq 2,\ \ |\Delta k| \leq 1.
\end{equation}
When a static electric or magnetic field is added, see section~\ref{MA},
some additional non-zero matrix elements exist, but sparsity is preserved.
The eigenvalues can then be calculated using an efficient diagonalization
routine such as the Lanczos 
algorithm \cite{lanczos,delande91,ericsson80,grimes94}.

Because -- in the presence of a microwave field -- the system is unbounded,
there are in general no exact bound states but rather resonances.
Using Sturmian functions, the
properties of the resonances can be calculated {\em directly} 
using the complex
rotation technique 
\cite{balslev71,graffi85,yajima82,nicolaides78,reinhardt83,ho83,moiseyev98}. 
The price
to pay is to diagonalize complex symmetric matrices, instead of
real symmetric ones. The advantage is that the resonances are obtained
as complex eigenvalues $E_n-i\Gamma_n/2$ of the complex rotated Hamiltonian,
$E_n$ being the position of the resonance, and $\Gamma_n$ its width.
All
essential properties of resonances can be obtained from complex
eigenstates \cite{abu94}.

\subsubsection{Simplified 1D and 2D models}

Because explicit calculations for the real 3D
 hydrogen atom may be
rather complicated, it is fruitful to study also simplified 
1D and 2D approximations of the real world.
Let us first consider the simplified
restriction of the atomic motion to one single dimension 
of configuration space,
\begin{equation}
H_0=\frac{p^2}{2} - \frac{1}{z}, \ \ \ \ {\mathrm with}\ z>0,
\label{h0_1d}
\end{equation}
with the external driving along $z$,
\begin{equation}
V = F z \cos \omega t.
\label{v_1d}
\end{equation}
The energy spectrum of $H_0$ is identical to the spectrum
in 3D~\cite{englefield72}:
\begin{equation}
E_n^{(1D)} = - \frac{1}{2n^2},\ \ \ \ {\mathrm with}\ n\geq 1.
\label{spectrum_1d}
\end{equation}
Such  a one-dimensional model allows to grasp
essential features of the driven atomic dynamics, and provides the simplest
example for the creation of non-dispersive wave-packets by a near-resonant
microwave field. The classical dynamics live on a three-dimensional
phase space, spanned by the single dimension of configuration space, the canonically conjugate momentum, and
by time. This is the lowest dimensionality for a Hamiltonian system to display
mixed regular-chaotic character \cite{lichtenberg83}.

For a circularly (or elliptically) polarized microwave, a 1D model
is of course inadequate. One can use a two-dimensional model
where the motion of the electron is restricted to the polarization
plane. The energy spectrum in two dimensions is:
\begin{equation}
E_n^{(2D)}= - \frac{1}{2\left(n+\half\right)^2},\ \ \ \ {\mathrm with}\ n\geq 0.
\label{spectrum_2d}
\end{equation}
It differs from the 3D (and 1D) energy spectrum by the additional
1/2 in the denominator, due to the specific
Maslov indices induced by  the Coulomb 
singularity.

\subsubsection{Action-angle coordinates}

\label{action-angle_hydrogen}

In order to apply the general theory of nonlinear resonances and
non-dispersive wave-packets derived in section~\ref{GM}, we need
the action-angle coordinates for the hydrogen atom.
For the simplified 1D model, the result is simple:
 the principal action $I$ and the canonically conjugate
angle $\theta$ are defined by \cite{goldstein80,jensen84}
\begin{eqnarray}
I & = & \sqrt{\frac{a}{2}}, \nonumber \\
\theta & = & \left\{
\begin{array}{lc}
\eta-\sin\eta, & \ \ p\geq 0, \\ 2\pi-\eta+\sin\eta, & \ \ p<0,
\end{array} \right. \nonumber \\
\eta & = & 2\sin^{-1}\sqrt{\frac{z}{a}},\ \ \  a=-E^{-1},
\label{aa_lin1d}
\end{eqnarray}
where $a$
is the maximum distance.
In celestial mechanics, $\theta$ and $\eta$ are known as
the {\em mean} and the {\em eccentric anomaly}, respectively \cite{marion}.
The Hamilton function depends on the action through:
\begin{equation}
H_0=-\frac{1}{2I^2},
\label{h0_I}
\end{equation}
and the classical Kepler frequency reads:
\begin{equation}
\Omega = \frac{dH_0}{dI} = \frac{1}{I^3}.
\label{omk}
\end{equation}
Due to the  Coulomb singularity at $z=0$,
 the Maslov index
of this system is $\mu=0$ instead of $\mu=2$, and the semiclassical 
energy spectrum,
 eq.~(\ref{semen})\footnote{As we are using atomic units, $\hbar$ is 
unity, and
the principal quantum number just coincides with the action.},
matches the exact quantum spectrum, eq.~(\ref{spectrum_1d}).

The classical equations of motion can be solved exactly and it is easy
to obtain the Fourier components of the dipole operator \cite{meerson82}:
\begin{equation}
z(\theta) = I^2 \left(\frac{3}{2} - 2\sum_{m=1}^{\infty}{ \frac{J'_m(m)}{m}
\cos (m\theta)} \right),
\label{v_1dfou}
\end{equation}
where $J'_n(x)$ denotes the  derivative of the usual Bessel function.
The strongly non-linear character of the Coulomb interaction is responsible
for the slow decrease of the Fourier components at high $m$.

For the 2D and 3D hydrogen atom, the action-angle variables
are similar, but more complicated because
of the existence of angular degrees of freedom.
The classical trajectories are ellipses with focus at the nucleus.
The fact that all bounded trajectories are periodic manifests the
degeneracy of the classical dynamics. As a consequence, although
phase space is six-dimensional with three angle and three action 
variables in 3D
-- four-dimensional with two angle and two action variables in 2D
-- the Hamilton function depends only on the total action $I$, precisely
like the 1D hydrogen atom, i.e. through eq.~(\ref{h0_I}).
In 3D, the Maslov index is zero, so that the energy spectrum is
again given by eq.~(\ref{spectrum}), and the semiclassical
approximation is exact. However, a different
result holds for the 2D hydrogen atom, where the
Maslov index is $\mu=2$ (still yielding exact agreement between the
semiclassical and   the quantum spectrum, cf.
eqs.~(\ref{ebkgen},\ref{spectrum_2d})).

The action-angle variables which 
parametrize a general Kepler ellipse  are well known
\cite{goldstein80}.
In addition to the action-angle variables $(I,\theta)$ which determine
the total action and the angular position
of the electron along the Kepler ellipse,
respectively, the orientation of the ellipse in space is defined by 
two angles: $\psi$, canonically conjugate to the total angular
momentum $L$, and the polar angle $\phi$, canonically conjugate to $M$, the
$z$-component of the angular momentum.
The angle $\psi$ conjugate to $L$ has a
direct physical meaning for $M=0$: it represents
the angle between the Runge-Lenz vector $\vec{A}$ (oriented along the major
axis) of the Kepler ellipse, and the
$z$-axis.
For the 2D hydrogen atom (in the $(x,y)$ plane), 
the orientation 
of the ellipse is defined by the angle $\psi$
(canonically conjugate to the total angular
momentum $L$)
 between the Runge-Lenz vector
and the $x$-axis.

Also the Fourier components of the unperturbed
classical position operator $\vec{r}(t)$
are well known \cite{casati88}. In the local coordinate
system of the Kepler ellipse (motion in the
$(x',y')$ plane, with major axis along $x'$), one gets:
\begin{eqnarray}
x'&=&-\frac{3e}{2}I^2+2I^2\sum_{m=1}^\infty
\frac{J_{m}^{'}(me)}{m}\cos m\theta \label{v_3d_x} \\
y'&=&2I^2\frac{\sqrt{1-e^2}}{e}\sum_{m=1}^\infty
\frac{J_{m}(me)}{m}\sin m\theta \label{v_3d_y} \\
z'&=&0
\label{v_3d_z}
\end{eqnarray}
where 
\begin{equation}
e = \sqrt{1-\frac{L^2}{I^2}}
\label{eccentricity}
\end{equation}
denotes the eccentricity of the ellipse.
$J_{m}(x)$ and $J^{'}_{m}(x)$ are the ordinary
Bessel function and
its derivative, respectively.

In the laboratory frame, the various components can be found
by combining these expressions with the usual Euler
rotations~\cite{marion,goldstein80}. The set of three Euler angles
describes the successive rotations required for the
transformation between the
laboratory frame and the frame $(x',y',z')$ linked to the classical
Kepler ellipse. We choose to rotate successively by an angle $\phi$ 
around the $z$ laboratory axis, an angle $\beta$ around the 
$y$-axis\footnote{Some authors define the second Euler rotation with
respect to the $x$-axis. The existence of the two definitions 
makes a cautious physicist's life much harder,
 but the physics does not -- or at least
should not -- depend
on such ugly details.}, and an angle $\psi$ around the $z'$ axis.
The physical interpretation of $\phi$ and $\psi$ is simple: 
$\phi$ corresponds to a rotation around the $z$-axis, and is thus 
canonically conjugate
to the $z$-component of the angular momentum, noted $M$. Similarly,
$\psi$ corresponds to a rotation around the axis of the total angular
momentum $\vec{L}$, and is thus canonically conjugate
to $L.$ 
By construction, the third angle $\beta$ is 
precisely the angle
between the angular momentum $\vec{L}$ and the $z$-axis. Thus:
\begin{equation}
\cos \beta = \frac{M}{L},
\label{beta}
\end{equation}
and $\beta =\pi/2$ for $M=0$.
Altogether, the coordinates in the laboratory frame are related
to the local coordinates through
\begin{eqnarray}
x&=&(\cos \psi \cos \beta \cos \phi - \sin \psi \sin \phi)\ x' +
(-\sin\psi \cos\beta \cos\phi -\cos\psi \sin\phi)\ y' + \sin \beta \cos\phi\ z',
\nonumber \\
y&=&(\cos \psi \cos \beta \sin \phi + \sin \psi \cos \phi)\ x' +
(-\sin\psi \cos\beta \sin\phi +\cos\psi \cos\phi)\ y' + \sin \beta \sin\phi \ z',
\nonumber \\
z&=& -\cos\psi \sin\beta \ x'  +  \sin\psi \sin\beta \ y' +  \cos\beta \ z',
\label{rot_euler}
\end{eqnarray}
which, combined with eqs.~(\ref{v_3d_x}-\ref{v_3d_z}), allows for a complete
expansion of the classical trajectories in terms of action-angle\
coordinates.

The situation is somewhat simpler for the 2D model of the hydrogen atom.
There, the angular momentum $\vec{L}$ is aligned along the $z$-axis,
which means that $L=M$ and $\beta=0.$ Also, the rotation around
$\vec{L}$ by an angle $\psi$ 
can be absorbed in a rotation by an angle $\phi$ around the $z$-axis.
Therefore, one is left with
two pairs $(I,\theta)$ and $(M,\phi)$ of action-angle variables.
The relation between the laboratory and the local coordinates
reads:
\begin{eqnarray}
x&=& \cos \phi \ x' - \sin\phi \ y' 
\nonumber \\
y&=&\sin \phi \ x' + \cos\phi\ y',
\label{twodrotfr} 
\end{eqnarray} 
which is nothing but a rotation of angle $\phi$ in the plane of the
trajectory.
Formally, the 2D result, eq.~(\ref{twodrotfr}), can be obtained from the 3D one,
eq.~(\ref{rot_euler}), by
specializing to $\beta=\psi=0$.
The eccentricity, eq.~(\ref{eccentricity}), of
the trajectory now reads:
\begin{equation}
e = \sqrt{1-\frac{M^2}{I^2}}.
\end{equation}

\subsubsection{Scaling laws}

\label{scaling_laws}

It is well known that the Coulomb interaction 
exhibits particular
scaling properties: for example, all 
bounded trajectories are similar
(ellipses), whatever the (negative) energy. Also, the
classical period 
scales in a well-defined way with the
size of the orbit (third Kepler law). This originates from
the fact that the Coulomb potential is a homogeneous function
-- of degree $-1$ -- of the radial distance $r$. Similarly,
the dipole operator responsible for the coupling between the Kepler 
electron and the 
external driving field is a homogenous function
-- of degree $1$ -- of $r.$ It follows that the
classical equations of motion of the hydrogen
atom exposed to an electromagnetic field
are invariant under the following
scaling transformation:
\begin{eqnarray}
\left \{
\begin{array}{l}
{\vec r}\rightarrow\alpha^{-1} {\vec r}, \\
{\vec p}\rightarrow\alpha^{1/2} {\vec p}, \\
H_0\rightarrow \alpha H_0, \\
t\rightarrow\alpha^{-3/2} t, \\
F\rightarrow\alpha^2 F,  \\
\omega\rightarrow\alpha^{3/2}\omega , \\
V \rightarrow \alpha V.
\end{array}
\right.
\label{scaling_law}
\end{eqnarray}
where $\alpha $ is an arbitrary, positive real number.
Accordingly, the action-angle variables
transform as
\begin{eqnarray}
\left \{
\begin{array}{l}
I \rightarrow \alpha^{-1/2} I,\\
L \rightarrow \alpha^{-1/2} L,\\
M \rightarrow \alpha^{-1/2} M,\\
\theta \rightarrow \theta,\\
\psi \rightarrow \psi,\\
\phi \rightarrow \phi.
\end{array}
\right.
\label{scaling_aa}
\end{eqnarray}
It is therefore useful to introduce the ``scaled" total angular momentum
and its component 
along the $z$-axis, by chosing 
\begin{equation}
\alpha =I^2,
\label{choosescale}
\end{equation}
what leads to
\begin{eqnarray}
\left \{
\begin{array}{l}
L_0 = \frac{L}{I},\\
M_0 = \frac{M}{I}.
\label{scaling_lm}
\end{array}
\right.
\end{eqnarray}
The eccentricity of the classical ellipse then reads:
\begin{equation}
e=\sqrt{1-L_0^2},
\end{equation}
and only depends -- as it should -- on scaled quantities. Similarly,
the Euler angles describing the orientation of the ellipse are
scaled quantities, by virtue of eq.~(\ref{scaling_aa}).

When dealing with non-dispersive wave-packets, it will be 
useful
to scale the amplitude and the frequency of the external field
with respect of the action $\hat{I}_1$ of the resonant orbit.
With the above choice of $\alpha$, eq.~(\ref{choosescale}), the scaling 
relation (\ref{scaling_law}) for $\omega$ defines the scaled frequency 
\begin{equation}
\omega_0=\omega I^3\,
\label{omega0}
\end{equation}
which turns into $\omega_0=\Omega \hat{I}_1^3$ with the resonance condition,
eq.~(\ref{resonance_condition}), and enforces
\begin{equation}
\hat{I}_1 = \omega^{-1/3},
\label{scaling_action}
\end{equation}
by virtue of eq.~(\ref{omk}).
Correspondingly, the scaled external field is defined as 
\begin{equation}
F_0=FI^4,
\label{scaling_F}
\end{equation}
which, with eq.~(\ref{scaling_action}), turns into $F_0=F\omega^{-4/3}$ at
resonance.
Hence, except for a global multiplicative factor $I^{-2},$
the Hamiltonian of a hydrogen atom in an external field
depends only on scaled quantities.

Finally, note that the {\em quantum} dynamics is not invariant
with respect to the above scaling transformations. Indeed, the Planck constant
$\hbar$ fixes an absolute scale for the various action variables. 
Thus, the spectrum
of the Floquet Hamiltonian will not be scale invariant,
while the underlying classical phase space
structure is. This latter feature 
will be used to identify in the quantum spectrum 
the remarkable features we are interested in.

\subsection{Rydberg states in linearly polarized microwave fields}
\label{LIN}

We are now ready to consider specific examples of non-dispersive
wave-packets. We consider first the simplest, one-dimensional,
driven hydrogen atom, as defined by eqs.~(\ref{h0_1d}),(\ref{v_1d}).

For a real 3D atom, this corresponds to driving the electron
 initially prepared in a one-dimensional 
eccentricity one orbit along the polarization axis of the field. 
In fact,
it turns out that the one-dimensionality of the dynamics is not
stable under the external driving: the Kepler ellipse (with
orientation fixed in configuration
space by the Runge-Lenz vector) slowly precesses off the
field polarization axis (see sections~\ref{LIN3D} and \ref{staticmw}).
Thus, the 1D presentation which follows has mostly
pedagogical value -- being
closest to the 
general case discussed later.
However, a one-dimensional model allows to grasp
essential features of the driven atomic dynamics and provides the simplest
example for the creation of non-dispersing wave-packets by a near-resonant
microwave field. 
A subsequent section will describe the dynamics of the real 3D atom under
linearly polarized driving, and amend on the flaws and
drawbacks of the one-dimensional model.

\subsubsection{One-dimensional model}
\label{LIN1D}

From eqs.~(\ref{h0_1d}),(\ref{v_1d}), the Hamiltonian of the driven 1D atom
reads
\begin{equation}
H=\frac{p^2}{2}-\frac{1}{z}+Fz\cos (\omega t),\ z>0.
\label{ham_lin1d}
\end{equation}
This has precisely the general form, eq.~(\ref{ham}), and we can therefore 
easily
derive explicit expressions for the secular Hamiltonian subject to the
semiclassical quantization conditions, eqs.~(\ref{ebkpt}),
(\ref{ebkires}),(\ref{ebkinres}), as well
as for the quantum mechanical eigenenergies, eq.~(\ref{prediction_mathieu}), in the
pendulum approximation.
With the Fourier expansion, eq.~(\ref{v_1dfou}), and
identifying $\lambda$ and $V(p,z)$ in eq.~(\ref{ham}) with $F$ and $z$ in
eq.~(\ref{ham_lin1d}), respectively, the Fourier coefficients in
eq.~(\ref{fourier_components}) take the explicit form
\begin{equation}
V_0=\frac{3}{2}I^2;\ \ \ \ V_m=-I^2\frac{J_m'(m)}{m},\ m\neq 0.
\label{coeff_lin1d}
\end{equation}
The resonant action -- which defines the position of the resonance island in 
fig.~\ref{pendulum} -- 
is given by eq.~(\ref{scaling_action}).
In a quantum 
description, 
the resonant action coincides with the 
resonant principal quantum number:
\begin{equation}
n_0=\hat{I}_1 = \omega^{-1/3},
\label{acteqqn}
\end{equation}
since the Maslov index vanishes in 1D, see section~\ref{RSEF}, and 
$\hbar\equiv 1$ in atomic units.
The resonant coupling is then given by:
\begin{equation}
 V_1=-\hat{I}^2J_1'(1),
\label{v1_LIN1D}
\end{equation}
and the secular Hamiltonian, eq.~(\ref{hsec_ap}), reads:
\begin{equation}
{\cal H}_{\mathrm sec} = \hat{P}_t - \frac{1}{2\hat{I}^2} 
-\omega \hat{I} -  J_1'(1)\hat{I}^2 F \cos\hat{\theta}.
\label{hsec_h1d}
\end{equation}
This Hamiltonian has the standard form of a secular
Hamiltonian with a resonance island centered around
\begin{equation}
\hat{I}=\hat{I}_1 = \omega^{-1/3}=n_0,\ \ \ \ \hat{\theta} = \pi,
\label{posres}
\end{equation}
sustaining librational motion  within its boundary.

Those energy values of ${\cal H}_{\mathrm sec}$ which
define contour lines (see fig.~\ref{pendulum}) 
such that the contour integrals, eqs.~(\ref{ebkires}),(\ref{ebkinres}),
lead to 
non-negative integer values of $N$, are the semiclassical quasienergies
of the 1D hydrogen atom under external driving. The non-dispersive wave-packet
eigenstate of this model atom in the electromagnetic field is represented by
the ground state $N=0$ of ${\cal H}_{\mathrm sec}$,
localized (in phase space) near the center
of the resonance island. A detailed comparison
of the semiclassical energies to the exact quantum solution of our problem
will be provided in the next subsection,
where we treat the
three-dimensional atom in the field. There it will turn out that the
spectrum of the 1D model is actually neatly 
embedded  in the 
spectrum of
the real 3D atom.

In the immediate vicinity of the resonance island, 
the secular Hamiltonian can be further simplified, leading to the
pendulum approximation, see section~\ref{CD} and eq.~(\ref{eqpend}).
The second derivative of the unperturbed Hamiltonian with respect
to the action is:
\begin{equation}
H^{''}_0=-\frac{3}{n_0^4},
\label{h0second}
\end{equation}
and the pendulum Hamiltonian reads:
\begin{equation}
{\cal H}_{\mathrm pend} = \hat{P}_t - \frac{3}{2n_0^2}
- J_1'(1)n_0^2 F \cos\hat{\theta} -\frac{3}{2n_0^4}(\hat{I}-n_0)^2.
\label{hpend_lin1d}
\end{equation}
Remember that $n_0$ is the resonant action, not necessarily an integer.
As we are interested in states deeply inside the resonance island,
we can employ the harmonic approximation around the
stable fixed point $(\hat{I}=n_0,\hat{\theta}=\pi)$, and finally
obtain the semiclassical energies of the non-dispersive
wave-packets:
\begin{equation}
{\cal E}_{N,k} = k\omega -\frac{3}{2n_0^2} + J_1'(1)n_0^2 F -
\left(N+\frac{1}{2}\right) \omega_{\mathrm harm},
\label{semiharm1dval}
\end{equation}
where, in agreement with eq.~(\ref{omega_harmonic})
\begin{equation}
\omega_{\mathrm harm}= 
\sqrt{3J_1'(1)} \frac{\sqrt{F}}{n_0} = \omega \sqrt{3J_1'(1)F_0}
\label{omega_h1d}
\end{equation}
is the classical librational frequency in the resonance island.
The quantum number $k$ reflects the global $\omega$ periodicity
of the Floquet spectrum, as a consequence of eq.~(\ref{ebkpt2}).

As already noted in section~\ref{scaling_laws}, the semiclassical quantization
breaks the scaling of the classical dynamics. Nontheless, 
the semiclassical
energy levels can be written in terms of the
scaled parameters introduced above, by virtue of
eqs.~(\ref{scaling_F},\ref{posres}):
\begin{equation}
{\cal E}_{N,k=0} = \frac{1}{n_0^2}
\left[ -\frac{3}{2} + J_1'(1) F_0 - \frac{N+1/2}{n_0} \sqrt{3J_1'(1)F_0}
\right].
\label{effective}
\end{equation}
Note that the term $(N+1/2)/n_0$ highlights 
the role of $1/n_0$ as an effective Planck constant.

As discussed in section~\ref{section_mathieu},
the fully ``quantum" quasienergies of the
resonantly driven atom can be obtained using the
very same pendulum approximation of the system, 
together with the solutions
of the Mathieu equation. In our case, the  characteristic exponent
in the Mathieu equation is given by eq.~(\ref{cexp})
and the Mathieu parameter is, according to 
eqs.~(\ref{map_mathieu3}),(\ref{v1_LIN1D}),(\ref{h0second}):
\begin{equation}
q=\frac{4}{3}Fn_0^6J'_1(1)=\frac{4}{3}F_0n_0^2J'_1(1).
\label{mathind_LIN1D}
\end{equation}
The quantum quasienergy levels are then given by eq.~(\ref{prediction_mathieu})
which reads:
\begin{equation}
{\cal E}_{\kappa,k=0} = - \frac{3}{2n_0^2} 
- \frac{3}{8n_0^4} \ a_{\kappa}(\nu,q).
\label{prediction_mathieu_lin1d}
\end{equation}

As discussed in section~\ref{QD}, the non-dispersive wave-packet
with maximum localization is associated with $\kappa=0$ and is well
localized inside the non-linear resonance island between the internal
coulombic motion and the external driving, provided the parameter $q$
is of the order of unity (below this value, the resonance island
is too small to support a localized state). The minimum scaled 
microwave amplitude is thus of the order of:
\begin{equation}
F_{0,\mathrm{trapping}}= F_{\mathrm{trapping}}n_0^4 \simeq \frac{1}{n_0^2}
\label{ftrapping}
\end{equation}
which is thus {\em much smaller} -- by a factor $n_0^2$, i.e.
three orders of magnitude in typical experiments --  
than the electric field created
by the nucleus. This illustrates that a well chosen weak perturbation
may strongly influence the dynamics of a non-linear system. From the
experimental point of view, this is good news, a limited
microwave power is sufficient to create non-dispersive wave-packets.

We have
so far given a complete description of the dynamics of the
resonantly driven, one-dimensional
Rydberg electron, from a semiclassical as well as from a quantum mechanical
point of view, in the resonant approximation. These approximate treatments
are now complemented by a numerical solution of the exact quantum mechanical
eigenvalue problem described by the Floquet equation (\ref{calhq}),
with $H_0$ and $V$ from eqs.~(\ref{h0_1d}),(\ref{v_1d}),
as well as by the numerical integration of the classical equations of motion
derived from eq.~(\ref{ham_lin1d}). Using this machinery, we 
illustrate some of the essential properties of nondispersive wave-packets
associated
with the principal resonance in this system, whereas we 
postpone the discussion of other primary resonances
to section \ref{ORH}.

\begin{figure}
\centerline{\psfig{figure=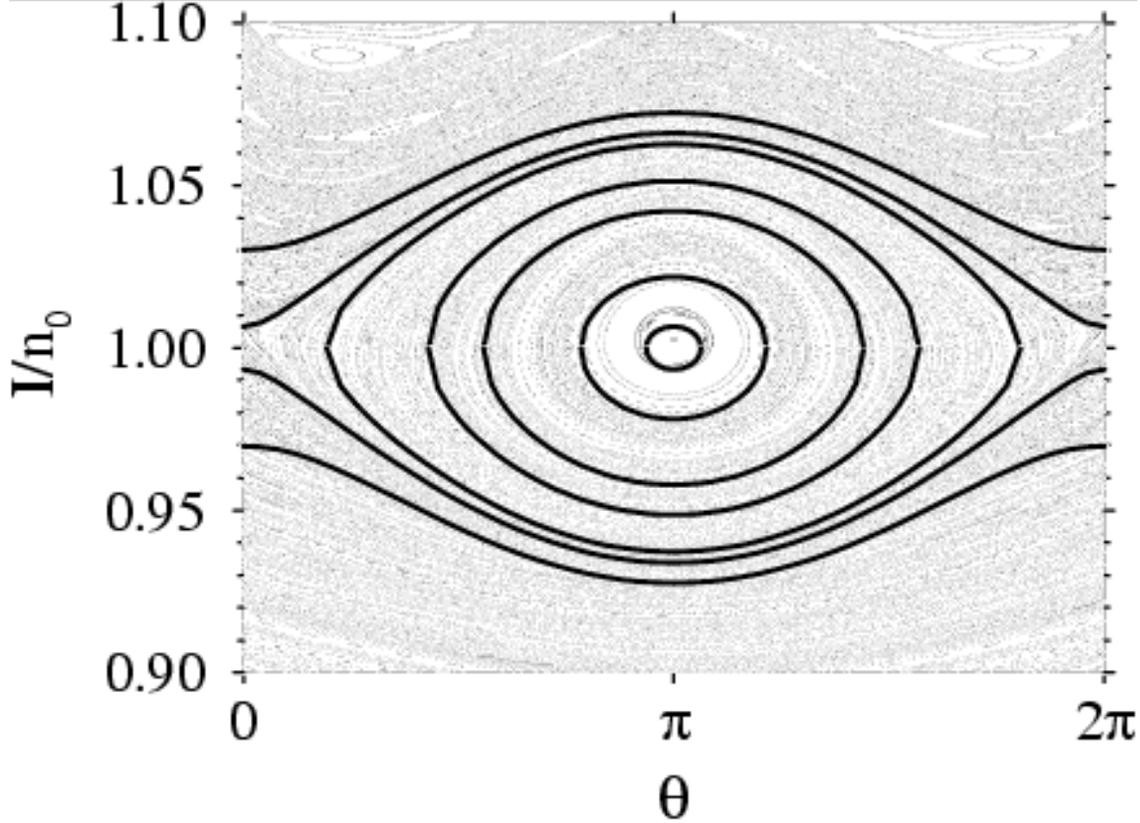,width=15cm,angle=-90}}
\caption{Poincar\'e surface of section for the dynamics of a 1D hydrogen
atom driven by an external oscillatory electric field, see eq.~(\protect\ref{ham_lin1d}).
The driving frequency is chosen as $\omega=1/60^3,$ such that the nonlinear
resonance island is centered at principal action (or effective
principal quantum number) $n_0=60.$ The scaled external field amplitude
is set to $F_0=Fn_0^4=0.01$. Although the driving field is much weaker
than the Coulomb field between the electron and the nucleus, 
it suffices to create a relatively
large resonance island which supports several non-dispersive
wave-packets.}
\label{lin1d_00}
\end{figure}

Fig.~\ref{lin1d_00} compares the phase space structure of the exact classical
dynamics generated by the Hamilton function (\ref{ham_lin1d}), and the isovalue
curves of the pendulum dynamics, eq.~(\ref{hpend_lin1d}),
for the case $n_0=60,$
at scaled field strength
$F_0=Fn_0^4=0.01$. The Poincar\'e surface of section is taken
at 
phases $\omega t=0$ (mod $2\pi$) and plotted in
$(I,\theta)$ variables which, for such times, coincide
with the $(\hat{I},\hat{\theta})$ variables, 
see eqs.~(\ref{rotframe_a}),(\ref{rotframe_b}).
Clearly, the pendulum approximation predicts the
structure of the invariant curves very well, 
with the resonance island surrounding
the stable periodic orbit at $(\hat{I}\approx 60,\hat{\theta}=\pi),$
the unstable fixed point at $(\hat{I}\approx 60,\hat{\theta}=0),$
the separatrix, and the rotational motion outside the resonance island.
Apparently, only tiny regions of stochastic motion invade the
classical phase space, which hardly affects the quality of the pendulum
approximation. It should be emphasized that -- because of the scaling laws,
see section~\ref{scaling_laws} -- the figure depends on
the scaled field strength $F_0$ only. Choosing a different microwave
frequency with the same scaled field leads, via eq.~(\ref{acteqqn}), 
to a change
of $n_0$ and, hence, of the scale of $I.$

\begin{figure}
\centerline{\psfig{figure=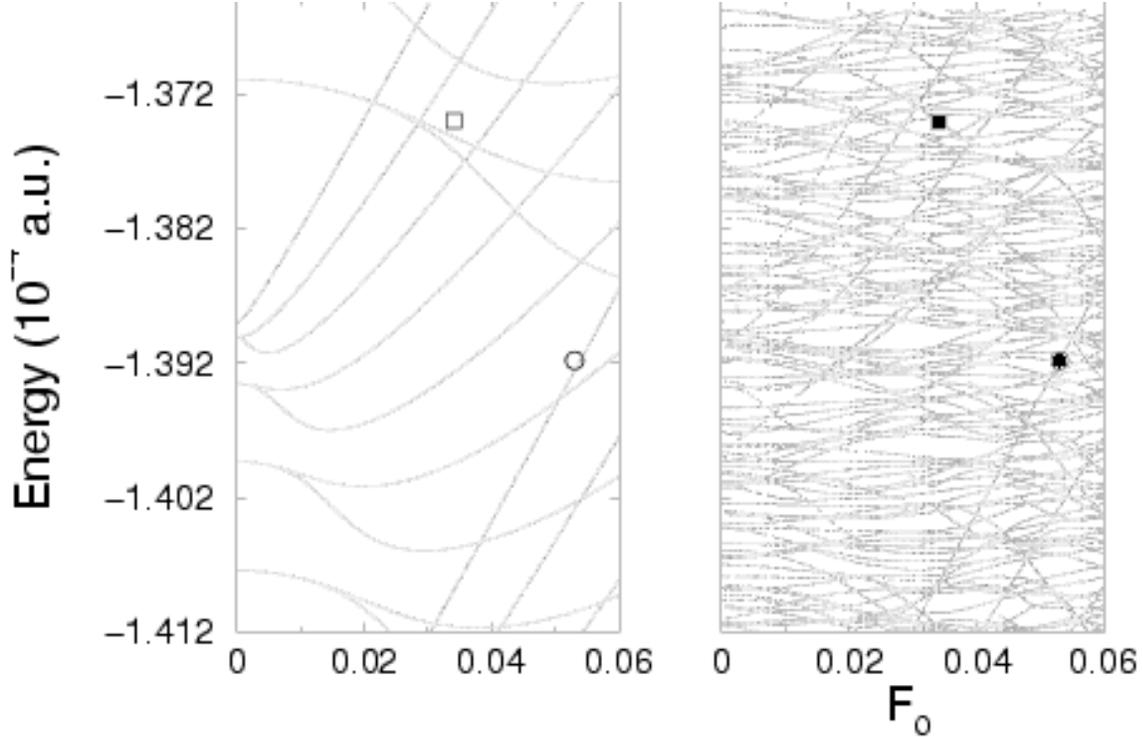,width=15cm,angle=-90}}
\caption{Comparison of the exact quasienergy spectrum of the 1D hydrogen atom
driven by a linearly polarized microwave field (right),
eq.~(\protect\ref{ham_lin1d}),
with the prediction of the pendulum approximation (left),
eq.~(\protect\ref{prediction_mathieu_lin1d}). Because of the $\hbar\omega$
periodicity of the Floquet spectrum, the energy levels
described by the pendulum approximation are folded inside one
Floquet zone. The agreement between the exact quantum result and the
pendulum approximation is very good. 
The filled circle shows the 
most localized non-dispersive wave-packet $N=0$ shown in
figs.~\protect\ref{lin1d_2}-\protect\ref{lin1d_4}, while
the filled square represents the hyperbolic non-dispersive wave-packet
partly localized in the vicinity of the unstable
equilibrium point of the pendulum, shown in
figs.~ \protect\ref{lin1d_5} and \protect\ref{lin1d_7}. 
The open circle and square compare the exact location of the
respective quasienergy values with the Mathieu prediction, which is
considerably better for the ground state as compared to the separatrix 
state.}
\label{lin1d_0}
\end{figure}

Fig.~\ref{lin1d_0} compares the prediction of the Mathieu approach,
eq.~(\ref{prediction_mathieu_lin1d}), for the quasienergy levels of the
Floquet Hamiltonian to
the exact numerical result obtained by diagonalization of the full Hamiltonian,
see section~\ref{RSEF}. Because of the $\hbar\omega$ periodicity
of the Floquet spectrum, the sets of energy levels of the pendulum,
see fig.~\ref{fig_mathieu},
are folded in one single Floquet zone.
For  states located inside or in the vicinity
of the resonance island -- the only ones plotted in 
fig.~\ref{fig_mathieu}a --
the agreement is very good for low and
moderate
field strengths.
Stronger
electromagnetic fields lead to deviations between the Mathieu and the exact
result. This indicates 
higer order corrections to the pendulum approximation.

\begin{figure}
\centerline{\psfig{figure=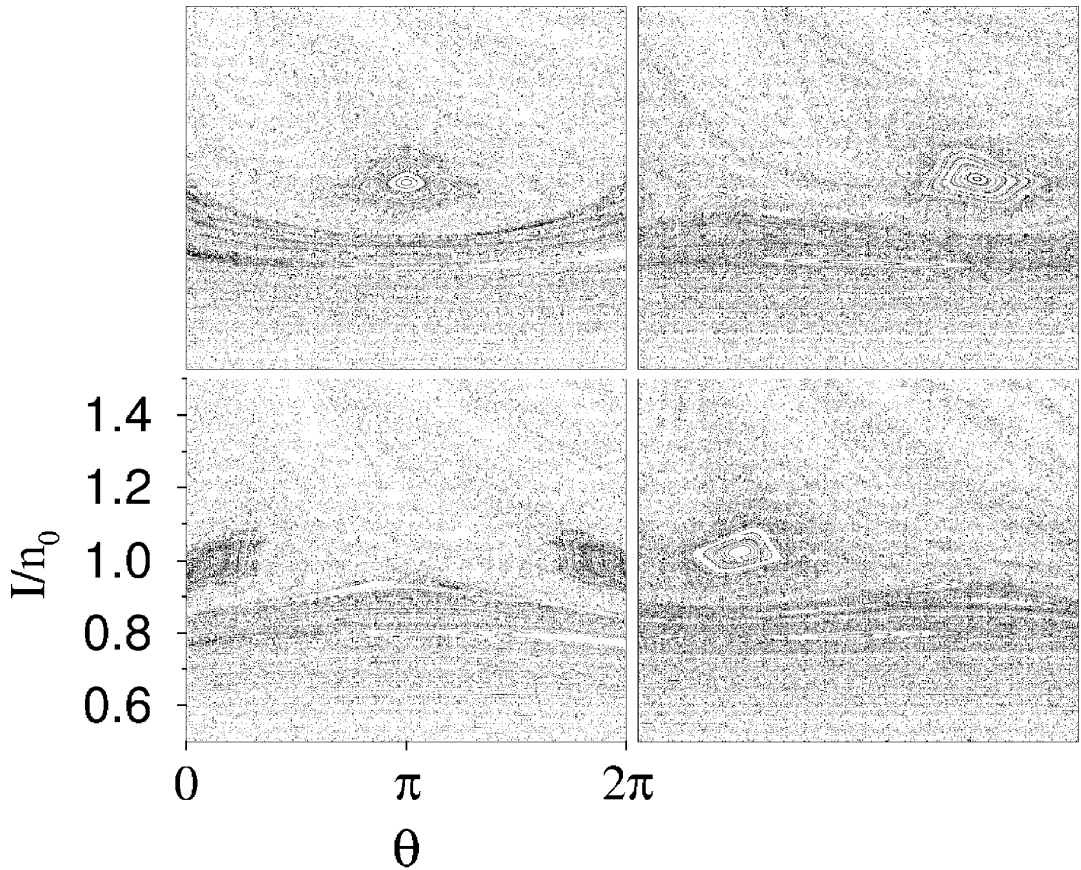,width=15cm,angle=-90}}
\caption{Surface of section of the classical phase space of a 1D hydrogen atom,
driven by a linearly polarized microwave field of amplitude $F_0=0.053$, 
for different values of
the phase: $\omega t=0$ (top left), $\omega t=\pi/2$ (top right), $\omega t=\pi$ (bottom left),
$\omega t=3\pi/2$ (bottom right). At this driving strength,
the principal
resonance remains as the
only region of regular motion of appreciable size,
in a globally chaotic phase space. The action-angle variables 
$I,\theta$ are defined by
eq.~(\protect\ref{aa_lin1d}), according to which a 
collision with
the nucleus occurs 
at $\theta=0$.}
\label{lin1d_1}
\end{figure}

Fig.~\ref{lin1d_1} shows a typical Poincar\'e surface of section of the
classical dynamics of the driven Rydberg electron, at different values
of the phase of the driving field. The field amplitude is chosen
sufficiently high to
induce largely chaotic dynamics,
with the
principal resonance as the only remnant of regular motion occupying
an appreciable volume of phase space.
The figure clearly illustrates
the temporal evolution of the elliptic island with the phase of the driving
field, i.e. the locking of the electronic motion on the external driving. The
classical stability island follows the dynamics of the unperturbed electron
along the 
resonant trajectory.
The distance from the nucleus is parametrized by the variable $\theta,$ 
see eq.~(\ref{aa_lin1d}). At $\omega t = \pi,$ the 
classical electron hits 
the nucleus (at $\theta=0$),
its velocity
diverges and changes sign discontinuously.
This explains the
distortion of the resonance island as it approaches $\theta=0,\pi$.

\begin{figure}
\centerline{\psfig{figure=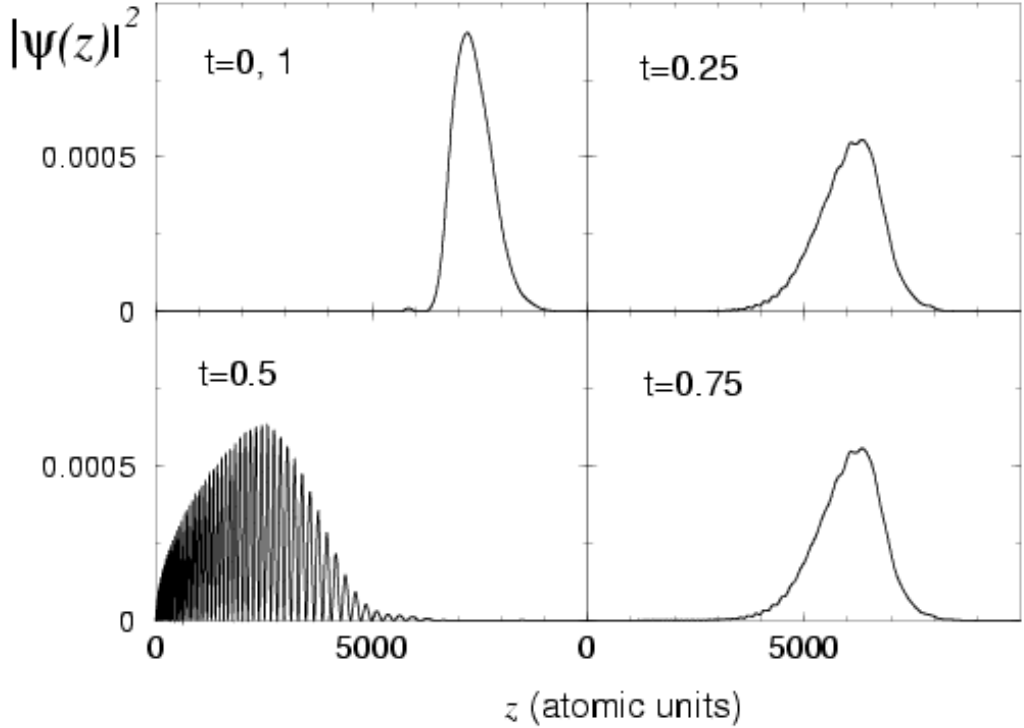,width=14cm,angle=-90}}
\caption{Configuration space representation of the electronic 
density of the non-dispersive wave-packet eigenstate
of a 1D hydrogen atom,
driven by a linearly polarized microwave field, for the same field amplitude
and phases as in fig.~\protect\ref{lin1d_1}. 
The eigenstate is 
centered on the principal resonance of classical phase space,
at action (principal quantum number) $n_0=60$.
In configuration space,
the wave-packet is localized at the outer
turning point (with zero average velocity) at time $t=0$, then propagates
towards the nucleus which it hits at $t/T=0.5$ (where $T=2\pi/\omega$ is the
microwave period). 
Afterwards, it propagates outward to the apocenter
which is reached at time $t/T=1$. After one period, the
wave-packet recovers {\em exactly} its initial shape, and
will therefore propagate along the classical trajectory 
forever, without spreading.
The wave-packet has approximately Gaussian shape(with time-dependent width) 
except at $t/T\simeq 0.5$.
At this instant, the head of the wave-packet, which 
already has been reflected off the nucleus, interferes with its tail,
producing interference fringes. 
}
\label{lin1d_2}
\end{figure}

\begin{figure}
\centerline{
\psfig{figure=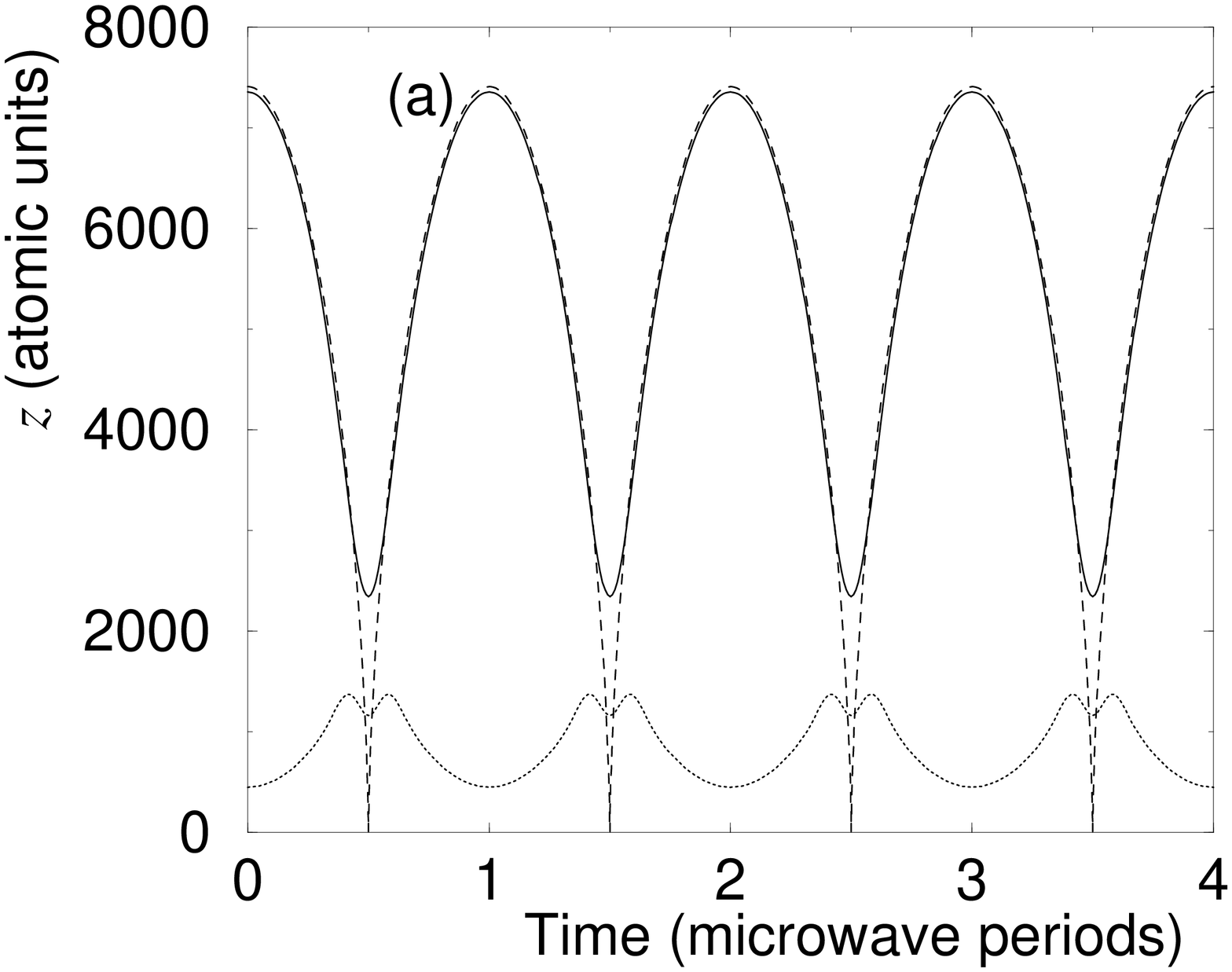,width=7cm,angle=-90}
\psfig{figure=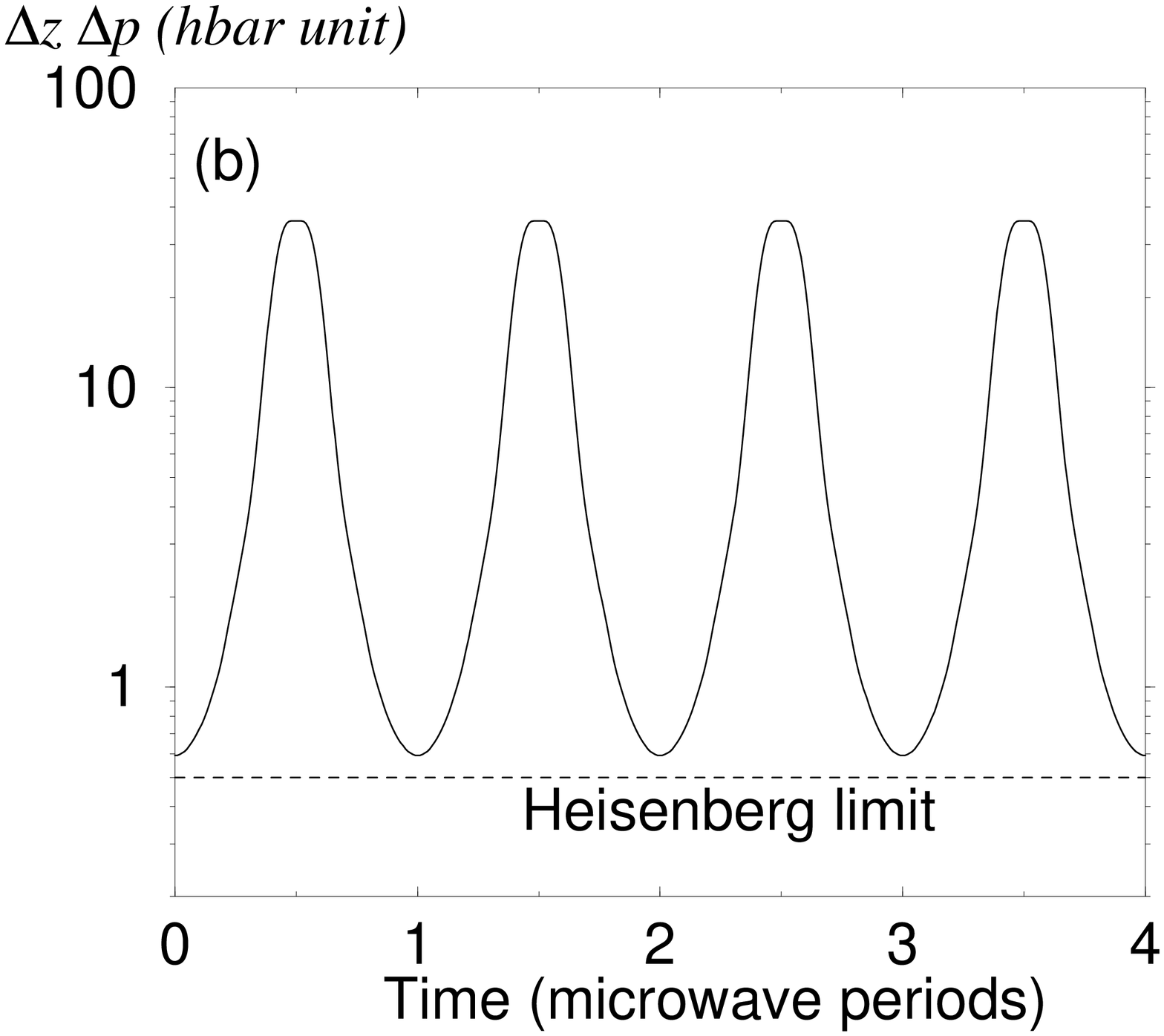,width=7cm,angle=-90}}
\caption{(a) Dashed line: 
Classical temporal evolution of the position of the Rydberg electron 
resonantly driven by a linearly polarized microwave field, in the 
one dimensional model, eq.~(\protect\ref{ham_lin1d}), of the hydrogen atom.
($n_0=60$; scaled field amplitude $F_0=Fn_0^4=0.053)$. 
Thick line: Expectation value $\langle z\rangle$ for the non-dispersive
wave-packet shown in fig.~\protect\ref{lin1d_2}, as a function of time.
It follows the classical trajectory remarkably well (except for collisions
with the nucleus).
Dotted line: The position uncertainty $\Delta z = \sqrt{\langle z^2\rangle -
\langle z \rangle ^2}$ 
of the wave-packet. 
$\Delta z$ being much smaller than
$\langle z \rangle$ (except near collisions with the nucleus)
highlights 
the efficient localization of the wave-packet.
(b) Uncertainty product $\Delta z\Delta p$ of the 
wave-packet. The periodically repeating maxima 
of this quantity indicate the collision of the electron 
with the atomic nucleus.
Note that the minimum uncertainty at
the outer turning point of the wave-packet is very close to the Heisenberg
limit $\hbar/2$. Although the wave-packet is {\em never}
a minimal one, it is nevertheless well localized and
an excellent 
approximation of a classical particle.}
\label{lin1d_3}
\end{figure}

Quantum mechanically, we expect a non-dispersive wave-packet eigenstate 
to be localized within the resonance island. 
The semiclassical prediction 
of its quasienergy, eq.~(\ref{prediction_mathieu_lin1d}),
facilitates to identify the nondispersive wave-packet within  the 
exact Floquet spectrum, after numerical diagonalization of the Floquet 
Hamiltonian (\ref{calhq}).
The wave-packet's configuration space
representation is shown 
in fig.~\ref{lin1d_2}, for the same phases of the field as
in the plots of the 
classical dynamics in fig.~\ref{lin1d_1}. Clearly, the wave-packet
is very well localized at the outer turning point of the Kepler electron
at phase $\omega t=0$
of the driving field, and is reflected off the nucleus half a period of the
driving field later. 
On reflection, the electronic density exhibits some
interference structure, as well as some transient spreading. This is a
signature of the quantum mechanical uncertainty in the angle $\theta$:
part of the wave-function, which still
approaches
the Coulomb singularity, interferes with 
the other part already reflected off the nucleus.  
The transient spreading is equally 
manifest in the temporal
evolution of the uncertainty product $\Delta z\Delta p$ itself, which is
plotted in fig.~\ref{lin1d_3}.
Apart from this singularity at $\omega t=\pi,$ the wave-packet
is approximately Gaussian at any time, with a time-dependent 
width
(compare $\omega t=0$ and
$\omega t=\pi/2$).

\begin{figure}
\centerline{\psfig{figure=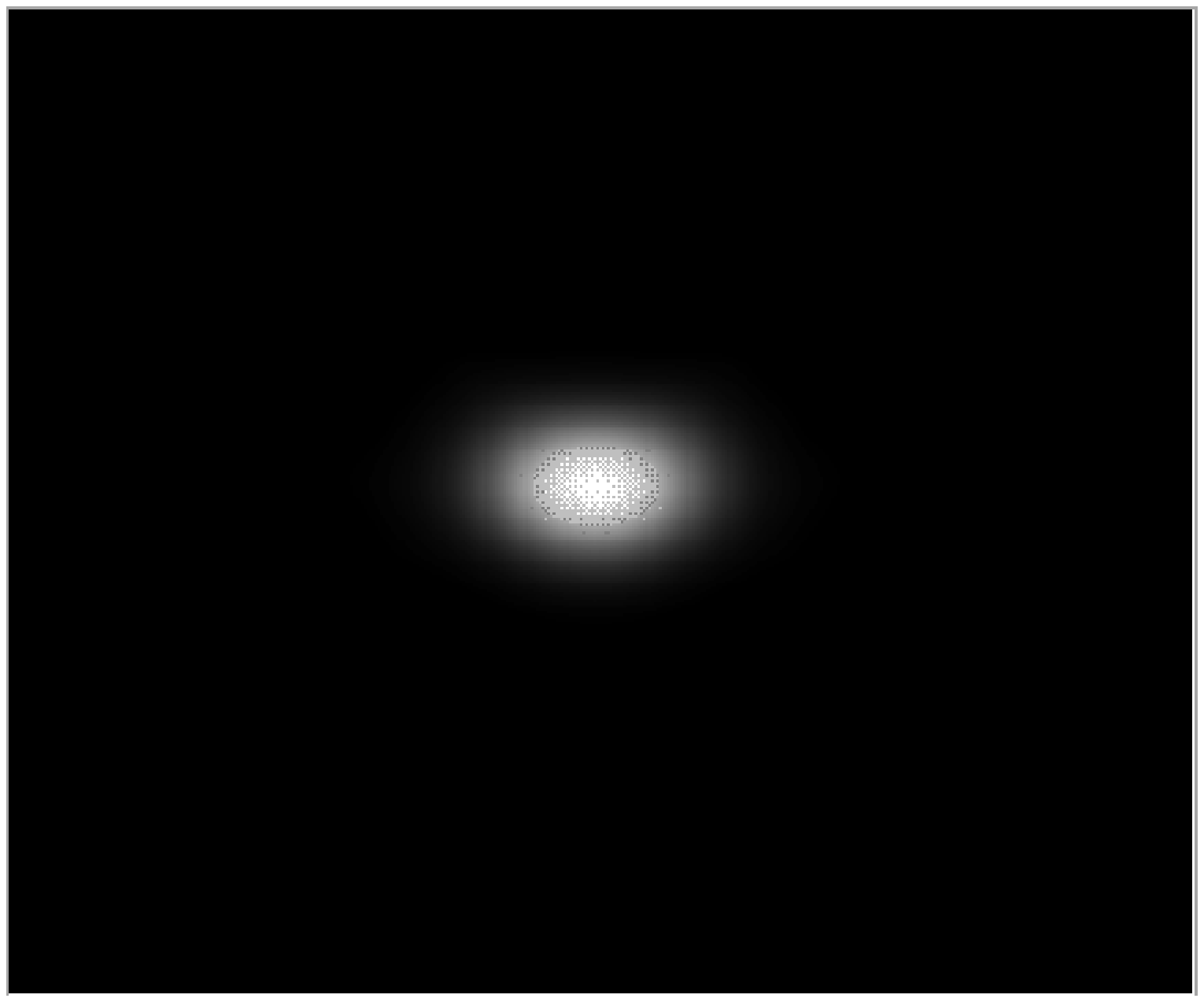,width=7cm}
\psfig{figure=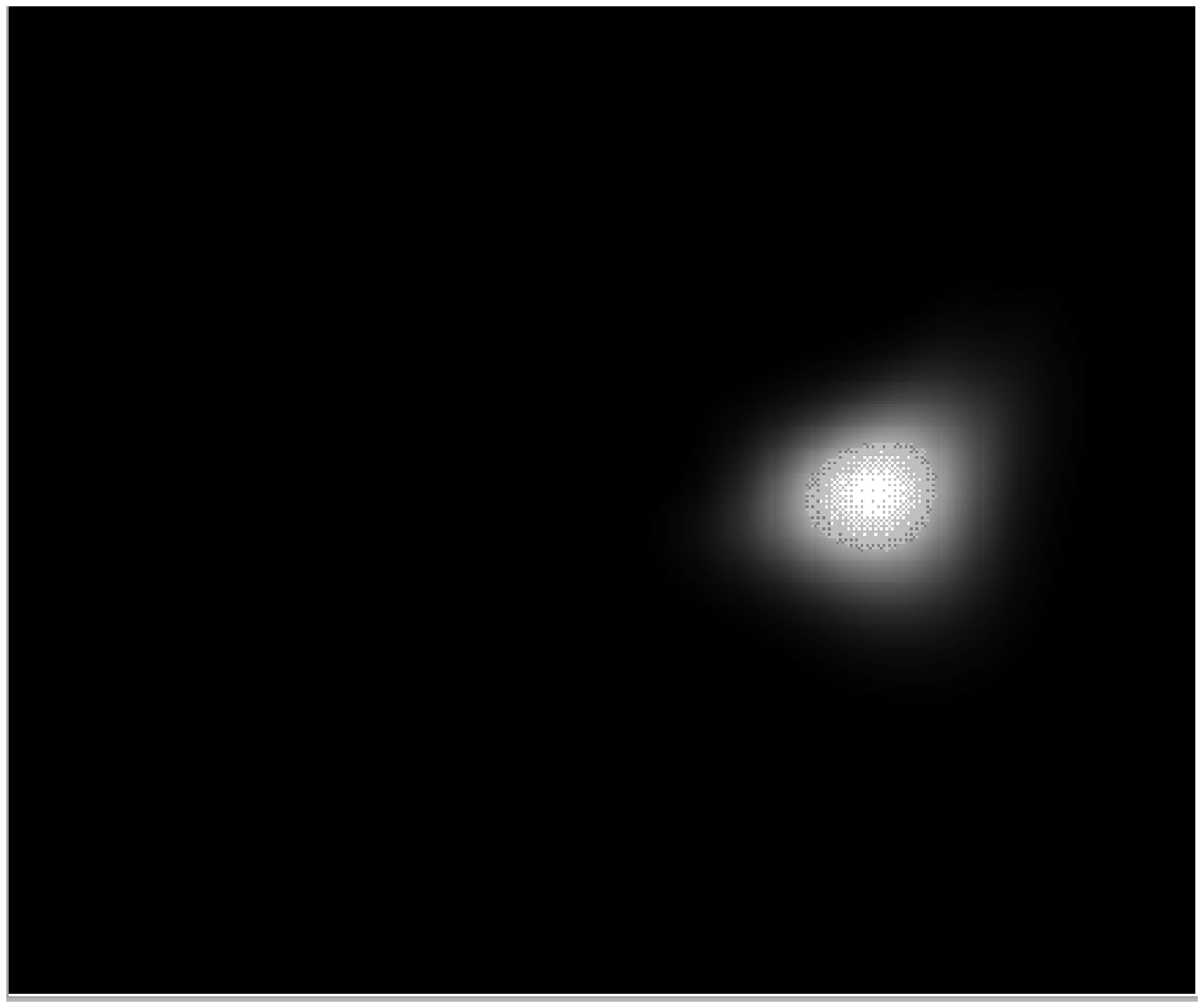,width=7cm}}
\smallskip
\centerline{\psfig{figure=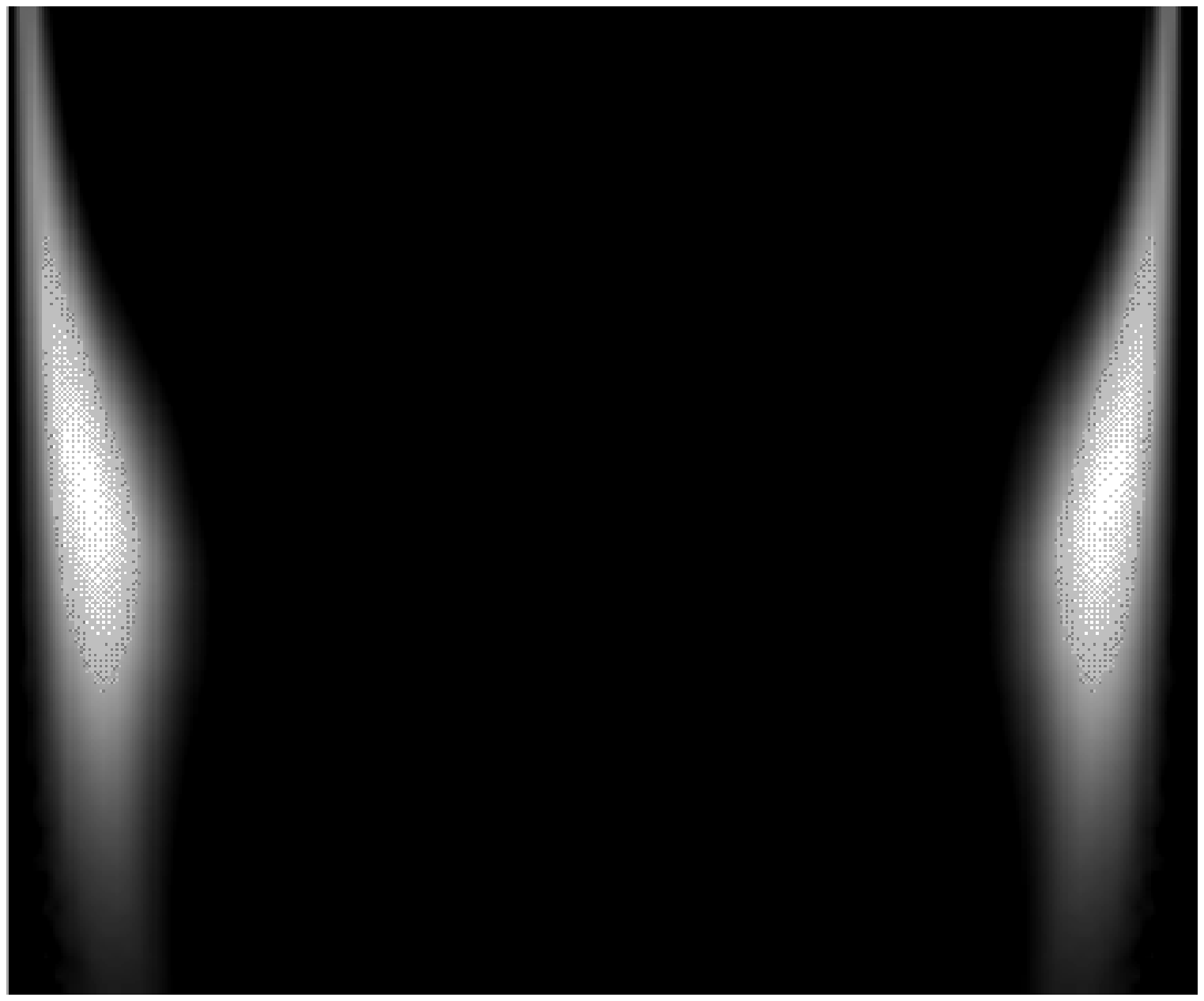,width=7cm}
\psfig{figure=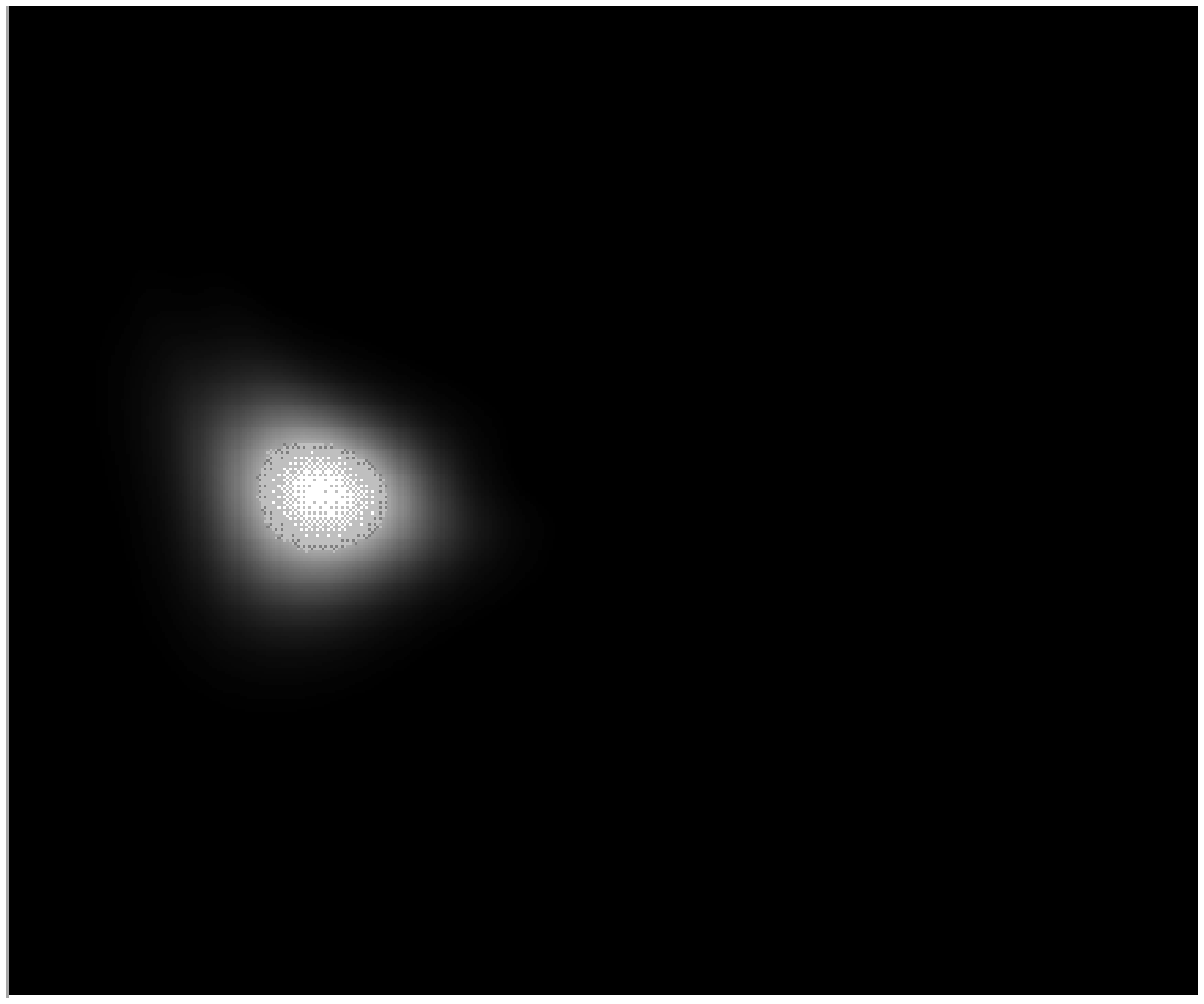,width=7cm}}
\caption{Husimi representation of the wave-packet eigenstate of
fig.~\protect\ref{lin1d_2} in classical phase space, for the same 
phases $\omega t$ and 
scales $(0\leq \theta \leq 2\pi, 30\leq I \leq 90)$ as employed for the
classical surface of section in fig.~\protect\ref{lin1d_1}. Clearly, the
quantum mechanical eigenstate of the atom in the field follows the classical
evolution without dispersion, except for its transient spreading when 
reflected off 
the nucleus (at $\omega t=\pi$, bottom left), due to the divergence
of the classical velocity at that position.
}
\label{lin1d_4}
\end{figure}

To complete the analogy between classical and quantum motion, we finally
calculate the Husimi distribution -- the
phase space representation of the wave-packet eigenstate
defined in section~\ref{coherent_states}, eq.~(\ref{husimi_def})  --
in order to obtain a direct comparison between classical and
quantum dynamics in phase space. Fig.~\ref{lin1d_4} shows the resulting 
phase
space 
picture, 
again
for different phases of the driving field. The association of the quantum
mechanical time evolution with the classical resonance island (see fig.~\ref{lin1d_1}) 
is
unambiguous. The transient spreading at the collision with 
the nucleus ($\omega t=\pi$) is due to the divergence
of the classical velocity 
upon reflection.

As discussed in section~\ref{CD} and visible in fig.~\ref{lin1d_00},
there is an hyperbolic fixed point (i.e. an unstable equilibrium point)
at $(\hat{I}=n_0,\hat{\theta}=0)$: it corresponds to the
unstable equilibrium position of the pendulum 
 when it points ``upwards". For the driven system,
it corresponds to an unstable periodic orbit resonant with the
driving frequency: it is somewhat similar to the stable
orbit supporting the non-dispersive wave-packets, except that is is
shifted in time by half a period. 

\begin{figure}
\centerline{\psfig{figure=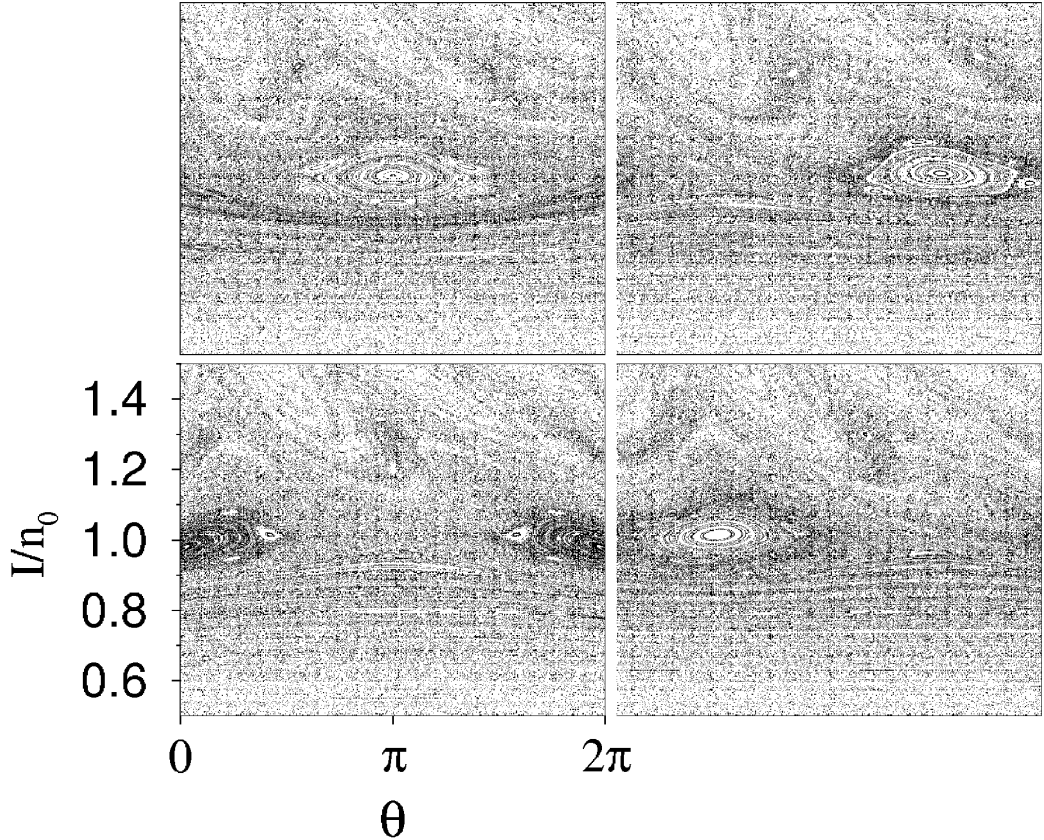,width=14cm,angle=-90}}
\caption{Surface of section of the classical phase space of a 1D hydrogen atom
driven by a linearly polarized microwave field, for different values of
the phase: $\omega t=0$ (top left), $\omega t=\pi/2$ (top right), $\omega t=\pi$ (bottom left),
$\omega t=3\pi/2$ (bottom right). The action angle variables $I,\theta$ are defined by
eq.~(\protect\ref{aa_lin1d}). At this value of the field amplitude,
$F_0=0.034$, the principal resonance island 
(and a small remnant of the $s=2$ resonance island) remain as the
only regions of regular motion, in a globally chaotic phase space.}
\label{lin1d_4a}
\end{figure}

\begin{figure}
\centerline{\psfig{figure=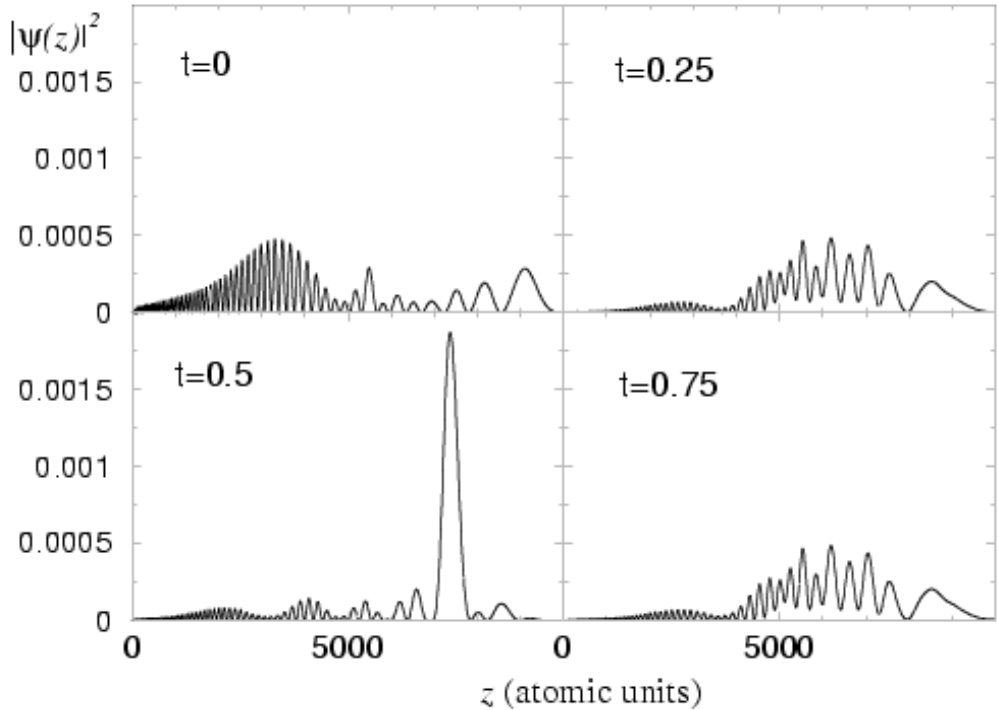,width=14cm,angle=-90}}
\caption{Wave-packet eigenstate anchored to the hyperbolic fixed point of the
principal resonance of the 1D hydrogen atom driven by a linearly polarized 
microwave field,
in configuration space,
for the same phases
of the driving field as in fig.~\protect\ref{lin1d_4a}. 
The wave-function is partly localized, especially close to
the outer turning point at $t=0.5\times 2\pi/\omega$, but the localization
is far from being perfect.
Comparison to figs.~\protect\ref{lin1d_1}, \protect\ref{lin1d_2} and
\protect\ref{lin1d_4a} shows
that the state evolves in phase opposition with the stable, non-dispersive
wave-packet, with significantly worse localization properties.}
\label{lin1d_5}
\end{figure}

\begin{figure}
\centerline{\psfig{figure=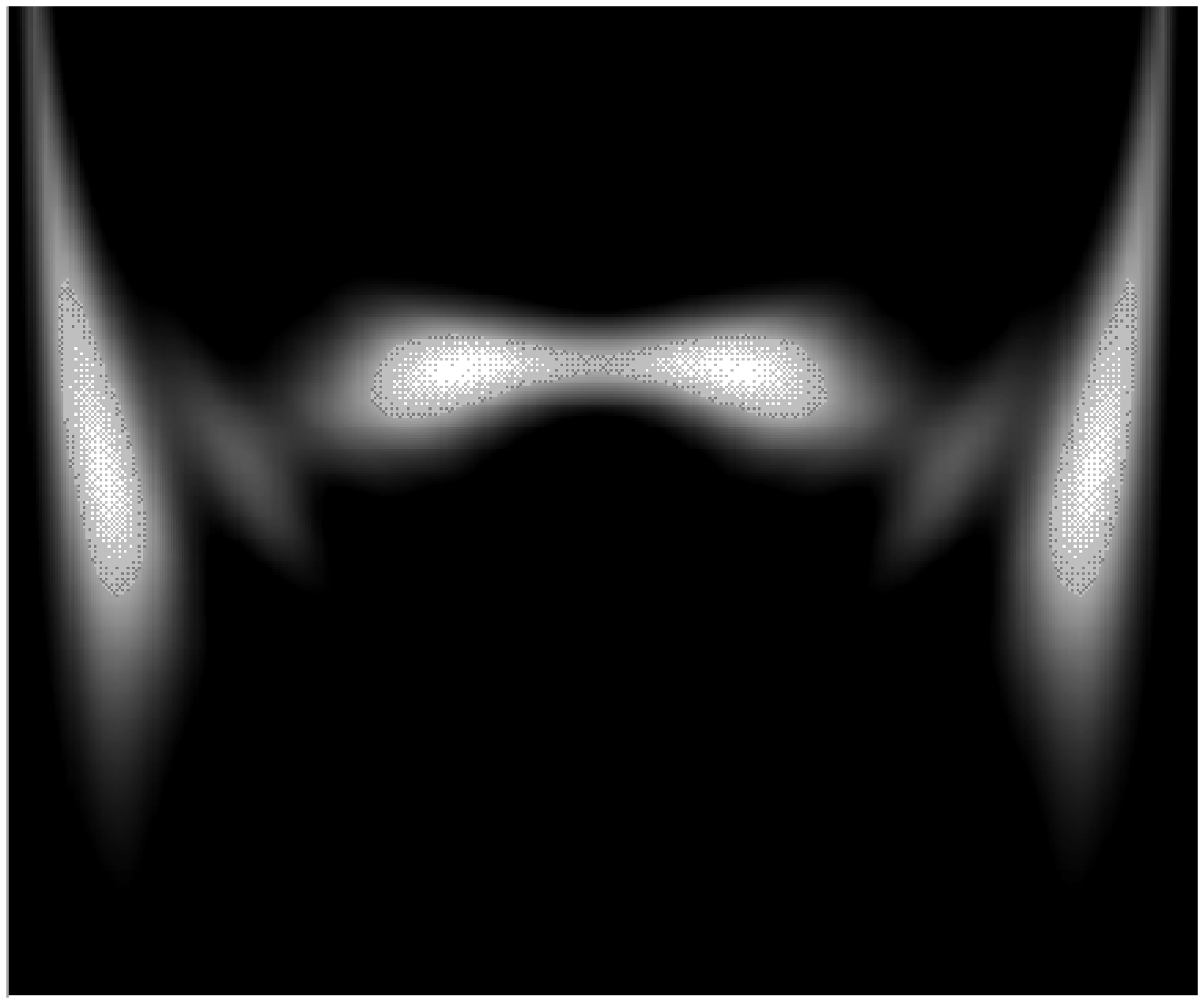,width=7cm}
\psfig{figure=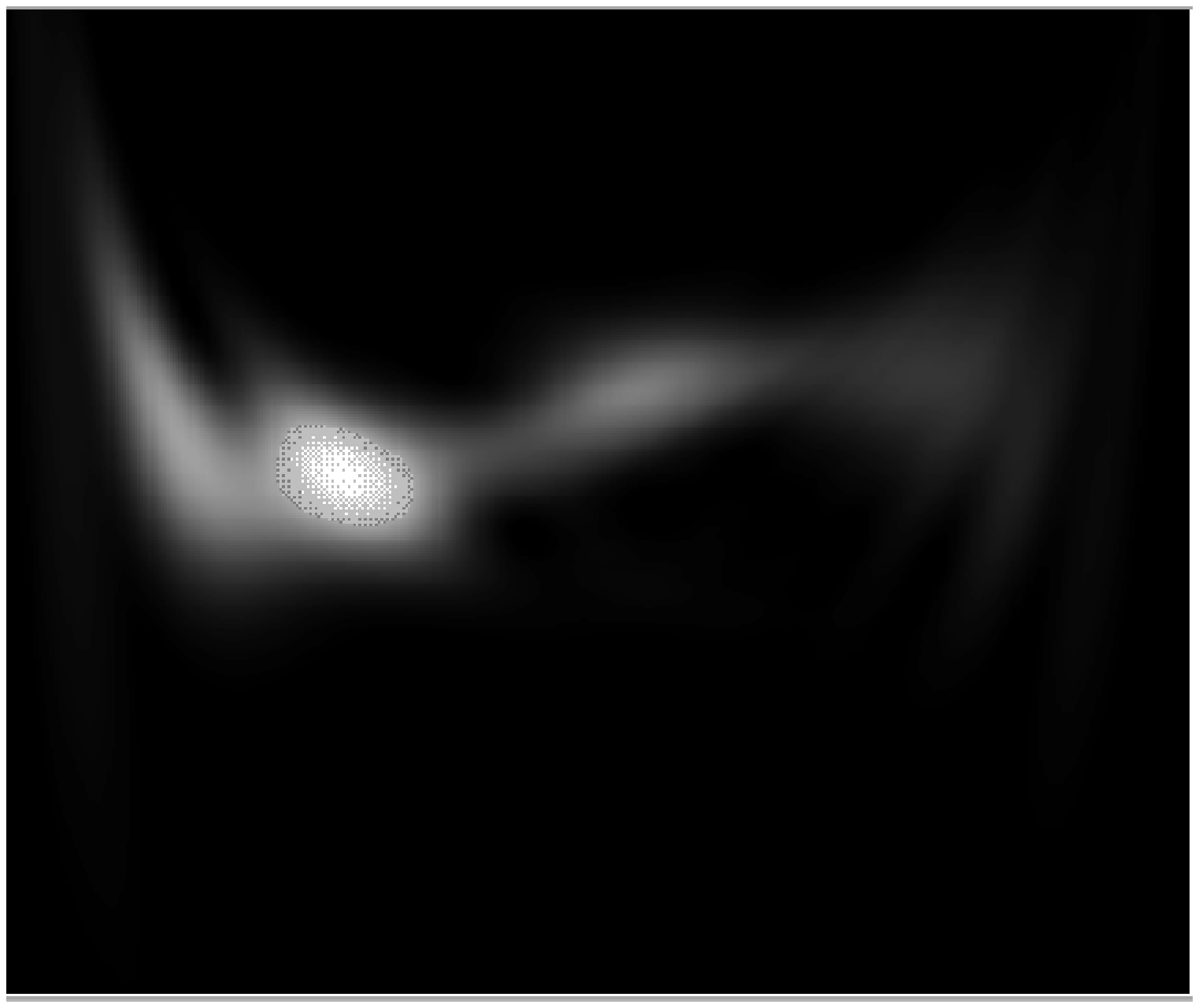,width=7cm}}
\smallskip
\centerline{\psfig{figure=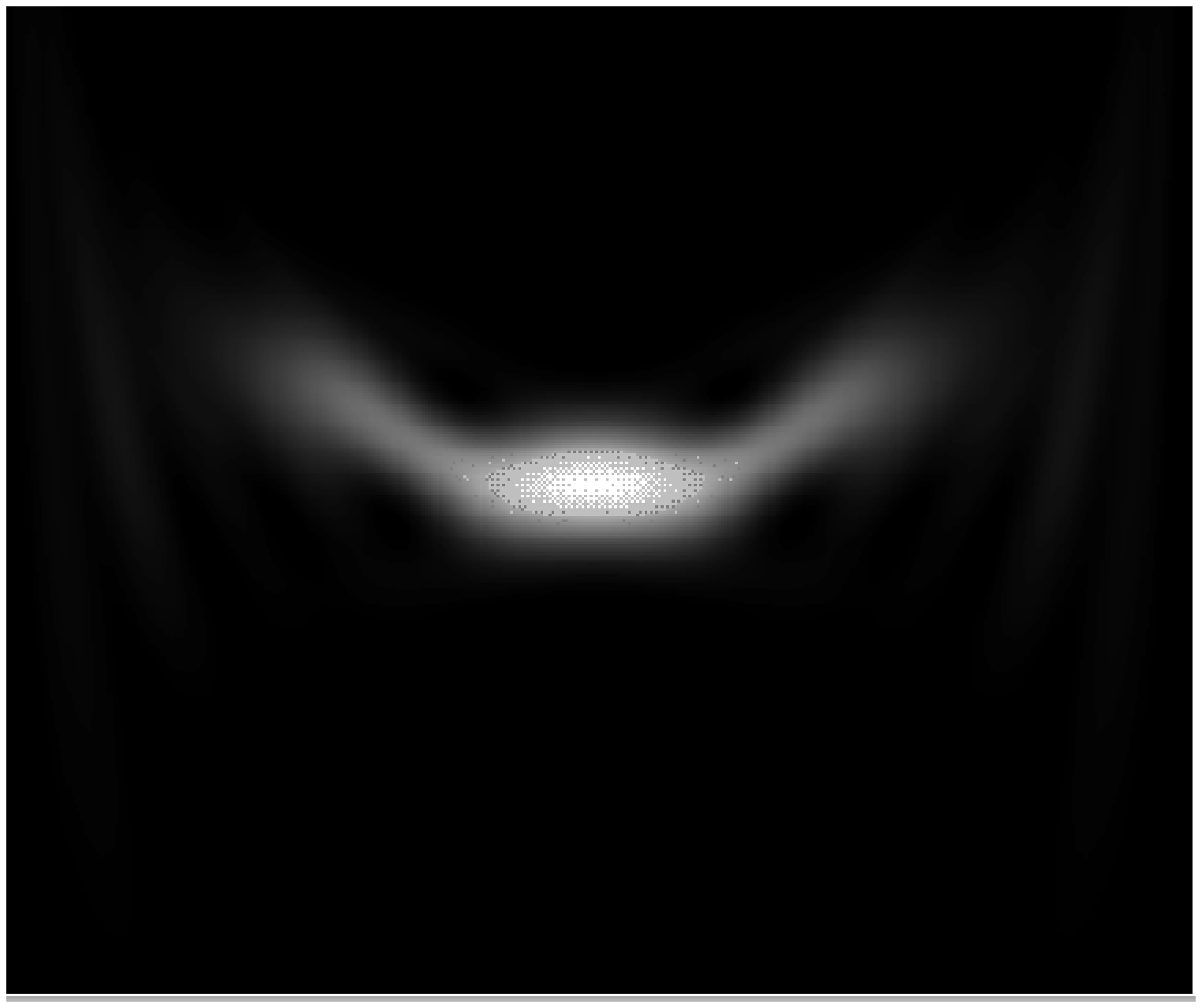,width=7cm}
\psfig{figure=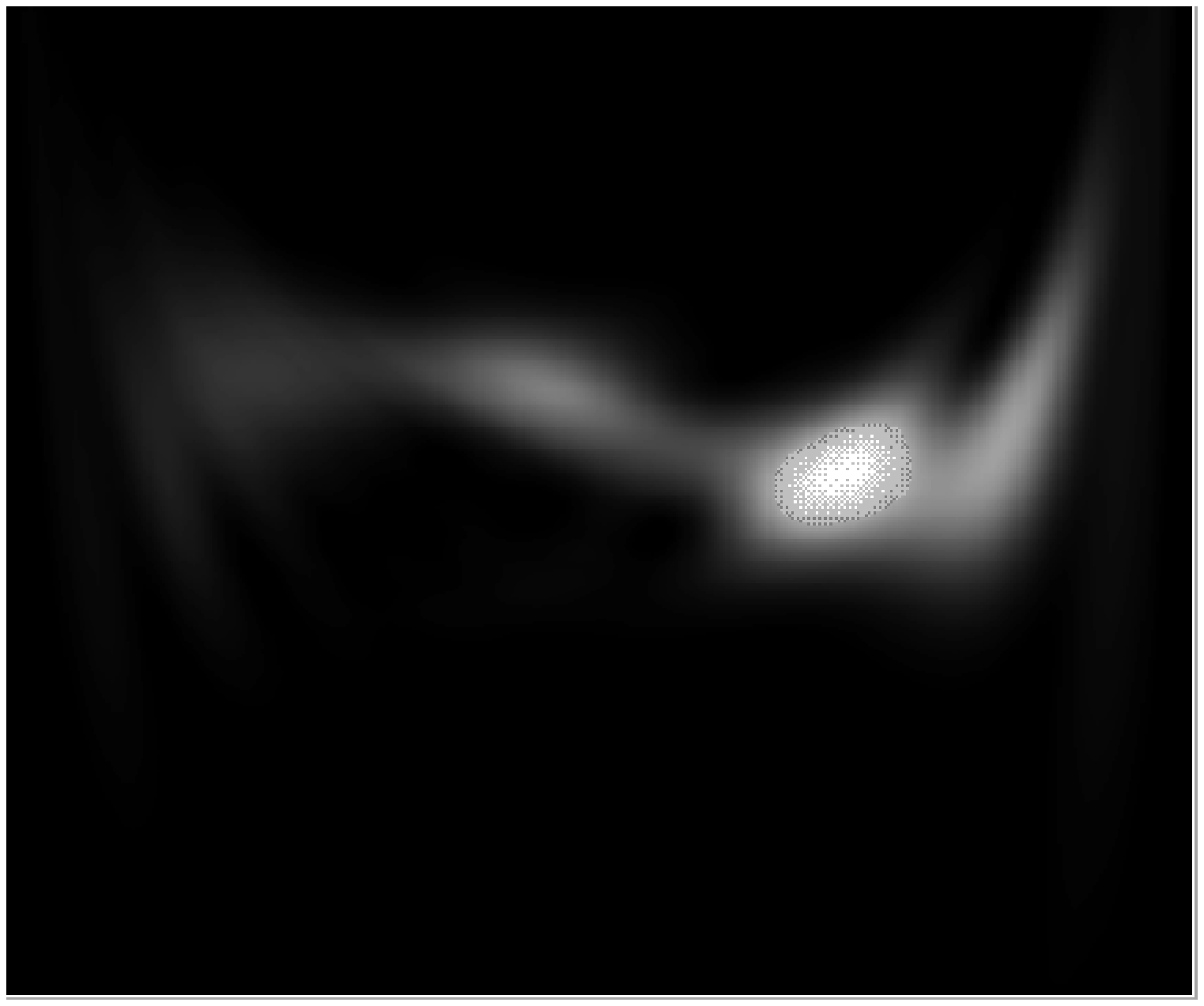,width=7cm}}
\caption{Husimi
representation of the ``hyperbolic" wave-packet eigenstate of
fig.~\protect\ref{lin1d_5}, for the same phases 
(scales as in fig.~\protect\ref{lin1d_4a}).
Clearly, the
quantum mechanical eigenstate of the atom in the field follows the classical
evolution. It is partly
localized on top of the hyperbolic fixed point, but also
spreads along the separatrix confining the principal
resonance.
The localization is more visible at $t=0$ (top left),
the spreading more visible  at
$t=0.5 \times 2\pi/\omega$ (bottom left).}
\label{lin1d_7}
\end{figure}

As discussed in section~\ref{scars}, the classical motion slows down 
at the hyperbolic fixed point
(the time to reach the unstable equilibrium point with zero velocity
diverges~\cite{marion}), and the eigenfunction must 
exhibit a maximum of the electronic density at this
position. 
In addition, due to the periodicity
of the drive, the corresponding (``hyperbolic'') wave-packet eigenstate
necessarily follows the dynamics of a classical particle 
which evolves along the unstable periodic orbit.
However, because the orbit is unstable, the quantum 
eigenstate cannot remain 
fully localized -- some probability
has to flow away along 
the unstable manifold of the classical flow in the vicinity of the 
hyperbolic fixed point.
Consequently,
such an eigenstate is partially localized along 
the separatrix between librational
and rotational motion.
For an illustration, first consider fig.~\ref{lin1d_4a}, which 
shows classical surfaces
of section of the driven (1D) hydrogen atom, at $F_0=0.034$, 
again for different phases $\omega t$.
Comparison with fig.~\ref{lin1d_1} shows a larger 
elliptic island at this slightly lower field amplitude, 
as well as remnants of the $s=2$ resonance island 
at slightly larger
actions $I/n_0\simeq 1.2\ldots 1.3$. 
The time evolution of the electronic density of the
eigenstate localized near the
hyperbolic fixed point is displayed in fig.~\ref{lin1d_5},
for different phases of the driving field.
Clearly, as compared to fig.~\ref{lin1d_2}, the wave-packet moves in phase
opposition to the driving field, and displays slightly irregular localization
properties.  Accordingly, the Husimi representation in
fig.~\ref{lin1d_7}
exhibits reasonnably good localization on top of the hyperbolic point at phase
$\omega t=\pi$, but the electronic probability spreads significantly
along the
separatrix layer at phase, as visible at $\omega t=0$.

In the above discussion of the localization properties of the wave-packet 
eigenstate we represented the wave-function in the $I$-$\theta$ phase space
of {\em classically bounded} motion (i.e., classical motion with negative 
energy). However, as we shall see in more detail in section~\ref{ION}, the 
microwave driving actually induces a nonvanishing overlap of {\em all} 
Floquet eigenstates \cite{graffi85,yajima82}, and, hence, of the wave-packet
eigenstates, with the atomic continuum. It suffices to say here that 
the associated finite decay rates induce finite life times of approx. 
$10^6$ unperturbed Kepler orbits for the quantum objects considered in this 
chapter, and are therefore irrelevant on the present level of our discussion.
In figs.~\ref{lin1d_2}, \ref{lin1d_4}, \ref{lin1d_5}, and \ref{lin1d_7}, a
finite decay rate would manifest as a slow reduction of the electronic 
density, without affecting its shape or localization properties, after 
$10^6$ classical Kepler periods.

\subsubsection{Realistic three-dimensional atom}
\label{LIN3D}

Extending our previous analysis to the three-dimensional hydrogen atom driven
by a linearly polarized microwave field, we 
essentially expand the accessible 
phase space. Since the Hamiltonian 
\begin{equation}
H_{\rm LP}=\frac{{\vec p}^2}{2}-\frac{1}{r}+Fz\cos(\omega t)
\label{ham_lin_3d}
\end{equation}
is invariant under rotations around the field polarization axis,  
the projection of the angular momentum is a conserved quantity and gives
rise to a good quantum number $M$. 
Hence, only two dimensions of configuration space are left, which, together
with the explicit, periodic time dependence, span a five-dimensional 
phase space. 

In the 1D situation described previously, the key ingredient for the existence
of non-dispersive wave-packets was the phase locking of the internal
degree of freedom on the external drive. In the 3D situation, there remains
one single drive, but there are 
several internal degrees of freedom. In the generic
case, not all internal degrees of freedom can be simultaneously locked
on the external drive, and one can expect only partial
phase locking, i.e. only partially localized wave-packets. The non-trivial
task is to understand how the phase locking of one degree of freedom
modifies the dynamics along the other degrees of freedom. In atomic systems,
the Coulomb degeneracy makes it possible to gain a full understanding
of this phenomenon. 

The starting point is similar to the 1D analysis in
sec.~\ref{LIN1D}, that is the expression of the Floquet
Hamiltonian  -- whose eigenstates are of interest --
as a function of action-angle coordinates
$(I,\theta)$, $(L,\psi)$,
$(M,\phi)$ introduced in section \ref{RSEF}.
Using eqs.~(\ref{beta},\ref{rot_euler})
and the Fourier expansion, eq.~(\ref{v_3d_x}-\ref{v_3d_z}),
of the position operator, one obtains:
\begin{equation}
{\mathcal H} = P_t -\frac{1}{2I^2}+ F \sqrt{1-\frac{M^2}{L^2}} 
\sum_{m=-\infty}^{+\infty}{\left[-X_m\cos\psi  \cos(m\theta-\omega t)
+  Y_m \sin\psi\sin(m\theta-\omega t)\right]},
\label{hamaa}
\end{equation}
with
\begin{eqnarray}
X_m(I)&=& I^2\frac{J_m'(me)}{m},\ \ \ \ m\neq 0, \label{xm} \\
Y_m(I)&=& I^2 \frac{\sqrt{1-e^2}J_m(me)}{me},\ \ \ \ m\neq 0,  \label{ym}\\ 
X_0(I)&=& -\frac{3}{2}eI^2, \label{x0}\\
Y_0(I)&=&0.
\label{y0}
\end{eqnarray}
where $e=\sqrt{1-L^2/I^2}$ is, as before, the eccentricity of the Kepler orbit
(see 
eq.~(\ref{eccentricity})).
The absence of $\varphi$ in the 
Hamiltonian reflects the azimuthal symmetry around the field axis and 
ensures the conservation of $M$.

Precisely as in the treatment of the 
one-dimensional 
problem, we now transform 
to slowly varying variables, given by 
eqs.~(\ref{rotframe_a})-(\ref{rotframe_c}): 
\begin{eqnarray}
\hat{\mathcal H} & = & \hat{P}_t -\frac{1}{2 \hat I^2}- \omega\hat I +
F \sqrt{1-\frac{M^2}{L^2}} \nonumber \\
\times & & \sum_{m=-\infty}^{+\infty} {\left[-X_m \cos\psi  
\cos(m\hat{\theta} + (m-1)\omega t)
+ Y_m \sin\psi \sin(m\hat{\theta} + (m-1)\omega t)\right]}.
\label{hamab}
\end{eqnarray}
Averaging over the fast variable $t$ (over the driving field period 
$T$) gives
the secular Hamiltonian of the three-dimensional problem
\begin{equation}
{\mathcal H}_{\rm sec}  = \hat{P}_t -\frac{1}{2\hat I^2}- 
\omega\hat{I}
 +F \sqrt{1-\frac{M^2}{L^2}}(-X_1(\hat{I})\cos\psi \cos\hat\theta+
Y_1(\hat{I})\sin\psi \sin\hat\theta).
\label{hamsc2}
\end{equation}
Its physical interpretation is rather simple: the $X_1$ term represents
the oscillating dipole (resonant with the frequency of the drive) 
along the major
axis of the classical Kepler ellipse, while the $Y_1$ term represents
the oscillating dipole along the minor axis. As these 
two components of the oscillating dipole are in quadrature, they interact
with two orthogonal components of the external drive, hence
the $\cos\hat\theta$ and $\sin\hat\theta$ terms.
Finally, both components can be combined to produce the compact form
\begin{equation}
{\mathcal H}_{\rm sec}=\hat{P}_t-\frac{1}{2\hat I^2}-\omega \hat{I}
+F \chi_1 \cos(\hat \theta + \delta_1),
\label{hamscfin}
\end{equation}
with
\begin{eqnarray}
\chi_1(\hat I,L,\psi) & := & \sqrt{1-\frac{M^2}{L^2}}\ \sqrt{X_1^2\cos^2\psi+Y_1^2\sin^2\psi} ,
\label{substc} \\
\tan\delta_1 (L,\psi)& := & \frac{Y_1}{X_1}\tan\psi = 
\frac{J_1(e)\sqrt{1-e^2}}{J'_1(e)e} \tan\psi.
\label{substb}
\end{eqnarray}
In this form, the secular Hamiltonian has the same structure as
the general 1D expression, eq.~(\ref{hsec_ap}), and its specialized
version for the 1D hydrogen atom, eq.~(\ref{hsec_h1d}).
The difference is that the additional action angle-variables
$(L,\psi)$, $(M,\phi)$ {\em only} enter in the amplitude and phase of 
the coupling defining the resonance island. 
This allows to separate various time scales 
in the system:
\begin{itemize}
\item The shortest time scale is associated with the Kepler
motion, which is also the period of the external drive.
In the resonant approximation discussed in detail
in section~\ref{CD}, this time scale is eliminated by passing to the
rotating frame. 
\item The time scale of the secular (or pendulum) motion
in the $(\hat{I},\hat{\theta})$ plane is significantly longer.
It is the inverse of the classical pendulum frequency,
eq.~(\ref{omega_harmonic}), of the order of
$1/\sqrt{F_0}$ Kepler periods. In the regime of weak external
driving we are interested in, $F_0 \ll 1$, it is thus much longer
than the preceding time scale.
\item The time scale of the ``transverse" (or angular) motion along the
$(L,\psi)$, $(M,\phi)$ variables. Because these 
are constant for the unperturbed Coulomb system, the time
derivatives like $dL/dt$ and 
$d\psi/dt$ generated by eqs.~(\ref{hamscfin})-(\ref{substb}) 
are proportional 
to $F$, and the resulting time scale is proportional
to $1/F$. More precisely, it is of the order
of $1/F_0$ Kepler periods, i.e., once again, significantly longer
than the preceding time scale.
\end{itemize}

From this separation of time scales, it 
follows that we can use the following, additional secular approximation:
for the motion in the $(\hat{I},\hat{\theta})$ plane, 
$\chi_1$ and $\delta_1$ are adiabatic invariants,
which can be considered as constant quantities. We then exactly
recover the Hamiltonian discussed for the 1D model of the atom,
with 
a resonance island confining trajectories
with librational motion in the $(\hat{I},\hat{\theta})$ plane, and
rotational motion outside the resonance island.
The center of the 
island is located at:
\begin{equation}
\hat{I}=\hat{I}_1 = \omega^{-1/3}=n_0,\ \ \ \ \ \ 
\hat{\theta} = -\delta_1.
\label{Kures}
\end{equation}
As already pointed out in section \ref{CD}, 
the 
size of the resonance island in phase space is determined
by the strength of the resonant coupling $\chi_1(\hat I,L,\psi).$
In the pendulum approximation, its extension in $\hat{I}$ 
scales as $\sqrt{\chi_1(\hat{I}_1,L,\psi)}$, i.e. with
$\chi_1$ evaluated at the center, eq.~(\ref{Kures}), of the island.

The last step is to consider the slow motion in the $(L,\psi)$ plane.
As usual, when a secular
approximation is employed, the slow motion
is due to an effective Hamiltonian which is 
obtained by averaging of the secular Hamiltonian over the fast motion.
Because the coupling $\chi_1(\hat{I},L,\psi)$ exhibits a simple scaling 
with $\hat{I}$ (apart from a global $\hat{I}^2$ dependence, it depends
on the scaled angular variables $L_0$ and $\psi$ only),
the averaging over the fast motion results in an effective
Hamiltonian for the ($L_0$,$\psi$) motion which depends
on $\chi_1(\hat{I},L,\psi)$ only.  
We deduce
that the slow motion 
follows curves of constant
 $\chi_1(\hat{I}_1,L,\psi),$ at a velocity which
depends on the average over the fast variables.
$\chi_1(\hat{I}_1,L,\psi)$ is thus a constant of motion, both for the fast
$(\hat{I},\hat{\theta})$  and the slow $(L,\psi)$ motion.  
This also implies that the order of the 
quantizations in the fast and slow variables
can be interchanged: using the dependence of $\chi_1(\hat{I}_1,L,\psi)$ 
on $(L,\psi),$ we obtain quantized values of $\chi_1$
which in turn can be used as constant values to quantize the
$(\hat{I},\hat{\theta})$ fast motion. 
Note that the separation of time scales is here essential
\footnote{If one considers non-hydrogenic atoms -- with
a core potential in addition to the Coulomb potential -- the classical
unperturbed ellipse precesses, adding an additional time scale,
and the separation of time scales is much less obvious. See also
section~\protect{\ref{SPEC}}.}.
Finally,
the dynamics in $(M,\phi)$ is trivial, since $M$ is a constant
of motion. In the following, we will consider 
the case $M=0$ for simplity. 
Note that, when the eccentricity of the classical
ellipse tends to $1$ -- i.e. $L\to 0$ -- and when
$\psi \to 0,$ the Hamiltonian (\ref{hamsc2}) coincides
exactly with the Hamiltonian of the 1D atom, eq.~(\ref{hsec_h1d}).
This is to be expected, as it corresponds to 
a degenerate classical Kepler ellipse along the $z$ axis.

The adiabatic separation of the 
radial and of the angular motion allows the separate WKB quantization of
the various degrees of freedom. In addition to the quantization conditions
in $(\hat{P}_t,t)$ and $(\hat{I},\hat{\theta})$, 
eqs.~(\ref{ebkpt}-\ref{ebkinres}),
already formulated in our general 
description of the semiclassical approach in 
section~\ref{sapp},
we
additionally need to quantize the angular motion, according to:
\begin{equation}
\frac{1}{2\pi}\oint_\gamma L d\psi =  \left(p+\frac{1}{2}\right)\hbar,
\label{ebkp}
\end{equation}
along a loop $\gamma$ of constant
$\chi_1$ in the $(L,\psi)$ plane.

\begin{figure}
\centerline{\psfig{figure=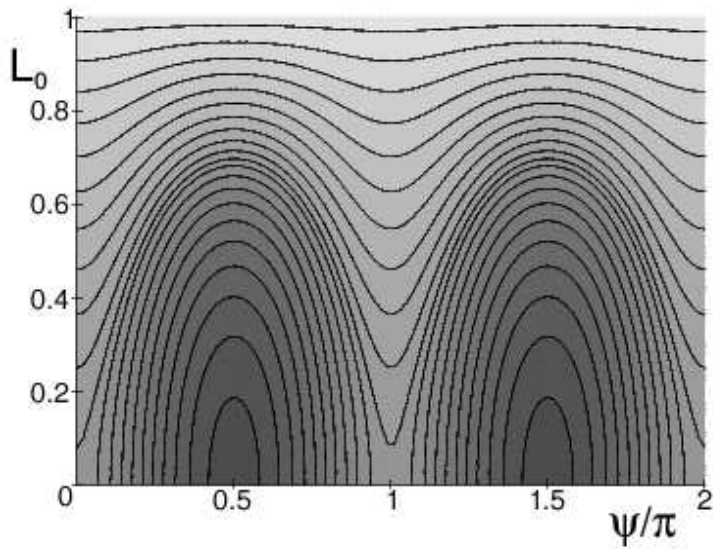,width=10cm,angle=-90}}
\caption{Isovalue curves of the angular part $\chi_1$, eq.~(\ref{substc}),
of the secular
Hamiltonian ${\mathcal H}_{\rm sec}$ represented in the plane
of the $L_0=L/\hat{I}_1$ and $\psi$ coordinates.
The slow evolution of the
Kepler ellipse of a Rydberg electron driven 
by a resonant, linearly polarized microwave field, takes place
along such isovalue curves.
$L_0=L/\hat{I}$ represents the total angular
momentum (a circular trajectory in a plane containing 
the field polarization axis has $L_0=1$), and
$\psi$ the [canonically conjugate] angle between
the field polarization axis and the major axis of the Kepler ellipse. 
The separatrix emanating from the unstable fixed point
($L_0=0,\psi=0$) separates rotational and librational motion, both 
``centered" around their respective stable
fixed points ($L_0=1$, $\psi$ arbitrary) and ($L_0=0$, $\psi=\pi/2$). 
The former corresponds to a circular orbit
centered around the nucleus. The latter
represents a straight linear orbit perpendicular to the field axis. The
unstable fixed point corresponds to linear motion along the
polarization axis. However, this initially degenerate Kepler ellipse will
slowly precess in the azimuthal plane. The equipotential curves shown here
satisfy the quantization condition (\protect\ref{ebkp}), 
for $n_0=21$
and $p=0\ldots 20$. At lowest order, the motion in the $(L_0,\psi)$ 
plane is independent of 
the microwave field strength and of the resonant principal quantum
number $n_0$.
}
\label{lin3d_1} 
\end{figure}

\begin{figure}
\centerline{\psfig{figure=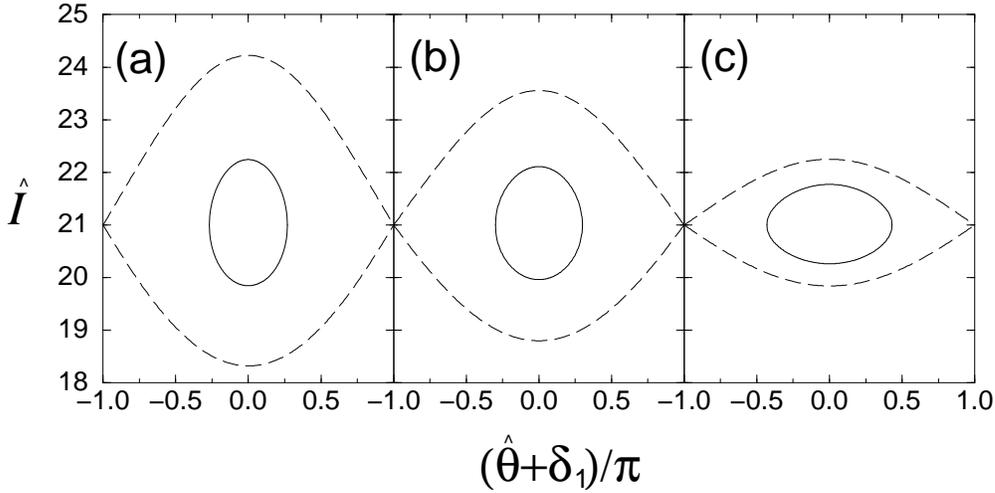,width=14cm,angle=-90}}
\caption{
Isovalue curves of the secular Hamiltonian ${\mathcal H}_{\rm sec}$,
eq.~(\ref{hamscfin}),
generating the ($\hat I,\hat \theta$)
motion of a Rydberg electron in a resonant microwave field. 
$\hat I$ and $\hat \theta$ correspond
to the atomic principal quantum number, and to the polar angle of the electron on
the Kepler ellipse, respectively. The scaled microwave amplitude is fixed at
$F_0=0.03$. Since the isovalues of 
${\mathcal H}_{\rm sec}$ depend on the 
transverse motion in $(L,\psi)$ via the constant value
of $\chi_1$, eq.~(\ref{substc}), 
contours (solid lines) are shown for three characteristic values of 
$\chi_1$, corresponding to fixed quantum numbers $p=0,10,20$, 
eq.~(\ref{ebkp}), 
of the angular
motion for the $n_0=21$ resonant manifold.
Only the ``ground state'' orbit satisfying eq.~(\protect\ref{ebkires}) 
with $N=0$ is 
shown, 
together with the separatrix (dashed lines) 
between librational 
and rotational motion in
the ($\hat I,\hat \theta$) plane. The separatrix encloses the  
principal
resonance island in phase space,
see also 
eqs.~(\ref{eqpend},\ref{area}). Panel (a) corresponds to the orbit 
with
$L_0=L/n_0\simeq 1$ (rotational orbit, $p=20$), panel (b) to the orbit
close to the separatrix of the angular motion ($p=10$), panel (c)
to the librational orbit close to the stable fixed point $L_0=0$,
$\psi=\pi/2$. Note that the resonance island  is smallest for librational, 
largest
for rotational, and of intermediate size for separatrix modes of the
angular motion.}
\label{lin3d_2}
\end{figure}

Importantly, the loops of constant $\chi_1$ are independent
of the microwave amplitude $F$ and scale simply with $\hat{I}.$
Thus, the whole quantization in the 
$(L,\psi)$ plane has to be done only once.
With
this prescription we can unravel the semiclassical 
structure of the quasienergy spectrum induced by the additional 
degree of freedom spanned by $(L,\psi)$, as an amendment to the spectral 
structure of the one-dimensional model discussed in section~\ref{LIN1D}.
Fig.~\ref{lin3d_1} shows the equipotential curves of $\chi_1$ in the
$(L,\psi)$ plane.
For a comparison with quantal data, the
equipotential lines plotted correspond to the quantized
values of $\chi_1$ for $n_0=21$.
Using the well-known properties of the Bessel functions~\cite{abramowitz72},
it is easy to show that $\chi_1(L,\psi)$ has the following
fixed points:
\begin{itemize}
\item ($L=\hat{I}_1$, arbitrary $\psi$). This corresponds to a Kepler
ellipse with maximum angular momentum, i.e. a circular orbit 
in a plane containing the microwave polarization axis along $\hat z$. 
As such a circle corresponds to a degenerate family of elliptical
orbits with arbitrary orientation of the major axis, $\psi$ is a
dummy angle. This fixed point corresponds to a global
maximum of $\chi_1(L=\hat{I}_1)= \hat{I}_1^2/2$,
and is surrounded by ``rotational"
trajectories in the $(L,\psi)$ plane.
An alternative representation of the ($L,\psi$) motion on the unit sphere,
spanned by $L$ and the $z$ and $\rho$-components of the Runge-Lenz vector,
contracts the line representing this orbit in 
fig.~\ref{lin3d_1} 
to an elliptic fixed point \cite{abu97}. 
\item ($L=0, \psi=\pi/2,3\pi/2).$ This corresponds to a 
degenerate straight line trajectory perpendicular to the
microwave field. Because of the azimuthal
symmetry around the electric field axis, the two points
actually correspond to the same physics.
The oscillating dipole clearly vanishes there,
resulting in a global minimum of $\chi_1(L=0,\psi=\pi/2,3\pi/2)=0.$
This stable fixed point is surrounded by ``librational" trajectories
in the $(L,\psi)$ plane.
\item ($L=0, \psi=0,\pi).$ This corresponds to a 
degenerate, straight line trajectory along the
microwave field, i.e. the situation already considered in the
1D model of the atom. $\psi=0$ and $\pi$ correspond
to the two orbits pointing up and down, which are of course
equivalent. This is a saddle point of
$\chi_1(L=0,\psi=0,\pi)=J'_1(1)\hat{I}_1^2.$
Hence, it is an unstable equilibrium point. As an implication,
in the real 3D world, the motion along the microwave axis, with the
phase of the radial motion locked on the external drive,
is angularly unstable (see also section \ref{staticmw}). This leads 
to a slow precession of the initially degenerate Kepler 
ellipse off the axis, 
and will manifest itself in the localization properties
of the 3D analog of the nondispersive wave-packet displayed in 
fig.~\ref{lin1d_2}. This motion takes place along
the separatrix between librational and rotational
motion.
\end{itemize}
Once the quantized values of $\chi_1$ (represented by 
the trajectories in fig.~\ref{lin3d_1})
have been determined,
we can quantize the (${\hat I},{\hat \theta}$) 
motion with 
these values fixed. 
Fig.~\ref{lin3d_2} shows the
equipotential lines of ${\mathcal H}_{\rm sec}$, 
for the three values of $\chi_1$
corresponding to the $p=0$, $10$ and $20$ states, see eq.~(\ref{ebkp}), 
of the 
$n_0=21$ manifold. In each case, the contour for the lowest state
$N=0$ has been drawn, together with the separatrix between the
librational and rotational (${\hat I},\hat{\theta}$) modes. 
The separatrix
determines the size of the principal resonance island
for the different substates of the transverse
motion. Note that the principal resonance is largest for the 
$p=20$ state, localized closest to the stable circular
orbit (hence associated with the
maximum value of $\chi_1$), whereas the smallest resonance
island is obtained for the $p=0$ state, localized
in the vicinity  
of (though not precisely at) 
the straight line orbit perpendicular to the field axis
(minimum value of $\chi_1$).
For the latter orbit itself, the first order coupling 
vanishes identically ($\chi_1=0$), which 
shows that the semiclassical results obtained 
from our first order approximation (in $F$) for the Hamiltonian may be quite
inaccurate in the vicinity of this orbit. Higher order corrections 
may become important.

\begin{figure}
\centerline{\psfig{figure=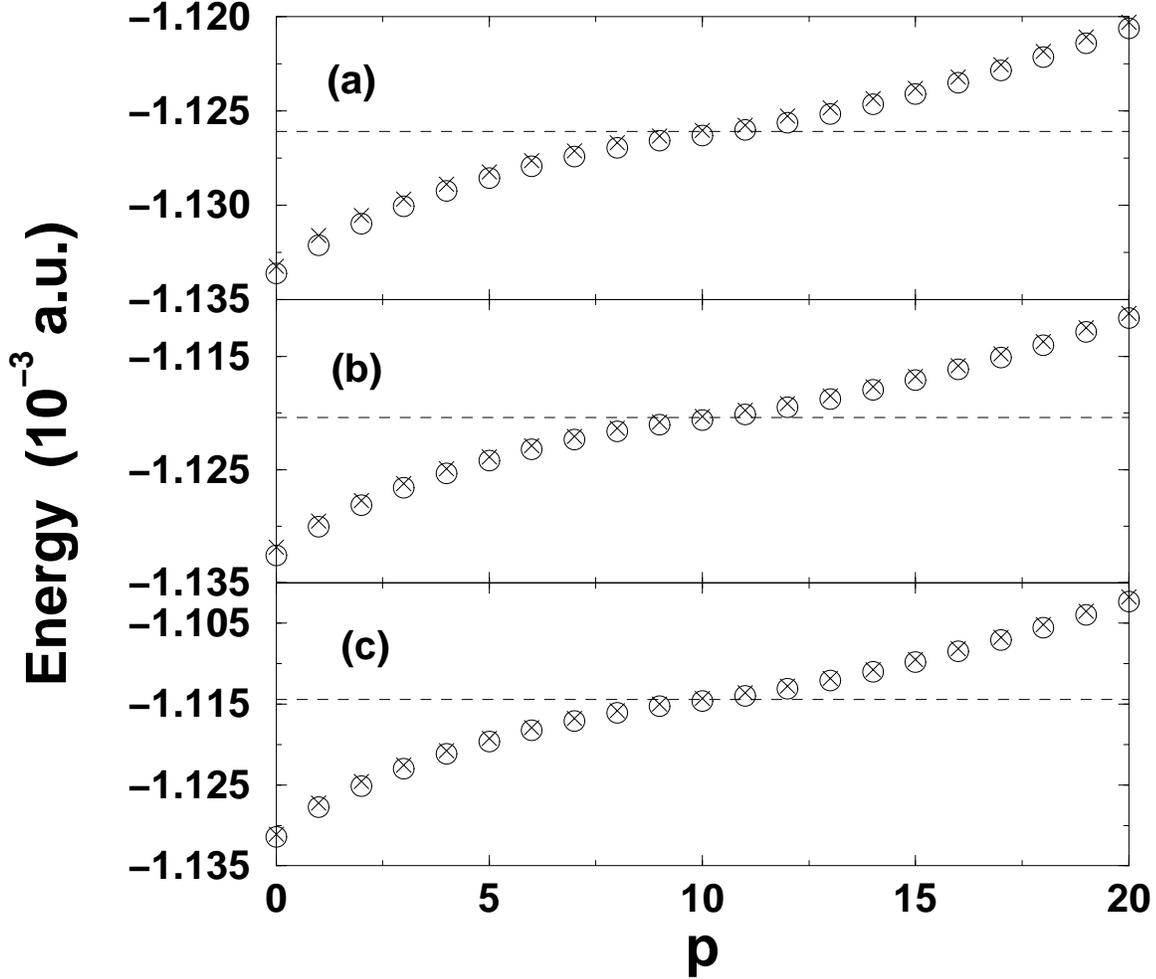,width=15cm,angle=-90}}
\caption{Comparison of the semiclassical quasienergies (circles; with
$N=0$, see eq.~(\protect{\ref{ebkires}})),
originating from the unperturbed $n_0=21$ manifold, 
to the exact quantum
result (crosses), at different values of the (scaled) driving field amplitude
$F_0=Fn_0^4=0.02$ (a), $0.03$ (b), $0.04$ (c). The agreement is excellent. The
quantum number $p=0\ldots 20$ labels the quantized classical
trajectories plotted in fig.~\protect{\ref{lin3d_1}}, starting from the
librational state $\mid p=0\rangle$ at lowest energy, rising through the
separatrix states $\mid p=10\rangle$ and $\mid p=11\rangle$, up 
to the rotational
state $\mid p=20\rangle$. The dashed line indicates the exact quasienergy of
the corresponding wave-packet eigenstate of the 1D model discussed in section
\ref{LIN1D}. The 1D dynamics is neatly embedded in the spectrum of
the real, driven 3D atom.}
\label{lin3d_3}
\end{figure}

As discussed above, the classical motion in the $(L,\psi)$ plane is 
slower than
in the (${\hat I},\hat{\theta}$) plane. In the semiclassical approximation,
the spacing between consecutive states corresponds to the
frequency of the classical motion (see also eq.~(\ref{cprinc})). 
Hence, it is to be expected that states
with the same quantum number $N$, but with successive quantum numbers $p$,
will lie at neighboring energies, building well-separated manifolds associated
with a single value of $N$. 
The energy spacing 
between
states in the same manifold should scale as $F_0$, while the spacing between
manifolds should scale as $\sqrt{F_0}$ (remember that $F_0\ll 1$ in the case
considered here). Accurate quantum calculations fully
confirm this prediction, with manifolds originating from
the degenerate hydrogenic energy levels at $F_0=0$, as we shall demonstrate
now. 
We first concentrate on the $N=0$ manifold, originating
from $n_0=21$. Fig.~\ref{lin3d_3} shows 
the comparison between the
semiclassical and the quantum energies, for different values of the 
scaled driving
field amplitude $F_0=Fn_0^4$. The agreement is excellent, except for 
the lowest lying states 
in the manifold for $F_0=0.02$. The lowest energy level ($p=0$)
corresponds to motion close to the stable fixed point $L=0$, $\psi =\pi/2$ 
in fig.~\ref{lin3d_1}; the highest energy level ($p=20$)
corresponds to rotational motion $L/n_0\simeq 1$. The levels with
the smallest energy difference ($p=10,11$) correspond to the librational and
the rotational trajectories closest to the separatrix, respectively. 
The narrowing
of the level spacing in their vicinity is just a consequence of the slowing 
down of the
classical motion~\cite{delande97}. 
In the same figure, we also plot (as a dashed line) the corresponding 
exact
quasienergy level for the 1D model of the atom (see section~\ref{LIN1D}).
As expected, it closely follows the separatrix state $p=10$. 
Such good agreement is a direct proof of the validity
of the adiabatic separation between the
slow motion in $(L,\psi)$, and the fast motion
in (${\hat I},\hat{\theta}$).

\begin{figure}
\centerline{\psfig{figure=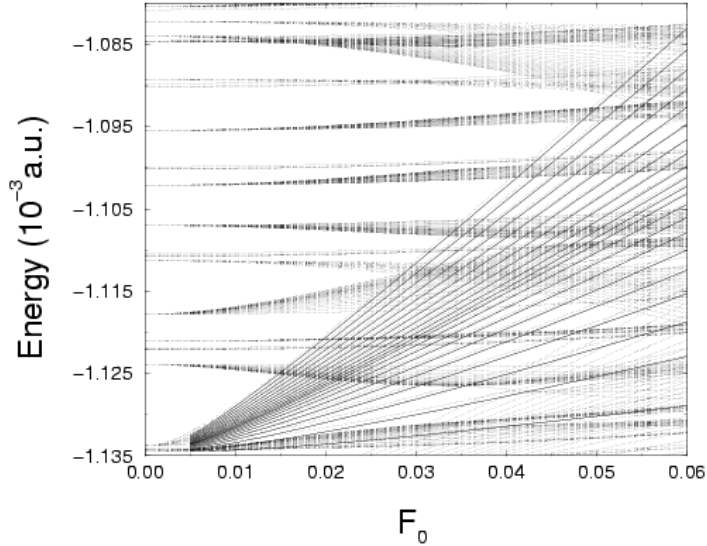,width=10cm,angle=-90}}
\caption{Level dynamics of the numerically 
exact quasienergies (dotted lines)
in the vicinity of the resonant manifold emerging from $n_0=21$ ($N=0$),
compared to the semiclassical prediction (full lines), for 
$F_0=Fn_0^4=0\ldots 0.06$.
Note that the maximum field amplitude exceeds the typical ionization
thresholds measured in current experiments for $\omega n_0^3 \simeq 1$
\protect\cite{bayfield89,koch95b,bellermann96}. 
Nontheless, the semiclassical prediction accurately 
tracks the exact
solution across a large number of avoided crossings
with other Rydberg manifolds.
}
\label{lin3d_4}
\end{figure}

Fig.~\ref{lin3d_4} shows a global comparison of the 
semiclassical prediction with
the exact level dynamics (energy levels vs. $F_0$), 
in a range from $F_0=0$ to $F_0=0.06$, which exceeds the
typical ionization threshold ($F_0\simeq 0.05$) observed in current 
experiments
\cite{bayfield89,koch95b,bellermann96}. We observe that the semiclassical 
prediction 
tracks the exact
quasienergies quite accurately, even
for large $F_0$-values, 
where the resonant 
$n_0=21$ manifold 
overlaps with other Rydberg manifolds, or with side bands of 
lower or higher lying
Rydberg states.
The agreement becomes unsatisfactory only 
in the region of very small $F_0$, where the size
of 
the resonance island in $(\hat I, \hat \theta)$ is very small.
This is not unexpected, as semiclassics should fail when the
area of the resonance island is comparable to $\hbar$, compare 
eq.~(\ref{number_of_trapped_states}).
In this weak driving regime, the pendulum approximation can be used
to produce more accurate estimates of the energy levels.
The fast $(\hat{I},\hat{\theta})$ motion is essentially identical to the one
of the 1D driven hydrogen atom: thus, the Mathieu approach used
in section~\ref{LIN1D} can be trivially extended to the 3D case.
The only amendment is to replace the factor $J_1'(1)n_0^2$ in
the expression of the Mathieu parameter $q$ by the various
quantized values of $\chi_1$ for $0\leq p \leq n_0-1,$ and to use
the same equation (\ref{prediction_mathieu_lin1d}) for the energy levels.

\begin{figure}
\centerline{\psfig{figure=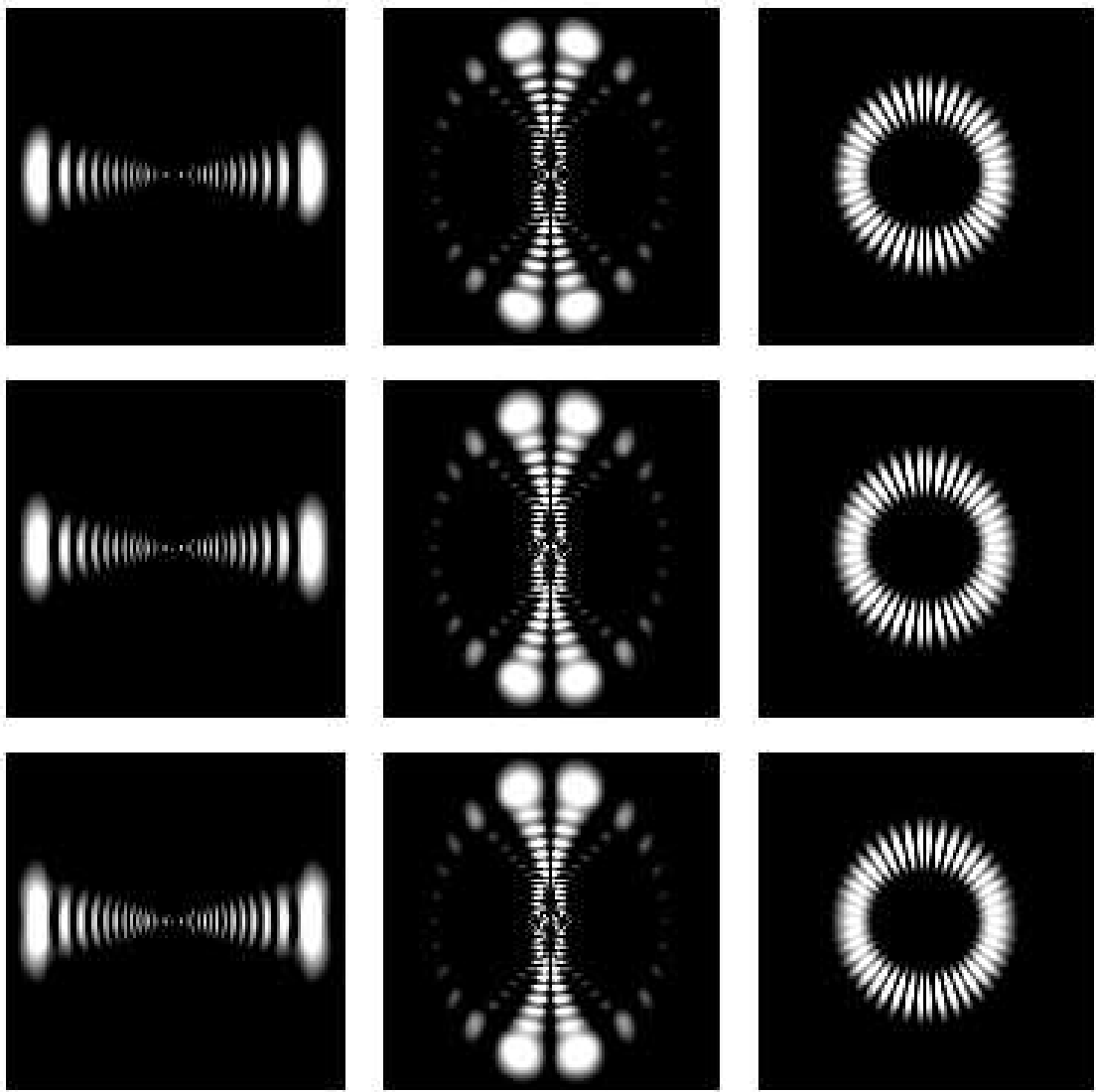,width=15cm}}
\caption{Electronic densities of the extremal librational ($p=0$, left), 
separatrix ($p=10$, center), and extremal rotational ($p=20$, right) 
quasienergy states of the $n_0=21$  manifold of a 3D hydrogen
atom exposed to a resonant microwave field
in cylindrical coordinates $(\rho,z)$, at different values of the
driving field amplitude $F_0=Fn_0^4=0.02$ (top), $0.03$ (middle), $0.04$ (bottom), 
averaged
over one period of the driving field. Note the clear localization along the
classical orbits corresponding to the respective contours in
fig.~\protect\ref{lin3d_1}, for {\em all} field amplitudes. The nodal lines
of the electronic densities clearly exhibit the direction of the underlying
classical motion. The field-induced finite decay rate of the eigenstates 
(see section~\protect\ref{ION}) is negligible on time scales shorter than 
approx. $10^6$ Kepler periods. Each box extends over 
$\pm 1000$ Bohr radii, in both $\rho$ (horizontal) and
$z$ (vertical) directions, with the nucleus at the center of the plot. The microwave polarization axis is oriented  
vertically along $z.$}
\label{lin3d_5}
\end{figure}

The semiclassical construction of the energy levels from classical
orbits is -- necessarily -- reflected in the localization properties of the
associated eigenstates, as demonstrated by the electronic densities of the
states $\mid p=0\rangle$, $\mid p=10\rangle$, and $\mid p=20\rangle$ in
fig.~\ref{lin3d_5}, for the same field amplitudes as in 
fig.~\ref{lin3d_3}. Note that, in this plot, the electronic densities 
are averaged over one field cycle,
hence display only the angular localization properties of the eigenstates.
Their localization along the classical orbits defined by the
stable or unstable fixed points of the
$(L,\psi)$ dynamics is obvious \cite{abu96,abu98a,abu97}.
Note in particular the nodal
structure of the state $\mid p=10\rangle$, associated with the unstable 
fixed point:
there are sharp nodal lines perpendicular to the $z$-axis, reflecting the
dominant motion along the $z$-axis, but also nodal lines of low visibility in
the angular direction. They are a manifestation of the slow classical
precession of the Kepler ellipse, i.e. the slow secular evolution in
the ($L,\psi$) plane. 
The quantum state, however, dominantly
exhibits the motion along the $z$-axis, as a signature of the effective
separation of time scales of radial and angular motion.
Finally, it should be realized from a comparison of the top to the middle and
bottom row of fig.~\ref{lin3d_5}
that the quasiclassical localization properties of the eigenstates are
essentially unaffected as $F$ rises, despite various avoided crossings which
occur at intermediate field values, see fig.~\ref{lin3d_4}. Especially,
the angular structure does not depend at all on $F$, as predicted by the
secular approximation.

\begin{figure}
\centerline{\psfig{figure=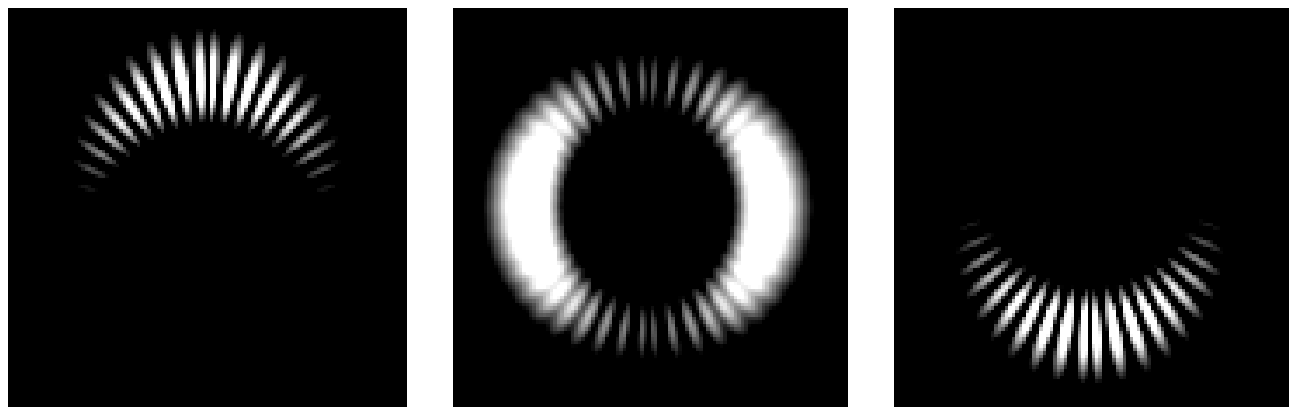,width=14cm}}
\caption{Temporal evolution of the electronic density
of the extremal rotational quasienergy state
$\mid p=20\rangle$ of the $n_0=21$ resonant manifold, 
for different phases $\omega t=0$ (left),
$\omega t=\pi/2$ (center), $\omega t=\pi$ (right) of the driving field, at
amplitude $F_0=0.03$, in cylindrical coordinates. 
Each box extends over $\pm 700$ Bohr radii, in both directions, $\rho$
(horizontal) and
$z$ (vertical). The microwave polarization axis is oriented 
along $z$. Because of the azymuthal symmetry of the problem, 
the actual 3D electronic
density is obtained by rotating the figure around the vertical axis. 
The state represents
a non-dispersive wave-packet shaped like a doughnut, moving periodically from
the north to the south pole (and back) of a sphere. For higher $n_0$, the
angular localization on the circular orbit should improve.}
\label{lin3d_8}
\end{figure}

The eigenstates displayed here are localized
along classical trajectories which are resonantly driven by the external
field. Hence, we should expect them to exhibit wave-packet like motion along
these trajectories, as the phase of the driving field is changed. 
This is indeed the case as illustrated  
in fig.~\ref{lin3d_8} for the state $p=20$ 
with maximal angular momentum 
$L/n_0\simeq 1$ \cite{kalinski95b,abu96,abu98a}. 
Due to the azimuthal symmetry of the problem, the actual 3D electronic
density is obtained by rotating the figure around the vertical axis.
Thus, the wave-packet is  actually a doughnut moving periodically
from the north to the south
pole (and back)
of a sphere, slightly deformed along the field direction. The interference
resulting from the contraction of this doughnut to a compact wave-packet at
the poles is clearly visible at phases $\omega t=0$ and $\omega t=\pi$ in the
plot. Note that the creation of unidirectional wave-packet
eigenstates moving along a circle in the plane containing the field
polarization axis is not possible for the real 3D atom \cite{abu98a},
 as opposed to the
reduced 2D problem studied in \cite{kalinski95b}, due to the 
abovementioned azimuthal
symmetry (see also section \ref{EP}).
For other states in the $n_0=21$ resonant manifold, the longitudinal
localization along the periodic orbit is less visible. The reason is that
$\chi_1$ is smaller than for the $p=20$ state, leading to a smaller
resonance island in $(\hat{I},\hat{\theta})$ (see fig.~\ref{lin3d_2}) 
and, consequently, to less
efficient localization. Proceeding to higher $n_0$-values should improve
the situation.

\begin{figure}
\centerline{\psfig{figure=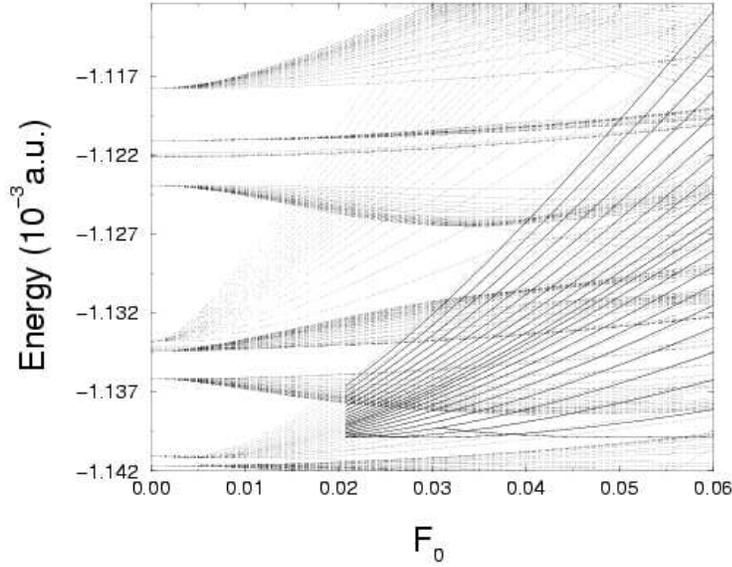,width=10cm,angle=-90}}
\caption{Comparison of the numerically 
exact level dynamics (dotted lines)
with the
semiclassical prediction (solid lines),
for the $n_0=22$ ($N=1$) manifold of a 3D hydrogen atom exposed to
a microwave field with frequency $\omega=1/(21)^3,$ resonant
with the $n_0=21$ manifold.
For sufficiently high $F_0,$ the quantum 
states originating at $F_0=0$ from the unperturbed $n_0=22$ level 
are captured by the principal resonance island and then represent 
the first excited state of the motion in the ($\hat I, \hat \theta$)
plane (i.e., $N=1$ in eq.~(\protect{\ref{ebkires}})). Since the island's 
size  
depends on the angular ($L,\psi$) motion (value of $p$ in 
eq.~(\protect{\ref{ebkp}}), see also fig.~\protect\ref{lin3d_2}),
states with large $p$ enter the resonance zone first. For these,
the agreement between quantum and semiclassical quasienergies 
starts to be satisfactory at lower $F_0$ values than for low-$p$ states.}
\label{lin3d_9}
\end{figure}

\begin{figure}
\centerline{\psfig{figure=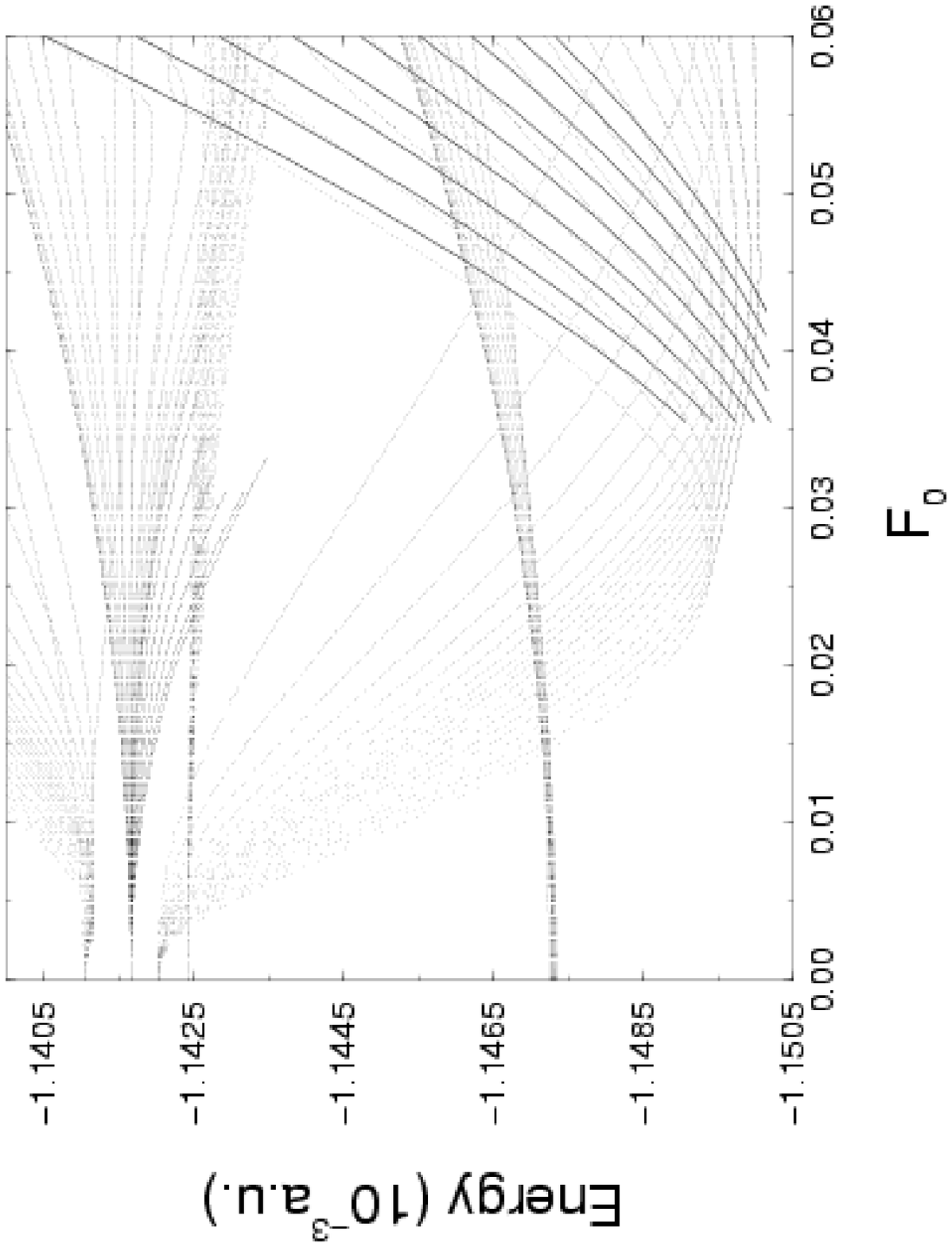,width=10cm,angle=-90}}
\caption{Same as fig.~\protect{\ref{lin3d_9}}, but for $N=2$. The quantum 
states originate
from the manifold $n_0=20$. The 
resonance island in the ($\hat I,\hat \theta$) 
coordinates
is too small to support any $N=2$ states for $F_0<0.03$, as seen from the
negative slope of the quasienergy levels. The quantum states 
``cross'' the separatrix as the amplitude is further increased,
and successively enter the resonance zone, starting from the 
largest value of $p$ in eq.~(\protect{\ref{ebkp}}) 
(the resonance island size increases with $p$).
Even for $F_0>0.04,$ only a minority of substates of 
the $n_0=20$ manifold is well represented by the resonant semiclassical 
dynamics (indicated by the solid lines).}
\label{lin3d_10}
\end{figure}

Let us briefly discuss ``excited'' states in the resonance island,
i.e. manifolds corresponding to $N>0$ in eq.~(\ref{ebkires}). 
Fig.~\ref{lin3d_9}
shows the exact level dynamics, with the semiclassical prediction for
$N=1$ superimposed \cite{abu98a}.
 The states in this manifold originate from $n_0=22$. We
observe quite good agreement between the quantum and semiclassical results for
{\it high} lying states in the manifold (for which the principal action
island is large, see fig.~\ref{lin3d_2}). 
For lower lying states the agreement is improved for higher
values of $F_0$. If $F_0$ is too low, the states are not fully localized 
inside 
the resonance island and, consequently,
are badly reproduced by the resonant semiclassical 
approximation.  This is further
exemplified in fig.~\ref{lin3d_10}, for $N=2$. 
Here, the
agreement is worse than for smaller values of $N$, and is observed only 
for
large $F_0$ and large $p$. This confirms the picture that the validity
of the semiclassical approach outlined here 
is directly related to the size,
eqs.~(\ref{area},\ref{number_of_trapped_states}), of the
resonance island in $(\hat I, \hat\theta)$ space (see also the discussion
in section~\ref{PTP}).

Finally note that, as already mentioned at the end of section~\ref{LIN1D}, all
wave-packet eigenstates have a finite decay rate which induces a slow, global
reduction of the electronic density localized on the resonantly driven
classical periodic orbit. However, the time scale of this decay is of the
order of thousands to millions of Kepler cycles, and therefore leaves our
above conclusions unaffected. However, some very intriguing consequences of
the nonvanishing continuum coupling will be discussed in section~\ref{ION}.

\subsection{Rydberg states in circularly polarized microwave fields}
\label{CP}

As shown in the preceding section, the use of a linearly polarized
microwave field is not sufficient to produce a non-dispersive
wave-packet fully localized in all three dimensions,
due to the azymuthal symmetry around the
microwave polarization axis. To get more flexibility, one may consider
the case of arbitrary polarization. It turns out that the results
are especially simple in circular polarization. They are the subject of this
section.

In most experiments on 
microwave driven Rydberg atoms,
linearly polarized (LP) microwaves have 
been used \cite{bayfield89,koch95b,bayfield74,bayfield96,galvez88}. 
For circular polarization (CP),
first experiments were performed for alkali
atoms in the late eighties \cite{fu90,cheng96}, 
with hydrogen atoms 
following only recently \cite{bellermann96}. The latter experiments 
also studied 
the general case of elliptic polarization (EP). While, 
at least theoretically, different
frequency regimes 
were considered for CP microwaves
(for a review see \cite{delande97b}) --
we shall restrict 
our discussion here to 
resonant driving.
Given a different 
microwave polarization, and thus a different form 
of the interaction Hamiltonian, eq.~(\ref{vcirc}),
Kepler trajectories which are distinct from those considered in the LP case
will be most efficiently locked on the external driving. Hence, in the sequal,
we shall launch nondispersive wave-packets along periodic orbits which are
distinct from those encountered above.

Historically, the creation of non-dispersive wave-packets in 
CP and LP microwave fields,
respectively, 
has been considered quite independently.
In particular, in the CP case, the notion of nondispersive 
wave-packets has been introduced \cite{ibb94} 
along quite 
different lines 
than the one adopted in this
review. The original work, as well as subsequent studies of the CP situation
\cite{farrelly95a,ibb95,kalinski95a,delande95,kuba95a,kalinski96b,kuba97a,kuba97b}
used 
the fact that, in this specific case, 
the time-dependence of the
Hamiltonian may be removed by a unitary transformation to the rotating frame
(see below). Thus, the stable periodic orbit 
at the center of the island 
turns into a stable equilibrium point in the rotating frame. This allows the
expansion of the Hamiltonian into a Taylor series in the vicinity of 
the fixed point,
and in particular a standard harmonic treatment using normal modes. 
We shall review 
this line of reasoning in detail below. It is, however, instructive 
to first 
discuss
the very same system using the general resonance approach
exposed in section~\ref{GM}.

\subsubsection{Hamiltonian}

With $\hat z$ the propagation direction of the microwave,
the electric field rotates
in the $x-y$ plane, and the Hamiltonian (\ref{h}) takes
the
following explicit form:
\begin{equation}
H_{\rm CP}=\frac{\vec{p}^2}{2}-\frac{1}{r}+
F\left\{x\cos(\omega t)+y\sin(\omega t)\right\}.
\label{ham_cp_3d}
\end{equation}

In contrast to the LP 
situation, there is no simplified one-dimensional
model in the CP case. However, a simplified two-dimensional 
model 
exists, where
the motion is restricted to the $x-y$ plane. As long as one is interested only
in the dynamics of non-dispersive wave-packets, this motion is 
stable (see below),
which means that a small deviation from the $z=0$ plane does not affect the
qualitative behavior. Hence, much physical insight can be obtained
from the simplified 2D model. It 
will be discussed in section \ref{2d_model}.

\subsubsection{Resonance analysis}
\label{reso_an}
We follow the general treatment exposed in sections~\ref{CD} and \ref{LIN1D}
for the 1D case,
and 
express the external perturbation as a function of the action-angle
variables 
$(I,\theta)$, $(L,\psi)$ and $(M,\phi)$ introduced in section~\ref{RSEF}. 
By inserting equations~(\ref{v_3d_x},\ref{v_3d_y}) in eq.(\ref{ham_cp_3d}), 
after appropriate account for the projection of the body-fixed frame 
$(x',y',z')$ onto the laboratory frame $(x,y,z)$, eq.~(\ref{rot_euler}),
we obtain for the Floquet Hamiltonian:
\begin{equation}
{\mathcal H}=
P_t
-\frac{1}{2I^{2}}+F   \sum_{m=-\infty}^{\infty}\left[
    V_m \cos(m\theta+\phi-\omega t)
    -U_m
    \sin(m\theta+\phi-\omega t)\right],
\label{h_circ_aa}
\end{equation}
where the Fourier coefficients are given by 
(see also eqs.~(\ref{xm}-\ref{y0})):
\begin{eqnarray}
 V_{m}(I,L,M,\psi) &=& \cos\psi \left(Y_m + \frac{M}{L} X_m\right), \label{Vb}\\
 U_{m}(I,L,M,\psi) &=& \sin\psi \left(X_m + \frac{M}{L} Y_m\right).
\label{Ub}
\end{eqnarray}
Once again, transformation 
to the ``rotating frame", eqs.~(\ref{rotframe_a}-\ref{rotframe_c}), and
averaging over one field period $T=2\pi/\omega$ (thereby neglecting all 
rapidly varying terms) leaves us with the explicit form of the
secular Hamiltonian:
\begin{equation}
{\mathcal H}_{\rm sec}=\hat{P}_t-\frac{1}{2\hat I^2}-\omega \hat I
+F\left[V_1(\hat{I},L,M,\psi)\cos (\hat{\theta}+\phi) - 
U_1(\hat{I},L,M,\psi)\sin (\hat{\theta}+\phi) \right],
\label{hamsccir}
\end{equation}
similar to eq.~(\ref{hamsc2}).
This can be rewritten as:
\begin{equation}
{\mathcal H}_{\rm sec}=\hat{P}_t-\frac{1}{2\hat I^2}-\omega \hat I
+F \chi_1 \cos (\hat{\theta}+\phi+\delta_1)  ,
\label{hamsccir2}
\end{equation}
with the effective perturbation
\begin{equation}
\chi_1(\hat{I},L,M,\psi)=\sqrt{V_1^2+U_1^2},
\label{widthcir}
\end{equation}
and
\begin{equation}
\tan \delta_1(\hat{I},L,M,\psi) = \frac{U_1}{V_1}.
\label{phasecirc} 
\end{equation}

This secular Hamiltonian, which has, once again, 
the same structure as the 1D secular Hamiltonian~(\ref{hsec_ap}),
governs the ``slow" dynamics of the system
in the vicinity of the resonance. Similarly to the LP case, the
various degrees of freedom evolve on different time scales:
\begin{itemize}
\item In the $(\hat{I},\hat{\theta})$ plane, the situation is exactly like
for a one-dimensional system. There is a resonance island around
the resonant action,
with a pendulum-like structure. Non-dispersive wave-packets
are associated with eigenstates localized at the center of this island,
at the point (see eqs.~(\ref{scaling_action},\ref{hamsccir2})):
\begin{equation}
\hat{I}_1 = \omega^{-1/3} = n_0,\ \ \ \ \hat{\theta} = -(\phi + \delta_1).
\label{center_island_CP}
\end{equation}
The period of the secular 
classical motion close to the resonance center scales as 
$1/\sqrt{F}$. 
It 
defines an
intermediate time scale, slower than
the Kepler frequency, but faster than the transverse motion in the
other coordinates $(L,\psi)$ and $(M,\phi)$. 
\item In the subspace spanned by $(L,M,\psi,\phi)$, the motion is much 
slower, with
a time scale proportional to $1/F$. The effective Hamiltonian
describing this motion is obtained by averaging the fast
motion in the perturbation which describes
this motion $(\hat{I},\hat{\theta})$ plane, which in turns
implies that $\chi_1$ itself is constant for both the motion in 
$(\hat{I},\hat{\theta})$ and $(L,M,\psi,\phi)$ space.
\end{itemize}
Note that
$\chi_1$ does not depend on the angle $\phi$. This,
in turn, implies that $M$ is a constant of the slow motion. 
This is because the circular polarization
does not define any preferred direction in the  $x-y$ polarization plane.

Once again, much alike our discussion in section~\ref{LIN3D},
the well-known properties of Bessel functions
\cite{abramowitz72}, together with eqs.~(\ref{Vb},\ref{widthcir}), imply
that, for given $M$, the maximum of $\chi_1$ occurs at 
$L=M$,
corresponding to the situation when the electronic motion is
restricted to the polarization plane. In this plane, 
the maximum $\chi_1=\hat{I}^2$
is reached for the circular orbit defined by $M=L=\hat{I}$ (i.e. 
$L_0=1$). 
This defines
a resonant periodic orbit locked on the 
external microwave driving, which maximizes the
effective Hamiltonian in each coordinate and is, therefore,
fully 
stable in all phase space directions. 
The orbit is a circular Kepler orbit in the polarization
plane, where the electron rotates around the nucleus with exactly
the angular velocity of the microwave. It is not really surprising
that this orbit maximizes the interaction energy with the external
field: indeed, along this orbit, the atomic dipole rotates 
exactly in phase with the polarization vector of 
the circularly polarized microwave field. 
As in the case of linear polarization discussed in 
section~\ref{LIN3D}, the angular motion in the
$(L,\psi,M,\phi)$ variables (which is trivial
in $(M,\phi)$, since $M$ is constant) could be studied in detail.
For the sake of brevity, we will not repeat such an analysis here. We rather
concentrate on the  wave-packets 
which are best localized in the resonance island
near the circular orbit.
The simplest approximation 
to describe these states is to replace
the largest quantized value of $\chi_1$ by its maximum value $\hat{I}^2$
estimated at the center of the resonance island, eq.~(\ref{center_island_CP}).
Then, the situation is similar to the 1D model of the atom, 
eq.~(\ref{hsec_h1d}),
except that the strength of the coupling is $\hat{I}^2$ instead of 
$-J_1'(1)\hat{I}^2.$ In complete analogy to the steps leading from 
eq.~(\ref{hsec_h1d}) to eqs.~(\ref{semiharm1dval},\ref{omega_h1d}) we employ 
the pendulum approximation with a subsequent harmonic expansion
around the pendulum's stable equilibrium point, 
deeply inside the resonance island.
The harmonic frequency of the motion in the $(\hat I,\hat\theta)$ plane is:
\begin{equation}
\omega_{\mathrm harm}= 
\frac{\sqrt{3F}}{n_0} = \omega \sqrt{3F_0},
\label{omega_harmonic_cp}
\end{equation}
and the quasi-energy levels are:
\begin{equation}
{\cal E}_{N,k} = k\omega -\frac{3}{2n_0^2} + n_0^2F - \left(N+\frac{1}{2}\right) \omega_{\mathrm harm}.
\label{estharm}
\end{equation}
Note that, by construction, 
$-3/2n_0^2 + F n_0^2$ is nothing but the energy at the center
of the resonance island, i.e. the energy of the resonant circular orbit.
For very small $F$, the resonance island shrinks and may
support only a small number of states, or even no state at all. 
In this regime, the harmonic
approximation, eq.~(\ref{spectrum_harmonic}), breaks down.
Alternatively, one can apply a quantum treatment
of the pendulum motion in the $(\hat{I},\hat{\theta})$ plane, as explained
in section~\ref{section_mathieu} and discussed in section~\ref{LIN1D} 
for the
1D model of the atom exposed to a linearly polarized
microwave. The analysis -- essentially identical to the one in
section~\ref{LIN1D} --  yields the following expression for the
energy levels:
\begin{equation}
{\cal E}_{k,N}= k\omega- \frac{3}{2n_0^2}-\frac{3a_N(\nu,q)}{8n_0^4},
\end{equation}
where $a_N(\nu,q)$ are the Mathieu eigenvalues (compare with 
eq.~(\ref{map_mathieu2})
for the general case), with
\begin{equation}
q=\frac{4}{3}Fn_0^6,
\end{equation}
and 
\begin{equation}
\nu=-2n_0 \ \ ({\mathrm mod}\ 2)
\end{equation}
the characteristic exponent.

These expressions are valid for the states localized close to the resonant
circular orbit.
For the other states, the calculation is essentially identical, the
only amendment being the use of the values 
of $\chi_1$
following from the quantization of the secular motion, instead of the maximum
value $n_0^2.$ 

Finally, as the center of the resonance island corresponds to a circular
trajectory in the $(x,y)$ plane, the Floquet states associated with 
the non-dispersive 
wave-packets
will be essentially composed of combinations of circular states 
$|n,\ L=M=n-1\rangle$,
with coefficients described by the solutions
of the Mathieu equation, as explained in section~\ref{section_mathieu}.
This Mathieu formalism has been rediscovered in this particular CP situation
via complicated approximations 
on the exact Schr\"odinger equation in \cite{kalinski96b}. 
We believe that the standard resonance analysis using the
pendulum approximation leads, at the same time, to simpler calculations, 
and to
a much more transparent
physical picture.
 
\subsubsection{The two-dimensional model}
\label{2d_model}

We shall now discuss the simplified 2D model of the CP problem, which 
amounts to restricting the motion to the
$(x,y)$ plane, but retains almost all the features of the full 3D problem. 
Instead of the six-dimensional phase space 
spanned by the 
action-angle variables $(I,\theta)$, $(L,\psi)$, $(M,\phi)$, one is left
with a four-dimensional submanifold 
with coordinates $(I,\theta)$, $(M,\phi),$ see section~\ref{action-angle_hydrogen}. 
The secular Hamiltonian  
then reads 
(compare eqs.~(\ref{hamsccir2},\ref{widthcir})):
\begin{equation}
{\mathcal H}_{\rm sec}=\hat{P}_t-\frac{1}{2\hat I^2}-\omega \hat I
+FV_1(\hat{I},M)\cos (\hat{\theta}+\phi)  ,
\label{hamsccir_2d}
\end{equation}
with (see eq.~(\ref{Vb}))
\begin{equation}
V_1(\hat{I},M)= \hat{I}^2 \left[ J_{1}^{'}(e) + {\mathrm sign}(M)
 \frac{\sqrt{1-e^2}}{e}J_{1}(e) \right] ,
\end{equation}
where (as in eq.~(\ref{eccentricity}))
\begin{equation}
e=\sqrt{1-\frac{M^2}{\hat{I}^2}}.
\end{equation}

However, the Maslov index for the $(\hat{I},\hat\theta)$ motion is different.
Indeed, the energy spectrum of 
the 2D atom is given by eq.~(\ref{spectrum_2d}).
Thus, quantized values of the action
are half-integer multiples of $\hbar$. The relation
between the resonant action $\hat{I}_1=\omega^{-1/3}$ and the
corresponding principal quantum number now reads (with $\hbar=1$):
\begin{equation}
n_0=\hat{I}_1-\frac{1}{2}.
\end{equation}
As explained in section~\ref{section_mathieu}, the optimal
case for the preparation of non-dispersive wave-packets --
where the states are the most deeply bound
inside the resonance island --  is for integer
values of $n_0,$ i.e. frequencies
(compare with eqs.~(\ref{resonance_condition},\ref{omk}))
\begin{equation}
\omega = \frac{1}{\left(n_0+\frac{1}{2}\right)^3}.
\end{equation}
For the energy levels of the non-dispersive wave-packets,
this also implies that the characteristic exponents in the Mathieu
equation -- see section~\ref{section_mathieu} -- are shifted
by one unit:
\begin{equation}
\nu=-2n_0 \ \ ({\mathrm mod}\ 2)\ \ = -2\hat{I}_1+1\ \ ({\mathrm mod}\ 2).
\end{equation}

\subsubsection{Transformation to the rotating frame}
\label{rotating_frame}

The resonance analysis developed above is restricted to first
order in the amplitude $F$ of the external drive. Extensions to higher
orders are possible, but tedious. For CP,
an alternative approach is possible,
which allows higher orders to be included quite easily.
It is 
applicable to CP only and thus
lacks the generality of the resonance approach we used so far. 
Still, it is rather
simple and deserves an analysis.

In CP, one may remove the time dependence of the
Hamiltonian (\ref{ham_cp_3d}) by a transformation 
to the noninertial frame rotating
with the external frequency $\omega$. The unitary fransformation
$U=\exp(i\omega L_z t)$ leads to \cite{bunkin64,grozdanov92}
\begin{equation}
H_{\rm rot}= U \ H_{\rm CP}\ U^{\dagger} + 
i U \frac{\partial U^{\dagger}}{\partial t}
=
\frac{\vec{p}^2}{2} -\frac{1}{r} +Fx -\omega L_z.
\label{hrot}
\end{equation}
Classically, such an operation corresponds to a time dependent rotation of the
coordinate frame spanned by $\overline{x} = x\cos \omega t +y\sin \omega t$, 
$\overline{y} = y\cos \omega t-x \sin \omega t$ 
(and dropping the bar 
hereafter)\footnote{Passing to the rotating frame implies  
a change of 
$\phi$ to $\overline{\phi}=\phi-\omega t$ in eq.~(\ref{h_circ_aa}). 
That is definitely
different from the change $\theta \to \hat{\theta}=\theta-\omega t$, 
eq.~(\ref{rotframe_a}), used in the resonance analysis. 
Both transformations are unfortunately
known under the same name of ``passing to the rotating frame". This is
quite confusing, but one has to live with it. Along the resonantly driven
circular orbit we are considering here, 
it happens that the azimuthal angle $\phi$ and the polar angle 
$\theta$ actually coincide.
It follows that the two approaches are equivalent {\em in the vicinity}
of this orbit.}.
The Hamiltonian~(\ref{hrot}), as a time-independent operator, has some energy
levels and corresponding eigenstates. Its spectrum is not 
$\omega$-periodic, although the unitary transformation assures that there is
a one-to-one correspondence between its spectrum (eigenstates) and the
Floquet spectrum 
of eq.~(\ref{ham_cp_3d})\footnote{In
fact, if $|\phi_i\rangle$ is an eigenstate of $H_{\rm rot}$ with energy $E_i$,
 then $U^{\dagger}|\phi_i \rangle$ is
a Floquet eigenstate with quasi-energy $E_i$, while states shifted in
energy by $k\omega$ are of the form $\exp (ik\omega t)U^{\dagger}|\phi_i
 \rangle.$ For a more detailed discussion of this point, 
see \cite{delande98}.}.
It was observed \cite{klar89} that the Hamiltonian~(\ref{hrot})
allows for
the existence of a stable fixed (equilibrium) 
point in a certain range of the microwave 
amplitude $F$. Later on, it was realized \cite{ibb94}
that wave-packets initially
localized in the vicinity of this fixed point will not disperse
(being bound by the fact that the fixed point is stable) for 
at least several Kepler periods. 
In the laboratory frame, these wave-packets (also
called ``Trojan states" \cite{farrelly95a,ibb95,ibb94,kalinski95a})
appear as wave-packets moving around the
nucleus along the circular trajectory, which is nothing
but the periodic orbit at the center of the resonance island discussed in 
section~\ref{reso_an}.
In the original formulation \cite{ibb94} and the discussion which followed
\cite{farrelly95a,ibb95,kalinski95a,kalinski96b},
great attention was paid to the accuracy of the harmonic
approximation (see below). This was of 
utmost importance for 
the non-spreading
character of {\it Gaussian-shaped} Trojan wave-packets considered in
\cite{farrelly95a,ibb95,ibb94,kalinski95a,kalinski96b}. 
As soon pointed out in \cite{delande95}, however, 
the accuracy of
this approximation
is immaterial for the very existence of
the wave-packets, which are to be identified, as shown above, with
well-defined Floquet states.

Let us 
recapitulate the fixed point analysis of \cite{ibb94,klar89}
in the rotating frame.
Inspection of the classical version of the  Hamiltonian $H_{\rm rot}$,
eq.~(\ref{hrot}), 
shows that, due to symmetry, 
one may seek the fixed point
at $z=y=0$. The condition for an equilibrium (fixed) point,
 i.e., $d\vec{r}/dt=0,\ d\vec{p}/dt=0$,
yields immediately that $p_{z,\rm eq}=p_{x, \rm eq}=0$, 
$p_{y,\rm eq}=\omega x_{\rm eq}$,
with the subscript ``$\rm eq$'' for 
``equilibrium''.
The remaining equation
for $d p_x/dt$ gives the condition
\begin{equation}
  -F+\omega^2x_{\rm eq}-\frac{|x_{\rm eq}|}{x_{\rm eq}^3}=0,
\label{wp0}
\end{equation}
that defines the position of the fixed point as a function of $F$.  
Following \cite{ibb94} let us introduce the 
dimensionless parameter
\begin{equation}
q=\frac{1}{\omega^{2}|x_{\rm eq}|^{3}}.
\label{q}
\end{equation}
One may easily express the fixed point position, the microwave field
amplitude, as well as the corresponding energy in terms of $q$ and $\omega$.
Explicitly:
\begin{equation}
x_{{\rm eq}1}=\frac{1}{q^{1/3}\omega^{2/3}}, 
\qquad F=\frac{1-q}{q^{1/3}}\omega^{4/3},
 \qquad  E_{\rm eq}=\frac{1-4q}{2}\left(\frac{\omega}{q}\right)^{2/3},
\label{wps}
\end{equation}
and
\begin{equation}
x_{{\rm eq}2}=-\frac{1}{q^{1/3}\omega^{2/3}}, 
\qquad F=\frac{q-1}{q^{1/3}}\omega^{4/3},
 \qquad  E_{\rm eq}=\frac{1-4q}{2}\left(\frac{\omega}{q}\right)^{2/3}.
\label{wpu}
\end{equation}
For $F=0$,  $q=1$ in eqs.~(\ref{wps},\ref{wpu}).
For $F<0,$ (i.e., $q>1$ in eq.~(\ref{wps})), $x_{{\rm eq}1}$ 
is an unstable fixed point, while for moderately
positive $F$ (i.e., $8/9<q<1$ \cite{ibb94,klar89}) it is stable.
Stability of the second equilibrium point  $x_{{\rm eq}2}$
is achieved by changing the sign of $F.$ For moderate
fields ($q$ close to unity), the stable and the unstable fixed points 
are located on
opposite sides of the nucleus, and at almost the same distance from it. 
As the whole analysis is classical, it has to obey the scaling laws
discussed in section~\ref{scaling_laws}. Hence, all quantities describing
the equilibrium points in the preceding equations scale as powers
of the microwave frequency. A consequence
is that there is a very simple correspondence
between the scaled microwave amplitude and the dimensionless parameter $q$:
\begin{equation}
F_0=F\omega^{4/3} = F n_0^4 = \frac{1-q}{q^{1/3}}.
\end{equation}
The parameter $q$ can thus be thought of as 
a convenient parametrization (leading to simpler algebraic formula)
of the scaled microwave amplitude.

A fixed point in the rotating frame corresponds
to a periodic orbit with exactly the period $T=2\pi/\omega$ of the
microwave driving field in
the original frame. The stable fixed point
(periodic orbit) thus corresponds to the center of the resonance island, and to
the stable equilibrium point of the pendulum in the secular approximation.
Similarly, the unstable fixed point corresponds to the
unstable equilibrium point of the pendulum.
Note that the stable fixed point approaches
$x_{\rm eq}=\omega^{-2/3}=\hat{I}_1^2$ when $F\rightarrow 0$, 
i.e. the radius of the circular classical Kepler 
trajectory with frequency $\omega$.
Thus,
the stable fixed point
smoothly reaches the location  of the circular state 
of the hydrogen atom, with a classical Kepler frequency equal
to the driving microwave frequency.  Its energy $E_{\rm eq}=-3\omega^{2/3}/2$
is the energy of the circular orbit in the rotating frame.

Since the non-dispersive wave-packets are localized in
the immediate vicinity of the stable fixed point in the rotating frame, an expansion of the
Hamiltonian around that position is useful. Precisely at the fixed point, all
first order terms (in position and momentum) vanish. At second order,
\begin{eqnarray}
H_{\mathrm rot} \simeq H_{\mathrm harmonic}&&=E_{\rm eq} +
\frac{{\vec{\tilde p}}^2}{2}
 - \omega(\tilde{x}\tilde{p}_y-\tilde{y}\tilde{p}_x)
+ \frac{\omega^2 q \tilde{y}^2}{2}
\nonumber\\ && - \omega^2 q \tilde{x}^2
+ \frac{\omega^2 q \tilde{z}^2}{2}\! ,
\label{harmonic} 
\end{eqnarray}
where 
 $(\tilde{x},\tilde{y},
\tilde{z})=(x-x_{\rm eq},y-y_{\rm eq},z-z_{\rm eq})$ 
(and accordingly for the
momenta) denotes the displacement with respect
to the fixed point. 
Thus, in the harmonic approximation
the motion in the $z$ direction
decouples from that in the $x-y$ plane and is an oscillation with
frequency $\omega\sqrt{q}.$ 
The Hamiltonian for the latter,
up to the additive constant $E_{\rm eq}$,  can be expressed in the
standard form for a 2D, rotating anisotropic oscillator 
\begin{equation}
H= \frac{\tilde{p}_x^2+\tilde{p}_y^2}{2} +\frac{\omega^2
(a\tilde{x}^2+b\tilde{y}^2)}{2}-\omega(\tilde{x}\tilde{p}_y-\tilde{y}\tilde{p}_x),
\label{rotaibb}
\end{equation}
where the two parameters $a$ and $b$ are equal to $-2q$ and $q$,
respectively. 
This standard form has been studied in textbooks~\cite{marion}.
It may be used to describe the stability of 
the Lagrange equilibrium points in
celestial mechanics (see \cite{ibb94,marion} and references therein).
Because this Hamiltonian mixes position and momentum coordinates,
it is not straightforward to determine the stability at the origin.
The result is that there are two domains of stability:
\begin{equation}
a,b \geq 1
\label{stability_region_1}
\end{equation}
and
\begin{equation}
-3 \leq a,b \leq 1,\ \ \ {\mathrm with}\ \ (a-b)^2+8(a+b) \geq 0
\label{stability_region_2}
\end{equation}

For the specific CP case, where $a=-2q$ and $b=q,$ only the second
stability region is relevant, and the last inequality
implies $8/9\leq q\leq 1$ for the fixed point to be stable.

Alternatively, one can ``diagonalize" the Hamiltonian~(\ref{rotaibb})
and construct 
its normal modes. The normal modes entangle position and momentum
 operators due
to the presence of crossed position-momentum terms in the
Hamiltonian.
Only along the $z$-mode (which is decoupled from the rest), the creation 
and
annihilation operators, $b_z^\dagger,b_z$ are
 the standard combinations of $\tilde z$
and $p_{\tilde z}$ operators. 
In the $(x,y)$ plane, the creation and annihilation operators in the
$\pm$ normal modes have complicated explicit formulae given in~\cite{delande98}. 
After some algebra, one ends up with 
the frequencies of the normal modes,
\begin{eqnarray}
 \omega_\pm &=& \omega \sqrt{\frac{2-q\pm Q}{2}},\label{omegas}\\
\omega_z &=& \omega \sqrt{q},
\label{ompmz}
\end{eqnarray}
where
\begin{equation}
Q=\sqrt{9q^2-8q}
\label{defQ}
\end{equation}
and $q\geq 8/9$ for $Q$ to be real.
In terms of  
creation/annihilation operators, the harmonic Hamiltonian,
eq.~(\ref{harmonic}),
takes the form \cite{ibb94,delande95,kuba95a,delande98}
\begin{equation}
H_{\rm harmonic}
 = E_{\rm eq} + \left(b^\dagger_+ b_+ +\frac{1}{2}\right)\omega_+ 
- \left(b^\dagger_- b_-+\frac{1}{2}\right)\omega_- 
 + \left(b^\dagger_z b_z+\frac{1}{2}\right)\omega_z .
\label{hh}
\end{equation}
A minus sign appears in front of the $\omega_-$ term. This is because
the fixed point is {\em not} a minimum of the Hamiltonian, although
it is fully stable\footnote{In the first stability region, 
eq.~(\ref{stability_region_1}),
only + signs appear.}. 
This is actually due to the momentum-position
coupling, hence the Coriolis force. It is the same phenomenon
which is responsible for the stability of the Trojan asteroids 
\cite{ibb94,yeazell00,marion}
and of an ion in a magnetic trap \cite{lee95} 
(in the latter case, the position-momentum
coupling is due to the magnetic field).

\begin{figure}
\centerline{\psfig{figure=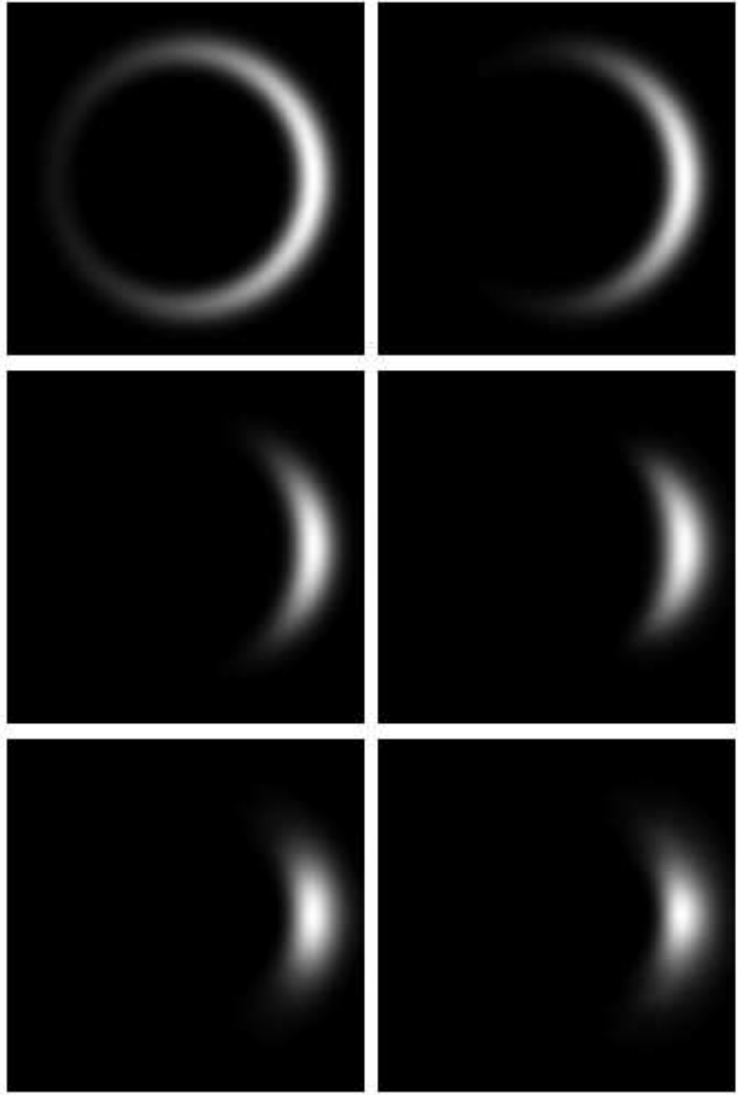,width=10cm}}
\caption{Non-dispersive wave-packets of the two-dimensional hydrogen atom
driven by a circularly polarized microwave field, for different values 
of the microwave amplitude $F$.
The microwave frequency is fixed at $\omega=1/(60.5)^3$,
corresponding to a resonance island centered at
$n_0=60.$ With increasing
microwave amplitude, more states are coupled and the wave-packet becomes
better localized. The scaled microwave amplitude $F_0=F\omega^{-4/3}$ 
is $0.0003$, $0.0011$,
$0.0111$, $0.0333$, $0.0444$, $0.0555$, from top left to bottom right.
The nucleus is at the center of the figure, which extends over
$\pm 5000$ Bohr radii in each direction. The microwave field is horizontal,
pointing to the right.}
\label{cp2df}
\end{figure}

Finally, with $n_\pm, n_z$ counting
the excitations in the corresponding modes, we obtain the
harmonic prediction for the energies of the eigenstates in the vicinity of the fixed
point: 
 \begin{equation}
E(n_+,n_-,n_z) = E_{\rm eq} + 
\left(n_++\frac{1}{2}\right)\omega_+
 - \left(n_-+\frac{1}{2}\right)\omega_-
+  \left(n_z+\frac{1}{2}\right)\omega_z .
\label{enharm}
\end{equation}
In particular, for $n_\pm=0,\ n_z=0$, we get a prediction for the ground 
state of the oscillator, a Gaussian localized on top of the fixed point, i.e.,
a Trojan wave-packet. In the following, we denote eigenstates in the 
harmonic approximation
as $|n_+,n_-,n_z\rangle$, thus the ground state non-dispersive wave-packet
as $|0,0,0\rangle$. 
In a 2D model, the $\omega_z$ term is dropped, the
corresponding eigenstates are denoted $|n_+,n_-\rangle$ and have energies:
\begin{equation}
E^{2D}(n_+,n_-) = E_{\rm eq} + 
\left(n_++\frac{1}{2}\right)\omega_+
 - \left(n_-+\frac{1}{2}\right)\omega_-.
\label{enharm2d}
\end{equation}
In fig.~\ref{cp2df}, we show the probability densities of the $|0,0\rangle$
wave-packets obtained by exact numerical diagonalization of the  
2D Hamiltonian (\ref{hrot}), for various
values of the microwave field amplitude.
Clearly, for sufficiently strong microwave amplitudes,
the wave-packets are well localized around
the classical stable fixed point, with banana-like shapes.
At very 
weak fields, the stability of the
fixed point gets 
weaker and weaker; for a vanishing microwave field, all points on the
circle with radius $\omega^{-2/3}$ are equivalent, and one has a ring of
equilibrium points. Thus, when $F$ tends to zero,
the non-dispersive wave-packet progressively extends along the angular direction
(with the radial extension almost unchanged), ending with a doughnut shape
at vanishing field. This means that, if $n_0$ is chosen
as an integer,
the non-dispersive wave-packet smoothly evolves into a circular
state $|n_0,M=n_0\rangle$ as $F\rightarrow 0$. 
The same is true for the 3D
atom, where the non-dispersive wave-packet smoothly evolves into the
circular state $|n_0,L=M=n_0-1\rangle.$

\begin{figure}
\centerline{\psfig{figure=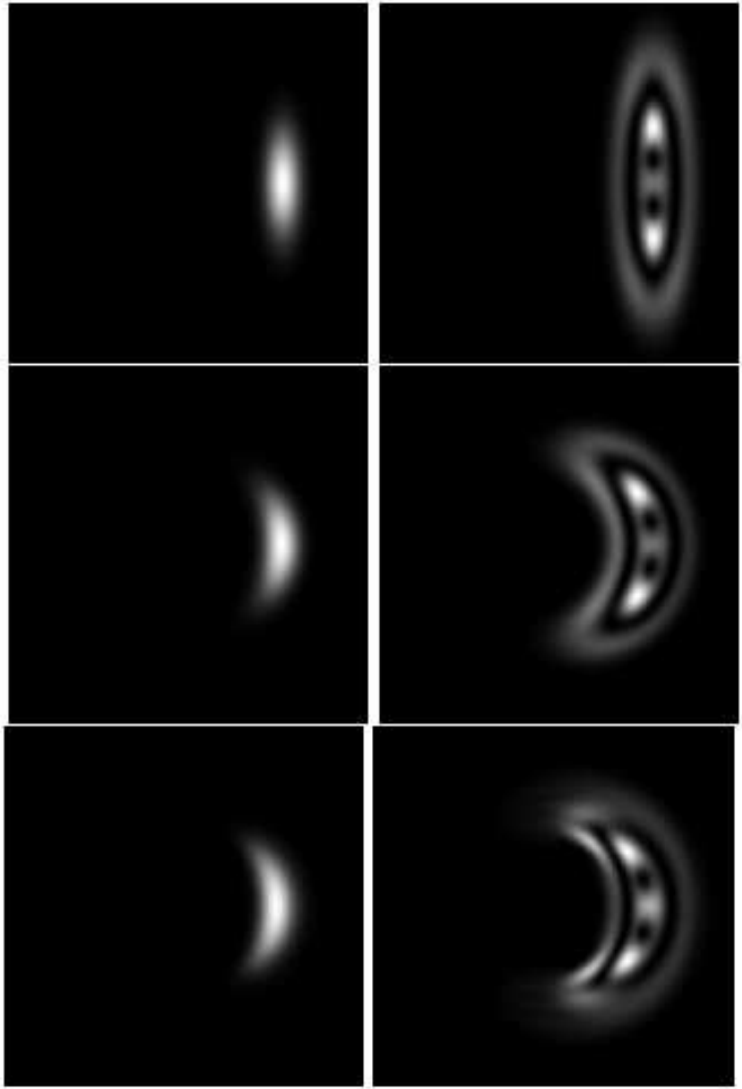,width=10cm} }
\caption{Comparison of exact Floquet eigenstates (bottom row) and of
their harmonic
approximations, for the ground state wave-packet $|n_+=0,n_-=0\rangle$ 
(left column),
and for the $|n_+=1,n_-=3\rangle$ excited state (right column), 
for the 2D hydrogen atom
driven by a circularly polarized microwave (field pointing to the
right of the figure), 
with amplitude $F_0=0.0333$ and resonant frequency
corresponding to $n_0=60$. The nucleus is located at the center of each plot
which extend over $\pm 5000$ Bohr radii. 
The top row represents eigenfunctions in the harmonic
approximation in $\tilde{x},\tilde{y}$ coordinates, 
eq.~(\protect\ref{rotaibb}).
The eigenfunctions in the middle row are obtained from the harmonic 
approximation to eq.~(\protect\ref{hrot}) in polar coordinates. 
They exhibit a clear bending of the electronic density along the circular 
trajectory.
The excited
wave-packet $|1,3\rangle$ appears in 
fig.~\protect{\ref{solit2}} as a straight 
line (modulo small avoided crossings) 
with a negative slope, meeting the state $|0,0\rangle$ in a broad avoided
crossing, around $F_0\simeq 0.036$.
}
\label{accharm}
\end{figure}

Provided the resonance island around the fixed point 
is large enough,
the harmonic approximation can also be used for studying properties
of ``excited" states inside the resonance island.
As an example, 
fig.~\ref{accharm} shows the $|n_+=1,n_-=3\rangle$ 
state calculated from the harmonic approximation, compared to
the state obtained by exact numerical diagonalization of the 2D Floquet
Hamiltonian. Obviously, the structure of the exact state is very similar to
the one obtained from 
its harmonic approximation. Because the creation and annihilation
operators in the $\omega_{\pm}$ modes entangle position
and momentum coordinates in a complicated way~\cite{delande98}, and 
although the system
is then completely integrable, the wave-function in the
harmonic approximation is {\em not} separable in any coordinate system 
(in contrast with the usual harmonic oscillator). Actually, the wave-function
can be written as a product of Gaussians and Hermite polynomials of 
the position
coordinates, but the Hermite polynomials have to be evaluated for 
complex values.
This results in the unusual pattern of the probability density displayed in
fig.~\ref{accharm}. An improvement over
the harmonic approximation is possible,
by bending the axis in the spirit of~\cite{kalinski95a}, in order to 
account for the spherical symmetry of the dominant Coulomb potential. 
With this
improvement, the probability density, shown in the middle row of 
fig.~\ref{accharm},
is almost indistinguishable from the exact result. Let us repeat that
this bending -- and consequently the deviation from Gaussian character of the
wave-function -- does not affect at all the non-dispersive character of the 
wave-packet.

\begin{figure}
\centerline{\psfig{figure=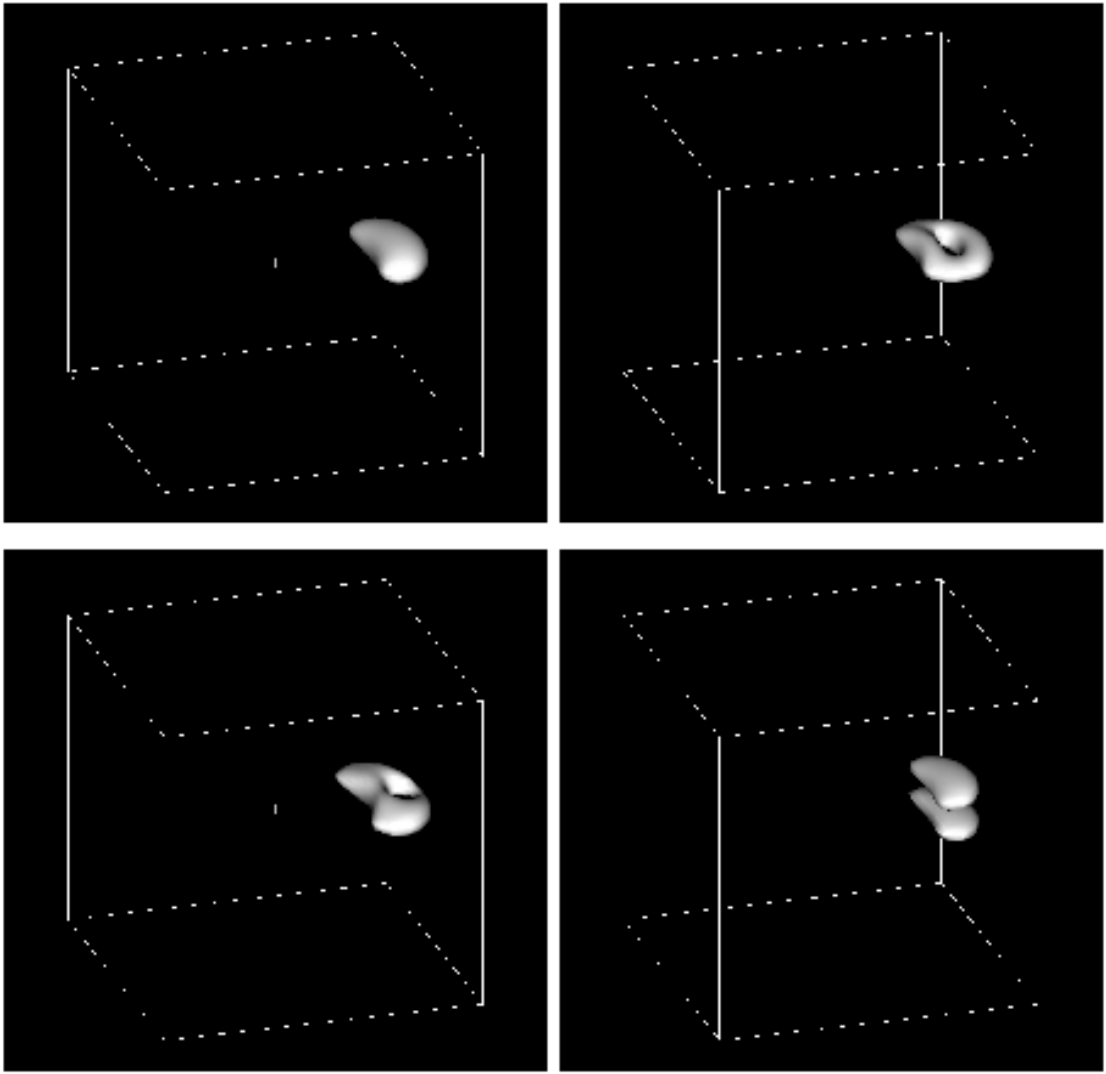,width=15cm}}
\caption{Isovalue plots (at 30\% of the maximum value) 
of non-dispersive wave-packets 
in the three-dimensional
hydrogen atom driven by a circularly polarized microwave field. 
Frequency of
the driving $\omega=1/60^3$, amplitude 
$F_0=0.04442$. 
In the laboratory frame, the wave-packets 
propagate -- without changing their shapes -- along a circular
trajectory 
centered around the nucleus indicated by a cross.  
The cube edges measure 10000 Bohr radii. The microwave polarization plane
is horizontal with the field pointing to the right. The
four wave-packets shown represent the ground state wave-packet $|0,0,0 
\rangle$ (top left), and the excited states $|1,0,0\rangle$ (bottom left),
$|0,1,0\rangle$ (top right), and $|0,0,1\rangle$ (bottom right).
Eventually, the microwave field will ionize such states, but their
lifetimes are extremely long, of the order of thousands to millions
of Kepler periods. 
}
\label{circwp3d}
\end{figure}

Let us now turn to the realistic 3D model of the atom.
Fig.~\ref{circwp3d} shows an isovalue contour of several 
non-dispersive wave-packets
for the hydrogen atom driven by a microwave
field with frequency
$\omega=1/60^3$, i.e., roughly resonant with 
the $n_0=60\to 59,61$ transitions 
(see eqs.~(\ref{cprinc},\ref{resonance_condition},\ref{omk})). 
The 
best localized wave-packet is the ground state $|0,0,0\rangle$, while
the three other states are excited by one quantum in either of 
the normal modes $\omega_{\pm,z}$, 
and are therefore significantly more extended in space.
Again, as already mentioned in sections~\ref{LIN1D} and \ref{LIN3D}, these
wave-packet eigenstates have finite, but extremely long life-times (several
thousands to millions of Kepler orbits), due to the
field induced ionization. For a detailed discussion of their decay
properties see section~\ref{ION}.

\begin{figure}
\centerline{\psfig{figure=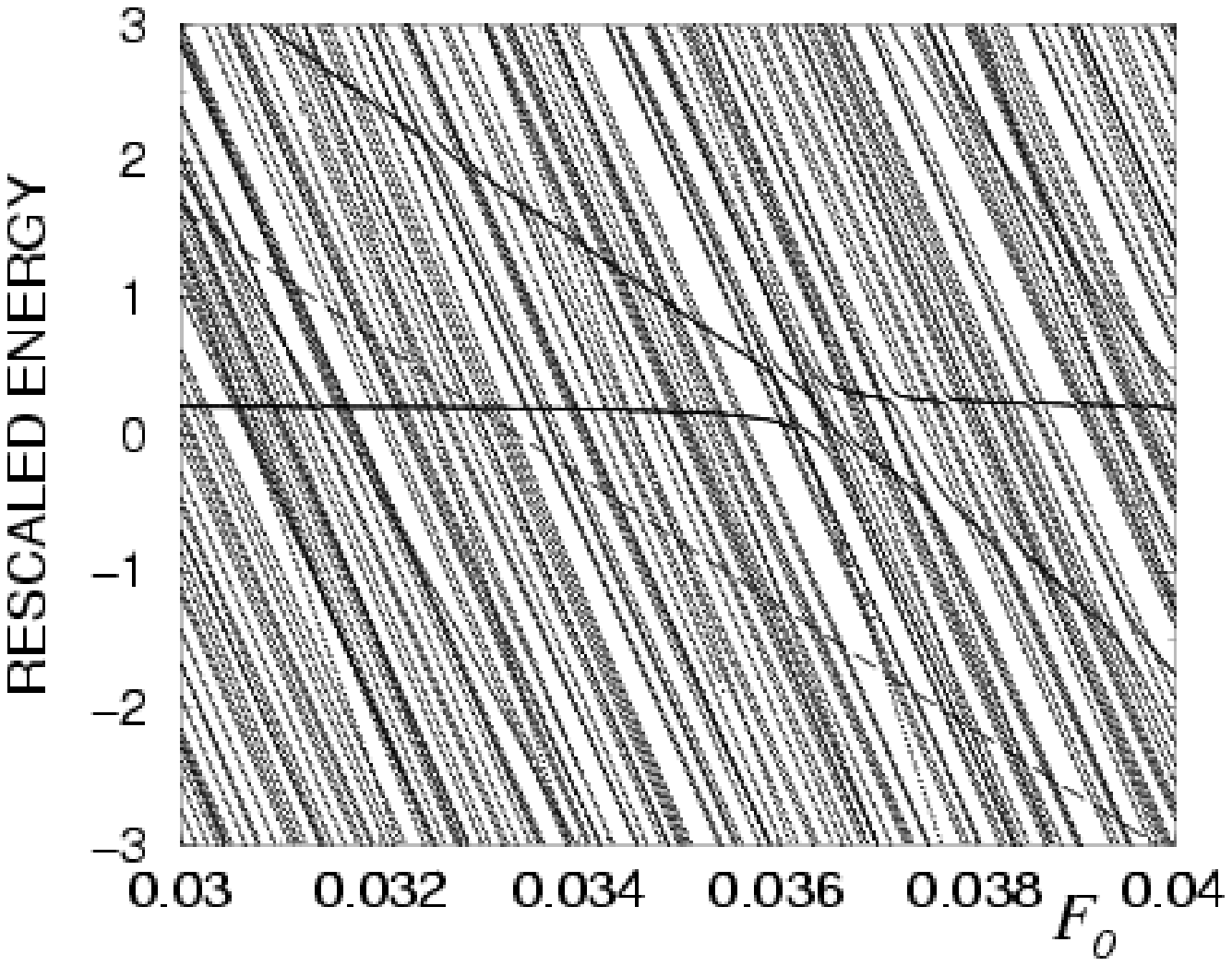,width=12cm}}
\caption{Spectrum of the two-dimensional hydrogen atom in
a circularly polarized microwave field 
of frequency $\omega=1/(60.5)^{3}$, 
as a function of the
scaled microwave amplitude $F_0=F\omega^{-4/3}$. 
In order to test the accuracy of the harmonic
prediction, 
we substract the semiclassical energy for the ground
state $|0,0\rangle$ wave-packet, eq.~(\protect{\ref{enharm}}), from the result
of the exact numerical diagonalization, and rescale the energy axis in units 
of the mean level spacing. The
almost horizontal line slightly above zero corresponds to the
non-dispersive ground state wave-packet, which typically undergoes small
avoided crossings with other Floquet states. A relatively large 
avoided crossing
occurs when two wave-packet-like states meet, as here 
happens around $F_0\simeq 0.036$.
The other (``colliding'') 
state is the excited wave-packet $|1,3\rangle$. The dashed
line indicates the harmonic prediction, eq.~(\ref{enharm}), for this state, 
which is obviously
less accurate. Note, however, that the slope is correctly predicted.
}
\label{solit2}
\end{figure}

As already demonstrated in 
the LP case (see figs.~\ref{lin3d_3},\ref{lin3d_4}), the semiclassical
prediction for the energies of the non-dispersive wave-packets
is usually excellent. In order to stress the (small) differences,
we plot in fig.~\ref{solit2} a part of the Floquet spectrum
of the two-dimensional model 
atom,
i.e. quasi-energy levels versus the (scaled) microwave amplitude,
after substraction of the prediction of the
harmonic approximation around the stable fixed point, eq.~(\ref{enharm}), 
for the 
ground state wave-packet $|0,0\rangle$.
The result is shown in units of the mean level spacing,
estimated\footnote{This estimate follows from the local energy splitting, 
$\sim n_0^{-3}$, divided by the number $n_0/2$ of photons needed to ionize the initial 
atomic state by a resonant driving field.} to be roughly $2/n_0^4$.
If the harmonic approximation was exact,
the ground state wave-packet would be represented by 
a horizontal line at zero.
The actual result is not very far from that, which proves that the
semiclassical method predicts the correct energy with an accuracy
mostly better than the mean level spacing. The other states of the system
appear as 
energy levels which rapidly evolve with $F_0$, and which  
exhibit extremely small
avoided crossings -- hence extremely small couplings -- with the wave-packet.
In the vicinity of such avoided
crossings, the energy levels are perturbed, 
the diabatic wave-functions mix (the wave-packet eigenstates 
get distorted),
and, typically, the lifetime of the
state decreases (induced by the coupling to the closest
Floquet state \cite{abuth,abu95a}, typically much less resistant against ionization, as we shall
discuss in detail in sec.~\ref{ION}).

Thus, strictly
speaking, when we speak of non-dispersive wave-packets as specific
Floquet states, we really have in mind a generic situation, {\em far} from
any avoided crossing. In particular, the examples of wave-packet states
shown in the figures above correspond to such situations.
The observed accuracy of the semiclassical approximation has 
important practical consequences:
in order to obtain the ``exact'' wave-packets numerically, 
we do not need many
eigenvalues for a given set of parameters. Using the
Lanczos algorithm for the partial diagonalization of a matrix, it is enough to extract
few (say five) eigenvalues only, 
centered on 
the semiclassical prediction. The accuracy of the latter
(a fraction of the mean level spacing) is
sufficient for a clear identification of the appropriate
quantum eigenvalue.
 Actually, in a real diagonalization of the Floquet Hamiltonian, we
identify the wave-packet states by both their vicinity 
to the semiclassical prediction
for the energy, 
and the large (modulus of the) slope of the
level w.r.t. changes of $F_0$, induced by its large dipole
moment in the rotating frame (by virtue of the Hellman-Feynman 
theorem \cite{cct92}).
The latter 
criterion is actually also very useful for the identification of excited
wave-packets in a numerically exact spectrum.
The state $|1,3\rangle$ (in the harmonic approximation)
presented in 
fig.~\ref{accharm} is precisely the excited
wave-packet which appears in fig.~\protect{\ref{solit2}} as a ``line''
with a negative slope, meeting the $|0,0\rangle$ state in a broad avoided
crossing around $F_0\simeq 0.036$. 
From that figure, it is
apparent that  the harmonic
prediction for the energy is not excellent for the  $|1,3\rangle$ state.
On the other hand, the slope of the Floquet
state almost matches the slope given by the harmonic
approximation, which confirms that the exact wave-function is still 
well approximated by its harmonic counterpart.
Note that also from the experimental point of view it is important to get accurate
and simple semiclassical estimates of the energies of the non-dispersive
wave-packets, since it may help in their preparation and unambiguous
identification. For a more detailed discussion, see section~\ref{SPEC}.

\begin{figure}
\centerline{\psfig{figure=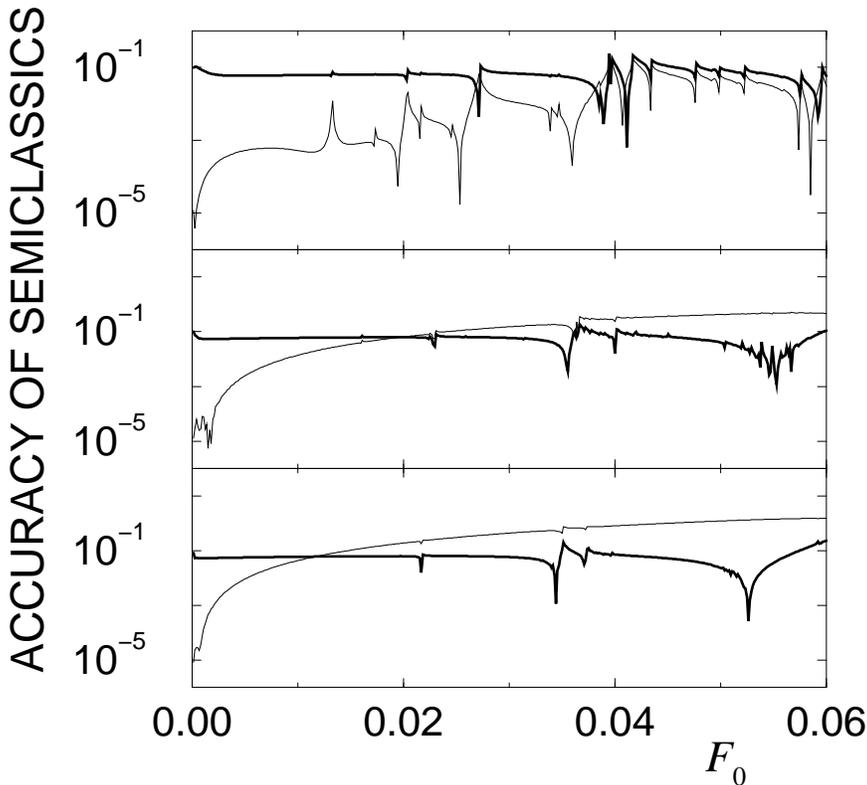,width=14cm,bbllx=0pt,bblly=50pt,bburx=600pt,bbury=400pt}}
\caption{
Difference between  the exact quantum energy of a non-dispersive
wave-packet of a 2D hydrogen atom in a 
circularly polarized microwave
field, and two different semiclassical predictions, as a function of 
the scaled microwave field amplitude, $F_0$.
The thick lines are obtained from 
a harmonic approximation 
of the motion around
the stable equilibrium point (in the rotating frame), eq.~(\ref{enharm2d}) 
and the 
thin lines use a quantum treatment (Mathieu approach, 
sec.~\protect\ref{section_mathieu})
of the motion in the 
resonance island,
combined with a semiclassical treatment of the secular motion. 
From top to bottom
$\omega=1/(30.5)^3,\ 1/(60.5)^3, \ 1/(90.5)^3$, corresponding to wave-packets
associated with Rydberg states of principal quantum number
$n_0=30,60,90$. The energy difference is expressed in units of the mean
level spacing, estimated by $n_0^4/2$.
The prediction of the Mathieu approach is
consistently better for low $F_0$, and the harmonic approximation
 becomes clearly superior for larger $F_0.$ For larger and larger $n_0,$
 the harmonic approximation is better and better. Note that both
 approximations make it possible to estimate the energy of the non-dispersive
 wave-packet with an accuracy better than the mean level spacing, allowing
 for its simple and unambiguous 
extraction from exact numerical data.}
\label{semicirc}
\end{figure}

It is interesting to compare the accuracy of 
the harmonic approximation
to the pendulum 
description outlined previously in sec.~\ref{reso_an}. The latter
results from lowest order perturbation theory in $F$.
Taking the small $F$ limit we get $\omega_+,\omega_z
\rightarrow
\omega$, and $\omega_-\rightarrow \omega\sqrt{3F_0}$, for the 
harmonic modes, see 
eqs.~(\ref{omegas}-\ref{defQ}). 
The latter result coincides -- as it should -- with the pendulum prediction,
eqs.~(\ref{omega_harmonic},\ref{omega_harmonic_cp}). Similarly, the energy of the stable
equilibrium point, eq.~(\ref{wps}), becomes at first order in $F$:
\begin{equation}
E_{\rm eq}= - \frac{3}{2n_0^2} + n_0^2\ F + O(F^2)
\end{equation}
which coincides with
the energy of the center of the resonance island, see section~\ref{reso_an}.
Thus, the prediction of the resonance analysis agrees with the harmonic
approximation in the rotating frame. 
For a more accurate estimate of the validity of both approaches, we have
calculated -- for the 2D model of the atom, but similar
conclusions are reached in 3D -- 
the energy difference between the exact quantum result  
and the prediction
using a semiclassical
quantization of the secular motion in the $(M,\phi)$ plane together with
the Mathieu method in the $(\hat{I},\hat\theta )$ plane on the one side,
and the prediction of the harmonic approximation around the fixed point,
eq.~(\ref{enharm}), on the other side. 
In fig.~\ref{semicirc}, we compare the results coming from both approaches.
As expected, the semiclassical approach
based on the Mathieu equation is clearly superior for very small microwave
amplitudes, as it is ``exact" at first order in $F$. On the other hand,
for the harmonic approximation to work well, the 
island around the fixed point has to be sufficiently large.
Since the size of the island increases as $\sqrt{F},$ the harmonic 
approximation
may become valid only for sufficiently large microwave amplitudes, when 
there is at least
one state trapped in the island. As seen, however,
in fig.~\ref{semicirc}, the harmonic
approximation yields a satisfactory prediction for the wave-packet energy
(within few \% of the mean spacing) almost everywhere. 
For increasing $n_0,$ the harmonic approximation is better and better
and the Mathieu approach is superior only over a smaller and smaller range
of $F_0=Fn_0^4$, 
close to 0. Still, both
approaches give very good predictions for the typical values of $F_0$ used in
the following, say $F_0\simeq 0.03$. The spikes visible in the figure are
due to the many small avoided crossings visible in fig.~\ref{solit2}.

While we have shown some exemplary  wave-packets for few values of
 $n_0$ and $F$
only, they generally look very similar provided that
\begin{itemize}
\item $n_0$ is sufficiently large, say $n_0>30$. For smaller $n_0$,
the wave-packet looks a bit distorted and one observes some
deviations from the harmonic approximation (for a more detailed 
discussion of this
point see secs.~\ref{ION} and \ref{SPO}, where 
ionization and spontaneous emission
of the wave-packets are discussed);
\item $F_0$ is sufficiently large, say $F_0> 0.001$, 
such that
the resonance island can support at least one state. For smaller
$F_0,$ the wave-packet becomes more extended in the angular coordinate, 
since less
 atomic circular states are significantly coupled, see fig.~\ref{cp2df};
\item $F_0$ is not too large, say smaller than $F_0\simeq0.065$. 
Our numerical data suggest that the upper limit is not given
by the limiting value of $q=8/9$,
for which the fixed point is still stable. The limiting
value appears to be rather linked to the  
$1:2$ resonance between the $\omega_+$ and 
$\omega_-$ modes, which occurs approx. at $F_0\simeq0.065$.
\item In particular, the value $q=0.9562$ (i.e. $F_0\simeq 0.04442$),
corresponding to 
optimal 
classical stability of the fixed point, advertised in~\cite{ibb94} as the
optimal one, is by no means favored. A much broader range of microwave
amplitudes is available (and equivalent as far as the ``quality'' of the
wave-packet is concerned). What is much more relevant, is the presence of
some accidental avoided crossings with other Floquet states.
\end{itemize}
Still, these are no very restrictive conditions, and we are left with 
a broad range
of parameters favoring the existence of nondispersive wave-packets, 
a range which
is experimentally fully accessible (see section~\ref{EXP} for a more
elaborate discussion of experimental aspects). 
\begin{figure}
\centerline{\psfig{figure=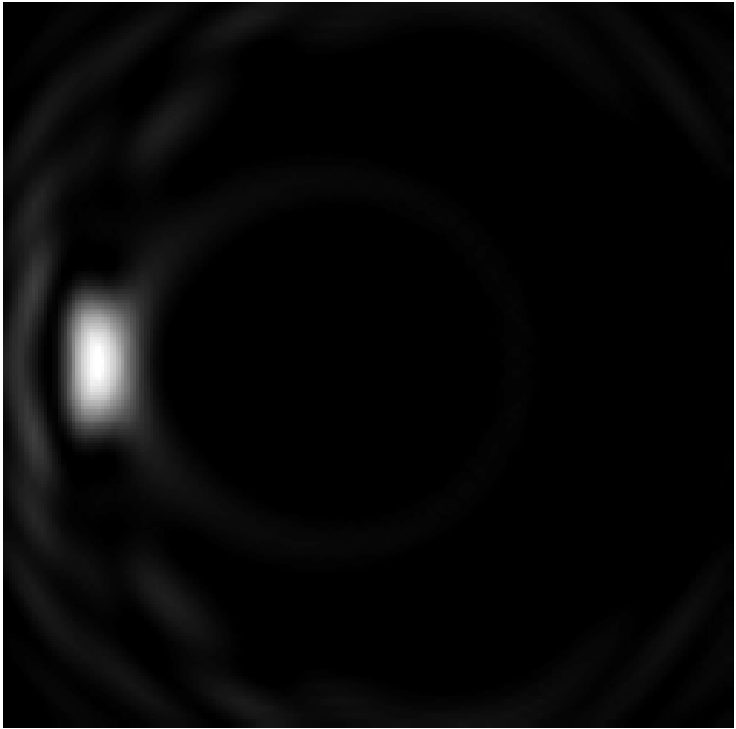,width=10cm}}
\caption{Floquet eigenstate of the two-dimensional hydrogen atom in a
circularly polarized microwave field. This state is 
partially localized on the unstable
equilibrium point (in the rotating frame), 
see eqs.~(\ref{wp0})-(\ref{wpu}), for $F_0=0.057$ and $n_0=60$.
This localization is of purely classical origin. The nucleus is at the center
of the figure which extends over $\pm 5000$ Bohr radii. The microwave
field points to the right.
}
\label{cp2unstb}
\end{figure}

Finally, in analogy with the LP case, we may consider 
Floquet states localized on the unstable fixed point associated 
with the principal resonance island. 
From
the discussion following eq.~(\ref{wpu}), this point is located 
opposite to the stable fixed point, on the other side of the nucleus. 
An example of
such a state is shown in fig.~\ref{cp2unstb}, for an amplitude of the microwave
field that ensures that most of the nearby Floquet states ionize rather 
rapidly.
The eigenstate displayed in the figure  
lives much longer (several thousands of Kepler periods). The
localization in the vicinity of an unstable fixed point, in analogy
to the LP case discussed previously, is of purely classical origin.
As pointed out in \cite{delande97},
such a localization must not be confused with scarring \cite{heller84} --
a partial localization on an unstable periodic orbit embedded in a chaotic
sea -- which disappears in the semiclassical limit \cite{heller84,bogomolny88a}.

\subsection{Rydberg states in elliptically polarized microwave fields}
\label{EP}

The origin of non-dispersive wave-packets being their localization inside
the resonance island (locking the frequency of the electronic motion
onto the external drive) suggests that
 such wave-packets are quite robust and 
should  exist not only for CP and LP, but
also for arbitrary elliptical polarization (EP). 

The possible existence
of nondispersive wave-packets for EP  was mentioned 
in \cite{lee97,Bth}, using the classical ``pulsating SOS'' approach. The
method, however, 
did not allow for quantitative predictions, and was restricted to
elliptic polarisations very close to the CP case.
However, the robustness of such wave-packets for arbitrary 
EP is obvious 
once the localization mechanism inside the resonance island 
is well understood \cite{sacha98b,sachath}.

Let us  consider an elliptically polarized driving field of constant
amplitude. With the ellipticity parameter $\alpha\in[0;1]$,
\begin{equation}
V=F(x\cos\omega t + \alpha y\sin\omega t)
\label{v_ep}
\end{equation}
establishes a continuous transition between linear ($\alpha =0$) and circular 
($\alpha =1$) polarization treated in the two preceding 
chapters\footnote{Note, however, that $\alpha =0$ defines a linearly polarized field 
along the $x$-axis, i.e., {\em in} the plane of elliptical polarization for 
$\alpha >0$. In sec.~\ref{LIN}, the polarization vector was chosen along
the $z$-axis. The physics is of course the same, but the algebraic expressions
are slightly different, requiring a rotation by an angle $\beta=\pi/2$
around the $y$-axis.}.
This general case is slightly more complicated than both  
limiting cases LP and CP. 
For LP microwaves (see section \ref{LIN3D}), the conservation of
the angular momentum projection onto the polarization axis, $M$, makes
the dynamics effectively two-dimensional. For the CP case,
the transformation (\ref{hrot}) 
to the frame rotating
with the microwave frequency removes the explicit 
time-dependence (see section \ref{CP}).
None of these simplifications is
possible in the general EP case, and the problem
is truly three dimensional {\em and} time-dependent.

To illustrate the transition from LP to CP via EP, the 
two-dimensional model of the atom is sufficient, and we shall restrain 
our subsequent treatment to this computationally less involved case.
The
classical resonance analysis for EP microwave ionization has been
described in detail in~\cite{sacha97,sacha98c}. It follows closely the lines described in detail
in sections~\ref{LIN} and \ref{CP}. By expanding the
perturbation, eq.~(\ref{v_ep}), in the action-angle coordinates $(I,M,\theta,\phi)$
of the two-dimensional atom (see sec.~\ref{action-angle_hydrogen}), 
one obtains the following secular Hamiltonian:

\begin{equation}
{\cal H}_{\rm sec}={\hat P_t}-\frac{1}{2{\hat I}^2}-\omega {\hat I}+
F\left[{\cal V}_1(\hat{I},M,\phi;\alpha)\cos\hat\theta-
{\cal U}_1(\hat{I},M,\phi;\alpha)\sin\hat\theta\right],
\label{EPr_0}
\end{equation}
with
\begin{eqnarray}
 {\cal V}_{1}(\hat{I},M,\phi;\alpha) &=& \cos\phi \left(X_1 + \alpha Y_1\right),\nonumber \\
 {\cal U}_{1}(\hat{I},M,\phi;\alpha) &=& \sin\phi \left(Y_1 + \alpha X_1\right).
\label{calUV}
\end{eqnarray}

This can be finally rewritten as:
\begin{equation}
{\cal H}_{\rm sec}={\hat P_t}-\frac{1}{2{\hat I}^2}-\omega {\hat I}+
F\chi_1(\hat{I},M,\phi;\alpha)\cos(\hat\theta+\delta_1).
\label{EPr}
\end{equation}
Both, $\chi_1$ and $\delta_1$,
depend on the shape and orientation of the   
electronic elliptical trajectory,
as well as on 
$\alpha$,
and are given by:
\begin{equation}
\chi_1(\hat{I},M,\phi;\alpha)= \sqrt{{\cal V}_1^2 + {\cal U}_1^2}, 
\end{equation}
and
\begin{equation}
\tan \delta_1(M,\phi;\alpha) = \frac{{\cal U}_1}{{\cal V}_1},
\end{equation}
which is once more the familiar form of a system with a resonance island 
in the $(\hat{I},\hat{\theta})$ plane. The expressions obtained are in fact
very similar to the ones we obtained for 
the three-dimensional atom exposed to a
circularly polarized microwave field, eqs.~(\ref{hamsccir2}-
\ref{phasecirc}), in section \ref{reso_an}. This is actually not surprising:
the relevant parameter for the transverse dynamics is the magnitude
of the atomic dipole oscillating with the driving field, i.e. the
scalar product of the oscillating atomic dipole with the polarization vector.
The latter can be seen either as the projection of the 
oscillating atomic dipole 
onto the polarization plane or as the projection of the polarization vector
on the plane of the atomic trajectory. If one considers a three-dimensional 
hydrogen
atom in a circularly polarized field, the projection of the polarization 
vector
onto the plane of the atomic trajectory is elliptically polarized
with ellipticity $\alpha=L/M.$ This is another method to rediscover 
the Hamiltonian~(\ref{EPr}) from Hamiltonian (\ref{hamsccir2}).
\begin{figure}
\centerline{\psfig{figure=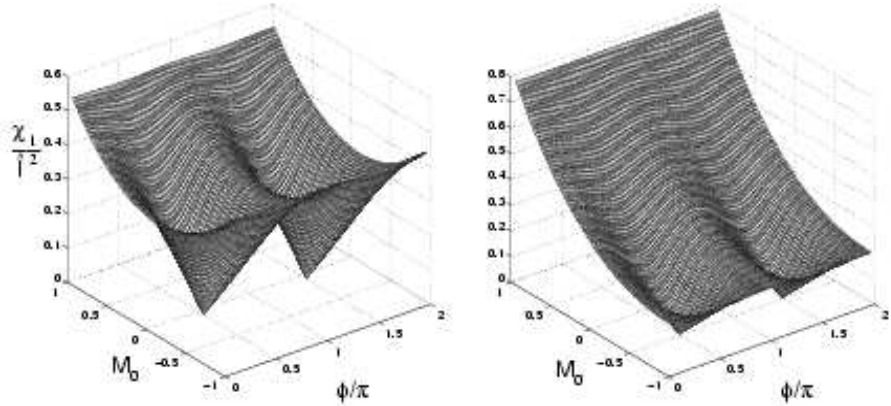,width=12cm}}
\caption{The scaled effective perturbation $\chi_1/\hat{I}^2$
driving the transverse/angular
motion of a two-dimensional hydrogen atom exposed to a resonant,
elliptically polarized microwave field, 
plotted as a function of
the scaled angular momentum, $M_0=M/\hat{I}$, and of the angle 
$\phi$ between the
Runge-Lenz vector and the major axis of the polarization ellipse.
Left and right panels correspond to
$\alpha=0.1$ and $\alpha=0.6$, respectively. Non-dispersive
wave-packets are localized around the maxima of this effective potential,
at $M_0=\pm 1,$ and are circularly co- and contra-rotating (with
respect to the microwave field) around the nucleus 
(see fig.~\protect\ref{ep2}).
}
\label{ep1}
\end{figure}

To obtain a semiclassical estimation of the energies of the nondispersive 
wave-packets 
 we proceed 
precisely in the same way as for LP and CP. Since the radial motion in 
$(\hat{I},\hat{\theta})$ is much faster than in the transverse/angular degree 
of freedom defined by $(M,\phi)$, we first quantize the effective perturbation
$\chi_1(\hat I,M,\phi;\alpha)$ driving the angular motion. 
Fig.~\ref{ep1} shows 
$\chi_1/\hat{I}^2$, as a function of $M_0=M/\hat{I}$ and 
$\phi$, for two different values of the driving field ellipticity $\alpha$. 
Note that $\chi_1$ becomes more symmetric as $\alpha\rightarrow 0$, since 
this limit defines the LP case, where the dynamics cannot depend on the 
rotational sense of the electronic motion around the nucleus. 
The four extrema 
of $\chi_1$ define the possible wave-packet eigenstates. Whereas the minima
at $\phi=\pi/2,3\pi/2$ correspond to elliptic orbits of intermediate 
eccentricity $0<e<1$ perpendicular to the driving field major axis, the 
($\phi$-independent)
maxima at $M_0=\pm 1$ define circular orbits which co- or
contra-rotate 
with the driving field. 
In the limit $\alpha\rightarrow 0$, the $M_0=\pm1$ maxima are associated with
the same value of $\chi_1$. 
Hence, the
actual Floquet eigenstates appear as tunneling doublets in the 
Floquet spectrum, 
each member of the doublet being a superposition of 
the co- and 
contra-rotating wave-packets.
 
The quantization of the fast motion in the $(\hat{I},\hat{\theta})$ plane 
is similar to the one already performed in the LP and CP cases.
As we already observed (see fig.~\ref{lin3d_2}), the size 
of the resonance island in the $(\hat{I},\hat{\theta})$ plane is 
proportional to 
$\sqrt{\chi_1}$. Correspondingly, also the localization properties of the 
wave-packet along the classical trajectory improve with increasing $\chi_1$.
We therefore conclude from fig.~\ref{ep1} that the eigenstates corresponding 
to the minima of $\chi_1$ cannot be expected to exhibit strong 
longitudinal localization,
whereas the eigenstates localized along the circular orbits at the maxima of 
$\chi_1$ can.

\begin{figure}
\centerline{\psfig{figure=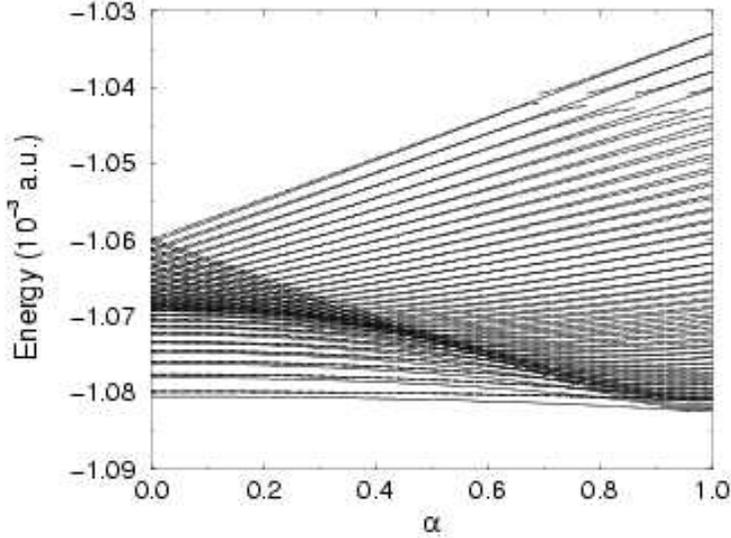,width=10cm}}
\caption{Energy levels of the two dimensional hydrogen atom driven by 
a resonant,
elliptically polarized microwave field of scaled amplitude $F_0=0.03$,
for $n_0=21$, as a function of the 
field ellipticity $\alpha$
(the resonant frequency of the microwave is $\omega=1/(21.5)^3)$. 
Full lines: semiclassical prediction; dotted lines: exact numerical result
for the states originating from the $n_0=21$ hydrogenic manifold. The 
non-dispersive wave-packets are the states originating from the upper doublet
at $\alpha=0.$ The ascending (resp. descending) energy level is associated
with the wave-packet co- (resp. contra-) rotating with the microwave field.
}
\label{ep3}
\end{figure}

Fig.~\ref{ep3} compares the semiclassical prediction obtained by quantization
of $\chi_1$ and ${\mathcal H}_{\rm sec}$ 
(following the lines already described in section~\ref{LIN3D},
for the ground state $N=0$ in the resonance island) 
to the exact quasienergies (determined by numerical diagonalization of the
Floquet Hamiltonian), for
$\alpha$ varying from LP to CP.
The agreement is excellent, with slightly 
larger discrepancies between the semiclassical and the exact results for the
states with smallest energy. For those states the resonance island is 
very small
(small $\chi_1$), what explains the discrepancy.
\begin{figure}
\centerline{\psfig{figure=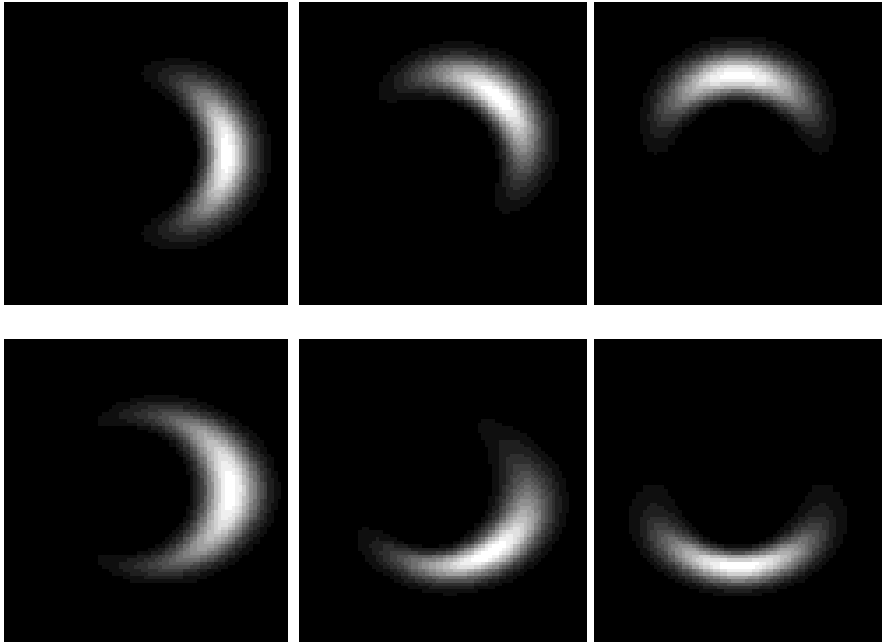,width=12cm}}
\caption{Non-dispersive wave-packets of the two dimensional hydrogen 
atom exposed to an elliptically polarized, resonant microwave field. 
Scaled microwave amplitude $F_0=0.03$
(for resonant 
principal quantum number  $n_0=21$), 
and ellipticity
$\alpha=0.4$. Top row: non-dispersive wave-packet moving on a 
circular orbit corotating with 
the microwave field, for phases $\omega t=0,\pi/4,\pi/2$ 
(from left
to right). This wave-packet evolves into the eigenstate represented in 
fig.~\protect\ref{cp2df}, under continuous increase of the ellipticity 
to $\alpha=1$.
Bottom row: 
non-dispersive wave-packet launched along the same circular orbit, but 
contra-rotating with  the driving field 
(for the 
same phases).
Note that, while the co-rotating wave-packet almost preserves its shape 
during the 
temporal evolution,
the contra-rotating one exhibits significant distortions during one 
field cycle, as a direct
consequence of its complicated level dynamics shown in 
fig.~\protect\ref{ep3}. 
Still being an exactly time-periodic Floquet
eigenstate, it regains its shape after
every period of the microwave.
The size of each box extends over 
$\pm 800$ Bohr radii, in both $x$ and $y$ directions,
with the nucleus in the middle. The major axis of the polarization
ellipse is along the horizontal $x$ axis and the microwave field points
to the right at $t=0.$  
}
\label{ep2}
\end{figure}

The highest lying state in fig.~\ref{ep3}, ascending with $\alpha$,
is a non-dispersive wave-packet state located on the circular orbit and
corotating with the EP field. It is shown in fig.~\ref{ep2} for $\alpha=0.4$.
As mentioned above, the corresponding counterrotating wave-packet 
is energetically
degenerate with the co-rotating one for $\alpha=0$. Its energy decreases with
$\alpha$ (compare fig.~\ref{ep3}). It is shown in the bottom row in
fig.~\ref{ep2} for $\alpha=0.4$. While the corotating wave-packet
preserves its shape for all $\alpha$ values (except at isolated avoided
crossings) the counter-rotating wave-packet undergoes a series of strong
avoided crossings for $\alpha >0.42$, progressively loosing its localized
character. This is related to a strong decrease of the maximum 
of $\chi_1$ at $M_0=-1$ with $\alpha$, clearly visible in fig.~\ref{ep1}.

While we have discussed the 2-dimensional case only, 
the CP situation (compare section \ref{CP}) indicates that for
sufficiently large $\alpha$, the important resonant motion occurs in the
polarization plane, being stable versus small deviations in the
$z$ direction.
Thus the calculations presented above are also
relevant for the real three dimensional world, provided $\alpha$ is not far from 
unity \cite{sachath,sacha98b}. For arbitrary $\alpha$, a full 3D
analysis is required. While this is clearly more involved, the general scenario
of a wave-packet anchored to a resonance island will certainly prevail.

\section{Manipulating the wave-packets}
\label{MA}

We have shown in the previous sections that non-dispersive wave-packets 
are genuine 
solutions of the Floquet eigenvalue problem, eq.~(\ref{calhq}), under resonant 
driving, for arbitrary polarization of the driving field. The semiclassical 
approximation used to guide our exact numerical approach directly demonstrates 
the localization of the electronic density in well defined regions of phase space,
which protect the atom against ionization induced by the 
external field (see, however, sec.~\ref{ION}). We have also seen that classical
phase space does not only undergo structural changes under changes of the 
driving field amplitude (figs.~\ref{lin1d_00}, \ref{lin1d_1},\ref{lin1d_4a}), 
but also under changes of the driving field ellipticity (fig.~\ref{ep1}).
Therefore, the creation of non-dispersive wave-packets can be conceived as an
easy and efficient means of quantum control, which allows the 
manipulation and the controlled transfer of quantum population accross phase space.
In particular, one may imagine the creation of a wave-packet moving along 
the polarization axis of a linearly
polarized microwave field. A subsequent,
smooth change through elliptical to finally
circular polarization allows to transfer the electron to a circular
orbit.

Adding additional static fields to the Hamiltonian (\ref{calhq}) provides us 
with yet another handle to control the orientation and shape of highly excited 
Rydberg trajectories, and, hence, to manipulate the localization properties 
of nondispersive wave-packet eigenstates in configuration and phase space.
The key point is that trapping inside the nonlinear resonance
island is a robust mechanism which protects the
non-dispersive wave-packet very efficiently from imperfections. This allows
to adjust the wave-packet's properties at will, just by adiabatically
changing the properties of the island itself. Moreover, when the strength of 
the external
perturbation increases, chaos generically invades a large part of 
classical phase space, but the resonance islands most often survives. 
The reason is that the phase locking phenomenon
introduces various time scales in the system, which have different
orders of magnitude.
That makes the system quasi-integrable (for example
through some adiabatic approximation {\em \`a la Born-Oppenheimer}) and
-- locally -- more resistant to chaos. 

Hereafter, we discuss two possible alternatives of manipulating the 
wave-packets.
One
is realized by adding a static
electric field to the 
LP microwave drive \cite{sacha98a}.
Alternatively, the addition of 
a static magnetic field to CP driving enhances the region of classical 
stability, and extends the range of applicability of the 
harmonic approximation \cite{lee97,farrelly95a,farrelly95,brunello96,cerjan97,lee95,brunello97}. 

\subsection{Rydberg states in linearly polarized microwave and static electric
fields}
\label{staticmw}
\label{LINF}

Let us first consider a Rydberg electron driven by a resonant, linearly
polarized microwave, in the presence of a static electric field.
We already realized 
(see the discussion in sec.~\ref{LIN3D}) that 
the classical 3D motion of the driven Rydberg electron is angularly unstable
in a LP microwave field. It turns out, however, 
that a stabilization
of the angular motion is possible by the addition of a
static
electric field $F_s$ parallel to the microwave polarization axis
\cite{sacha98a,leopold86,leopold87}.
The corresponding Hamiltonian reads:
\begin{equation}
H=\frac{p_x^2+p_y^2+p_z^2}{2}-\frac{1}{r}+Fz\cos\omega t +F_s z,
\label{hcart}
\end{equation}
which we examine 
in the vicinity of the $s=1$ resonance. As in sec.~\ref{LIN3D}, the 
angular momentum projection $M$ on the $z$ axis remains a constant of 
motion, and we shall assume $M=0$ in the following.
Compared to the situation of a pure microwave field, there is an
additional time scale, directly related to the static field.
Indeed, in the presence of a perturbative static field alone, it is known 
that the Coulomb degeneracy of the hydrogenic energy levels (in $L$)
is lifted. The resulting eigenstates are combinations of the $n_0$ substates
of the $n_0$ manifold (for $0 leq L \leq n_0-1$)\footnote{These states 
are called ``parabolic" states, since the
eigenfunctions are separable in parabolic coordinates \protect\cite{landau2}.}.
The associated energy 
levels are equally spaced by a quantity proportional to $F_s$ 
($3n_0F_s$ in atomic units). Classically, the trajectories are no longer
closed but rather Kepler ellipses which slowly librate around the static 
field axis,
periodically changing their shapes, at a (small) frequency  
$3n_0F_s\ll \omega_{\rm Kepler}=1/n_0^3.$
Thus, the new time scale associated with the static electric field
is of the order of $1/F_{s,0}$ Kepler periods, where
\begin{equation}
F_{s,0} = F_s n_0^4
\end{equation}
is the scaled static field. This is to be compared to the time scales $1/F_0$
and $1/\sqrt{F_0}$, which characterize the secular time evolution in the
$(L,\psi)$ and $(\hat{I},\hat{\theta})$ coordinates, respectively 
(see discussion in section~\ref{LIN3D}), in the
presence of the microwave field alone. To achieve confinement of the
electronic trajectory in the close vicinity of the field polarization axis, we
need $1/F_0\simeq 1/F_{s,0}$, with both, $F_s$ and $F$ small enough 
to be treated at first order.

If we now consider the $s=1$ resonance, we deduce the secular Hamiltonian
by keeping only the term which does not vanish after
averaging over one Kepler period. For the microwave field, this 
term was already identified in eq.~(\ref{hamscfin}).
For the static field, only the static Fourier component of the atomic dipole,
eqs.~(\ref{x0},\ref{y0}),
has a non-vanishing average over one period.
Altogether, this finally leads to
\begin{equation}
{\mathcal H}_{\rm sec}=\hat P_t -\frac{1}{2\hat{I}^2} -\omega \hat{I}
+ F_s X_0(\hat{I},L) \cos \psi  
 + F \chi_1(\hat{I},L,\psi) \cos(\hat \theta+\delta_1).
\end{equation}
Since the last two terms of this Hamiltonian depend differently on
$\hat{\theta}$, it is no more possible, as it was in the pure LP case (see
section \ref{LIN3D}), to perform the quantization of the slow LP motion
first. Only the secular approximation \cite{cct92} which consists in
quantizing first the fast variables $(\hat{I},\hat{\theta})$, and subsequently
the slow variables $(L,\psi)$, remains an option for the general
treatment. However, since we are essentially interested in the wave-packet
eigenstate with optimal localization properties, we shall focus on the ground
state within a sufficiently large resonance island induced by a microwave
field of an appropriate strength. This motivates the harmonic expansion of the
secular Hamiltonian around the stable fixed point at 
\begin{equation}
\hat{I}=\hat{I}_1=\omega^{-1/3},\ \ \ \ \hat\theta = -\delta_1.
\end{equation}
with the characteristic
frequency, see eq.~(\ref{omega_harmonic}):
\begin{equation}
\Omega(\hat{I}_1L,\psi) = \frac{\sqrt{3F\chi_1(\hat{I}_1,L,\psi)}}{\hat{I}_1^2}.
\end{equation} 
Explicit evaluation of the ground state energy of the locally harmonic
potential yields 
the effective Hamiltonian for the slow motion in 
the $(L,\psi)$ plane:
\begin{equation}
H_{\rm eff}= -\frac{\Omega(L,\psi)}{2} +F\chi_1(L,\psi)
+F_sX_0(L) \cos \psi ,
\label{slow}
\end{equation}
where all quantities are evaluated at $\hat{I}=\hat{I_1}.$
For the determination of the angular localization properties of the wave-packet
it is now sufficient to inspect the extrema of $H_{\rm eff}$.

\begin{figure}
\centerline{\hbox{
\psfig{figure=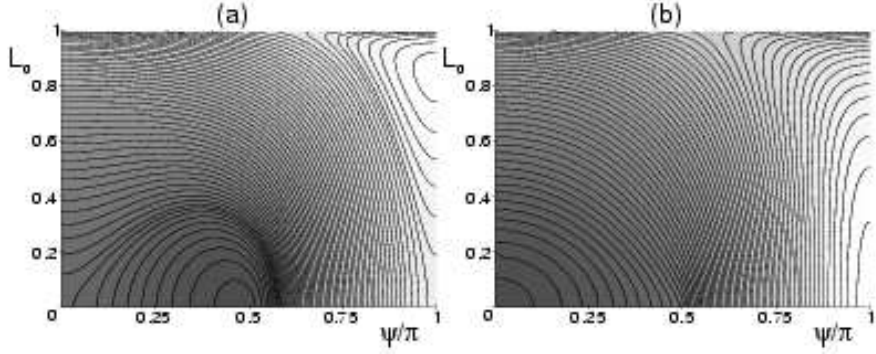,width=12cm}}}
\caption{Contours of the effective Hamiltonian,
eq.~(\protect{\ref{slow}}), in the
$(L_0,\psi)$ plane (with $L_0=L/n_0$ the scaled angular momentum,
and $\psi$ the angle between the major axis of the elliptical trajectory
and the field axis). 
The potential surface generates the slow 
evolution
of the angular coordinates of the 
Kepler trajectory of a Rydberg electron exposed to collinear,
static and resonant microwave electric fields.  
Initial atomic principal quantum number $n_0=60$; 
scaled microwave amplitude $F_0=Fn_0^4=0.03$, and scaled static field amplitude
$F_{s,0}=0.12F_0<F_{s,c}$ (a),
$F_{s,0}=0.25F_0>F_{s,c}$ (b), with $F_{s,c}$ the critical static field 
amplitude defined in eq.~(\protect\ref{ec}). 
The lighter the background, the higher
the effective energy. 
The contours are plotted at the semiclassical energies which 
quantize $H_{\rm eff}$
according to eq.(\ref{ebkp}), and thus represent the
60 eigenenergies shown in fig.~\protect\ref{f2}.
Observe the motion of the stable island along the
$\psi=\pi$ line (corresponding to the energetically highest 
state in the manifold), under changes of $F_s$.
 }
\label{f1}
\end{figure}
For $F_{s,0}=0$, 
we recover the pure LP case with a maximum along the line $L_0=1$ 
(circular state), 
and a minimum at $L_0=0$, $\psi=\pi/2$ (see fig.~\ref{lin3d_1}), 
corresponding to 
a straight line orbit perpendicular to the field. 
For increasing
$F_{s,0}$, the maximum moves towards lower values of $L_0$, and contracts
in $\psi$, whereas the minimum approaches $\psi=0$ for constant $L_0=0$, 
see fig.~\ref{f1}. It is easy to show that there exists
a critical value $F_{s,c}$ of the static field, depending on $\hat{I}_1$,
\be
F_{s,c}=\frac{2}{3}\left\vert F_0J'_1(1)
-\frac{\sqrt{3F_0J'_1(1)}}{4\hat{I}_1}\right\vert
   \simeq 0.217F_0 -0.164\frac{\sqrt{F_0}}{\hat{I}_1},
\label{ec}
\ee
above which both fixed
points reach $L_0=0.$ Then, in particular, the maximum 
at $L_0=0$, $\psi=\pi$, corresponds to a straight line orbit {\em parallel} 
to
$\vec{F}_s$. Note that in the classical limit, $\hat{I}_1\rightarrow\infty$,
eq.~(\ref{ec}) recovers the purely
classical value \cite{leopold87} for angular stability of the 
straight line orbit 
along the polarization axis, as it should.
\begin{figure}
\centerline{\psfig{figure=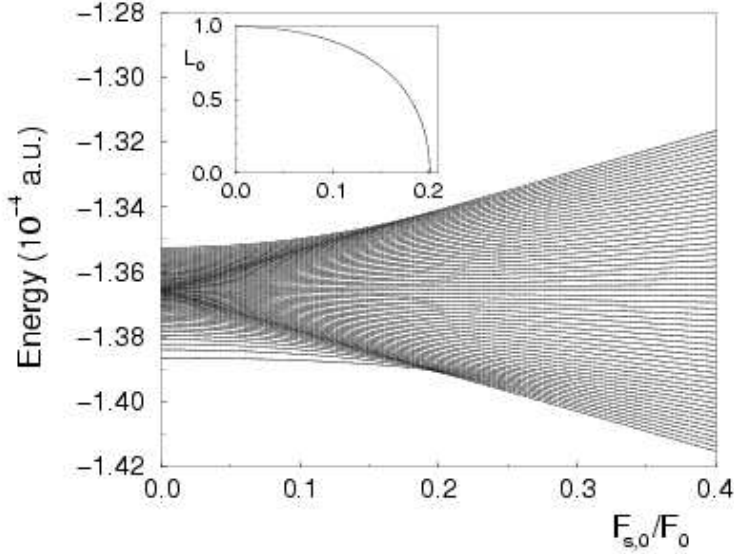,width=10cm}}
\caption{Semiclassical energy levels of the resonantly driven manifold 
$n_0=60$ of the hydrogen atom, as a function of the ratio of the scaled static 
electric field strength $F_s$ to the scaled 
microwave amplitude $F_0$, for fixed $F_0=Fn_0^4=0.03$. The insert shows 
the scaled angular momentum $L_0=L/n_0$ of the stable fixed point 
$(L_0,\psi=0)$, see fig.~\protect\ref{f1},
as a function of the same variable. The corresponding trajectory evolves
from a circular orbit coplanar with the polarization axis to a straight
line orbit stretched along this axis, via orbits of intermediate eccentricity.
For $F_{s,0}>F_{s,c}$, eq.~(\ref{ec}), the stable fixed point is stationary
at $L_0=0$. The corresponding wave-packet state, localized in
the vicinity of the fixed point, is the energetically 
highest state in the spectrum.
For $F_{s,0}>F_{s,c}$, it is a completely localized wave-packet
in the three dimensional space, propagating back and forth along the
polarization axis, without spreading (see fig.~\ref{linf_wp}).
}
\label{f2}
\end{figure}
Therefore, by variation of 
$F_{s,0}\in[0,F_{s,c}]$, we are able to continuously tune 
the position of the maximum in 
the $(L_0,\psi)$
plane.
Consequently, application of an additional static electric field gives us 
control over the trajectory traced by 
the wave-packet.
This is further illustrated in fig.~\ref{f2}, through the 
{\em semiclassical} level dynamics of the
resonantly driven manifold originating from the $n_0=60$ energy shell, as a 
function of $F_s$. In the limit $F_s=0$, the spectrum is equivalent to the one
plotted in fig.~\ref{lin3d_3}(b). 
As the static field is ramped up, the highest lying state
(maximum value of the semiclassical quantum number $p$, for $F_s=0$, see 
fig.~\ref{lin3d_3}, and the right column of fig.~\ref{lin3d_5}) gets stretched 
along the static field direction and finally, for $F_{s,0}>F_{s,c}$, collapses onto
the quasi one dimensional wave-packet eigenstate bouncing off the nucleus along
a straight line Kepler trajectory. Likewise, the energetically lowest 
state of the manifold (at $F_s=0$, minimum value of the semiclassical 
quantum number $p$, left column of fig.~\ref{lin3d_5}) is equally rotated 
towards the direction defined by $\vec{F_s}$, but stretched in the opposite 
direction.
 
\begin{figure}
\centerline{\psfig{figure=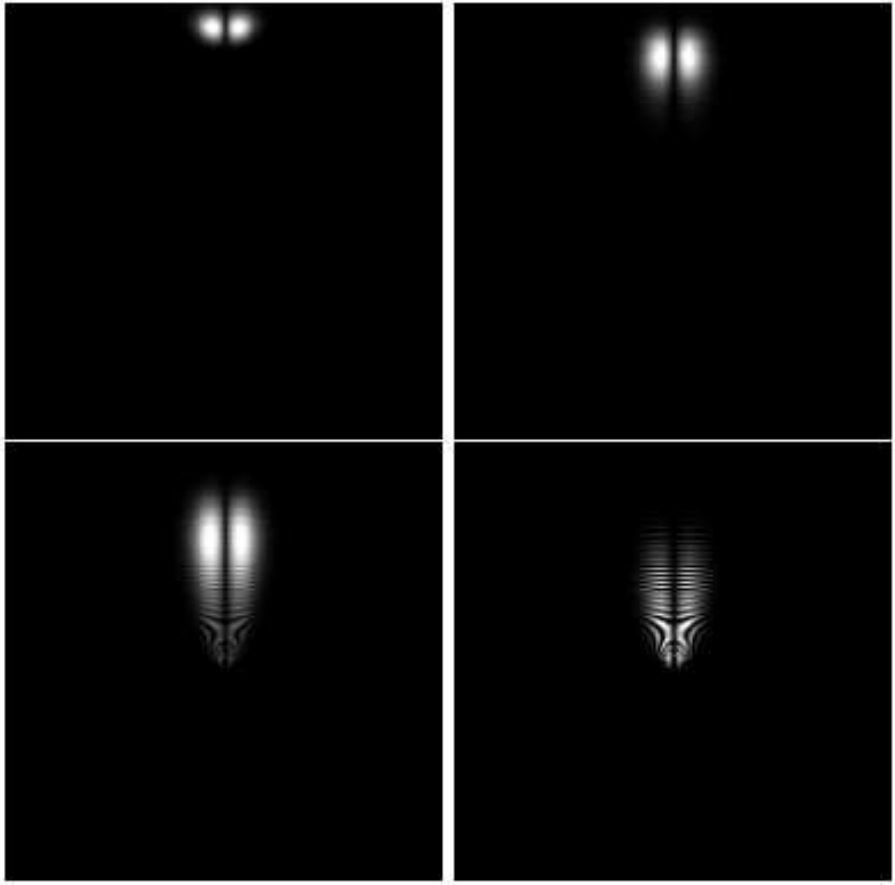,width=12cm}}
\caption{Temporal dynamics of the non-dispersive wave-packet of a
three-dimensional hydrogen atom exposed to a linearly polarized,
resonant microwave field, in the presence of a parallel static electric field. 
$F_0=0.03$, $F_{s,0}=0.009$, $F_{s,0}/F_0=0.3$, $n_0=60$. 
Driving field phases at the different stages of the wave-packet's evolution:
$\omega t= 0$ (top left), $\pi/2$ (top right),
$3\pi/4$ (bottom left), $\pi$ (bottom right). 
It is well localized in the three dimensions of space and repeats its shape
periodically (compare fig.~\protect\ref{lin1d_2} for the analogous dynamics in 
the restricted 1D model, where no additional static electric field is needed). 
The nucleus is at the center of the plot which extends
over $\pm 8000$ Bohr radii. The microwave polarization axis and the
static field are oriented along the vertical axis.
}
 \label{linf_wp}
\end{figure}

The existence of a non-dispersive wave-packet localized in all three
dimensions of space is confirmed by a pure quantum calculation, using
a numerically exact  
diagonalization of the Floquet Hamiltonian. Figure~\ref{linf_wp}
shows the electronic density of a single Floquet eigenstate (the
highest one in fig.~\ref{f2}, for $F_{s,0}/F_0=0.3$), 
at various phases of the driving field.
The wave-packet is clearly localized along the field axis, and
propagates along a straight line classical 
trajectory, repeating its shape periodically. Its dynamics precisely reproduces 
the dynamics of the 1D analogue illustrated in fig.~\ref{lin1d_2}. 
Once again, as for previous
examples, the finite ionization rate (see section \ref{ION}) of the 
3D wave-packet is of
the order of some million Kepler periods.

\subsection{Wave-packets in the presence of a static magnetic field}
\label{MAG}

Similarly to a static electric field which may stabilize an angularly
unstable wave-packet, 
the properties of non-dispersive wave-packets
under circularly polarized driving, in the presence of an additional 
static magnetic field normal to 
the polarization plane,
has been studied in a series of papers
\cite{lee95,farrelly95a,farrelly95,brunello96,cerjan97,lee97,brunello97}.
The Hamiltonian of the system in the coordinate frame corotating with the CP
field 
reads (compare with eq.~(\ref{hrot}), for the pure CP case)
\be
H=\frac{p_x^2+p_y^2+p_z^2}{2}-\frac{1}{r}-
(\omega-\omega_{\mathrm c}/2)L_z +Fx
+\frac{\omega_{\mathrm c}^2}{8}(x^2+y^2),
\label{fareq1}
\ee
where $\omega_{\mathrm c}$ is the cyclotron frequency. In atomic 
units, the cyclotron frequency $\omega_{\mathrm c}=-qB/m$
equals  the magnetic field value. It can be both positive or negative,
depending on the direction of the magnetic field\footnote{A different
convention is used (quantization axis defined by the orientation of
the magnetic field) in many papers on this subject. It leads to
unnecessarily complicated equations.}.
This additional parameter modifies the dynamical properties which characterize 
the
equilibrium points, the analysis of which may be carried out alike
the pure CP case treated in sec.~\ref{CP}. 
A detailed stability analysis can be found in 
\cite{lee97,farrelly95,rakovic98} and we summarize here
the main results only.
  
Since changing the sign of $F$ in eq.~(\ref{fareq1}) is equivalent to changing
the sign of $x$ 
from positive to negative, we only consider the 
equilibrium position at $x_{\rm eq}>0$ (compare eqs.~(\ref{wp0}-\ref{wpu})).
For nonvanishing magnetic field, its position is given by
\be
\omega(\omega-\omega_{\mathrm c})x_{\rm eq}-\frac{1}{x_{\rm eq}^2}-F=0.
\label{fareq2}
\ee
Redefining the dimensionless parameter $q$ (see eq.~(\ref{q})) via 
\be
q=\frac{1}{\omega(\omega-\omega_{\mathrm c})x_{\rm eq}^3},
\label{mq}
\ee
we obtain for the
microwave amplitude 
\be 
F=[\omega(\omega-\omega_{\mathrm c})]^{2/3}(1-q)/q^{1/3},
\label{fm}
\ee
and  
\be 
E_{\mathrm eq}=[\omega(\omega-\omega_{\mathrm c})]^{1/3}(1-4q)/2q^{2/3}
\label{me}
\ee
for the equilibrium energy.

Harmonic expansion of
eq.~(\ref{fareq1}) around 
the equilibrium point $(x_{\rm eq},y_{\rm eq}=0,z_{\rm eq}=0)$ allows for 
a linear stability analysis
in its vicinity.
Alike the pure CP case,
the $z$ motion decouples from the motion in
the $(x,y)$ plane. For the latter, we recover 
the generic harmonic Hamiltonian 
discussed
in section \ref{CP}, eq.~(\ref{rotaibb}), provided we substitute
\be
\tilde{\omega}=\omega-\omega_{\mathrm c}/2.
\label{om}
\ee

When expanded at second order around the equilibrium point, the
Hamiltonian (\ref{fareq1}) takes the standard form
of a rotating anisotropic oscillator, eq.~(\ref{rotaibb}), with
$\tilde{\omega}$ replacing $\omega$, and with the stability parameters:
\bea
a&=&\frac{1}{\tilde{\omega}^2}\left(\frac{\omega_{\mathrm c}^2}{4}-\frac{2}
{x_{\rm eq}^3}
\right), \nonumber \\
b&=&\frac{1}{\tilde{\omega}^2}\left(\frac{\omega_{\mathrm c}^2}{4}+\frac{1}
{x_{\rm eq}^3}
\right).
\label{ab_mag}
\eea

\begin{figure}
\centerline{\psfig{figure=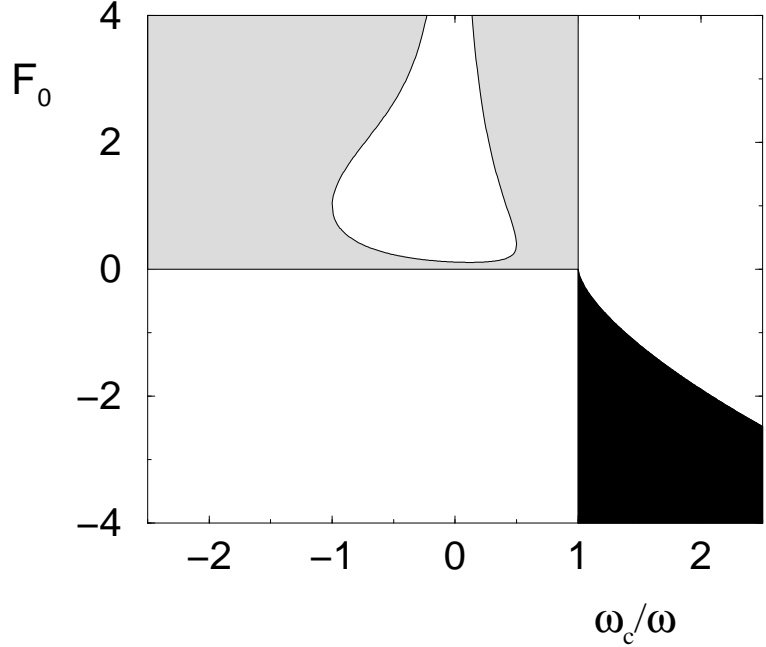,width=10cm,angle=-90}}
\caption{(Shaded) Regions of stability of the equilibrium point 
$(x_{\rm eq},y_{\rm eq},z_{\rm eq})$ for circularly polarized 
driving (amplitude $F$, 
frequency $\omega$) of a hydrogen atom, in the presence of a magnetic
field (corresponding cyclotron frequency $\omega_{\mathrm c}$). The black and grey regions correspond 
to the two regions of stability, described by 
eqs.~(\protect{\ref{stability_region_1}}) and (\protect{\ref{stability_region_2}}),
respectively.
}
\label{farfig1}
\end{figure}

The regions of 
stability of the equilibrium point $(x_{\rm eq},y_{\rm eq},z_{\rm eq})$
are thus obtained from the domains of stability of the
2D rotating anisotropic oscillator, given by 
eqs.~(\ref{stability_region_1},\ref{stability_region_2}). They
are visualized in terms of the 
physical parameters $F$ and $\omega_{\mathrm c}$,
(using the standard scaled electric field $F_0=Fn_0^4=F\omega^{-4/3}$)
in fig.~\ref{farfig1}, with the black region corresponding
to eq.~(\ref{stability_region_1}), and the grey region to eq.~(\ref{stability_region_2}).
Observe that the presence of the magnetic field tends to enlarge the 
region
of stability in parameter space; 
for $\omega_{\mathrm c}=0$ (pure CP case, no magnetic field) the
stability region is quite tiny, in comparison to large values of 
$|\omega_{\mathrm c}|$
\footnote{As long as we are interested in long-lived wave-packets,
the region of small $F_0$ is of interest only. At higher $F_0$ and for
$\omega_{\mathrm c}<\omega$, the
strong driving field will ionize the atom rather fast -- see section \ref{ION}.
 This makes the gray
region $F_0>0.1$ of little practical
interest.}.
On the other hand, the stability diagram does not provide us any detailed 
information on the actual size of the resonance island surrounding the 
equilibrium point.
However, 
it is precisely the size of the resonance island which is crucial for 
anchoring
nondispersive quantum wave-packets close to the classical equilibrium point
(see sec.~\ref{sapp}). 

An alternative approach to characterize the stability properties of
the classical motion near $(x_{\rm eq},y_{\rm eq},z_{\rm eq})$ has been 
advertized in 
\cite{lee97,farrelly95a,farrelly95,brunello96,cerjan97,lee95,brunello97}: the 
concept of zero velocity surfaces (ZVS).
In order to construct a ZVS, 
the Hamilton function is expressed in terms of velocities  
rather than canonical momenta. For the harmonic Hamiltonian
(\ref{rotaibb}), the calculation yields
\begin{equation}
H=\frac{v_x^2+v_y^2}{2}+\frac{\omega^2}{2}\left [(a-1)x^2+ (b-1)y^2
\right ].
\label{ZVS1}
\end{equation}
Thus, the ``kinetic energy'' becomes a positive function of velocities, and one
can define the ZVS as
\begin{equation}
S=H-\frac{v_x^2+v_y^2}{2},
\label{ZVS2}
\end{equation}
the generalization of an effective potential for interactions which mix 
position and momentum coordinates.
Note that, when the 
velocities coincide with the canonical momenta, $S$ is nothing 
but the potential energy surface. We prefer to denote it $S$ instead of $V$,
to stress the difference. As discussed in detail in \cite{lee97}, a ZVS
may be used to locate the equilibrium points. However, 
their stability properties are not obvious 
({\em contrary} to the potential surface, where 
minima define stable
fixed points, while maxima and saddle points are unstable). For a ZVS, saddles
are also unstable, but maxima may either be stable {\em or} 
unstable.
For example, the first stability region, eq.~(\ref{stability_region_1}), 
of the rotating 2D anistropic
Hamiltonian is associated with
a stable minimum of the ZVS. The second region of stability,
eq.~(\ref{stability_region_2}), corresponds to $a,b\leq 1$ and thus
to a maximum of the ZVS. However, the ZVS does not show
any qualitative change whether $(a-b)^2+8(a+b)$ is positive or negative,
i.e. whether the equilibrium point is stable or unstable.
Thus, a ZVS is clearly inappropriate, or at least potentially
dangerous, for the discussion of the classical motion close to
equilibrium.
As a matter of fact, this difficulty with the ZVS is crucial
in our case, even for the pure CP case, without additional magnetic field.
Indeed, the ZVS
becomes
\begin{equation}
S=\left(-\frac{2q+1}{2}x^2+\frac{q-1}{2}y^2\right) \omega^2,
\label{ZVS3}
\end{equation}
where we use the single parameter $q$ to parametrize $S$. At $q=1$
(i.e. $F=0$, see eq.~(\ref{fm})), the equilibrium point 
turns from a saddle
(for $q>1$) into a maximum. Consequently, the ZVS correctly reflects 
the change of the equilibrium
point from unstable ($q>1$) to stable ($q<1$).
However, for any  $q\in(0,1)$, the equilibrium remains a maximum of the ZVS,
which completely misses the change of stability at $q=8/9$.
Thus, the very same maximum
may change its stability (which fundamentally affects the classical 
motion in its
vicinity) without being noticed by inspection of the 
ZVS. The latter
evolves very smoothly around $q=8/9$. Thus, the ZVS contours provide
{\em no} information on the nature of the classical 
motion in the vicinity of the
equilibrium point, in {\em disaccord} with   
\cite{lee97,cerjan97,farrelly95a,farrelly95,brunello96,lee95,brunello97}.
Similarly, the isovalue contours of the ZVS (which are ellipses in the
harmonic approximation) have no relation
with the isovalue  contours of the ground-state wave-packet
localized
around the equilibrium point (these contours
are also ellipses in the harmonic approximation where the
wave-packet is a Gaussian),
contrary to what is stated in \cite{lee97,farrelly95a}.
For example, the aspect ratio (major axis/minor axis)
of the ZVS contour lines is $\sqrt{(2q+1)/(1-q)}$ which varies smoothly 
around $q=8/9$, while the aspect ratio of the isocontours 
of the ground state wave-packet diverges when $q\to 8/9.$ 
\footnote{While this argument has been presented here
for the simplest case of the harmonic
oscillator hamiltonian (\ref{rotaibb}), it carries over to 
the full, nonlinear model,
eq.~(\ref{fareq1}).}

Nonwithstanding, a ZVS may be used for other purposes \cite{lee95}, e.g., 
to show the existence of an ionization threshold 
for the Hamiltonian (\ref{fareq1}), when $\omega_{\mathrm c}>\omega$
(area coded in black in fig.~\ref{farfig1}). 
Clearly, due to the parabolic confinement in the 
$x-y$-plane, ionization is only possible along the 
$z$ direction.
The threshold  is given by
\cite{lee95}
\be
E_{\mathrm ion}=F^2/2\omega(\omega-\omega_{\mathrm c}),
\label{thresh_m}
\ee
which 
lies above the equilibrium energy $E_{\rm eq}$. Thus, for parameters 
in that region,
the electron -- initially placed close to the stable fixed point --
cannot ionize. One may expect, therefore, that wave-packets built
around the equilibrium point for $\omega_{\mathrm c}>\omega$ lead to
{\em discrete} Floquet states. In other cases, e.g., for pure
CP driving, non-dispersive wave-packets are rather represented by 
long-living resonances (see section \ref{ION}).

Finally, it has been often argued  
\cite{lee97,farrelly95a,farrelly95,brunello96,cerjan97,lee95,brunello97}
that the presence of the magnetic field is absolutely necessary for the
construction of non-dispersive wave-packets. The authors consider 
non-dispersive
wave-packets 
as equivalent 
to Gaussian shaped wave-functions (using
equivalently the notion of coherent states). Then it is vital that the
motion in the vicinity of the fixed point is locally harmonic within a
region of size $\hbar$. This leads the authors to conclude that
non-dispersive wave-packets may not exist for the pure CP case except in the
extreme semiclassical regime. 
As opposed to that, the diamagnetic term in
eq.~(\ref{fareq1}) gives a stronger weight to 
the harmonic term, which is the basis of the above claim\footnote{Note
 that the non-harmonic terms, being entirely due
to the Coulomb field, are not removed or decreased by the addition
of a magnetic field. They are just hidden by a larger harmonic term.}.
From our point of view, which, as already stated above, 
attributes 
the non-dispersive character of the wave-packet 
to a classical nonlinear
resonance, the accuracy of the harmonic approximation 
(which, anyway, always
remains an approximation) is 
irrelevant for the existence of
non-dispersive wave-packets. 
The best proof is that \cite{farrelly95a}
concludes, on the
basis of the validity of the harmonic approximation,
that non-dispersive wave-packets should not exist for
$n_0\simeq 60$ in CP field, in complete
contradiction to numerically exact experiments showing their
existence down to $n_0=15$ \cite{delande95}.
On the other hand, it is an interesting question
how good 
the harmonic approximation actually is in the pure CP case. The interested
reader may find a more quantitative discussion of this point in section
\ref{SPO}.

\section{Other resonances}
\label{HOR}

\subsection{General considerations}
\label{ORG}

We have so far restricted our attention to nondispersive wave-packets anchored
to the principal resonance of periodically driven Hamiltonian systems. In
section~\ref{GM}, we already saw that {\em any} harmonic of the unperturbed 
classical motion can dominate the harmonic expansion (\ref{hamfou}) of the 
classical Hamilton function, provided it is resonantly driven by the external
perturbation, i.e.
\begin{equation}
s\theta -\omega t\simeq {\mathrm const},\ s>0\ {\mathrm integer}.
\label{rescond}
\end{equation}
This is the case when the $s^{\mathrm th}$ 
harmonic of the classical internal frequency $\Omega$ is resonant with
the external driving $\omega$. As $\Omega$ depends on the classical unperturbed
action, the corresponding classical
resonant action is defined by
\begin{equation}
\Omega(I_s) = \frac{\partial H_0}{\partial I}(I_s) = \frac{\omega}{s}.
\label{Is}
\end{equation}
At this action, 
the period of the classical motion is $s$ times the period 
of the external drive.
Precisely like in the $s=1$ case (the principal resonance), for any integer
$s>1,$ Floquet eigenstates of the driven system exist which are 
localized on the 
associated classical stability islands in phase space. The energy of 
these eigenstates 
can again be estimated through the semiclassical quantization of the secular
dynamics. 
To do so, we start from eqs.~(\ref{hamfou},\ref{hamext}) and transform to 
slowly varying variables (the ``rotating frame") defined by
\begin{eqnarray}
\hat{\theta}&=&\theta-\frac{\omega t}{s}, \label{trafo_s_a} \\
\hat{I}&=&I, \nonumber \label{trafo_s_b} \\
\hat{P_t}&=&P_t+\frac{\omega I}{s} \nonumber.
\label{trafo_s_c}
\end{eqnarray}
The Floquet Hamiltonian in this rotating frame now reads:
\begin{eqnarray}
\hat{\mathcal H} =  \hat{P}_t+H_0(\hat{I})-\frac{\omega\hat{I}}{s} 
+ \lambda
\sum_{m=-\infty}^{+\infty}V_m(\hat{I})\left\{\cos\left(m\hat{\theta}+
\left(\frac{m-s}{s}\right)\omega t\right)\right\}, 
\label{ham_trafo_lin1d}
\end{eqnarray}
which is periodic with period $\tau=sT$ with $T=2\pi/\omega$.
Passing to the rotating frame 
apparently destroys the $T$-periodicity of the original Hamiltonian.
This, however, is of little importance, the crucial point
being to keep the periodicity  $sT$ of the internal motion.
If we now impose the resonance condition (\ref{rescond}), the major 
contribution to the sum in eq.~(\ref{ham_trafo_lin1d}) will come 
from the slowly evolving resonant term $m=s$, while the other 
terms vanish upon averaging the fast variable $t$ over one period $\tau$, 
leading to the secular Hamilton function
\begin{equation}
{\cal H}_{\rm sec}=\hat{P}_t+H_0(\hat{I})-\frac{\omega\hat{I}}{s}
+\lambda V_s(\hat{I})\cos(s\hat{\theta}).
\label{sec_s}
\end{equation}
This averaging procedure eliminates the explicit time dependence of 
$\hat{\mathcal H}$, and is tantamount to restricting the validity of 
${\mathcal H}_{\rm sec}$ to the description of those classical 
trajectories 
which comply with eq.~(\ref{rescond})
and, hence, exhibit a periodicity with period $\tau$. This will have an 
unambiguous signature in the quasienergy spectrum, as we shall see further
down.
The structure of the secular Hamiltonian is simple and reminds us of the
result for 
the principal $s=1$ resonance, eq.~(\ref{hsec_ap}). However,
due to the explicit appearance of the factor $s>1$ in the argument of the
$\cos$ term,  a juxtaposition of $s$ resonance
islands close to the resonant action, eq.~(\ref{Is}), is created.
It should be emphasized that these $s$ resonance islands
are actually $s$ clones of the same island. Indeed, a trajectory trapped
inside a resonance island will successively visit all the islands:
after one period of the drive, $\theta$ is approximately increased
by $2\pi/s,$ corresponding to a translation to the next island. 
At the center of the islands, there is a single, stable resonant
trajectory whose period is exactly $\tau=sT.$

At lowest order in $\lambda,$ all the quantities of interest can
be expanded in the vicinity of $I_s$,
exactly as for the principal resonance in section~\ref{CD}.
$V_s$ is consistently evaluated at the resonant action. 
The pendulum approximation of the secular Hamiltonian then reads:
\begin{equation}
{\cal H}_{\mathrm pend} = \hat{P}_t + H_0(\hat{I_s}) 
-\frac{\omega}{s}\hat{I}_s 
+ \frac{1}{2}
H^{''}_0(\hat{I_s})\ (\hat{I}-\hat{I_s})^2 + \lambda
V_s(\hat{I_s}) \cos s\hat{\theta},
\label{eqpend_s}
\end{equation}
with the centers of the islands located at:
\begin{equation}
\hat{I}=\hat{I}_s=I_s,
\end{equation}
\begin{equation}
\left\{
\begin{array}{l}
\displaystyle \hat\theta = k \frac{2\pi}{s},\ \ \ \ \ {\mathrm if}\ \lambda V_s(\hat{I}_s)H^{''}_0(\hat{I}_s)<0,\\
\displaystyle \hat\theta = k \frac{2\pi}{s} + \frac{\pi}{s},\ \ \ \ \ {\mathrm if}\ \lambda V_s(\hat{I}_s)H^{''}_0(\hat{I}_s)>0,
\end{array}
\right.
\end{equation}
where $k$ is an integer running from 0 to $s-1.$
For $H^{''}_0(\hat{I}_s)$ -- see eq.~(\ref{second_derivative}) -- positive,
these are minima of the secular Hamiltonian, otherwise they are
maxima. 
The extension of each resonance island is, as a direct generalization of the
results of section \ref{CD}:
\begin{eqnarray}
\Delta\hat\theta& = &\frac{2\pi}{s},\\
\Delta\hat{I} & = & 4\sqrt{\left|\frac{\lambda V_s(\hat{I_s})}{ H^{''}_0(\hat{I_s})}\right|,}
\end{eqnarray}
and its area
\begin{equation}
A_s(\lambda) = \frac{16}{s} \sqrt{\left|
\frac{\lambda V_s(\hat{I_s})}{ H^{''}_0(\hat{I_s})}\right|}.
\label{area_s}
\end{equation}
Again, the dependence of $A_s(\lambda)$ on $\sqrt{|\lambda|}$ implies that even
small perturbations may
induce
significant changes in the phase space structure, provided the perturbation
is resonant with a harmonic of the unperturbed classical motion\footnote{
The situation is very different in the opposite case, when $\omega$
is the $s^{\mathrm th}$ SUB-harmonic \cite{marion} of the internal frequency. A resonance
island may then exist but it is typically much smaller as it comes into
play only at order $s$ in perturbation theory, with a
size scaling as $|\lambda|^{s/2}.$}.

The construction of a non-dispersive wave-packet is simple once the
$s$-resonance structure is understood: indeed, any set of initial
conditions trapped in one of the $s$-resonance islands will classically remain
trapped forever. Thus, a quantum wave-packet
localized initially inside a resonance island is a good candidate for
building a  non-dispersive wave-packet. There remains, however, a difficulty:
the wave-packet can be initially placed in any of the $s$ resonance islands.
After one period of the driving, it will have jumped to
the next island, meaning that it will be far from its initial position.
On the other hand, the Floquet theorem guarantees the existence
of states which are strictly periodic with the period of the drive 
(not the period of the resonant internal motion). 
The solution to this
difficulty is to build eigenstates which simultaneously occupy
{\em all} 
$s$-resonance islands, that is, which are composed
of $s$ wave-packets each localized on a different resonance
island. After one period of the drive, each individual wave-packet replaces
the next one, resulting in globally $T$-periodic motion of this ``composite''
Floquet state.
 If the system has a macroscopic size (i.e. in the
semiclassical limit), 
individual wave-packets will appear extremely well localized
and lying far from the other ones
while maintaining a well-defined phase coherence with them.
For $s=2$, the situation
mimics a symmetric double well potential, where even and odd solutions
are linear combinations of nonstationary states, each localized in either one
well \cite{landau2}.

In order to get insight in the structure of the Floquet quasi-energy spectrum,
it is useful to perform the semiclassical 
EBK quantization of the secular Hamiltonian (\ref{sec_s}).
Quantization of the motion in 
$(\hat{I},\hat{\theta})$,
see section~\ref{sapp}, provides 
states trapped within the
resonance islands (librational motion),
and states localized outside 
them (rotational motion). As usual, the number of
trapped states is given 
by the size, eq.~(\ref{area_s}) of the
resonance island:
\begin{equation}
{\mathrm Number\ of\ trapped\ states} \simeq \frac{8}{\pi \hbar s}
\sqrt{\left|\frac{\lambda V_s(\hat{I_s})}{ H^{''}_0(\hat{I_s})}\right|}.
\label{number_of_trapped_states_s}
\end{equation}
The quantization can be performed along the contours of 
any of the $s$ clones of the resonance island,
giving of course the same result. However, this does {\em not}
result in a $s$-degeneracy of the spectrum: indeed, the $s$ clones
belong to the same torus in phase space (see above) and do
not generate $s$ independent states.

If the number of trapped states is sufficiently large,
the harmonic approximation to the pendulum (or secular) Hamiltonian
can be used, with the frequency of the harmonic motion
around the stable resonant orbit given by:
\begin{equation}
\omega_{\mathrm harm} = s\sqrt{|\lambda V_s(\hat{I_s})H^{''}_0(\hat{I_s})|}.
\label{omega_harmonic_s}
\end{equation}

In order to get the complete semiclassical Floquet spectrum,
we additionally have to perform the semiclassical quantization in the 
$(t,\hat{P}_t)$ plane,
giving:
\begin{equation}
\frac{1}{2\pi}\int_0^{\tau}\hat{P}_tdt=\frac{\hat{P}_t\tau}{2\pi}
=\frac{s\hat{P}_t}{\omega} = \left(
j+\frac{\mu}{4}\right)\hbar,
\label{wkb_k1d}
\end{equation}
with $j$ integer.
This finally yields the semiclassical Floquet levels (in the 
harmonic approximation):
\begin{equation}
{\cal E}_{N,j} = H_0(\hat{I_s}) - \frac{\omega}{s} \hat{I}_s + \left(j+\frac{\mu}{4}\right)\hbar \frac{\omega}{s}
- \ {\mathrm sign}(H^{''}_0(\hat{I_s}))
\left[ |\lambda V_s(\hat{I_s})| -
\left(N+\frac{1}{2}\right) \hbar \omega_{\mathrm harm} \right],
\label{spectrum_harmonic_s}
\end{equation}
with $N$ a non-negative integer.
The wave-packet with optimum localization in the
$(\hat{I},\hat{\theta})$ plane, i.e. optimum localization
along the classical unperturbed orbit, is the $N=0$ state.
According to eq.~(\ref{spectrum_harmonic_s}), the
semiclassical quasi-energy spectrum has
a periodicity $\hbar\omega/s,$ whereas the ``quantum" Floquet theory
only enforces 
$\hbar\omega$ periodicity.
Thus, inside a Floquet zone of width $\hbar\omega,$
each state appears $s$ times (for $0\leq j < s$), at energies
separated by  $\hbar\omega/s.$ Note that this property is a direct
consequence of the possibility of eliminating the time dependence of
$\mathcal H$ in eq.~(\ref{ham_trafo_lin1d}) by averaging over $\tau$,
leading to the time independent expression (\ref{sec_s}) for
${\cal H}_{\rm sec}$. 
Therefore, it will 
be only approximately valid for the exact quantum Floquet
spectrum. In contrast, the $\hbar\omega$ periodicity holds exactly,
as long as the system Hamiltonian is time-periodic.

We will now recover the $\hbar\omega/s$ 
periodicity in a quantum description of our problem,
which will provide us with the formulation of an eigenvalue problem 
for the wave-packet eigenstates anchored to the $s$-resonance, in terms of a 
Mathieu 
equation. In doing so, we shall extend the 
general concepts outlined in section \ref{section_mathieu} above.

Our starting point is eq.~(\ref{coupled_equations}), which we again consider
in the regime where the eigenenergies $E_n$ of the 
unperturbed Hamiltonian $H_0$ are locally approximately spaced 
by $\hbar\Omega.$
The resonance condition (\ref{rescond}) implies
\begin{equation}
\left. \frac{dE_n}{dn}\right|_{n=n_0}=\hbar\frac{\omega}{s},
\label{deriven}
\end{equation}
where, again, $n_0$ is not necessarily an integer, and is related
to the resonant action and its associated Maslov index through:
\begin{equation}
\hat{I}_s = \left(n_0+\frac{\mu}{4}\right) \hbar.
\label{def_n0_s}
\end{equation}
When the resonance condition is met, the only efficient
coupling in eq.~(\ref{coupled_equations}) connects 
states with the same value of $n-sk.$ In other words, 
in the secular approximation, a given state $(n,k)$ only 
couples to $(n+s,k+1)$ and $(n-s,k-1).$
We therefore consider a given ladder of coupled states labeled 
by $j=n-sk.$ Because of the overall $\omega$ perodicity
of the spectrum, changing $j$ by $s$ units (i.e., shifting all $k$-values
by 1) is irrelevant, so that it is enough to consider the $s$ independent
ladders $0\leq j \leq s-1.$
Furthermore,
in analogy to section~\ref{section_mathieu}, eq.~(\ref{matel_mat}), 
we can replace the coupling matrix
elements in eq.~(\ref{ham_trafo_lin1d}) by the resonantly driven Fourier
coefficients of the classical motion, 
\begin{equation}
\langle \phi_n|V|\phi_{n+s}\rangle\simeq\langle \phi_{n-s}|V|\phi_n\rangle\simeq V_s(\hat{I}_s).
\label{coupling_s}
\end{equation}
With these approximations, and the shorthand notation $r=n-n_0$,
eq.~(\ref{coupled_equations}) takes the form of $s$ independent
sets of coupled equations, 
identified by the integer $j$:\footnote{For $s=1$, this equation reduces of course
to eq.~(\protect\ref{tridiag}). We here use $n_0$ instead of $\hat{I}_s$; 
the two quantities differ only by the Maslov index, eq.~(\protect\ref{def_n0_s}).}
\begin{equation}
\left[
{\cal E}-E_{n_0}+\frac{n_0-j}{s}\hbar\omega-\frac{\hbar^2 r^2}{2}H_0^{''}
(\hat{I}_s)\right]
d_{r}=\frac{\lambda}{2}V_s[d_{r+s}+d_{r-s}],
\label{coupled_s}
\end{equation}
where 
\begin{equation}
d_{r}=c_{n_0+r,\frac{n_0+r-j}{s}},
\label{d_s}
\end{equation}
as a generalization of the notation in eq.~(\ref{diag_mat}). Again, $r$ is
not necessarily an integer, but the various $r$ values involved
in eq.~(\ref{coupled_s}) are equal modulo $s$.
Precisely as in the case of the principal resonance, eq.~(\ref{coupled_s}) 
can be
mapped on its dual space expression, via eq.~(\ref{dual}):
\begin{equation}
\left[
-\frac{\hbar^2}{2}H_0^{''}(\hat{I}_s)\frac{d^2}{d\phi^2}+E_{n_0}
-\frac{n_0-j}{s}\hbar\omega+\lambda
V_s\cos(s\phi)\right] f(\phi) = {\cal E} f(\phi),
\label{mathieu_s}
\end{equation}
and identified with the Mathieu equation (\ref{mat_eq}) through
\begin{eqnarray}
s\phi & = & 2v, \\
a & = & \frac{8}{\hbar^2s^2H_0^{''}(\hat{I}_s)}\left[{\cal E}-E_{n_0}+(n_0-j)\hbar
\frac{\omega}{s}\right],\ j=0,\ldots, s-1, \nonumber \\
q & = & \frac{4\lambda V_s(\hat{I}_s)}{s^2\hbar^2H_0^{''}(\hat{I}_s)}. \nonumber
\label{ident_s}
\end{eqnarray}
The quasienergies associated with the $s$ resonance in the pendulum
approximation then follow immediately as
\begin{equation}
{\cal E}_{\kappa,j}=E_{n_0}-(n_0-j)\hbar\frac{\omega}{s}+\frac{\hbar^2s^2}{8}
H_0^{''}(\hat{I}_s)a_{\kappa}(\nu,q),
\label{qu_mat_s}
\end{equation}
where the index $j$ runs from $0$ to $s-1$, and $\kappa$	labels the 
eigenvalues of the Mathieu equation \cite{abramowitz72}.
Again, the boundary condition for the solution
of the Mathieu equation is incorporated via the characteristic exponent, which
reads
\begin{equation}
\nu=-2\frac{n_0-j}{s}\ ({\mathrm mod}\ 2),\ \ \ \ \ j=0,\ldots s-1.
\label{char_s}
\end{equation}
The structure of this quasi-energy spectrum apparently displays the
expected $\hbar\omega/s$ periodicity. However, the characteristic 
exponent $\nu$
-- and consequently the $a_{\kappa}(\nu,q)$ eigenvalues --
depend on $j$, what makes the periodicity approximate only. It is
only far inside
the resonance island that the $a_{\kappa}(\nu,q)$ eigenvalues are almost
independent
of $\nu$ and the periodicity is recovered. Deviations from this
periodicity are further discussed in section~\ref{OGB}.
Finally, the asymptotic expansion of the $a_{\kappa}(\nu,q)$ for large $q$,
eq.~(\ref{math_asy}),
gives again (compare section \ref{section_mathieu}) 
the semiclassical estimate of the energy levels in the
harmonic approximation, eq.~(\ref{spectrum_harmonic_s}), where the
indices $\kappa$ and $N$ coincide.

\subsection{A simple example in 1D: the gravitational bouncer}
\label{OGB}

As a first example of non-dispersive wave-packets localized on $s>1$ primary
resonances, 
we consider
the particularly simple 1D model of
a particle
moving vertically in the gravitational field, and bouncing off 
a periodically driven horizontal plane.
This system is known as the Pustylnikov model \cite{lichtenberg83} (or,
alternatively, the ``gravitational bouncer", or the ``bubblon model"
\cite{holthaus95,benvenuto91,oliveira94,flatte96})  
and represents a standard example of chaotic motion. Moreover,
despite its simplicity, it may find possible 
 applications in the dynamical manipulation of cold atoms
\cite{steane95}.
A gauge transformation shows its equivalence to a periodically
driven particle moving in a triangular potential well, with the Hamiltonian
\begin{equation}
H=\frac{p^2}{2}+V(z)+\lambda z \sin(\omega t),
\label{boueq1}
\end{equation}
where 

\begin{equation}
V(z)=\left\{ 
\begin{array}{ll}
z & {\rm for}\ z\geq 0 , \\
\infty & {\rm for}\ z<0 .
\end{array}
\right.
\end{equation}
The strength $\lambda$ of the periodic driving is proportional 
to the maximum
excursion 
of
the oscillating surface.
Classically, this system is 
well approximated by the standard map \cite{lichtenberg83}
(with the momentum and the phase of the driving
 at the moment of the bounce as variables), with kicking amplitude
$K=4\lambda$. 

Apart from a (unimportant) phase shift $\pi/2$ in $\omega t$, 
eq.~(\ref{boueq1}) 
is of the general type of eq.~(\ref{h_gen}) 
and the scenario for the creation of 
non-dispersive wave-packets described in sections~\ref{CD} and
\ref{QD} is applicable. 
As a matter
of fact, a  careful analysis of the problem 
using the Mathieu approach described
in sec.~\ref{section_mathieu}, as well as the semiclassical
quantization of the Floquet Hamiltonian 
were already outlined in \cite{holthaus95}, 
where the associated Floquet eigenstates were baptized ``flotons".
We recommend \cite{holthaus95,flatte96} for 
a detailed discussion of the problem,
reproducing here only the main results,
with some minor modifications.

Solving the classical equations of motion for the unperturbed Hamiltonian
is straightforward 
(piecewise uniformly accelerated
motion alternating with 
bounces off the mirror) 
and it is easy to express the unperturbed Hamiltonian and the
classical internal frequency in terms of action-angle variables:
\begin{eqnarray}
\displaystyle H_0&=&\frac{(3\pi I)^{2/3}}{2},\\
\displaystyle \Omega &=& \frac{\pi^{2/3}}{(3I)^{1/3}},
\end{eqnarray}
while the full, time-dependent Hamiltonian reads:
\begin{equation}
 H=\frac{(3\pi I)^{2/3}}{2}+\frac{\lambda\pi I^{2/3}}{(3\pi)^{1/3}}
\sin(\omega t)
-\frac{2\lambda (3I)^{2/3}}{\pi^{4/3}}\sin(\omega t)\sum_{n=1}^\infty
 \frac{\cos(n\theta)}{n^2}.
\label{boueq2}
\end{equation}

\begin{figure}
\centerline{\psfig{figure=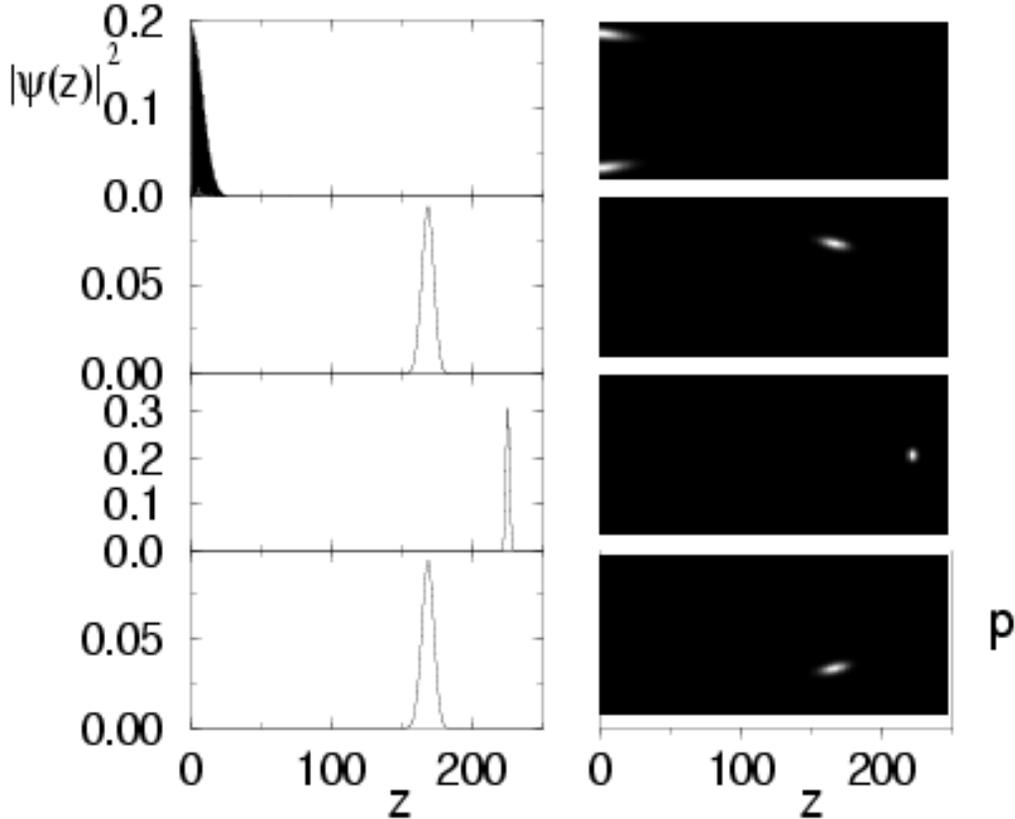,width=14cm}}
\caption{A non-dispersive $s=1$ wave-packet of 
the gravitational bouncer, eq.~(\protect\ref{boueq1}). The quantum number 
of the resonant state is chosen as $n_0=1000$, to match typical 
experimental dimensions \protect\cite{steane95}.
The left
column shows the time evolution of the wave-packet 
for $\omega t=0,\pi/2,\pi,3\pi/2$ (from top to bottom).
The right column shows the
corresponding phase space (Husimi, see eq.~(\ref{husimi_def})) 
representation ($z$ axis horizontal as in the left column,
momentum $p$ on the vertical axis).
The parameters are 
$\omega\simeq 0.1487$, $\lambda=0.025$. 
The periodic, nondispersive dynamics of 
the wave-packet bouncing off the mirror in the gravitational field is 
apparent.}
\label{boufig2}
\end{figure}

Thus, the resonant action (\ref{Is}) is given by
\begin{equation}
\hat{I}_s = \frac{\pi^2s^3}{3\omega^3},
\label{act-s-res}
\end{equation}
with the associated strength of the effective coupling
\begin{equation}
V_s = \frac{(3I)^{2/3}}{s^2\pi^{4/3}}.
\end{equation}
Using the framework of
sections~\ref{CD}, \ref{QD} (for $s=1$),
and \ref{ORG} (for $s>1$), the reader
may easily compute the various  properties  of non-dispersive wave-packets
in this system\footnote{There is, however, a tricky point:
the Maslov index in this system
is 3, with a contribution  1 coming from the outer turning point,
and 2 from $z=0$, since the oscillating plane 
acts as a hard wall. Hence, the relation
between the principal quantum number and the action is 
$I_1=n+3/4$, see eq.~(\protect\ref{WKB}).}.  
An example for $s=1$
is presented in fig.~\ref{boufig2},
for the resonant principal quantum number $n_0=1000,$ (i.e., 
$\hat{I}_1=1000.75$) where both,
the (time-periodic) probability densities in 
configuration and phase space 
are shown. Note that such 
high $n_0$ values (or even higher) correspond to typical experimental 
falling heights (around 0.1 mm for $n_0=1000$) in experiments on cold atoms \cite{steane95}.
Therefore, the creation of an atomic wave-packet in such an experiment would 
allow to store 
the atom in a quasi classical ``bouncing mode" over arbitrarily long times, 
and might find some application in the field 
of atom optics \cite{oberthaler99}.

\begin{figure}
\centerline{\psfig{figure=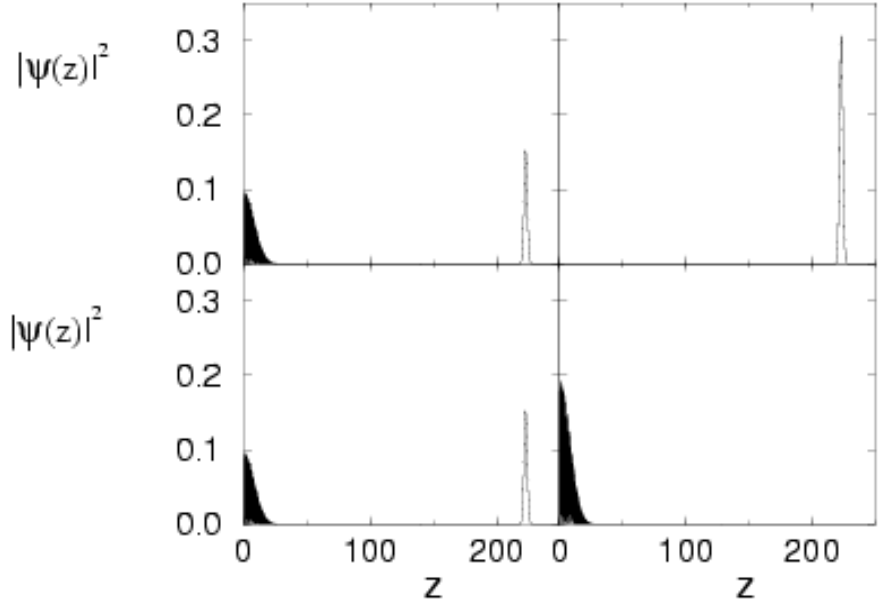,width=12cm}}
\caption{Floquet eigenstates anchored to the $\omega=2\Omega$ 
resonance in the gravitational bouncer (left column), in configuration space,
at $\omega t=0$. Each eigenstate exhibits two wave-packets shifted by a phase
$\pi$ along the classical trajectory, to abide the Floquet
periodicity imposed by eq.~(\ref{flstate}). Symmetric and antisymmetric 
linear combinations 
of these states isolate either one 
of the wave-packets, which now evolves precisely like a classical particle
(right column), periodically bouncing off the wall at $z=0$.
The parameters are 
$\omega\simeq 0.2974$,
(corresponding to 
$n_0=1000$, for the $2:1$ resonance), 
$\lambda=0.025$.}
\label{othfig1}
\end{figure}

\begin{figure}
\centerline{\psfig{figure=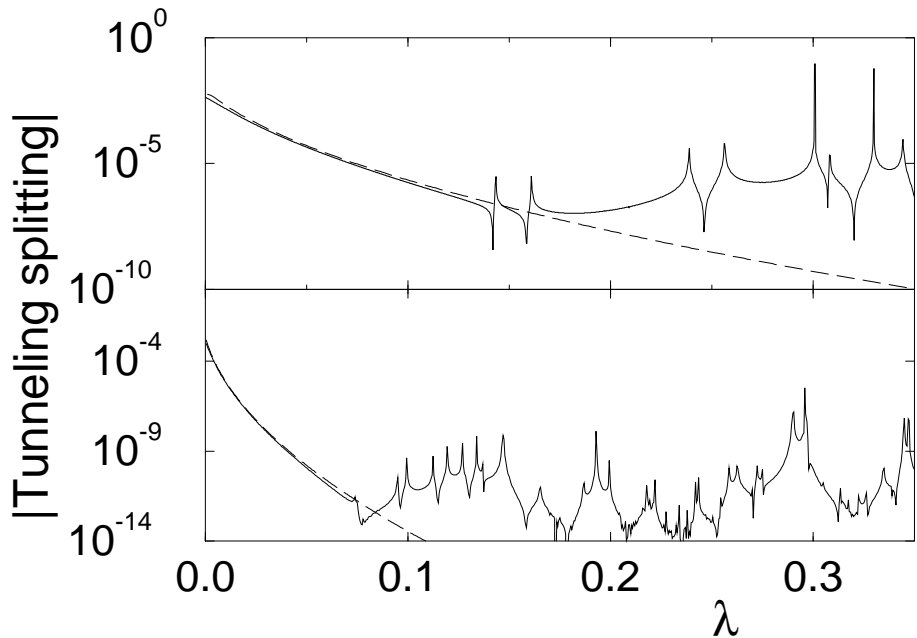,width=14cm,bbllx=0pt,bblly=50pt,bburx=600pt,bbury=400pt}}
\caption{Tunneling splitting between the energies (modulo
$\omega/2$) of two Floquet states of the gravitational bouncer, anchored to the
$s=2$ resonance island, as a function of the driving amplitude 
$\lambda$.
Driving frequency 
$\omega\simeq 1.0825$ (top) and $\omega\simeq 0.8034$ (bottom),
corresponding to resonant states $n_0=20$ and $n_0=50$, respectively. 
The dashed line reproduces   
the prediction of Mathieu theory \protect\cite{holthaus95}, 
eq.~(\protect{\ref{otheqn}}).
Observe that the latter fits the exact 
numerical data only for small values of $\lambda$. At larger 
$\lambda,$ small avoiding crossings between one member of the
doublet and eigenstates originating from other manifolds dominate
over the pure tunneling contribution and the Mathieu prediction 
is not accurate.}
\label{othfig2}
\end{figure}

For $s=2$, we 
expect, following the general discussion in section \ref{ORG},
 two quasi-energy levels, separated 
by $\omega/2$, according to the semiclassical result, 
eq.~(\ref{spectrum_harmonic_s}), which are
{\em both} associated with the $s=2$ resonance. 
As a matter of fact, such states are born out from an exact numerical 
diagonalization of the Floquet Hamiltonian derived from eq.~(\ref{boueq1}). 
An exemplary situation is shown in fig.~\ref{othfig1}, for $n_0=1000$ (i.e., 
$\hat{I}_2=1000.75$, in eq.~(\ref{act-s-res})). The tunneling coupling between the 
individual wave-packets shown in the right column of fig.~\ref{othfig1} is 
given by the tunneling splitting $\Delta$ between the energies of both 
associated Floquet states (left column of fig.~\ref{othfig1}) modulo 
$\omega/2$. From the Mathieu approach -- 
eqs.~(\ref{qu_mat_s},\ref{char_s}) --
this tunneling coupling is directly related to the variations of the
Mathieu eigenvalues when the characteristic exponent is changed.
In the limit where the resonance island is big enough, $q\gg 1$ in
eq.~(\ref{qu_mat_s}), asymptotic expressions \cite{abramowitz72}
allow for the following estimate~\cite{holthaus95}:
\be
\Delta=\frac{8\sqrt{2}\lambda^{3/4}}{\pi\sqrt{\omega}}
\exp\left(-\frac{16\pi\sqrt{\lambda}}{\omega^3}\right)=
\frac{8[3(n_0+3/4)]^{1/6}\lambda^{3/4}}{\pi^{4/3}}
\exp\left(-6(n_0+3/4)\sqrt{\lambda}/\pi\right),
\label{otheqn}
\ee
which we can test with our numerically exact quantum treatment. 
The result is shown in fig.~\ref{othfig2}, for two different values of $n_0$.

Observe the 
excellent agreement for small $\lambda$, 
with an almost
exponential decrease of the splitting with $\sqrt{\lambda}$, as expected from
eq.~(\ref{otheqn}). However,
for larger values of $\lambda$ (nontheless still in the regime of predominantly
regular classical motion) the splitting saturates and then starts to fluctuate
in
an apparently random way. While the phenomenon has not been completely 
clarified so far, 
we are inclined to attribute it
 to tiny avoided crossings with Floquet states localized is 
 some other resonance islands
(for a discusion
 of related phenomena see
\cite{bonci98,brodier01}).
Comparison of the two 
panels of fig.~\ref{othfig2} additionally indicates that the region of 
$\lambda$ values where 
avoided crossings become important 
increases in the semiclassical limit, and that eq.~(\ref{otheqn})
remains valid for small $\lambda$ only. This is easily understood: in the semiclassical
limit $n_0\rightarrow \infty,$ 
the tunneling splitting decreases exponentially, while the density of states
increases.

\begin{figure}
\centerline{\psfig{figure=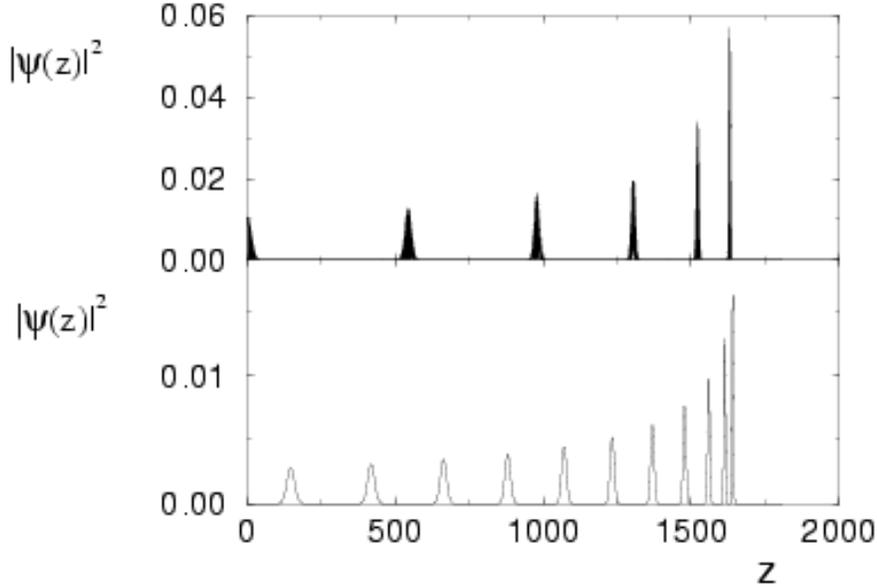,width=12cm}}
\caption{Single Floquet state of the gravitational bouncer at phases 
$\omega t=0$ (top) and $\omega t=\pi/2$ (bottom), anchored to
the $11:1$ resonance island chain in classical phase space.
Driving frequency $\omega=11\Omega\simeq 0.6027$, 
$n_0=20000$, and
$\lambda=0.025$. In the upper plot,
among 11 individual
wave-packets which constitute the eigenstate,
five pairs (with partners moving in opposite
directions) interfere at distances $z>0$ from the mirror, whereas the 
11th wave-packet bounces off the wall and interferes with itself. At
later times (bottom), the 11 wave-packets are well separated in space.}
\label{othfig3}
\end{figure}

Finally, fig.~\ref{othfig3} shows a Floquet eigenstate anchored to the $s=11$
resonance island chain. 
It may
be thought of as a linear combination of 11 non-dispersive
wave-packets which, at a given time, may interfere with each other, or,
at another time (bottom panel), are spatially well separated. 
For a Helium atom bouncing off an atom mirror in the earth's gravitational
field (alike the setting in \cite{steane95}),
the $z$ values for such a 
state reach 5 millimeters. This non-dispersive wave-packet is
thus a macroscopic object composed of 11 individual components
keeping a well-defined phase coherence.

\subsection{The $s=2$ resonance in atomic hydrogen under linearly 
polarized driving}
\label{ORH}

\begin{figure}
\centerline{
\psfig{figure=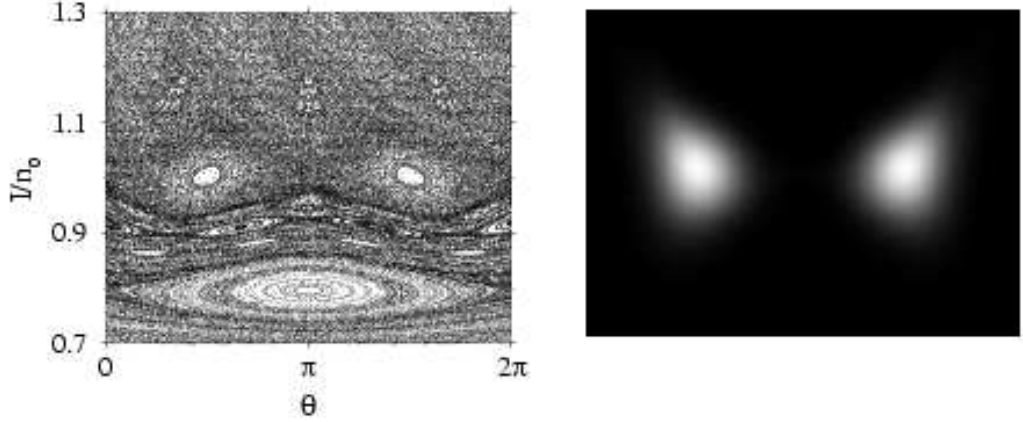,width=14cm}
}
\caption{Left: Poincar\'e surface of section of the one-dimensional hydrogen atom 
under linearly polarized driving, eq.~(\protect\ref{ham_lin1d}), for resonant
driving at twice the Kepler frequency.
The scaled field strength is $F_0=0.03$, 
and the phase is fixed at $\omega t=0.$
The $s=2$ resonance islands are 
apparent, embedded in the chaotic sea, and separated from the $s=1$ resonance
by invariant tori. Right: Husimi representation \protect\cite{abu95a}
of a Floquet eigenstate (for $n_0=60$)
anchored to the $s=2$ resonance displayed on the left.}
\label{orh_0} 
\end{figure}

Let us now return to the  hydrogen atom driven by LP microwaves.
The highly nonlinear character of the Coulomb interaction favours 
non-dispersive wave-packets anchored to the $s:1$ resonance island, since the
Fourier components $V_s$ of the coupling between the atom and the
microwave decay 
slowly\footnote{The very same behavior characterizes the gravitational
bouncer discussed in the previous section. For the bouncer the slow
inverse square dependence of $V_s$ on $s$ is due to a hard 
collision with the oscillating surface.
 For the Coulomb problem, the singularity at the origin
is even stronger.}
 with $s$ -- compare eqs.~(\ref{v_1dfou},\ref{v_3d_x}).
Consider first the simpler
1D model of the atom. We  discuss the 
$s=2$ case only, since similar conclusions can be obtained
for higher $s$ values.
The left panel in fig.~\ref{orh_0} shows the classical phase space 
structure (Poincar\'e surface of section) 
for $F_0=0.03$,
 with the $s=2$
resonance completely embedded in the chaotic sea, and well separated from the 
much larger principal resonance island.
From our experience with the principal 
resonance, and from the general considerations on $s:1$ resonances above, 
we expect to find 
Floquet eigenstates which are localized on this classical phase space
structure and mimic the temporal evolution of the corresponding 
classical trajectories.
Indeed, the right plot in fig.~\ref{orh_0} 
displays a Floquet eigenstate obtained by 
``exact" numerical diagonalization, which precisely exhibits 
the desired properties.

Again, this observation has its direct counterpart in the realistic 
3D atom, where the $2:1$ resonance allows for the
construction of non-dispersive wave-packets along elliptic trajectories,
as we shall demonstrate now. 
We proceed as for the $s=1$ case (sec.~\ref{LIN3D}): 
the secular Hamiltonian is obtained by averaging the full Hamiltonian,
eq.~(\ref{hamaa}), after transformation to the ``rotating frame",
eq.~(\ref{trafo_s_b}),
over one period $\tau=sT$ of the resonantly driven classical
trajectory:
\begin{equation}
{\mathcal H}_{\rm sec}=\hat{P}_t-\frac{1}{2\hat I^2}
-\frac{\omega \hat I}{s}
+F\sqrt{1-\frac{M^2}{L^2}}\left[-X_s(\hat{I}) \cos\psi\cos\hat{\theta} + Y_s(\hat{I}) \sin\psi\sin\hat{\theta}\right],
\label{hamscfin_s}
\end{equation}
where $X_s$ and $Y_s$ are given by eqs.~(\ref{xm},\ref{ym}). 
This can be condensed into 
\begin{equation}
{\mathcal H}_{\rm sec}=\hat{P}_t-\frac{1}{2\hat I^2}
-\frac{\omega \hat I}{s}
+F\chi_s\cos(s\hat{\theta}+\delta_s) ,
\label{hamscfin2}
\end{equation}
with
\begin{eqnarray}
\chi_s(\hat{I},L,\psi) & := & \sqrt{1-\frac{M^2}{L^2}}\sqrt{X_s^2\cos^2\psi+Y_s^2\sin^2\psi},
\label{substc2} \\
\tan\delta_s (L,\psi)& := & \frac{Y_s}{X_s}\tan\psi = \frac{J_s(se)\sqrt{1-e^2}}{eJ_s'(se)} \tan\psi.
\label{substb2}
\end{eqnarray}

\begin{figure}
\centerline{\psfig{figure=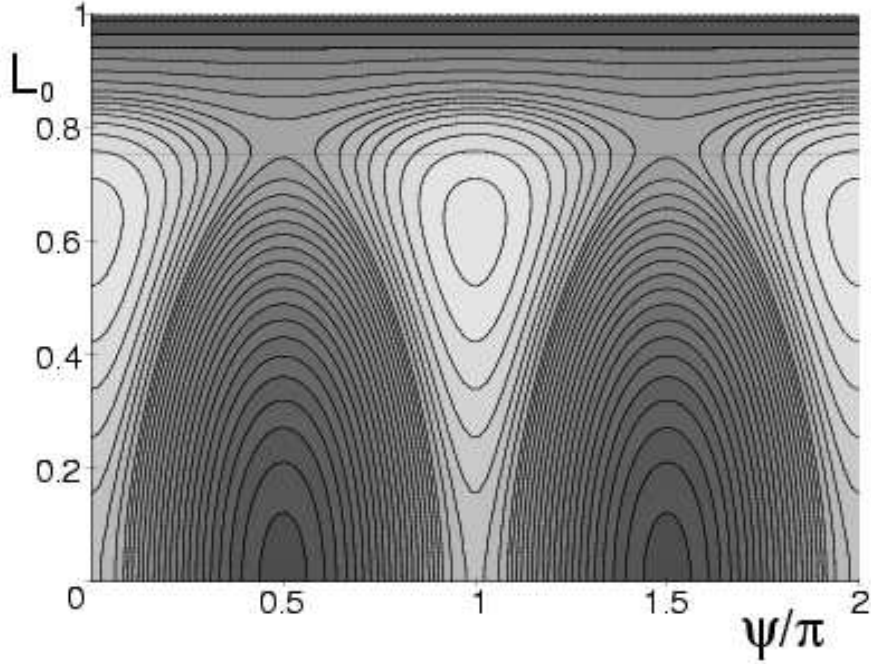,width=12cm,angle=-90}}
\caption{Contour plot of the effective perturbation $\chi_2$, 
eq.~(\protect\ref{substc2}), generating the slow evolution of the 
electronic trajectory in the $(L_0=L/\hat{I},\psi)$ plane. The secular
motion in this case is topologically different from that corresponding to
the $s=1$ resonance (compare with fig.~\protect\ref{lin3d_1}). 
In particular, 
there appear new fixed points at $L_0\simeq 0.77,\psi=\pi/2,3\pi/2$ (unstable,
corresponding to unstable elliptic
orbits with major axis perpendicular to the polarization axis) and at
$L_0\simeq 0.65,\ \psi=0,\pi$ (stable, corresponding
to stable elliptic orbits with major axis parallel and antiparallel
to the polarization axis). The resonance island in the $(\hat I, \hat \theta)$ 
plane is quite large for the latter stable orbits, and the associated
eigenstates are
non-dispersive wave-packets localized both longitudinally along
the orbit (locked on the microwave phase), and in the transverse direction, 
see figs.~\protect\ref{orh_5}, \protect\ref{orh_6}, \protect\ref{orh_7}.}
\label{orh_2} 
\end{figure}

For simplicity, we will now discuss the case $M=0,s=2.$
Fig.~\ref{orh_2} shows the equipotential lines of $\chi_2$
in the $(L,\psi)$ plane, calculated from 
eqs.~(\ref{hamscfin2}-\ref{substb2})\footnote{Since $\chi_s$ scales globally 
as $\hat{I}^2$,
the equipotential lines in fig.~\protect\ref{orh_2} do not depend
on $\hat{I}.$}.
For a comparison with quantum data, the equipotential lines represent
the values of $\chi_2$ for $n_0=42$, quantized from the WKB prescription
in the $(L,\psi)$ plane, exactly as done for the principal
$s=1$ resonance in section~\ref{LIN3D}.
 
One immediately notices that the secular
motion is in this case topologically different from that corresponding to
the $s=1$ resonance (compare to fig.~\ref{lin3d_1}), 
with the following features:
\begin{itemize}
\item three different types of motion coexist, with separatrices
originating from the straight line orbits parallel 
($L_0=0,\psi=0,\pi)$ to the
polarization axis.
\item The straight line orbits perpendicular to the polarization axis
($L_0=0,\psi=\pi/2,3\pi/2)$ lie at minima -- actually zeros -- of
$\chi_2.$ At lowest order, they exhibit vanishing coupling to 
the external field,
as for the $s=1$ resonance. Hence, the resonance island in the
$(\hat{I},\hat{\theta})$ plane will be small, and the
wave-packets  localized along the corresponding orbits 
are not expected to exist for moderate excitations.
\item The circular orbit (in the plane containing the polarization axis,
$L_0=1,$ arbitrary $\psi$) also exhibits vanishing coupling (since the 
circular motion is purely harmonic, no coupling is possible
at $\omega=2\Omega$).
\item There are ``new" fixed points at $L_0\simeq 0.77,\psi=\pi/2,3\pi/2$
(unstable), and at $L_0\simeq 0.65,\psi=0,\pi$ (stable), corresponding
to elliptical orbits with major axis perpendicular and parallel to the
polarization axis, respectively. The latter ones correspond
to {\em maxima} of $\chi_2$, and are associated with a large
resonance island in the $(\hat{I},\hat{\theta})$ plane. The motion in
their vicinity is strongly confined,
{\it both} in the angular ($L_0,\psi$) and 
in the ($\hat I,
\hat \theta$) coordinates: the corresponding eigenstates can be characterized
as non-dispersive wave-packets, localized both longitudinally along
the orbit (locked on the microwave phase), and in the transverse direction.
\end{itemize}

\begin{figure}
\centerline{\psfig{figure=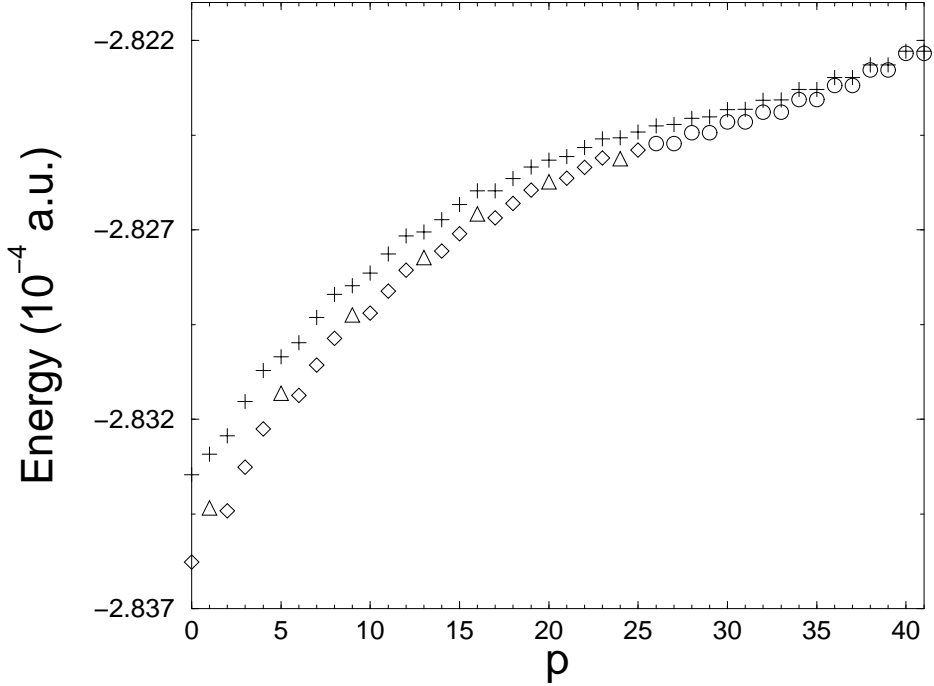,width=12cm,angle=-90}}
\caption{Comparison of numerically 
exact quasienergies originating from the 
$n_0=42$ manifold (depicted by pluses) to the semiclassical prediction 
(open symbols) based
on the quantization of the $s=2$ resonance island
(microwave frequency $=2 \times$ Kepler frequency), for 
scaled microwave field $F_0=0.04$. Circles
correspond to doubly degenerate states localized in the vicinity of maxima of 
$\chi_2$, around the elliptic fixed points at $(L_0\simeq 0.65,\psi=0,\pi)$
in fig.~\protect\ref{orh_2}. Triangles correspond to almost 
circular states in the vicinity of the stable minimum at 
($L_0=1,\psi\ \rm arbitrary)$, while diamonds correspond to states localized 
around the stable minima at $L_0=0,\psi=\pi/2,3\pi/2$. 
The agreement between 
the semiclassical and quantum energies is very good, provided the size of the 
resonance island in the $(\hat{I},\hat{\theta})$ plane 
is sufficiently large
(high lying states in the manifold). For low 
lying states in the manifold, the discrepancies between quantum and 
semiclassical 
results are significant, due to the insufficient size of the island.
}
\label{orh_3} 
\end{figure}

In order to separate quantum states localized in different regions
of the $(L_0,\psi)$ space, we
show in fig.~\ref{orh_3} a comparison between the semiclassical prediction and
the numerically exact 
Floquet energies (obtained as
in section \ref{LIN3D} for the $s=1$ resonance) originating from 
this manifold, with  
$N=0$ in eqs.~(\ref{ebkires},\ref{ebkinres}), at $F_0=0.04$.
Observe that the 16 upmost states appear in eight quasi-degenerate pairs
differing by parity. 
Exact degeneracy does not happen because of 
tunneling effects: the lower the doublet in energy, the larger its
tunneling splitting. The tunneling process 
involved here is a ``transverse" tunneling in the $(L,\psi)$ plane, 
where the electron jumps from the elliptic $(L_0\simeq 0.65,\psi=0)$ 
Kepler 
trajectory to its image under $z$-parity, the $(L_0\simeq 0.65,\psi=\pi)$ 
trajectory (compare fig.~\ref{orh_2}). This tunneling process is entirely due 
to the specific
form of $\chi_2$,
with two {\em distinct} maxima.

The energetically highest doublet in fig.~\ref{orh_3} 
corresponds to states
localized as close as possible 
to the fixed points $L_0\simeq 0.65,\ \psi=0,\pi$.
For these states (large resonance island in the ($\hat I, \hat \theta$)
plane),
semiclassical quantization nicely
agrees with the quantum results. On the other hand, the agreement
between quantum and semiclassical results progressively degrades 
for lower energies,
as the size of the  island
in the ($\hat I, \hat \theta$) plane becomes smaller. Still, 
the disagreement between
semiclassical and quantum results is at most of the order of
the spacing between adjacent levels\footnote{
A quantum approach based on the pendulum approximation and 
the Mathieu equation would give a much better prediction
for such states.}.
Below the energy of the unstable fixed points at 
$L_0\simeq 0.77,\psi=\pi/2,3\pi/2$,
there are no more pairs of classical trajectories in the $(L,\psi)$ plane
corresponding to distinct classical dynamics related by $z$-parity.
Hence, the doublet structure 
has to disappear, as confirmed by the
exact quantum results shown in fig.~\ref{orh_3}. On the other hand, there
are two disconnected regions in the $(L,\psi)$ plane which can give rise to 
quantized values of
$\chi_2$ (and, consequently, to quasienergies) within the same quasienergy 
range: the neighbourhood of
the stable fixed points $(L_0=0,\psi=\pi/2,3\pi/2)$, and the region close to
$L_0=1$. In the semiclassical quantization 
scheme, these regions are completely decoupled  
and induce two independent, 
non-degenerate series of quasienergy levels. 
Consequently, the complete spectrum exhibits a rather
complicated structure, caused by the interleaving of these two series.

\begin{figure}
\centerline{\psfig{figure=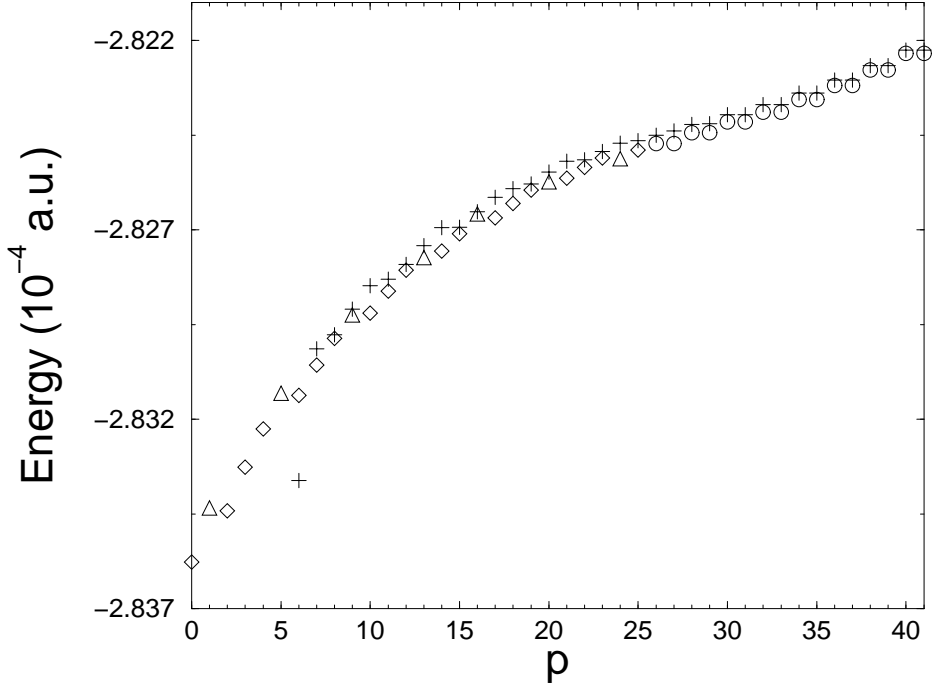,width=12cm,angle=-90}}
\caption{Same as fig.~\protect\ref{orh_3}, but for the mirror manifold 
shifted in energy by $\omega/2$. While, for most states, the agreement with 
semiclassics is of the same quality as in fig.~\protect\ref{orh_3}, no quantum
data are plotted at the bottom of the manifold. Indeed, at those energies, 
another Rydberg manifold strongly perturbs the spectrum due to close 
accidental degeneracy. Consequently, the unambiguous identification of 
individual states is very difficult.}
\label{orh_4} 
\end{figure}

As discussed in section~\ref{ORG}, for a $s:1$ resonance, in a Floquet 
zone of width
$\omega=s\Omega$, there is
not a single manifold of states, but rather a set
of $s$ different manifolds approximately identical and
separated by $\Omega.$
Deviations from the exact $\omega/s$ periodicity are due
to tunneling~\cite{holthaus95}. This tunneling process is however
{\em completely} different from the 
``transverse" one in the $(L,\psi)$ plane described above. 
It is a case of 
``longitudinal" tunneling, where the electron jumps from one location
on a Kepler orbit to another, shifted along 
the {\em same} orbit. This longitudinal tunneling is 
similar in origin to the tunneling described in section \ref{OGB}. 
It has to be stressed that it represents a  general
phenomenon in the vicinity of a $s:1$ resonance 
(with $s\ge2)$, due to the
phase space structure in the $(\hat I, \hat \theta$) plane, see section
~\ref{ORG},
 in contrast
to the ``transverse" quasi-degeneracy (discussed in fig.~\ref{orh_3})
due to the specific form of $\chi_2.$
Inspecting the numerically exact quantum quasienergy spectrum,
we indeed find the manifold 
shown in fig.~\ref{orh_4} (compared to the semiclassical prediction). 
Observe that the agreement between quantum and
semiclassical quasienergies is similar to that observed in fig.~\ref{orh_3},
except for the low lying states. Here, incidentally, the states
anchored to the resonance island are strongly perturbed by another
Rydberg manifold; proper identification of the individual 
quantum states is very difficult in this region, and therefore no quantum 
data are shown at low energies.

Finally, let us consider the localization properties of the wave-functions 
associated with 
the upmost states of the manifolds in figs.~\ref{orh_3} and \ref{orh_4}. 
These wave-functions should localize in the vicinity of 
stable trajectories of period 2, i.e. they should be 
strongly localized, both in angular and orbital coordinates, 
along an elliptic Kepler orbit of intermediate eccentricity. 
However, because of the longitudinal quasi-degeneracy, we expect the
associated Floquet eigenstates to be composed of
two wave-packets on the ellipse, exchanging their positions
with period $T.$ 
Furthermore, due to the transverse quasi-degeneracy,
we should 
have combinations of the elliptic orbits labeled by 
$\psi=0$ and $\psi=\pi.$ Altogether, this makes four individual
wave-packets represented by each Floquet state. 
Due to the azimuthal symmetry of the problem around the field polarization
axis, each wave-packet actually is doughnut-shaped (compare fig.~\ref{lin3d_8}
for the simpler $s=1$ case).

\begin{figure}
\centerline{\psfig{figure=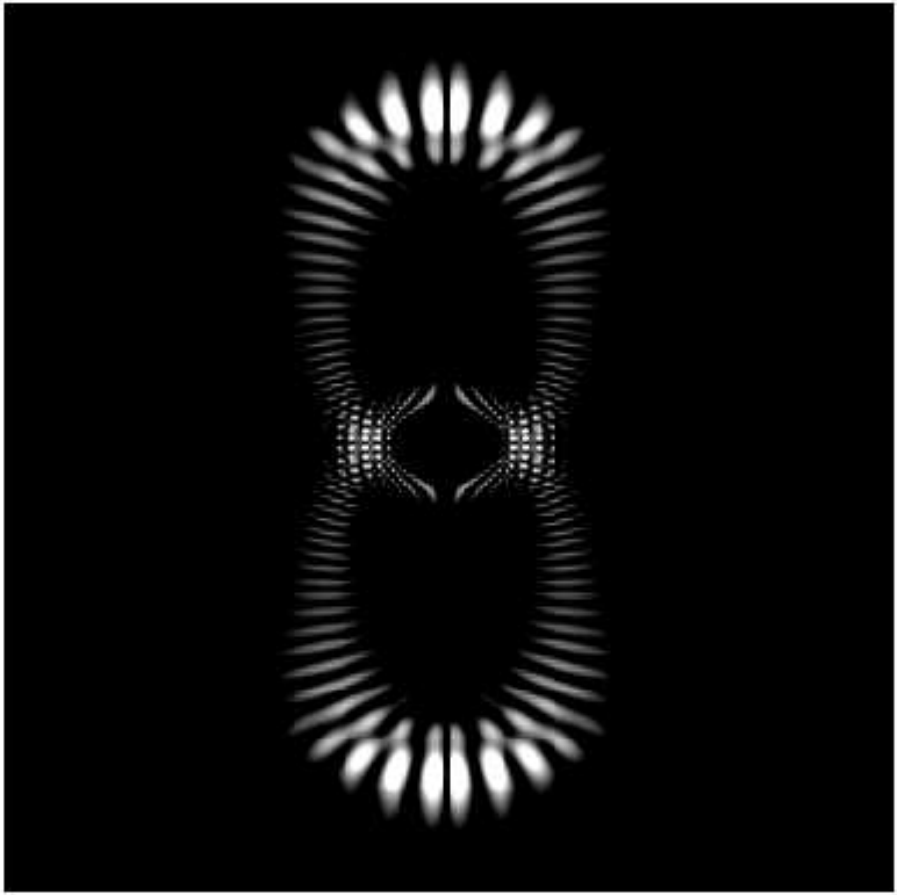,width=12cm,angle=0}}
\caption{Electronic density of the upmost eigenstate of the $n_0=42$ manifold
of fig.~\protect\ref{orh_3}, averaged over one microwave period. This state 
presents localization along a pair of Kepler ellipses oriented along the field
polarization axis. The box measures $\pm 3500$ Bohr radii in both $\rho$ and 
$z$ directions, with the nucleus at the center. 
The microwave polarization axis along $z$ is 
parallel to the vertical 
axis of the figure. The orientation and eccentricity of the ellipse
are well predicted by the classical resonance analysis.}
\label{orh_5} 
\end{figure}

Exact quantum calculations fully confirm this prediction. Fig.~\ref{orh_5}
shows the electronic density of the upmost Floquet state in the
$n_0=42$ manifold (fig.~\ref{orh_3}), averaged over one field period. 
As expected,
it is localized along two symmetric Kepler ellipses ($\psi=0$ and $\psi=\pi$,
respectively), but
longitudinally delocalized because of the time average.
In fact, there are four such Floquet states displaying very similar electronic
densities. These are the energetically highest doublet 
in the $n_0=42$ manifold,
and the upmost doublet in the ``mirror" manifold displayed in fig.~\ref{orh_4}.

\begin{figure}
\centerline{\psfig{figure=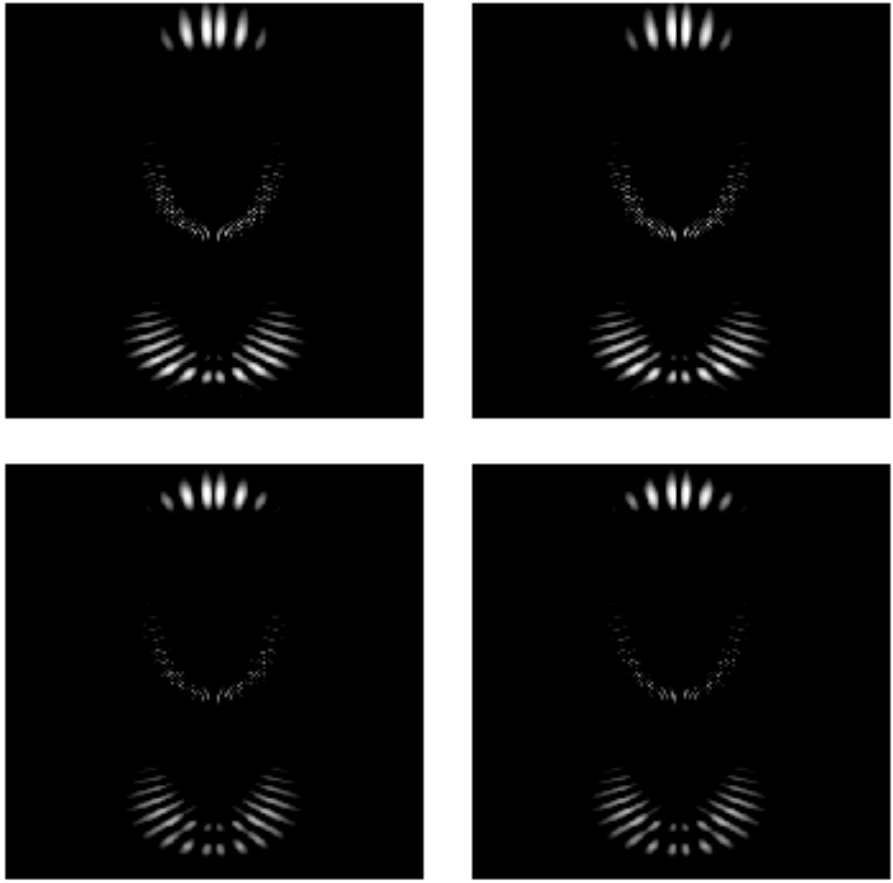,width=12cm,angle=0}}
\caption{Electronic densities of the eigenstates of the upmost doublet states 
(top) of the $n_0=42$ manifold of fig.~\protect\ref{orh_3}, and of their mirror
states (bottom), shifted in energy by $\omega/2$ (fig.~\ref{orh_4}), 
at driving field phase $\omega t=0$. 
The longitudinal localization
on the Kepler ellipses (similar for all states) is apparent. On each ellipse,  
four different individual wave-packets (or rather, due to azimuthal symmetry,
two doughnut wave-packets) 
can be distinguished, 
propagating along the Kepler ellipse.
Notice the phase shift of $\pi$ in the temporal evolution on the two ellipses,
implied by $z$-inversion. The microwave polarization axis along $z$ is 
given by the
vertical 
axis of the figure, with the nucleus at the center of the figure.}
\label{orh_6} 
\end{figure}

\begin{figure}
\centerline{\psfig{figure=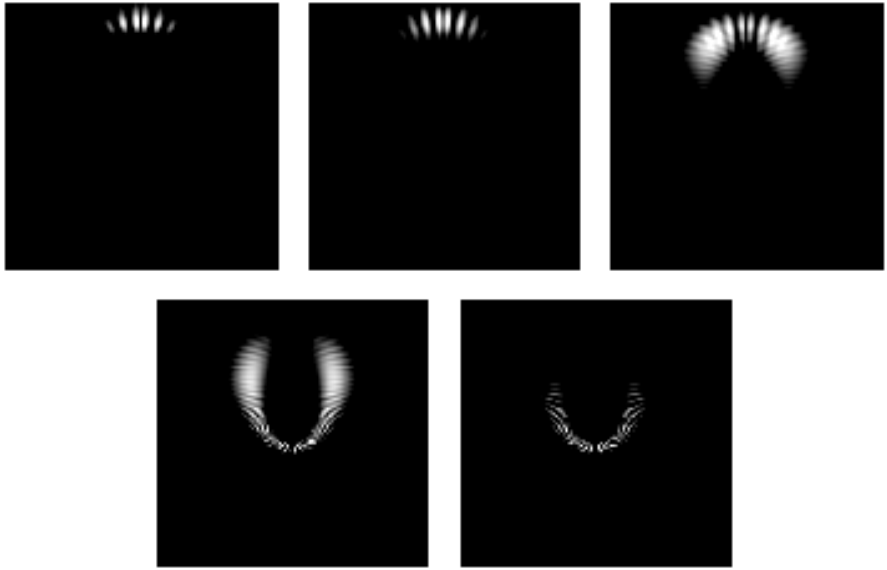,width=12cm,angle=0}}
\caption{Temporal evolution of a convenient linear combination of the four 
eigenstates 
of fig.~\protect\ref{orh_6}, for phases $\omega t=0$ (top left), 
$\pi/2$ (top center), $\pi$ (top right), $3\pi/2$ (bottom left), $2\pi$ (bottom
right) of the driving field. Clearly, a single doughnut propagating along a 
single trajectory has been selected by the linear combination. This wave-packet
essentially repeats its periodic motion with period $2T=4\pi/\omega$. 
It slowly 
disperses, because the four states it is composed of 
are not exactly degenerate 
(tunneling effect), and because it ionizes (see sec.~\ref{ION}). 
The microwave polarization axis 
along $z$ is 
parallel to the
vertical 
axis of the figure, with the nucleus at the center of the plot.}
\label{orh_7} 
\end{figure}

Fig.~\ref{orh_6} shows the electronic densities of these four
Floquet eigenstates at phase $\omega t=0$ of the driving field:
the four doughnuts  are now clearly visible,
as well as the orbital and radial 
localizations along the two elliptic trajectories. 
Very much in the same way as for a double well potential
(or for the bouncer discussed in section \ref{OGB}, compare 
fig.~\ref{othfig1}),
a linear combination 
of these four states allows for the selection of one single doughnut, 
localized along one single classical Kepler ellipse. 
This wave-packet then evolves along this 
trajectory without dispersion, as demonstrated in fig.~\ref{orh_7}. 
Note, however, that 
this single wave-packet is {\em not} a single Floquet state, and
thus does not exactly
repeat itself periodically. It slowly disappears
at long times, for 
at least two 
reasons: firstly, because of
longitudinal and transverse tunneling, the phases of the four Floquet
eigenstates accumulate small differences as time evolves, what induces
complicated oscillations between the four possible
locations of the wave-packet, and secondly, 
the ionization rates of the individual
Floquet states lead to ionization and loss of phase coherence,
especially if the ionization rates (see sec.~\ref{ION})
of the four states are not equal.

\section{Alternative perspectives}
\label{ALTP}

There are several known systems where 
an oscillating field is used to
stabilize a specific mode of motion, such as 
particle accelerators \cite{lichtenberg83}, Paul traps 
\cite{paul92} for ions, etc. 
In these cases, the 
stabilization is a completely
classical phenomenon based on 
the notion of nonlinear resonances. What 
distinguishes our
concept of non-dispersive wave-packets discussed in the preceding chapters 
from those situations is
the necessity to use quantum (or semi-classical) mechanics to describe a given
problem, due to relatively low quantum numbers. Still, the principle of 
localization 
remains the same, and consists in locking the motion of the system
on the external drive.
However, it is not essential that the drive be provided externally, it 
may well be supplied 
by a (large) part of the system to the (smaller) remainder. Note that, 
rather formally, also 
an atom exposed to a microwave field can be understood as one large 
quantum system -- a 
dressed atom, see sec.~\ref{SQ} -- where the field-component 
provides the drive for 
the atomic part \cite{cct92}.
In the present section, we shall therefore briefly recollect 
a couple of related phase-locking phenomena in slightly more complicated 
quantum systems, which open additional perspectives for creating
non-dispersive 
wave-packets in the microscopic world.

\subsection{Non-dispersive wave-packets in rotating molecules}
\label{MOL}

A situation closely related to atomic hydrogen exposed to CP microwaves 
(sec.~\ref{CP}) is met when considering the dynamics of a single, 
highly
excited Rydberg electron in a
rotating molecule~\cite{benvenuto94}. 
In \cite{ibb96}, the following  
model Hamiltonian has been proposed:
\be
H=\frac{\vec{p}^2}{2}-\frac{1}{|\vec{r} + \vec{a}(t)|},
\label{molibb0}
\ee
where $\vec{a}(t)$ denotes the position of the center of the Coulomb field
w.r.t. the molecular center of mass, and is assumed to rotate in the 
$x-y$ plane 
with constant frequency $\omega$:
\begin{eqnarray}
{\vec a}(t) & = & \left (
\begin{array}{lc}
\cos\omega t & \ \ -\sin\omega t \\
\sin\omega t & \ \ \cos\omega t
\end{array} \right ){\vec a}.
\label{rotvec}
\end{eqnarray}
In the rotating frame, one obtains the Hamiltonian \cite{benvenuto94} 
\be
H=\frac{p_x^2+p_y^2+p_z^2}{2}-\frac{1}{r} +a\omega^2 x -\omega L_z +
\frac{a^2\omega^2}{2},
\label{molibb}
\ee
which, apart from the constant term $a^2\omega^2/2$, is equivalent to the one 
describing an atom driven by a CP field
(compare eq.~\ref{hrot})).  Note that the
role of the microwave amplitude (which can be arbitrarily tuned in the 
CP problem)
is taken by $a\omega^2$, i.e., a combination of molecular parameters, 
what, of course, 
restricts the
experimental realization of 
non-dispersive wave-packets in the molecule to 
properly selected molecular species \cite{ibb96}. 

With the help of the stability analysis outlined in sec.~\ref{rotating_frame}, 
eqs.~(\ref{wp0}-\ref{wpu}), the 
equilibrium position $x_{\mathrm eq}$ 
of the molecular Rydberg electron is easily estimated according to 
(assuming a small value of $a$, limited by the size 
of the molecular core)
\be
x_{\mathrm eq}\simeq ({\cal I}/{\cal J})^{2/3},
\label{x0mol}
\ee
 with ${\cal I}$ 
the molecular momentum of inertia, 
${\cal J}=\omega{\cal I}$
the rotational quantum number.
To optimize the angular localization of the wave-packet, 
it is necessary that $x_{\mathrm eq}$ 
be sufficiently large 
(from sec.~\ref{RSEF}, $x_{\mathrm eq}\sim n_0^2$, where
$n_0$ is the electronic principal quantum number). Thus, for given ${\cal I}$,
${\cal J}$
should be 
small. In 
\cite{ibb96}, 
a hydrogen-tritium molecule is considered, which yields 
$n_0\simeq 18$ for ${\cal J}=1$. 

Note, however, that such reasoning 
is {\em not} justified. 
The effective Hamiltonian (\ref{molibb0})
implies a {\em classical} description of 
the molecular rotation (much as the {\em classical} treatment 
of the periodic drive in 
eq.~(\ref{h_gen}), with a well-defined phase) defined by the position vector
${\vec a}(t).$
For such an approach to
be valid, ${\cal J}$ must be sufficiently 
large. The molecular rotation plays
the role of the microwave field in the analogous CP problem,
the number of rotational quanta is just equivalent to the average number of
photons defining the amplitude of the (classical) coherent state 
of the driving 
field. Clearly, if ${\cal J}$ is too small, the  
effect  of an exchange
of angular momentum between the Rydberg electron and the core
on 
the {\em quantum state} 
of the core (and, hence, on $a\omega^2$ assumed to be constant
 in eq.~(\ref{molibb})) cannot be
neglected and, therefore, 
precludes any semiclassical treatment, see also \cite{delande98,shirley65}. 
In other words, if ${\cal J}$ is too small, the number of rotational 
states of the core which are coupled 
via the interaction is too small to mimic a quasi-classical evolution as 
suggested by eq.~(\ref{rotvec}).

Nontheless, this caveat does not completely
rule out the existence of 
 molecular non-dispersive wave-packets,
provided a fast
rotation of a core  
with large momentum of inertia (to render $x_{\mathrm eq}$
sufficiently large, eq.~(\ref{x0mol}),
such that the electronic wave-packet gets localized 
far away from the molecular core) can be 
realized, as also suggested
in \cite{ibb96}. 

\subsection{Driven Helium in a frozen planet configuration}
\label{HE}

In the previous examples of non-dispersive wave-packets,
the key point has been the generic appearance of a nonlinear
resonance 
for periodically driven quantum
systems whose  unperturbed dynamics is integrable.
A natural question to ask is whether the concept of non-dispersive wave-packets 
can be generalized to systems which exhibit mixed regular-chaotic dynamics 
even in the absence of the external perturbation. In the atomic realm, 
such a situation is realized for the helium atom, where 
electron-electron interactions provide 
an additional source of 
nonlinearity.
The corresponding Hamiltonian writes in atomic units
\begin{equation}
H_{\rm He} = \frac{{\vec p}_1^2}{2}+\frac{{\vec p}_2^2}{2}-\frac{2}{r_1}
-\frac{2}{r_2}+\frac{1}{|{\vec r}_1-{\vec r}_2|}.
\label{he_free}
\end{equation}
As a matter of fact, the classical and quantum 
dynamics of the three-body Coulomb problem generated by Hamiltonian~(\ref{he_free}) 
has been a largely unexplored
``terra incognita'' until very recently \cite{tanner00}, since the dimensionality 
of the phase space dynamics increases from effectively two to effectively eight 
dimensions when a second electron is added to the familiar Kepler problem.
Furthermore, the exact quantum mechanical treatment of the helium atom 
remains a formidable task since the early days of quantum mechanics, and 
considerable advances could be achieved only very recently, with the 
advent of modern semiclassical and group theoretical methods 
\cite{wintgen93,gremaud97,gremaudth,puttner01}.
Already the classical dynamics of this system exhibits 
a largely chaotic phase space structure, which typically 
leads to the rapid autoionization 
of the associated {\em doubly excited} 
quantum states of the atom. One of the major 
surprises in the analysis of the three body Coulomb problem during 
the last decades has therefore been the discovery of a new, highly correlated
and classically globally stable electronic configuration, the 
``frozen planet'' \cite{eichmann90,richter90}. 
The appeal of this configuration resides in its 
counterintuitive, asymmetric character where both electrons are located on the
same side of the nucleus. Furthermore, this configuration turns out to 
be the most robust of all known doubly excited two-electron configurations, in the 
sense that it occupies a large volume in phase space. Its stability is 
due to the strong coupling of the two electrons by the $1/|\vec{r}_1-\vec{r}_2|$ 
term in the Hamiltonian (\ref{he_free}), which enforces their highly correlated motion.

The frozen planet is an ideal candidate to test the prevailance of the 
concept of nondispersive wave-packets in systems with intrinsically
mixed dynamics.
In a recent study~\cite{schlagheck98a,schlagheck99,schlagheck99b,schlagheckth} 
the response of this highly correlated two-electron 
configuration to a periodic force has been investigated from a classical
and from a quantum mechanical point of view. 
The Hamiltonian for the driven problem writes, in the length gauge,
\begin{equation}
H=H_{\rm He}+F\cos(\omega t)(z_1+z_2).
\label{he_driven}
\end{equation}
Guided by the experience 
on non-dispersive wave-packets in one electron Rydberg states,
the driving frequency $\omega$ 
was chosen near resonant with the natural frequency
$\Omega_{\rm FP}\approx 0.3n_i^{-3}$ of the frozen planet, where $n_i$ denotes the 
principal
quantum number of the inner electron. It was found that, for a suitably chosen 
driving field amplitude $F$, a nonlinear resonance between the correlated 
electronic motion and the external drive can be induced in the classical 
dynamics, at least for 
the collinear frozen planet where the three particles (two electrons and the
nucleus) are aligned 
along the polarization
axis of the driving field. 

However, contrary to the situation for the 
driven hydrogen atom discussed in sections \ref{LIN1D} and \ref{LIN3D}, 
there is a fundamental difference between the one 
dimensional model of the driven three body Coulomb problem
and the full 3D problem. For the one-electron system,
we have seen that the classical Kepler ellipse performs
a slow precession in the angular variables, though
remains bounded and does not ionize. In contrast, if one permits deviations
from collinearity in the driven frozen planet dynamics, it is found that
the transverse
direction is generally unstable and leads to rapid ionization. This transverse
ionization is simply due to the fact that the external field destroys the 
intricate electron electron correlation which creates the unperturbed 
frozen planet. 
Notwithstanding, 
it has been shown that the application of an additional, weak 
static electric field allows to compensate 
for the transverse instability, and to establish a classically globally 
stable dynamical 
situation for the frozen planet. The transverse confinement 
through the static field again justifies the collinear model, 
and first quantum calculations performed 
for this restricted model show the existence of a 
wave-packet associated with the principal resonance between the frozen planet orbit and 
the driving field,
which faithfully traces the classical trajectory at
the period of the drive. 
As for driven one-electron systems, 
these nondispersive two-electron 
wave-packets exhibit life times of typically $10^6$ driving field periods
\footnote{Again, in contrast to the driven one electron 
problem, nothing guarantees that the life times obtained for the 1D model 
carry over
to the real 3D object. On the contrary, first results on the bare 3D Coulomb 
problem 
\protect\cite{schlagheckth} indicate a strong dependence 
of the life times on 
the dimension of the accessible configuration space.}.

Hence, there is strong evidence that a resonant external forcing allows
for the creation of quantum eigenstates with a quasi-classical
temporal evolution, even in the presence of strong two-particle correlations.

\subsection{Non-dispersive wave-packets in isolated core excitation of
multielectron atoms}
\label{LAMBRO}

Another example of non-dispersive wave-packets in a two-component 
atomic system has 
recently been proposed for two-electron 
atoms \cite{hanson95,zobay96,mecking98}. 
The scheme
uses an isolated-core excitation in which one of the electrons is transfered
to a Rydberg trajectory 
by a {\em short} laser pulse, forming 
an initially well-localized wave-packet.
A second 
source {\em continuously} drives a transition between two discrete states 
of the remaining atomic 
core. The latter
induces Rabi oscillations (or a coherent superposition) between 
two 
Rydberg series to which the
first electron is
excited. If the 
Rabi 
frequency (controlled by the continuous drive of the core) is matched with the
Kepler frequency of the orbit of  
the 
outer electron,
the autoionization rate of the latter
may be strongly suppressed, provided the respective
phases are also matched properly: 
if 
the electron approaches its 
inner turning radius (where 
the configuration-interaction between Rydberg electron and core -- 
leading to autoionization --
is 
strongest) while 
the core is in its ground state, 
autoionization becomes impossible since the  
configuration-interaction 
does not compensate for the ionization potential of the Rydberg electron. 
On the other
hand, when the electron is far from the nucleus 
(and electron-electron
interaction is weak), the core may be in its excited state, 
without ejecting the Rydberg 
electron.

Consequently, autoionization is supressed for the center of the 
Rydberg wave-packet.
During 
time evolution, however, the wave-packet spreads, 
its head and
its tail desynchronize with the Rabi evolution of the core, 
and eventually approach   
the region close
to the nucleus (where configuration-interaction is most pronounced)  
when the core is not
in its ground state. Then these parts of the wave-packet autoionize, 
and the remaining Rydberg population 
is reshaped into a localized wave-packet, since the spreading tails have been 
chopped off.
Hence, these wave-packets exhibit a rather rapid ``melting'' (on a time scale
of at most some hundred Kepler periods) --
to be compared 
to hundreds of thousands or even more Kepler cycles performed 
by non-dispersive 
wave-packets in microwave driven hydrogen atoms studied above (which also
ionize, however {\em very slowly}, see sec.~\ref{ION}).

The present scenario is in some sense reminiscent of the one 
 in sec.~\ref{MOL}, with a (quantum) two-level core replacing
the rotating molecular core.
As mentioned above, a two-level system alone can only exchange one quantum
with the the outer electron and thus cannot provide an
 exact phase locking mechanism for the highly excited Rydberg electron.
However, the two-level core
is here {\em driven} by an external electromagnetic field and consequently
gains an additional degree of freedom which can be used for the
phase locking mechanism. The drawback is that this phase locking
implies losses (through autoionization). Nevertheless,
the quasi-classical evolution over $\sim 100$ Kepler cycles 
is still quite impressive, and presumably stems from the relatively sharp 
confinement of efficient configuration-interaction within a spatial region close
to the inner turning point of the Rydberg wave-packet.

\section{Characteristic properties of non-dispersive wave-packets}
\label{CHPROP}

After presenting several examples of non-dispersive wave-packets in the previous
chapters, we now study their specific properties in more detail. 
Especially, several important physical
processes which may affect the existence of wave-packets 
have so far been hidden under the carpet~\cite{carpetmen}. 
The two most important ones, at least for
driven atoms, are ionization and spontaneous emission, and they will 
be discussed
in detail below. First, let us briefly discuss the general properties of
wave-packet eigenstates 
under the variation of various parameters of 
the driven system 
(e.g.,
microwave amplitude 
and frequency, the strength of 
an external static 
field, etc.).

\subsection{Ionization rates and chaos assisted tunneling}
\label{ION}

Atoms driven by microwaves will eventually ionize. Therefore, the
non-dispersive wave-packet states discussed up till now cannot be,
rigorously speaking, discrete states, 
they are rather resonances~\cite{goldberger50} with
some finite life-times. Importantly, as we shall discuss in detail below,
these life-times may be extremely long, of the order of millions of microwave
periods. 
In that sense, they are comparable to 
those of highly 
excited atomic Rydberg states,
which also decay, by spontaneous emission, 
on time scales of few millions of classical periods.
Even more importantly, the life-times of the 
non-dispersive wave-packets are typically orders of
magnitude larger than the life-times of other 
states in the Floquet spectrum: the wave-packets
are particularly resistant
to ionization. This is due to the 
classical confinement
of the
electron inside the regular island. 
To ionize, the electron has no other option but
to tunnel
out of the classically confining island,
before gaining energy by diffusive excitation~\cite{casati88}. 
The resonance island is strictly confining only
for a one-dimensional system. For multi-dimensional systems, the 
tori in the resonance islands are not fully isolating and a very
slow classical diffusion process might eventually lead to ionization. 
This, however, takes place on extremely long time scales and is
completely negligible in atomic systems. 
In practice, ionization of the wave-packet is essentially mediated by a pure
quantum process, exponentially unlikely 
in the
semiclassical limit. As we shall see below, this tunneling
process has quite interesting properties which may be quantitatively
described 
for microwave driven atoms. 
More details 
can be found in \cite{kuba95a,kuba98,kuba98a}. 

Due to the initial tunneling step, the life-times of 
non-dispersive wave-packets will typically be 
much longer than those of 
Floquet states 
localized in the chaotic
sea surrounding the island \cite{abu95a}. Moreover, 
since the ionization mechanism involves 
chaotic
diffusion, many quantum mechanical paths link the initial wave-packet to
the final continuum. Thus, the life-time of the wave-packet will reflect 
the interferences between those different possible paths,
and will sensitively depend on parameters such as the microwave frequency or
amplitude, that affect the interfering paths 
through the chaotic sea. 
These fluctuations,
reported first in \cite{kuba95a}, are perfectly deterministic
and
resemble the conductance fluctuations observed in mesoscopic systems 
\cite{stone85} . 
\begin{figure}
\centerline{\psfig{figure=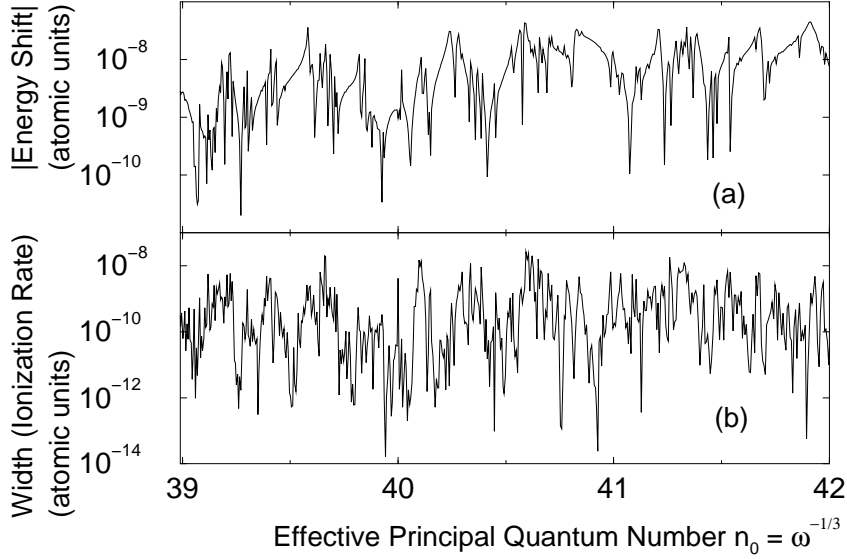,width=8cm,bbllx=100pt,bblly=150pt,bburx=520pt,bbury=600pt,angle=-90}}
\caption{Typical fluctuations of the width (ionization rate)
and of the energy (with respect to its averaged, smooth behavior)
of the non-dispersive wave-packet of a two-dimensional
hydrogen atom in a circularly polarized microwave field.
 The data presented
are obtained for small variations
of the effective principal quantum number $n_0=\omega^{-1/3}$ around 40,
and a scaled
microwave electric field $F_0=0.0426$.
To  show that the fluctuations 
cover
several orders of magnitude, we use a logarithmic vertical scale, and
plot the absolute value of the shift.}
\label{ionfig1}
\end{figure}
In fig.~\ref{ionfig1}, we show the fluctuations of the
ionization rate (width) of the non-dispersive wave-packet
of the two dimensional hydrogen atom in a circularly
polarized microwave field. The energy levels
and widths
are obtained as explained in sec.~\ref{RSEF}, by numerical
diagonalization of the complex rotated Hamiltonian. 
All the data presented in this section have been obtained 
in the regime where the typical ionization rate is smaller than the
mean energy spacing between consecutive levels, so that the
ionization can be thought as a small perturbation acting
on bound states.
The width (although 
very small) displays
strong fluctuations over several orders of magnitude.
Similarly, the real part of the energy (i.e., the center of the
atomic resonance) displays wild fluctuations. The latter
can be observed only if the smooth variation
of the energy level with the control parameter (following 
approximately the semiclassical prediction given
by eq.~(\ref{estharm}))
is substracted.
Therefore, we fitted the numerically obtained energies by a smooth function
and substracted this fit to obtain the displayed fluctuations. Note that
these fluctuations are so small that an accurate fit is 
needed\footnote{In particular, 
the semiclassical expression is not sufficiently
accurate for such  a fit.}. This can be easily seen in fig.~\ref{solit2}
where, on the scale of the mean level spacing, these fluctuations
are invisible by eye (the level appears as a straight horizontal line).

The explanation  
for the fluctuations is the following: 
in a quantum language, they are due to the coupling
between the localized wave-packet and states
localized in the chaotic sea surrounding the resonance island. 
While the energy 
of the wave-packet is a smooth function of the
parameters $F$ and $\omega,$  the energies
of the chaotic states  display a complicated behavior characterized
by level repulsion and large avoided crossings. It happens often
that -- for some parameter values -- there is a quasi-degeneracy
between the wave-packet eigenstate and a chaotic state, see the numerous
tiny avoided crossings in fig.~\ref{solit2}. 
There, the two
states are efficiently mixed, the wave-packet captures
some part of the coupling of the chaotic state to the continuum
and its ionization
width increases (see also \cite{abu95a}). 
This is the very origin
of the observed fluctuations. Simultaneously, the 
chaotic state repels the wave-packet state leading to a deviation of the 
energy from its smooth behavior,
and thus to the observed fluctuations. 
This mechanism is similar 
to 
``chaos assisted tunneling'', described in the literature
\cite{LB90,LB92,grossmann91,GDJH91b,GDJH93,plata92,bohigas93,bohigas93b,tomsovic94,shudo95,leyvraz96}
for both, driven one-dimensional and two-dimensional autonomous systems.
There, the tunneling rate between two symmetric
islands -- which manifests itself through the splitting between
the symmetric and antisymmetric states of a doublet -- 
may be strongly enhanced by the
chaotic transport between the islands. We have then a ``regular''
tunneling escape 
from one island, a chaotic diffusive transport from the vicinity of
one island to the other (many paths, leading to interferences and resulting
in large fluctuations of the splitting),
 and another ``regular'' tunneling penetration into the
second island. In our case, the situation is even simpler -- we have a
``regular'' tunneling escape supplemented by a chaotic diffusion and
eventual ionization. Thus, instead of the level splitting, we observe 
a shift of the energy level and a finite width.

Since these fluctuations stem from the coupling between the regular
wave-packet state and a set of chaotic states, it is quite natural
to model such a situation via a Random Matrix model \cite{kuba98}, 
the
approach being directly motivated by a similar treatment of the
tunneling splitting in \cite{tomsovic94}. For details,  
we refer the 
reader to the original work~\cite{kuba98}. It suffices to say here that
the model is characterized by three real parameters: $\sigma$ -- which
characterizes the mean strength of the coupling between the regular state 
and the chaotic levels, $\gamma$ -- which measures the decay of the chaotic
states (due to ionization;  direct ionization transitions from
the wave-packet state to the continuum are negligible),
and $\Delta$ -- which is the mean level spacing of chaotic levels.
The two physically relevant, dimensionless 
parameters are  $\gamma/\Delta$ and $\sigma/\Delta$.
In the perturbative regime $(\gamma/\Delta,\sigma/\Delta\ll 1)$ it is possible
to
obtain analytical \cite{kuba98} 
predictions for the statistical distribution of the energy
shifts $P(s)$ (of the 
wave-packet's energy from its unperturbed value)
and 
for the distribution of its widths $P(\Gamma)$. $P(s)$ turns out to
be a Cauchy distribution (Lorentzian), similarly to the tunneling
splitting distribution found in \cite{tomsovic94}. The distribution of the 
widths 
is a bit more complicated (it is the square root of $\Gamma$ which
is approximately 
Lorentzian distributed). 
The perturbative approach fails for the 
asymptotic behavior of the 
tails of the distributions, where an exponential cut-off is expected and
observed in numerical studies \cite{kuba98,tomsovic94}. By fitting 
the predictions of the Random Matrix model to the numerical data of 
fig.~\ref{ionfig1}, we may finally extract the values of 
$\gamma/\Delta$, the strength of the decay, and of $\sigma/\Delta,$  
the coupling between the regular and the chaotic states.
\begin{figure}
\centerline{\psfig{figure=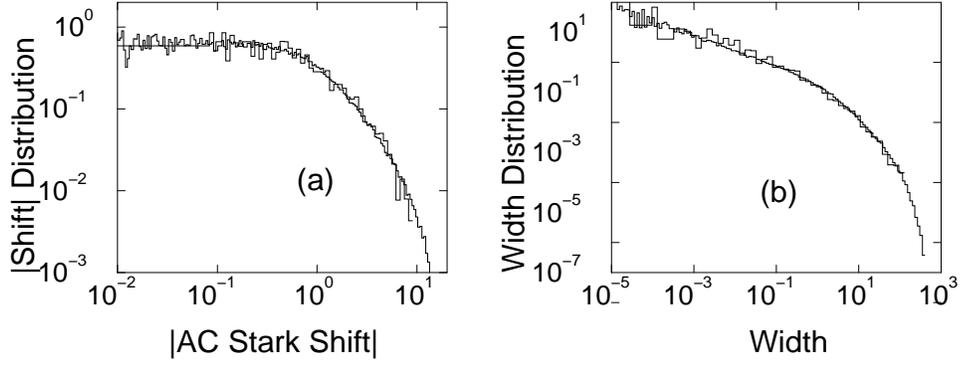,width=8cm,bbllx=100pt,bblly=120pt,bburx=320pt,bbury=500pt,angle=-90}}
\caption{The distribution of energy shifts
(a) and 
ionization widths 
(b) for the non-dispersive wave-packet of a two-dimensional hydrogen atom
in a circularly polarized field, obtained by numerical diagonalization
of the Hamiltonian (large bins), compared to 
the random matrix model (small bins). 
Both distributions
are shown on a double logarithmic scale to better visualize the behavior
over a large range of shift and width values. 
Since the energy shift 
may be 
positive or negative, we show the distribution of its modulus. 
The random matrix model fits very well the numerical results,
with both distributions showing 
regions of algebraic behavior followed by an exponential cut-off.}
\label{ionfig2}
\end{figure}
An example of such a fit is shown in fig.~\ref{ionfig2}. 
The numerical data
are collected around some mean values of $n_0$ and $F_0$, typically
1000 data points were used for a single fit \cite{kuba98}. This
allowed us to study the dependence of the parameters 
$\gamma/\Delta,\sigma/\Delta$ on $n_0$ and $F_0.$
\begin{figure}
\centerline{\psfig{figure=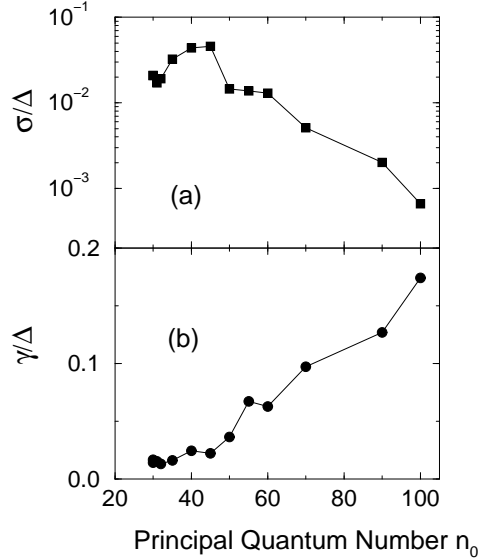,width=8cm,bbllx=100pt,bblly=150pt,bburx=520pt,bbury=600pt,angle=-90}}
\caption{ 
Effective tunneling rate $\sigma/\Delta$  
of the wave-packet (a), 
as a function of the
effective quantum number $n_0=\omega^{-1/3}$
(the inverse of the effective Planck constant), 
for fixed classical dynamics, $F_0=0.0426$. Note the
exponential decrease for sufficiently high $n_0$ (the vertical scale is
logarithmic). The corresponding effective chaotic 
ionization rate $\gamma/\Delta$
(b)
smoothly increases with $n_0$, approximately as $n_0^2$.}
\label{ionfig3}
\end{figure}
The dependence 
on $n_0$
is shown in fig.~\ref{ionfig3}.
Clearly, the tunneling rate $\sigma/\Delta$ decreases 
{\em exponentially} with $n_0.$
Since $n_0$ is the inverse of the effective Planck constant in our problem
(see the discussion in section~\ref{LIN1D} and eq.~(\ref{effective})), 
this shows that
\begin{equation}
\sigma/\Delta \propto \exp \left (-\frac{S}{\hbar_{\mathrm eff}}\right),
\label{dec_exp_tun}
\end{equation}  
where $S$, corresponding to some effective imaginary action 
\cite{shudo95}, is found to be given for our specific choice of
parameters by 
 $S\simeq 0.06 \pm 0.01$ (as fitted from the plot). Such an exponential
 dependence is a hallmark of a tunneling process, thus confirming that
 the wave-packets are strongly localized in the island and communicate
 with the outside world via tunneling.
The $n_0$ dependence of the dimensionless chaotic ionization
rate $\gamma/\Delta$ is very different: it shows a slow, algebraic 
increase with $n_0$. A simple analysis based on a Kepler map
\cite{casati88} description 
would yield a linear increase with $n_0$, whereas our 
data seem to suggest a quadratic function of $n_0$. This discrepancy
is not very surprising, bearing in mind the simplicity of the
Kepler map approach.

\begin{figure}
\centerline{\psfig{figure=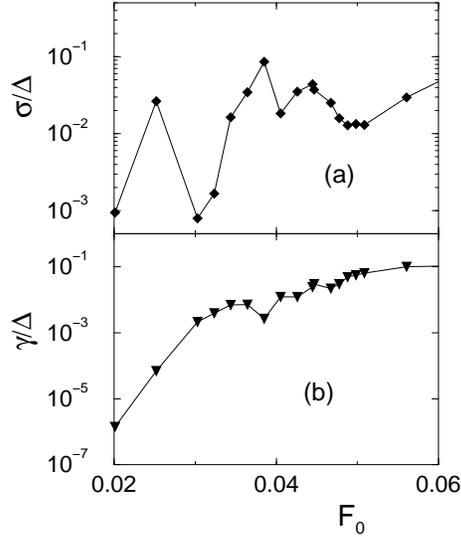,width=8cm,bbllx=100pt,bblly=150pt,bburx=520pt,bbury=600pt,angle=-90}}
\caption{The tunneling rate $\sigma/\Delta$ (panel (a)) and the chaotic
ionization rate $\gamma/\Delta$ (panel (b)), as a function of the scaled
microwave amplitude $F_0$, for wave-packet eigenstates of a two-dimensional 
hydrogen atom in a circularly polarized
microwave field. 
Observe the oscillatory behavior of the tunneling rate. The bumps
are due to secondary nonlinear resonances in the classical dynamics of the
system.}
\label{ionfig4}
\end{figure}

Similarly, we may study, for fixed $n_0$, the dependence of $\gamma/\Delta$
and $\sigma/\Delta$ on $F_0$, i.e. on the microwave field strength. Such
studies, performed for both linear and circular polarizations, have 
indicated that, not very surprisingly, the chaotic 
ionization rate
$\gamma/\Delta$ increases rather smoothly with the microwave amplitude
$F_0$. On the other hand, the tunneling rate $\sigma/\Delta$ shows
pronounced non-monotonic variations with $F_0,$ see fig.~\ref{ionfig4}.
This unexpected behavior can nontheless be explained \cite{kuba98}. 
The bumps in $\sigma/\Delta$ occur at microwave field
strengths where secondary nonlinear resonances emerge 
within the 
resonance island in
classical phase space. 
For circular polarization, this 
corresponds to some resonance
between two eigenfrequencies $\omega_+$ and $\omega_-$ (see
sec.~\ref{CP} and figs.~\ref{solit2},\ref{solfig}) of the dynamics in the classical resonance island. 
Such resonances 
strongly perturb
the classical dynamics and necessarily affect the
quantum transport 
from 
the island.

Let us
stress finally that, even for rather strong microwave fields (say $F_0=0.05$),
where most of the other Floquet states have life-times of few tens or hundreds
of microwave
periods, and irrespective of the polarization of the driving field or of the 
dimension of the accessible configuration space (1D, 2D or 3D), the life-time 
(modulo fluctuations) 
of a non-dispersive wave-packet is typically of the order of
$10^5$ Kepler periods, for $n_0\simeq 60$. This may be used for their possible
experimental detection, see section~\ref{LIFE}.

\subsection{Radiative properties}
\label{SPO}

So far, we have considered the interaction of 
the atom with 
the coherent driving field
only. However, this is not the full story. 
Since the driving field couples
excited atomic states, it remains to be seen to which extent 
spontaneous emission 
(or, more precisely, the coupling to other, initially unoccupied modes of
the electromagnetic field) affects the wave-packet properties. 
This is very important, since the non-dispersive wave-packets
are supposed to be long living objects, and spontaneous
emission obviously limits their life-time.
Furthermore, we have here an example of decoherence effects due to 
interaction with
the environment.
More generally, the interaction of non-dispersive wave-packets
with an additional weak external electromagnetic field may
provide 
a useful tool to probe 
their properties.
In particular, 
their localization within the resonance island
implies that 
an external probe will couple them efficiently 
only to neighboring states within the island. In turn, that   should
make their experimental characterization easy and unambiguous.
Of course, 
external drive (microwave field) and probe must 
not be treated
on the same footing. 
One should rather consider the 
atom 
dressed by the external 
drive
as a strongly coupled system,
or use the Floquet picture described above,
and 
treat the additional mode(s) of the probe (environment) 
as a perturbation. 
We first start with the simplest situation, where a single mode
of the environment is taken into account.

\subsubsection{Interaction of a non-dispersive wave-packet
with a monochromatic probe field}
\label{spofloq}
Let us first consider the addition of a monochromatic probe field
of frequency $\omega_p.$ The situation is very similar to the 
probing of a time-independent system by a weak monochromatic field,
with the only difference  
that the Floquet Hamiltonian replaces
the usual time-independent Hamiltonian. Thus, the
weak probe field may induce a transition between two Floquet states
if it is {\it resonant} with this transition, i.e. if
$\hbar \omega_p$ is equal to the quasi-energy difference between
the two Floquet states. According to Fermi's Golden Rule,
the transition probability is proportional to the
square of the matrix element coupling the initial Floquet state
$|{\cal E}_i\rangle$ to the final one $|{\cal E}_f \rangle .$

Using 
the Fourier representation of Floquet states,
\begin{equation}
|{\cal E}(t)\rangle = \sum_k{\exp (-ik\omega t) |{\cal E}^k\rangle},
\end{equation}
and averaging over one driving field cycle $2\pi/\omega$ we get
\begin{equation}
\langle {\cal E}_f | {\cal T} | {\cal E}_i\rangle =
\sum_{k=-\infty}^{\infty}
{\langle {\cal E}_f^k | {\cal T} | {\cal E}_i^k \rangle}.
\label{matelm_floquet}
\end{equation}
$\cal T$ denotes the transition operator, usually some
component of the dipole operator depending on the polarization
of the probe beam.
If the quasi-energy levels are not bound states but
resonances with finite life-time (for example because of
multiphoton transition amplitudes to the continuum) 
this approach is easily extended \cite{cct92}, yielding
the following expression for the photoabsorption cross-section of the
probe field at frequency $\omega_p$:
\begin{equation}
\sigma(\omega_p) = \frac{4 \pi \omega_p \alpha}{c} 
{\mathrm Im} \sum_{f}{|\langle {\cal E}_f | {\cal T} | {\cal E}_i \rangle |^2
\left [\frac{1}{{\cal E}_f-{\cal E}_i-\omega_p}+\frac{1}{{\cal E}_f-{\cal E}_i+\omega_p} \right ]}
\label{sigma_probe}
\end{equation}
where $\alpha$ is the fine structure constant, 
and where ${\cal E}_f$ and ${\cal E}_i$ are the complex energies 
of the initial and final
Floquet states. The sum extends over all the Floquet states of the system.
Although the Floquet energy spectrum is itself $\omega$-periodic
(see sec.~\ref{QD}), this is {\em not} the case for the photoabsorption
cross-section. Indeed, the Floquet states at energies ${\cal E}_f$ and
${\cal E}_f+\omega$ have the same Fourier components, but shifted
by one unit in $k$, resulting in different matrix elements. 
The photoabsorption spectrum is thus
composed of series of lines separated by $\omega$ with unequal intensities.
When the driving is weak, each Floquet state has a dominant Fourier component.
The series then appears as a dominant peak accompanied by side bands
shifted in energy by an integer multiple of the driving frequency $\omega.$
In the language of the scattering theory \cite{cct92,goldberger50,faisal87}, 
these side bands
can be seen as the scattering of the probe photon assisted
by one or several photons of the drive. In any case, the Floquet
formalism is well suited, since it contains 
this weak driving regime as a limiting case, as well as the strong 
driving regime
needed to 
generate a non-dispersive wave-packet.
   
\subsubsection{Spontaneous emission from a non-dispersive wave-packet}
\label{spospo}
We now 
address the situation where no probe field is added to
the microwave field. 
Still, photons of the driving field can be scattered in the 
(initially empty) 
remaining modes of
the electromagnetic field. This is thus some kind of spontaneous emission
or rather resonance fluorescence of the atom under coherent driving. 
It can be seen as spontaneous
emission of the dressed atom, where 
an initial
Floquet state decays spontaneously to another Floquet 
state with a lower quasi-energy,
the energy difference being carried by the spontaneous photon.
As an immediate consequence, the spectrum of the emitted photons
is composed of the resonance frequencies of the Floquet system,
the same that are involved in eq.~(\ref{sigma_probe}).
The decay rate along a transition depends
on the dipole matrix element connecting the initial and the final states,
but also on the density of modes for the emitted photons.
If we consider, for simplicity, the case of free atoms,
one obtains:
\begin{equation}
\Gamma_{if} = \frac{4\alpha^3 ({\cal E}_i-{\cal E}_f)^3}{3} 
|\langle {\cal E}_f|{\cal T}|{\cal E}_i\rangle|^2
\label{gamma}
\end{equation}
where ${\cal E}_i-{\cal E}_f$ is the positive energy difference 
between the initial 
and final Floquet states. As the matrix element of the dipole 
operator $\cal T$
is involved, clearly the localization properties of the Floquet states 
will be of 
primordial importance for the spontaneous emission process.

The total decay rate (inverse of the life-time) of a 
state $|{\cal E}_i\rangle$
is obtained by summing the partial rates
$\Gamma_{if}$ connecting the initial state to all states with lower
energy. It is not straightforward to determine
which Floquet states contribute most to the decay rate --
the two factors in eq.~(\ref{gamma}) compete:
while $|\langle {\cal E}_f|{\cal T}|{\cal E}_i\rangle|^2$ tends to favor states localized close
to the initial state (maximum overlap), the factor $({\cal E}_i-{\cal E}_f)^3$
(due to the density of modes in free space)
favors transitions to much less excited states. Which factor wins depends
on the polarization of the driving field.

\subsubsection{Circular Polarization}
\label{spocp} 
Consider first a circularly polarized microwave field. 
A first analysis of spontaneous emission
has been given in \cite{ibb97b}, where the rotating frame 
(see sec.~\ref{CP}) approach was used. The driven problem becomes
then time-independent, and the analysis of spontaneous emission 
appears to be simple.
This is, however, misleading, and it is quite easy to omit some
transitions with considerable 
rate. 
The full and correct analysis, both in the rotating and in the standard
frame \cite{delande98},  discusses this problem extensively. The
reader should consult the original papers for details.
 
\begin{figure}
\centerline{\psfig{figure=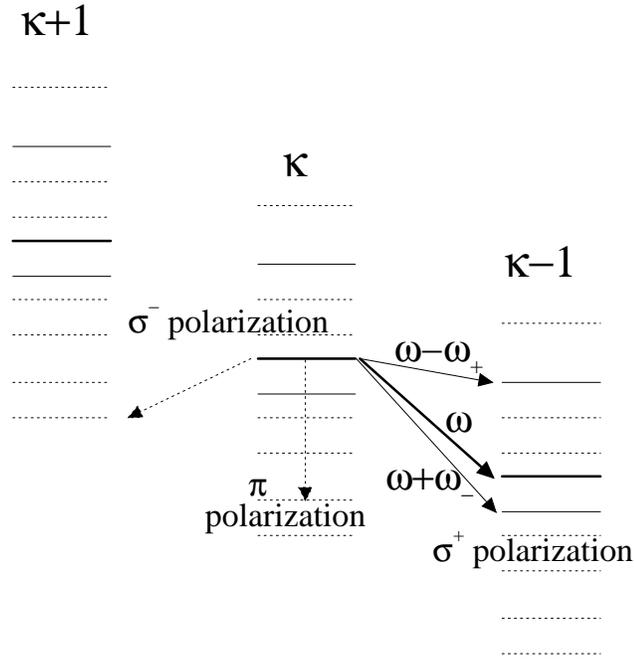,width=9cm,bbllx=100pt,bblly=100pt,bburx=420pt,bbury=400pt}}
\caption{Spontaneous emission transitions for the non-dispersive wave-packet
of an atom 
in
a circularly polarized microwave field. The quasi-energy levels of the
Floquet Hamiltonian can be split in series labelled by $\kappa$
(total angular momentum of the atom and of the microwave field). 
The various series
are identical, except for an energy shift equal to an integer multiple
of the microwave frequency $\omega.$
The arrows indicate possible spontaneous transitions 
leaving the initial state $|0,0,0\rangle$.
Only arrows drawn with solid lines are allowed in the harmonic approximation.
The position of the $|0,0,0\rangle$ wave-packet in each Floquet ladder is
indicated by 
the fat lines.}
\label{spon1}
\end{figure}

A crucial point is to realize that the Floquet spectrum of the Hamiltonian
in CP splits into separate blocks, all of  them being identical,
except for a shift by an integer multiple of the driving frequency $\omega.$ 
Each block corresponds to a fixed
quantum number $\kappa=k+M$ where $k$ labels the photon block 
(Fourier component) in the Floquet approach, while $M$ is the azimutal 
quantum number. This merely signifies that the absorption
of a driving photon of circular polarization $\sigma^+$ increases $M$ by
one unit. In other words, $\kappa$ is nothing but the total
angular momentum (along the direction of propagation of the microwave field)
of the 
entire system comprising the atom and the driving field.
The separate $\kappa$ blocks 
are coupled by spontaneous emission.
Since, again, the spontaneously emitted photon carries 
one quantum of angular
momentum, spontaneous emission couples states within the 
same $\kappa$-block (for
$\pi$ polarization of the emitted photon w.r.t. the $z$ axis, 
which leaves $M$ invariant) or 
in neighboring $\kappa$-blocks $\kappa'=\kappa \pm 1$ (see fig.~\ref{spon1}).
$\sigma^+$ polarization of the emitted photon gives rise
to higher frequency photons
since -- for the same initial and final Floquet states --
the energy difference in the $\sigma^+$ channel is larger by
$\hbar \omega$ than in the $\pi$ channel (and by
$2\hbar \omega$ than in the $\sigma^-$ channel), as immediately
observed in fig.~\ref{spon1}. As the emission
rate, eq.~(\ref{gamma}), changes with the cubic power of the energy difference,
spontaneous photons with $\sigma^+$ polarization
are expected to be dominant.

In the absence of any further approximation, the spontaneous emission
spectrum is  
fairly complicated - it consists of three series with
different polarizations, $\sigma^{\pm}$ and $\pi$.
We may use, however, the harmonic approximation,
discussed in detail in sec.~\ref{CP}. The Floquet
states localized in the vicinity of the stable fixed point
may be labelled
by three quantum numbers $|n_+,n_-, n_z\rangle$, 
corresponding to the various excitations in
the normal modes. The non-dispersive wave-packet we are most interested in
corresponds to the ground state $|0,0,0\rangle$. The dipole operator (responsible for the spontaneous transition)
may be expressed as a linear combination of the creation and annihilation 
operators in these normal modes. Consequently, we obtain strong selection
rules for dipole transitions between $|n_+,n_-,n_z\rangle$ states belonging
to different ladders (at most $\Delta n_i=0,\pm 1$ with not all possibilities
allowed -- for details see \cite{delande98}). The situation is even simpler
for 
$|0,0,0\rangle$,
which may 
decay 
only via three transitions, all 
$\sigma^+$ polarized 
(i.e., from the $\kappa$ block 
to the $\kappa-1$ block):
\begin{itemize}
\item a transition to the 
$|0,0,0\rangle$ state 
in the $\kappa-1$
block. By definition, this occurs precisely at the microwave frequency 
of the drive. One
can view this process as elastic scattering of the microwave photon;
\item a transition to the state $|1,0,0\rangle$,  
at frequency $\omega - \omega_+$;
\item a transition to the state $|0,1,0\rangle$, 
at frequency $\omega + \omega_-$.
\end{itemize}
In the harmonic approximation, 
explicit analytic expressions can be obtained for
the corresponding transition rates \cite{delande98}. It suffices to say 
here that in the semiclassical
limit the elastic component becomes dominant, since its intensity scales as 
$\omega^{5/3}$,
while the intensities of the other two components 
are proportional to $\omega^2,$ i.e., are typically
weaker by a factor $n_0=\omega^{-1/3}$.
This implies that the non-dispersive wave-packet decays exclusively
(in the harmonic approximation) to its immediate neighbor
states, 
emitting a photon with frequency in the microwave range, comparable
to the driving frequency. Direct decay to the atomic $|n=1,L=M=0\rangle$
ground state
or to weakly excited states of the system is forbidden by the selection rules
of the dipole operator. This is easily understood: the CP nondispersive ground
state wave-packet $|0,0,0\rangle$
is built essentially from states with large angular momentum (of the
order of $n_0$), and as it can lose only one unit of angular
momentum per spontaneous emission event, it can decay only to similar
states. When the harmonic approximation breaks down, additional lines
may appear, but, for the same reason, in the microwave range only. 
Another important observation is that the inelastic component at
$\omega + \omega_-$ is by far stronger than the one at $\omega - \omega_+$.
This is entirely due to the cubic power of the transition frequency entering
the expression for the rate (\ref{gamma}). Note the sign difference, due to
the sign difference between $\pm$ modes 
in the harmonic hamiltonian, eq.~(\ref{hh}). 

In the semiclassical limit $\omega=n_0^{-3} \to 0,$ the decay is dominated
by the elastic component, and the total decay rate is \cite{delande98}:
\begin{equation}
\Gamma = \frac{2 \alpha^3 \omega^{5/3} q^{-2/3}}{3},
\label{jacks}
\end{equation}
what, multiplied by the energy $\omega$ of the 
spontaneous photon, gives the energy loss 
due to spontaneous emission:
\begin{equation}
\frac{dE}{dt} = \frac{2 \alpha^3 \omega^{8/3}
q^{-2/3}}{3}=\frac{2\alpha^3\omega^4|x_{\rm eq}|^2}{3},
\label{jacks2}
\end{equation}
where we used eq.~(\ref{q}).
This is nothing but the result 
obtained from classical electrodynamics 
\cite{jackson} for a point charge moving
on a circular orbit of radius $|x_{\rm eq}|$ with frequency $\omega$.
Since the charge loses energy, it cannot survive on a circular orbit and
would eventually fall onto the nucleus following a spiral trajectory. 
This model
stimulated Bohr's original 
formulation
of quantum mechanics. 
Let us notice that the non-dispersive
wave-packet is the first physical realization of the Bohr model. There is no
net loss of energy since, in our case, the electron is driven by the microwave
field and an emission at frequency $\omega$ occurs in fact
as an elastic scattering of a microwave photon. Thus
the non-dispersive wave-packet is a cure of the long-lasting
Bohr paradox.

\begin{figure}
\centerline{\psfig{figure=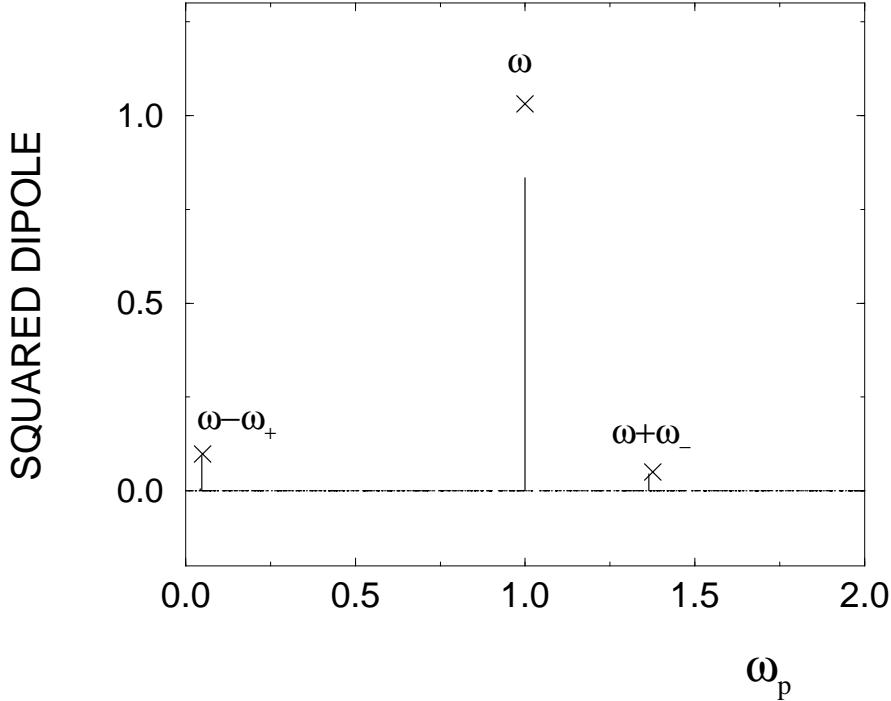,width=12cm,angle=-90}}
\caption{Square of the dipole matrix element (scaled w.r.t. $n_0$, i.e.
divided by $n_0^2$)
connecting the $|0,0,0\rangle$ non-dispersive 
wave-packet of a three-dimensional hydrogen atom
in a circularly polarized microwave field with
other Floquet states, as a function of the
energy difference between
the two states. The stick spectrum  is the exact result obtained
from a numerical diagonalization 
with $\omega=1/60^3$, corresponding to a principal
quantum number $n_0=60$, and scaled amplitude
$F_0=0.04446$; in natural units, the microwave frequency $\omega/2\pi$ 
is 30.48 GHz,
and the
microwave amplitude 17.6 V/cm. The crosses represent  the
analytic prediction within the harmonic approximation \protect\cite{delande98}.
There are three dominant lines ($\sigma^+$ polarized) discussed in the text,
other transitions (as well as transitions with
$\sigma^-$ or $\pi$ polarizations) are negligible, what proves
the validity of the harmonic approximation. 
If a weak probe field (in the microwave domain) is applied to the system 
in addition
to the driving field, its absorption spectrum should therefore
show the three dominant lines, allowing an unambiguous
characterization of the non-dispersive wave-packet.
}
\label{spon2}
\end{figure}

Figure \ref{spon2} shows the square of the dipole matrix elements
connecting the non-dispersive wave-packet, for
$n_0=60$ (i.e., microwave frequency $\omega=1/60^3$) and
scaled microwave field 
$F_0=0.04446$, to other Floquet states
with lower energy. These are the results 
of an exact numerical diagonalization of the full Floquet Hamiltonian.
They are presented as a 
stick spectrum because the widths of the important lines are very narrow
on the scale of the figure, which is given as a function of 
the energy difference
between the initial and the final state, that is the frequency of the 
scattered 
photon. Thus, this figure shows the lines that could be observed
when recording the photoabsorption of a weak microwave probe field.
As expected, there is a dominant line at the frequency $\omega$ of the
microwave, and two other lines at frequencies $\omega-\omega_-$
and $\omega+\omega_+$ with comparable intensities, while
all other lines are at least 10 times weaker. This means that
the harmonic approximation works here very well; its predictions,
indicated by the crosses in the figure, are in good quantitative
agreement with the exact result (apart from tiny shifts  
recognizable in the figure,
which correspond to the mismatch between the exact and the 
semiclassical energies 
observed in figs.~\ref{solit2} and \ref{solfig}). This is not completely
surprising as the energy levels themselves are well reproduced
by this harmonic approximation, see sec.~\ref{rotating_frame}.
However, the photoabsorption  spectrum probes the {\em wave-functions} 
themselves 
(through the overlaps) which are well known to be much more sensitive
than the energy levels. The good agreement for both the energy spectrum
and the matrix elements is a clear-cut proof of the
reliability of the harmonic approximation for physically accessible principal
quantum numbers, say $n_0<100$; in fact, it is good down to
$n_0 \simeq 30$, and the non-dispersive wave-packet exists even
for lower $n_0$ values (e.g. $n_0=15$ in \cite{delande95}) although
the harmonic approximation is not too good at such low quantum numbers. 
There were repeated claims in the literature 
\cite{lee97,farrelly95a,farrelly95,brunello96,cerjan97,lee95}
that the stability island 
as well as the effective potential 
are necessarily
unharmonic in the vicinity of the equilibrium point, and that the
unharmonic terms will destroy the stability of the non-dispersive wave-packets.
The present results prove that these claims are doubly wrong:
firstly, as explained in sec.~\ref{rotating_frame}, harmonicity is
{\em not} a requirement for non-dispersive wave-packets to exist 
(the only condition is the existence of a sufficiently large
resonance island); secondly, the harmonic approximation
is clearly a very good approximation even for moderate values
of $n_0.$

\begin{figure}
\centerline{\psfig{figure=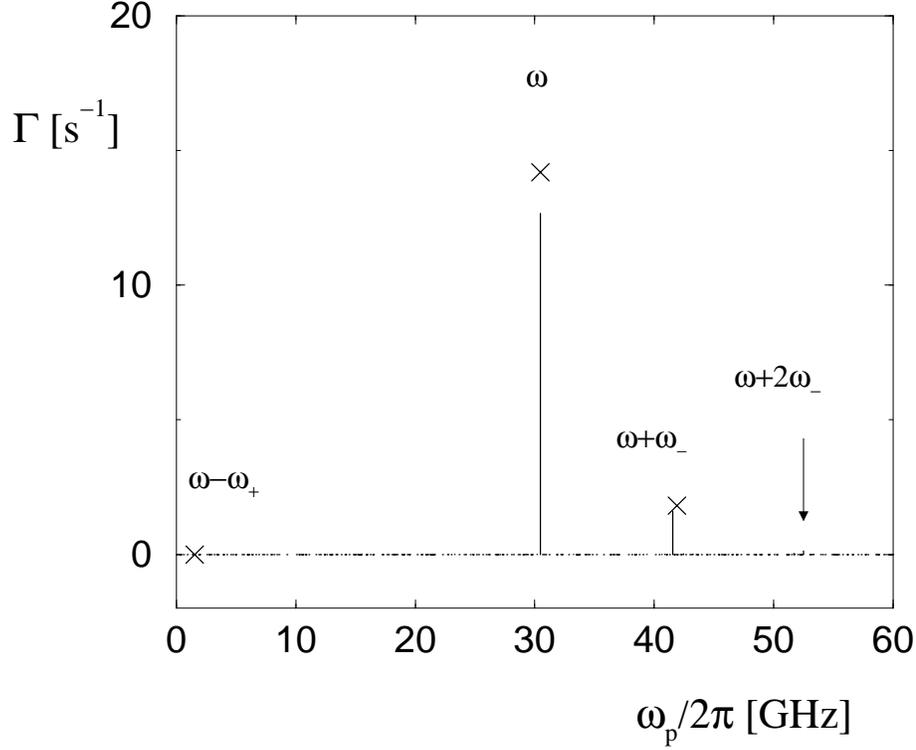,width=12cm,angle=-90}}
\caption{
Same as figure \protect{\ref{spon2}}, but for the
decay rates along the different transitions and in natural units.
The density of modes
of the electromagnetic field completely kills the transition at 
frequency $\omega-\omega_+$,
invisible in the figure. One can see a  small line at frequency approximately
$\omega+2\omega_-$ (see arrow), an indication of a weak breakdown of the harmonic
approximation.
}
\label{spon3}
\end{figure} 

Multiplication by the free space density of states transforms fig.~\ref{spon2}
in fig.~\ref{spon3}, which shows that the corresponding 
spontaneous decay rates 
are very low, of the order of 100 Hz at most. They are
few orders of magnitude smaller than the ionization rates and thus may be
difficult to observe. With increasing $n_0,$ the spontaneous rate decreases
algebraically while the ionization rate decreases {\it exponentially}, 
see sec.~\ref{spolp}. 
Thus for large $n_0,$ the spontaneous emission may be the
dominant process. For $F_0\simeq 0.05$ the cross-over may be expected around
$n_0=200$. However, for smaller $F_0,$ the ionization rate decreases
considerably, and for $F_0\simeq 0.03$ both rates become comparable around
$n_0=60$. Still, a rate of few tens of photons (or electrons
in the case of ionization) per second 
may be quite hard to observe experimentally.

To summarize, resonance fluorescence of 
non-dispersive wave-packets in 
circularly polarized microwave occurs only in the microwave range
(close to the driving frequency). In 
particular, the elastic component (dominant in the semiclassical limit)
does not destroy the wave-packet, the wave-packet merely converts the 
microwave photon into a photon emitted with the same polarization, but in
a different direction.
Let us stress that we assumed the free space density of modes in this
discussion. Since 
the microwave field may be also 
supplied to the atom by
putting the latter in a microwave cavity, it should be 
interesting to 
investigate how the density of modes in such a cavity affects the
spontaneous emission rate either by increasing
or decreasing it (see \cite{haroche92} for a review)
or, for special cavities (waveguides), even
invalidates the concept of a decay rate \cite{lewenstein88,lewenstein88b}. 
  
\subsubsection{Linearly polarized microwave}
\label{spolp}
  
Let us now discuss the spontaneous emission of non-dispersive wave-packets
driven by a linearly polarized microwave field. 
The situation becomes complicated since we should consider
different wave-packets corresponding to (see fig.~\ref{lin3d_5})
extreme librational states ($p=0$, located perpendicularly to the polarization
axis), separatrix states elongated along the polarization axis, and 
extreme rotational (maximal $p$,  doughnut shaped) states
of the resonantly driven manifold. Clearly, all these wave-packet states have
different localization properties and spontaneous emission will couple
them to different final states. No systematic analysis of the effect has
been presented until now, only results based on the simplified 
one-dimensional model are available \cite{hornberger98}. 
Those are of relevance 
for the spontaneous emission of the separatrix based wave-packet
and are reviewed below. Quantitatively we may, however, expect that
the spontaneous emission properties of the extreme rotational wave-packet
will resemble those of the non-dispersive wave-packet in circular
polarization. Indeed, the linearly polarized wave 
can be decomposed into two circularly
polarized waves, and the extreme
rotational state (fig.~\ref{lin3d_8}) is a coherent superposition of
two circular wave-packets -- each locked on one 
circularly polarized component -- moving in the opposite sense. 
The decay of each
component can then be
obtained from the preceding discussion.
 
A completely different picture emerges for other wave-packets. Their electronic
densities averaged over one period are concentrated along either the $\rho$
(extreme librational) or $z$ axes and do not vanish close to the 
nucleus (fig.~\ref{lin3d_5}).
They have non-negligible dipole elements with Floquet states 
built on low lying
atomic states (i.e. states practically unaffected by the driving field).
Because of the cubic power dependence of the decay rate, eq.~(\ref{gamma}),
on the
energy of the emitted photon, these will dominate the spontaneous emission. 
Thus, in contrast to the CP case, 
spontaneous decay will lead to the destruction of the wave-packet. 
A quantitative analysis confirms this qualitative picture.
To this end, 
a general master equation formalism can be developed \cite{hornberger98}, which
allows 
to treat 
the
ionization process induced by the driving field exactly,
while 
the spontaneous emission is treated perturbatively, 
as in the preceding section. 
Applied 
to the one-dimensional model of the atom (sec.~\ref{LIN1D}), 
with the density of
field modes of the real three-dimensional world, it is possible 
to approximately model 
the behaviour of the separatrix states of the three-dimensional atom.
 
\begin{figure}
\centerline{\psfig{figure=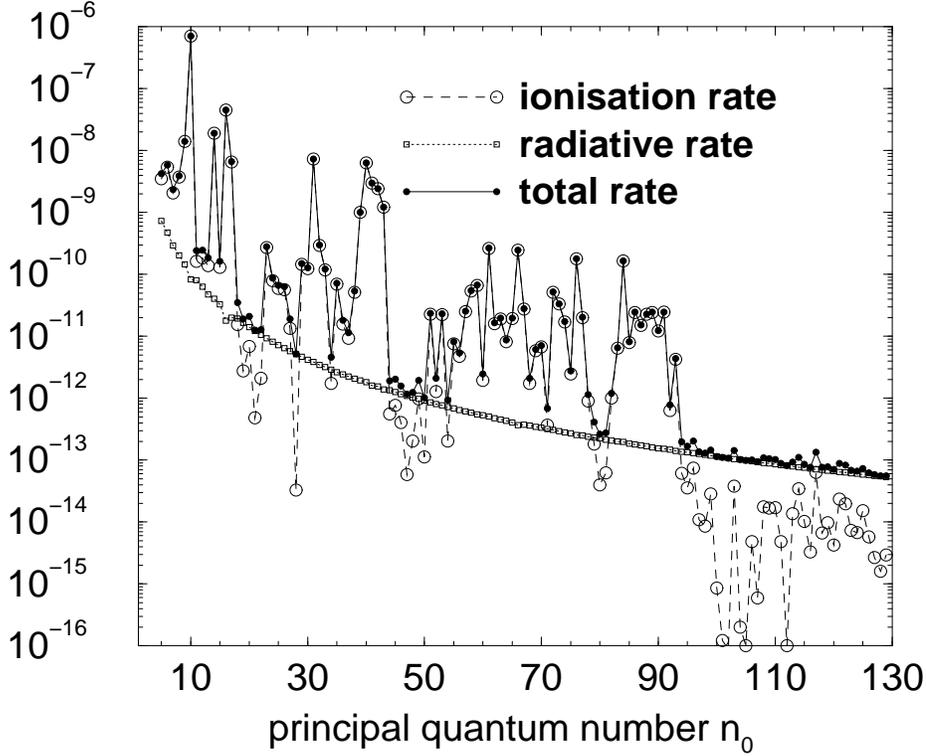,width=12cm,angle=-90}}
\caption{
Comparison of the spontaneous decay rate, the 
ionization rate, and their sum, for a
non-dispersive wave-packet in a linearly polarized microwave field,
as a 
function of the principal quantum number $n_0$. 
Microwave amplitude $F_0=0.04442,$ decay rates in atomic units. 
The full decay rate exhibits a cross-over 
from a dominantly coherent (ionization)
to a dominantly incoherent (spontaneous emission) regime. 
The fluctuations of the
rate present in the coherent regime are suppressed in the incoherent regime.
The data presented here are obtained by an exact numerical calculation
on the one-dimensional model of the atom \protect\cite{hornberger98}, 
see sec.~\ref{LIN1D}.
}
\label{spon4}
\end{figure}

A non-dispersive wave-packet may decay
either by ionization or by spontaneous emission, the total 
decay rate being the sum of the two rates
\cite{hornberger98}. 
Like
in the CP case, the  
decay rate to the atomic continuum decreases exponentially with $n_0$ 
(since it is essentially a tunneling process, see sec.~\ref{ION}, eq.~(\ref{dec_exp_tun}))
while the spontaneous decay rate depends 
algebraically on $n_0$ \cite{bethe77}. The wave-packet is a coherent
superposition of
atomic states with principal quantum number close to $n_0=\omega^{-1/3}$, and
the dipole matrix element between an atomic state $n$ and a weakly 
excited state scales as $n^{-3/2}$\cite{bethe77}. 
Since the energy of the emitted
photon is of order one (in atomic units), eq.~(\ref{gamma})
shows
that the spontaneous emission rate should decrease like 
$\alpha^3/n_0^3.$
The numerical results, presented in fig.~\ref{spon4}, fully confirm
this $1/n_0^3$ prediction.
However, the spontaneous decay of real, 3D wave-packets with near 1D
localization properties (see fig.~\ref{lin3d_5}, middle column, and 
fig.~\ref{linf_wp}) is
certainly slower. 
Indeed, these states are combinations of atomic states
with various total angular momenta $L$; among them, only the
low-$L$ values decay rapidly to weakly excited states, 
the higher $L$
components being coupled only to higher excited states. 
In other words, they are dominantly composed by extremal parabolic Rydberg
states, which have well-known decay properties \cite{bethe77}.
Altogether,
their decay rate is decreased by a factor of the order of $n_0,$  
yielding a $n_0^{-4}$ law instead of  $n_0^{-3}$. 

On the other hand, since the ionization process is dominated by tunneling in
the direction of the microwave polarization axis
\cite{abuth,abu98a,abu97,abu97b}, the ionization rate in 3D remains globally
comparable to the ionization rate in 1D, for the wave-packet launched along
straight line orbits. This remains true even if the generic
fluctuations of the ionization rate (see section~\ref{ION}) may induce locally
(in some control parameter) large deviations between individual 3D and 1D
decay rates\footnote{A similar behaviour is observed
in circular polarization for the 2D and 3D non-dispersive wave-packets: they
exhibit comparable ionization rates, but distinct fluctuations
\protect{\cite{kuba95a}}.}. 
Therefore, the transition from dominant ionization to dominant spontaneous
decay will shift to slightly higher values of $n_0$ in 3D.
As in the CP case studied above, this cross-over may be moved to smaller 
values of $n_0$  by reducing the
ionization rate, i.e., by decreasing $F_0$.

\subsection{Non-dispersive wave-packet as a soliton}
\label{SOLI}

The non-dispersive character of the wave-packets discussed in this review
brings to mind solitons, i.e. solutions of {\it nonlinear} wave equations
that propagate without deformation: the non-linearity is there
essential to overcome the spreading of the solution. The non-dispersive
wave-packets discussed by us are, on the other hand, solutions of
the {\it linear} Schr\"odinger wave equation, and it is not some
non-linearity of the wave equation which protects them from spreading, but
rather the periodic driving. Thus, at first glance,
there seems to be no link  between both phenomena. This is
not fully correct. One may 
conceive non-dispersive wave-packets as solitonic
solutions of particular nonlinear equations, 
propagating not in time, but in parameter 
space \cite{kuba97b}.
The 
evolution of energy levels in such a space, 
called ``parametric level dynamics'',
has been extensively studied (see 
\cite{haake90,nakamura93} for 
reviews), both for time-independent
and  for periodically time-dependent systems. In the latter case, 
the energy levels are the quasi-energies 
of the Floquet Hamiltonian (see sec.~\ref{QD}). 

For the sake of simplicity, 
we consider here the two-dimensional hydrogen atom 
exposed to a circularly polarized microwave (sec.~\ref{2d_model}),
where the explicit time dependence can be removed by 
transforming to
the rotating frame (see sec.~\ref{rotating_frame}), but completely
similar results are obtained for the Floquet Hamiltonian of any 
periodically time-dependent system.
The Hamiltonian, given by eq.~(\ref{hrot}),
\be
H=\frac{{\vec p}^2}{2}-\frac{1}{r}+Fx-\omega L_z,
\label{solieq0}
\ee
may be thought of as an example of a generic system of the form
\be
H(\lambda)=H_0+\lambda V,
\label{solieq1}
\ee
where $\lambda$ is a parameter. In our case, 
for example, the
microwave amplitude may be tuned, leading to $V=x$ and $\lambda=F$. 
The interesting quantities are then 
the eigenvalues $E_i(\lambda)$ and the eigenfunctions
$|\psi_i(\lambda)\rangle$ of eq.~(\ref{solieq1}). 
Differenciating the
Schr\"odinger equation with respect to $\lambda$,
one shows (with some algebra)
\cite{haake90} that the behavior of $E_i(\lambda)$ with
$\lambda$ may be viewed as the motion of $N$ fictitious
classical particles
(where $N$ is
the dimension of the Hilbert space) with positions $E_i$ and momenta
$p_i=V_{ii}=\langle\psi_i|V|\psi_i\rangle$, governed by the
Hamiltonian
\begin{equation}
{\cal H}_{\rm cl}=\sum_{i=1}^N\frac{p_i^2}{2}+\frac{1}{2}\sum_{i=1}^N\sum_
{j=1,j\neq i}^N
\frac{\mid{\cal L}_{ij}\mid^2}{(E_j-E_i)^2},
\label{hclas}
\end{equation}
where
${\cal L}_{ij}=(E_i-E_j)\langle\psi_i|V|\psi_j\rangle$
are additional independent variables obeying the general Poisson brackets
for angular momenta. The resulting dynamics,
although nonlinear, is integrable \cite{haake90}.

Let us now consider the parametric motion of some eigenstate 
$\mid n_+,n_- \rangle$, 
for example of the ground state wave-packet $|0,0\rangle$.
Its coupling to other states is quite weak -- because
of its localization in a well defined region of phase space -- 
and the corresponding ${\cal L}_{ij}$ are consequently very small.
If we 
first suppose
that the wave-packet state is well isolated (in energy)
from other wave-packets (i.e., states with low values of $n_+,n_-$), 
the fictitious particle associated with 
$|0,0\rangle$ 
basically ignores the other 
particles and propagates freely at constant
velocity. It preserves its properties across the
successive interactions with neighboring states, 
in particular its shape:
in that sense, it is a solitonic solution of the
equations of motion generated by Hamiltonian~(\ref{hclas}).

\begin{figure}
\centerline{\psfig{figure=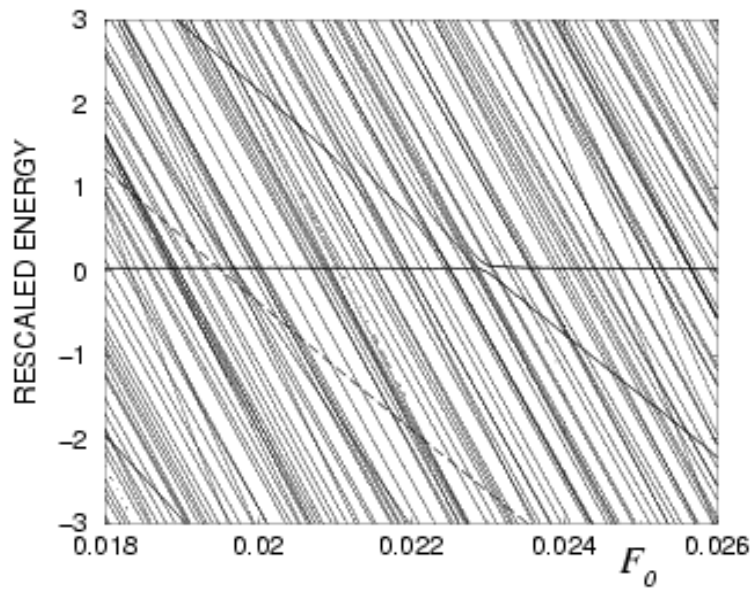,width=10cm}}
\caption{
The quasi-energy spectrum of a two-dimensional hydrogen atom 
in a circularly polarized microwave field,
as a function of the scaled microwave amplitude $F_0$ (for
$n_0=60$). In order to emphasize the dynamics of wave-packet states,
the semiclassical prediction, eq.~(\protect\ref{estharm}) for the
ground state wave-packet energy is substracted from the
numerically calculated energies. Consequently, the
ground state wave-packet $|0,0\rangle$ is represented by 
the almost horizontal line. 
The dashed line represents the
semiclassical prediction for the $|1,4\rangle$ wave-packet. Although it is 
rather far from the exact result, the slope
of the energy level is well reproduced. The size
of the avoided crossing between the ``solitonic"
levels $|0,0\rangle$ and  $|1,4\rangle$ is a direct measure
of the failure of the harmonic approximation.
 } 
\label{solfig}
\end{figure}

Suppose that, in the vicinity of some $F$ values, 
another wave-packet 
state (with low $n_+,n_-$ quantum numbers) becomes quasi-degenerate 
with $|0,0\rangle$. 
In the harmonic approximation, see sec.~\ref{rotating_frame}, the two states
are completely uncoupled; 
it implies that the corresponding
${\cal L}_{ij}$ vanishes and the two levels cross. 
The coupling
between
the two solitons 
stems from the {\em difference} between the exact hamiltonian
and its harmonic approximation, 
i.e. from third order or higher terms, beyond the harmonic
approximation. Other
$|n_+,n_-\rangle$ states having different slopes w.r.t. $F$ induce 
``solitonic collisions'' 
at some other values of $F$.  
To illustrate the effect,
part of the spectrum of the two-dimensional hydrogen atom in 
a CP microwave
is shown in fig.~\ref{solfig}, as a function of the scaled microwave
amplitude $F_0$.
For the sake of clarity, the
energy of the ground state wave-packet 
$|0,0\rangle$ calculated
in the harmonic approximation, eq.~(\ref{estharm}) is substracted,
such that it appears as an almost
horizontal line.
Around $F_0=0.023$, it is crossed by another solitonic solution, 
corresponding to
the $|1,4\rangle$ wave-packet\footnote{Similarly to the 
wave-packet $|1,3\rangle$ discussed in fig.~\protect\ref{solit2},
the semiclassical harmonic
prediction for the energy of $|1,4\rangle$ is not satisfactory.
However, the slope of the energy level is well reproduced as a 
function of $F_0$.}, 
what represents the collision
of two solitons. Since this
avoided crossing 
is narrow and well isolated from
other avoided crossings, the wave-functions before and after 
the crossing preserve
their shape and character, as typical for an isolated two-level system.
This may be further verified by wave-function plots before and after
the collision (see 
\cite{kuba97b} for more details). 
The avoided crossings become larger 
(compare fig.~\ref{solit2}) with increasing $F_0$. In fact, as mentioned 
in sec.~\ref{rotating_frame},
we have numerically verified that the solitonic character of the ground
state wave-packet practically disappears at
the  $1:2$ resonance, close to $F_0\simeq 0.065$ \cite{kuba97b}. For larger
$F_0,$ while one may still find nicely
localized wave-packets for {\em isolated} values of $F_0,$ 
the increased size of the avoided crossings makes it difficult
to follow the wave-packet 
when sweeping $F_0.$ 
For such strong fields, the ionization rate of
wave-packet states becomes appreciable, comparable to the level
spacing between consecutive states and the simple solitonic model breaks down.
In order to understand the variations of the 
(complex) energies of the resonances with $F$, a slightly
more complicated model -- level dynamics in the complex plane --
should be used~\cite{haake90}.

\section{Experimental preparation and detection of non-dispersive wave-packets}
\label{EXP}

In the preceding chapters, we have given an extensive
theoretical description of the characteristic 
properties of non-dispersive wave-packets in driven Rydberg
systems. We have seen that these surprisingly robust ``quantum particles''
are ubiquitous in the interaction of electromagnetic radiation with 
matter. However, any theoretical analysis needs to be confronted with 
reality, and we have
to deal with the question of creating and identifying non-dispersive
wave-packets in a laboratory experiment.
In our opinion, none of the currently operational experiments on the
interaction of Rydberg atoms with microwave fields 
allows for an unambiguous identification of non-dispersive wave-packets,
although some of them \cite{koch95b} 
certainly have already populated such states.
In the following, we shall therefore start out with 
a brief description of the typical approach of state of the art experiments, 
and subsequently extend on various alternatives to create and to probe  
non-dispersive wave-packets in a real experiment.
We do not aim at a comprehensive 
review on the interaction 
of Rydberg atoms with microwave fields, but rather refer to 
\cite{abuth,delande94,sachath,koch95b,casati87b}
for a detailed treatment of various aspects of this intricate
problem. Here, we strictly focus on
issues pertinent to our specific purpose.

\subsection{Experimental status}
\label{EXSQ}

The theoretical interest in the interaction of Rydberg states of atomic
hydrogen
with low frequency electromagnetic
fields has been triggered by early experiments \cite{bayfield74}
which showed a surprisingly
efficient excitation and subsequent ionization of the atoms
by the field. More precisely, a microwave field
of frequency $\omega$
comparable to the energy
difference between the initial atomic state and its nearest neighbor
was observed to induce appreciable ionization, 
for atom-field interaction times of 
approx. $100$ driving field cycles, and for 
field amplitudes 
beyond a certain threshold value
(of the order of $5-10\%$ of the
Coulomb field experienced by the Rydberg electron on its unperturbed
Kepler orbit). This threshold behavior of the
ionization probability as a function of the driving field amplitude
 rather than
of the driving frequency -- in apparent contradiction to the photoeffect --
motivated a theoretical analysis of the classical dynamics of the Rydberg
electron under external driving. It turned out that
the ionization threshold marks the transition
from  regular
to  chaotic {\em classical dynamics} of the driven electron \cite{leopold78}.

The microwave ionization of atomic Rydberg states was thus identified
as
an experimental testing ground for quantum transport
under the conditions of classically mixed regular chaotic
dynamics, where the transport was simply measured by the experimentally
observed ionization yield, or -- with some additional experimental effort --
by the time dependent redistribution of the atomic population over the
bound states \cite{bayfield88a,bluemel89a,bluemel91a,abu91}. 
Depending on the precise value of the scaled frequency $\omega_0$ -- the ratio
of the microwave frequency $\omega$ 
to the Kepler frequency $\Omega_{\rm Kepler}$
of the initially excited Rydberg atom, eq.~(\ref{omega0}) --
of the driving field, theory soon predicted essentially
``classical'' ionization yields ($\omega_0<1.0$), or some quantum suppression
of chaotic ionization ($\omega_0>1.0$) 
\cite{casati84}, 
mediated by the quantum mechanical
interference effect known as 
{\em dynamical localization}, 
analogous to 
Anderson localization in the electronic transport
through disordered solids  
\cite{fishman82,grempel84,brenner96,abu98b,wimbergerda,wimberger01}.
The physical process involved in chaotic ionization is 
classically deterministic
diffusion, 
therefore essentially statistical in nature, and 
insensitive to the details of the transport process.
Correspondingly, the mere ionization probability
condenses all details of the ionization process in one single number,
without revealing details on 
individual local
structures in phase space. It reflects the statistical characteristics
of the excitation process, rather than the population of some well defined
individual atomic  states in its course \cite{abuth,abu95a}.
Hence, state of the art experiments are ``blind'' for the details of the
atomic excitation process on the way to ionization, and therefore not
suitable for the unambiguous identification of individual eigenstates of the
atom in the field, notably of non-dispersive wave-packets. 
The case is getting
worse with additional complications which are unavoidable in a real experiment,
such as the unprecise definition of the initial state the atoms are prepared
in 
\cite{koch95b,galvez88,sirko96,sirko93a,sauer92b,leeuwen85,koch95a,koch92b,koch92a,koch89}, 
the experimental uncertainty on the envelope of the amplitude of the
driving field experienced by the atoms as they enter the interaction
region with the microwave 
(typically a microwave cavity or wave guide) \cite{bluemel91a,sauer92b,noel00},
stray electric fields due to contact potentials in the interaction region,
and finally uncontrolled noise sources which may affect the
coherence effects involved in the quantum mechanical transport process
\cite{bayfield91}.
On the other hand, independent experiments on the microwave ionization
of Rydberg states of atomic hydrogen \cite{bayfield89,galvez88}, 
as well as on hydrogenic initial states 
of lithium \cite{noel00}, did indeed provide hard evidence
for the relative stability of the atom against ionization when driven by a
resonant field of scaled frequency $\omega_0\simeq 1.0$.
Furthermore, in the hydrogen experiments, this stability was observed 
to be insensitive
to the polarization of the driving field, be it linear,
circular or elliptical \cite{bellermann96}.
These experimental findings suggest that some atomic dressed states 
anchored to
the principal resonance island in the classical phase space
are populated by switching on the microwave field,
since these states  tend to be more stable
against ionization than  states localized in the chaotic sea~\cite{abu95a}, 
see section \ref{ION}.

Consequently, for an unambiguous preparation and identification of
non-dispersive wave-packets launched along well defined classical trajectories
the experimental strategy has to be refined. 
We suggest
two 
techniques for their preparation:
\begin{itemize}
\vskip -10pt
\item The direct, selective optical excitation from
a low lying state {\em in the presence of the microwave field}. 
This approach actually 
realizes some kind of  
``Floquet state absorption spectroscopy'' \cite{abu96}.
\item The preparation of the appropriate atomic initial 
state -- optionally in the presence of a static field --
followed by  
switching  
the microwave field on the appropriate time scale 
to the desired maximum field amplitude \cite{kuba95b}.
\end{itemize}

Two, possibly complementary methods should allow 
for an efficient detection of such wave-packets: 
\begin{itemize}
\item Floquet spectroscopy  -- this time 
involving either microwave or optical transitions between states dressed
by the microwave field;
\item Measurements of the time-dependence of the ionization yield 
of the non-dispersive
wave-packet. This 
requires the ability to vary the
interaction time between the atoms and the microwave by more 
than one order of magnitude
\cite{abuth,abu95a,noel00,abu95b,arndt91,benson95}.
\end{itemize}

All these techniques 
are experimentally well-developed and actually realized in different, 
currently
operational experimental settings \cite{noel00,abu95b,noel00b}.
The only prerequisite for an unambiguous identification of non-dispersive 
wave-packets therefore remains an experimental set-up which allows 
to follow these
complementary strategies simultaneously.

\subsection{Direct preparation}
\label{SPEC}

The 
most straightforward way to populate a non-dispersive
wave-packet state is its
direct optical excitation
in the presence of the driving field,
from a 
weakly excited state of the system at energy $E_0$.
In section \ref{spofloq}, we have discussed how a weak electromagnetic probe 
field can induce transitions between Floquet states.
This is particularly easy if  
one of the states involved is in an energetically low
lying state. 
Such a state
is practically unaffected by the driving
microwave field (which is weak as compared to the Coulomb field 
experienced by a deeply 
bound state, and very  far
from any resonance), such that the corresponding Floquet state is almost
exactly 
identical with the time-independent atomic state. In other words,
all the Fourier components of the Floquet state vanish,
except the $k=0$ component, 
which represents the 
unperturbed atomic state
$|\phi_0\rangle$.
In such a case, the photoexcitation cross-section (\ref{sigma_probe}) becomes
(neglecting the anti-resonant term):

\begin{equation}
\sigma(\omega_p) = \frac{4 \pi \omega_p \alpha}{c}
{\mathrm Im} \sum_{f}{|\langle {\cal E}_f^0 | {\cal T} | \phi_0 \rangle |^2
\frac{1}{{\cal E}_f-E_0-\omega_p}},
\label{sigma_probe_simple}
\end{equation}
where the sum extends over all Floquet states with energy ${\cal E}_f$ and
involves only their $k=0$ Fourier component
\footnote{Alternatively, the sum could be rewritten as a sum
over one Floquet zone only, with all the Fourier
components $|\phi_f^k\rangle$ involved, with the denominator
replaced by ${\cal E}_f+k\omega-E_0-\omega_p.$}.

\begin{figure}
\centerline{\psfig{figure=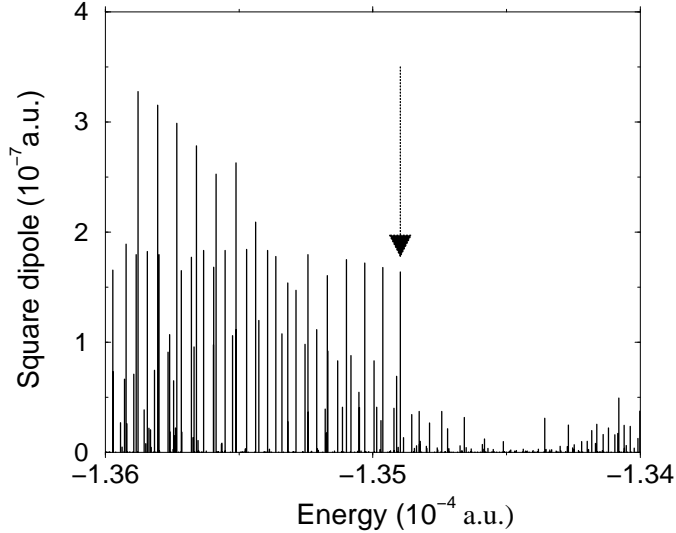,width=9cm,angle=-90}}
\caption{Photo-excitation of highly excited Rydberg states of the
hydrogen atom in the presence of a linearly polarized microwave field
of frequency $\omega/2\pi=30.48$ GHz and amplitude
$F= 7.93$ V/cm and a parallel static electric field $F_s=2.38$ V/cm.
The initial state is the ground state of the atom, and the
polarization of the probe beam is parallel to the static
and microwave fields. The spectrum 
is displayed
in the region where Floquet states are mainly composed of Rydberg states
with principal quantum number around $n_0=60$, while the microwave driving 
is resonant with the Kepler frequency of such states. Hence, some of
the Floquet states are trapped in the non-linear resonance island
and behave as non-dispersive wave-packets. Of special interest
is the state marked with an arrow, which is similar to the non-dispersive
wave-packet displayed in fig.~\protect{\ref{linf_wp}}, localized
in all three dimensions of space (the field amplitude is
slightly different). Its large photo-excitation probability
should make a direct experimental preparation possible. At the scale of the figure,
the width of the various lines is very small, and the spectrum is
almost a pure stick spectrum.
}
\label{abs_spec}
\end{figure}

Eq.~(\ref{sigma_probe_simple}) shows that the excitation probability 
exhibits a maximum any
time the laser is scanned across a frequency which is resonant with
the transition from the 1s ground state $|\phi_0\rangle$ 
to a specific dressed state of the atom
in the field. Fig.~\ref{abs_spec} shows
an example for the photoabsorption probability from the ground state
of atomic hydrogen in the presence of 
a microwave field and a parallel static electric field. The microwave
frequency is resonant with
atomic transitions in the region of $n_0\simeq 60$.
Clearly, the
cross-section shows extremely narrow peaks each of which corresponds to
a Floquet eigenstate of the atom in the field. As a matter of fact, the
state marked by the arrow is similar to the dressed state of the 3D atom
displayed in fig.~\ref{linf_wp}, 
a wave-packet periodically moving along the
field polarization axis. As obvious from the figure, this state can be
efficiently reached by direct excitation from the ground state.
Furthermore, due to its sharp signature in
$\sigma(\omega_p)$, it is easily and unambiguously identified.
On the other hand, this kind of preparation of the wave-packet is
obviously reserved to those dressed states 
which
have nonvanishing overlap with the deeply bound atomic states. 
For wave-packets tracing circular or elliptical orbits far from the
nucleus, another strategy is needed, that
is discussed below.

A slightly adapted spectroscopic approach should 
be equally 
useful for the unambiguous identification of wave-packet eigenstates.
Instead of probing the dressed spectrum from a weakly excited
state using a laser field, one may equally well probe the local structure
of the dressed spectrum in the vicinity of the wave-packet eigenstate
by inducing transitions from the wave-packet to neighboring states
by a second, weak microwave field of linear or circular 
polarization \cite{bluemel90a}.
Such stimulated transitions will be mediated by the dipole matrix elements
given in equation~(\ref{matelm_floquet}), and allow to measure
the energy spacings in the
immediate vicinity of the wave-packet state directly. Hence, microwave probe
spectroscopy should be an extremely
sensitive probe, since it allows for the unambiguous
identification of the wave-packet via the characterization of its local
spectral environment.

Given the spectroscopic
resolution which is nowadays available in the optical as well as in the
microwave domain, the spectroscopic approach outlined above seems to be the method
of choice for an unambiguous identification, and -- where possible --
for an efficient launch of nondispersive wave-packets along a periodic
orbit of the classical dynamics. What it requires, however, is a
precise determination of Floquet spectra from the accompanying
quantum calculation. Fortunately, both for hydrogen and for alkali
atoms, the necessary theoretical quantum data may be obtained from 
already existing
software \cite{krug00,krug01}.

Although we elaborated in this review paper only the case of the hydrogen
atom, the general concepts are also fruitful for non-hydrogenic atoms.
Indeed, the major difference between the Rydberg electron in a hydrogen
atom and in a non-hydrogenic atom is the existence in the latter case
of an ionic core which affects the classical and quantum dynamics of the
Rydberg electron. On the scale of a Rydberg atom, the ionic core is 
a extremely small object which will thus induce a very local perturbation.
As long as the Rydberg electron does not approach the ionic core,
it behaves completely similarly in hydrogen or non-hydrogenic atoms.
Thus, the properties of non-dispersive wave-packets tracing circular
or elliptical classical orbits are essentially independent of the ionic core,
and the hydrogenic analysis holds. 
For orbits which come close to the nucleus, the ionic core may scatter
the Rydberg electron. Thus, instead of being indefinitely trapped on a
torus inside a resonance island, it may happen that the Rydberg electron
hops from a torus to another one when it gets close to the nucleus.
This of course will affect the long time classical and quantum dynamics.
Nevertheless, it remains true that most of the time the classical
dynamics -- and consequently the phase locking phenomenon responsible
for the existence of non-dispersive wave-packets -- is identical
to the hydrogenic dynamics. From the quantum point of view, the ionic
core is responsible for the existence of non-zero quantum defects
in the low angular momentum channels. The energy levels, mixed by the microwave
driving, will thus be significantly shifted from their
hydrogenic positions. However, the {\em structure} of the energy levels --
grouped in manifolds -- will essentially survive, see \cite{krug00}.
It is likely that some non-dispersive wave-packets also exist in
non-hydrogenic atomic species.

\subsection{Preparation through tailored pulses}
\label{PTP}

Another, indirect method for preparing  non-dispersive wave-packets
is also available. This will be the method of choice for wave-packets
moving along classical orbits of large angular momentum (small
eccentricity). Such states, obviously, are not accessible to a direct
optical excitation from weakly excited, low angular momentum states.
The same general scheme may be also applicable to high eccentricity
wave-packets although in that case we expect that the direct
excitation may be more efficient and flexible.
The method to be discussed here
consists of two stages.
We first prepare the atom in a well chosen and well defined initial
highly excited state, and then turn the microwave field on relatively slowly,
from zero amplitude to its plateau value $F_{\rm max}$.

A non-dispersive wave-packet is a single eigenstate $|{\cal E}\rangle$
of the Floquet Hamiltonian
describing the driven system at fixed driving field amplitude
$F.$
As shown in sections  \ref{LIN3D}, \ref{CP} and \ref{SOLI},
 the evolution of the quasienergies
of the driven atom with an external control parameter like the driving field
amplitude is rather complicated.
 It reflects the dramatic transformation
of the structure of classical phase space, manifesting in an
abundance of avoided crossings of various sizes in the level dynamics.
Still, as exemplified in figs.~\ref{fig_mathieu}, \ref{solit2}, and \ref{solfig},
the wave-packet states may be followed 
rather easily 
under changes of $F$ (parametrized by $t$, during the switching of the pulse)
in the level dynamics, 
in agreement with their ``solitonic''
character (see section \ref{SOLI}).
Nontheless, the very same figures illustrate clearly that the targeted 
wave-packet state undergoes
many avoided crossings as the  microwave amplitude is 
swept. To remain in 
a single eigenstate, the avoided crossings should be passed
either adiabatically or diabatically, with a
branching ratio at an individual crossing
being described by the well known Landau-Zener scenario
\cite{landau2,breuer89b}.
Consequently, if we want to populate an individual wave-packet eigenstate
from a field free atomic state $|\psi_0\rangle$,
we need some knowledge
of the energy level dynamics. Then it is possible to identify those
field free states which are connected to the wave-packet via
adiabatic and/or diabatic transitions in the network of energy levels,
and subsequently
to design $F(t)$ such as to transfer population
from $|\psi_0\rangle$ to $|{\cal E}\rangle$ most efficiently.
A precise experimental preparation of $|\psi_0\rangle$
is the prerequisite of any such approach.

When the driving field is increased from zero, the major modification
in the classical phase space is the emergence
 of the resonance
island (see figs.~\ref{lin3d_4}, \ref{lin3d_9}, \ref{lin3d_10}). 
Quantum mechanically, the states with initial
principal quantum number close to $n_0=\omega^{-1/3}$ will
enter progressively inside the resonance island.
For a one-dimensional system, the Mathieu equation, discussed
in section \ref{section_mathieu}, fully describes the evolution of the
energy levels in this regime. As shown for example
in figure~\ref{fig_mathieu}, the non-dispersive wave-packet
with the best localization, i.e. $N=0,$ 
is -- in this simple situation -- adiabatically connected 
to the field-free state with principal quantum number closest
to $n_0,$ i.e. the eigenstate $\kappa=0$ of the
Mathieu equation. When the Mathieu parameter $q,$ eq.~(\ref{map_mathieu3}), 
is of the order
of unity, the state of interest is trapped in the resonance island, which
happens at field amplitudes given by eq.~(\ref{ftrapping})
for the one-dimensional atom, and for 
linear polarization of the microwave field.
A similar scaling
is expected for other polarizations, too.
In the interval
$F\leq F_{\rm trapping}$, the field has to be increased slowly enough such
as to avoid losses from the ground state to the excited states of
the Mathieu equation, at an energy separation of the order of $n_0^{-4}$.
The most favorable situation is then the case of
``optimal'' resonance (see section~\ref{QD}),
when $n_0$ is an integer, the situation in figure~\ref{fig_mathieu}.
The wave-packet state is always separated from the other
states by an energy gap comparable to its value at $F=0$,
i.e. of the order of $3/(2n_0^4)$.
The situation is less favorable if $n_0$ is not an integer,
because the energy gap between the wave-packet
of interest and the other states is smaller when $F \to 0.$
The worst case is met when $n_0$ is half-integer: the free
states $n_0+1/2$ and $n_0-1/2$ are quasi-degenerate, and selective
excitation of a single wave-packet is thus more difficult.

The appropriate time scale for switching on the field is given 
by the inverse of the
energy splitting, i. e. for ``optimal" resonance
\begin{equation}
\tau_{\rm trapping}\sim n_0^4=n_0\times 2\pi/\omega,
\label{t_smallF}
\end{equation}
or $n_0$ driving field periods.

Once trapped in the resonance island, the coupling to states localized
outside the island will be residual -- mediated by quantum mechanical
tunneling, see section \ref{ION} -- and the size of the
avoided crossings between the trapped and the untrapped states is exponentially
small. After adiabatic switching into the resonance island on a time 
scale of $n_0$
Kepler orbits, we now have to switch diabatically from $F_{\rm trapping}$ to
some final
$F$ value, in order to avoid adiabatic
losses from
the wave-packet into other states while passing through the avoided crossings.

The preceding discussion is based on a one-dimensional model
and the Mathieu equation. Taking into account the other
``transverse'' degrees of freedom is not too difficult. Indeed,
as noticed in sections \ref{LIN3D}, \ref{CP}, 
the various time scales of the problem
are well separated. The transverse motion is slow and can be
adiabatically separated from the fast motion in $(\hat{I},\hat\theta)$. 
Instead of getting
a single set of energy levels, one gets a family of sets,
the various families being essentially uncoupled. An example
for the 3D atom in a linearly polarized microwave field is shown
in fig.~\ref{lin3d_4}.
 It follows that the estimate for the trapping field
and the
switching time are essentially the same as for 1D systems.
Inside the resonance island, the situation becomes slightly more
complicated, because there is not a single frequency for
the secular motion, but several frequencies along the
transverse degrees of freedom. For example, in CP,
it has been shown that there are three eigenfrequencies, 
eqs.~(\ref{omegas},\ref{ompmz}), in the harmonic
approximation -- see section \ref{rotating_frame}.
 This results in a large number of excited energy levels
which may have avoided crossings with the ``ground state", i.e.
the nondispersive wave-packet we want to prepare.
As shown in section \ref{SOLI}, most of these avoided
crossings are extremely small and can be easily crossed diabatically.
However, some of them are rather large, especially when there
is an internal resonance between two eigenfrequencies. Examples are
given in figs.~\ref{solit2} and \ref{solfig}, where $\omega_+=3\omega_-$ and 
$\omega_+=4\omega_-$, respectively. 

These
avoided crossings are large and dangerous, because the states
involved lie {\em inside} the resonance island, which thus
loses its protective character.
They are mainly due to the unharmonic character of the
Coulomb potential.
Their size  may be qualitatively analyzed as we do below on the
CP example, 
expecting similar
sizes of the avoided crossings for any polarization.

The unharmonic corrections to the
harmonic approximation around the stable fixed point
$x_e,p_e$ in the center of the nonlinear resonance -- as outlined in
section \ref{CP} -- are due to the higher order terms
$(-1)^j\tilde{x}^j/j!x_e^{j+1}$ in the
Taylor series of the Coulomb potential, where
$\tilde{x},\tilde{y}$ are excursions from
the equilibrium position.
$\tilde{x}$ and $\tilde{y}$
can be expressed as linear combinations of $a_{\pm}^{\dagger}$ and
$a_{\pm}$ operators \cite{delande98} giving
$\tilde{x},\tilde{y}\sim \omega^{-1/2}\sim n_0^{3/2}$. Furthermore,
the equilibrium distance from the nucleus scales as the size of the atom,
$x_e\sim n_0^2$, and, therefore,
\begin{equation}
\tilde{x}^j/x_e^{j+1}\sim n_0^{\left(-2-\frac{j}{2}\right)}.
\label{abu_ptp_scaleac}
\end{equation}

A state $|n_+,n_-\rangle$ is obtained
by the excitation of
$N_+$ quanta in the $\omega_+$-mode and of $N_-$ quanta in the
$\omega_-$-mode, respectively, i. e.
 by the application of the
operator product $(a_+^{\dagger})^{n_+}(a_-^{\dagger})^{n_-}$ on
the wave-packet state $|0,0\rangle$, which -- by virtue of
eq.~(\ref{abu_ptp_scaleac}) -- will be subject to an unharmonic correction
scaling like
\begin{equation}
\Delta E_{\rm unharmonic}\sim n_0^{\left(-2-\frac{n_++n_-}{2}\right)}.
\label{abu_ptp_unharm}
\end{equation}
Hence, the size of the
avoided crossings between the wave-packet eigenstate and
excited states of the local potential around the stable fixed point
decreases with the number of quanta in the excited modes.

In addition, we can determine the width $\Delta F_{\rm unharmonic}$
of such avoided crossings in
the driving field amplitude $F$, by differentiation of the energy
(\ref{enharm})
of $|n_+,n_-\rangle$ with respect to $F$.
Then, the difference between the energies of two eigenstates localized
in the resonance island is found to scale like $Fn_0$. Defining
$\Delta F_{\rm unharmonic}$ by the requirement that $Fn_0$ be of the order
of $\Delta E_{\rm unharmonic}$, we find
\begin{equation}
\Delta F_{\rm unharmonic}\sim n_0^{\left(-3-\frac{n_++n_-}{2}\right)}.
\label{abu_ptp_fun}
\end{equation}
Since we want to switch the field to a maximum value
$F_{\rm max}\sim n_0^{-4}$, the Landau-Zener formula
\begin{equation}
\tau\sim \frac{F_{\rm max}}{\Delta E \Delta F}
\label{abu_ptp_lz}
\end{equation}
yields
\begin{equation}
\tau_{\rm unharmonic}\sim n_0^{n_-+n_++1}\sim n_0^{n_-+n_+-2}\ \ \
{\mathrm microwave\ periods} 
\label{abu_ptp_tunharm}
\end{equation}
for the scaling behavior of the time scale which guarantees diabatic
switching through avoided crossings of the wave-packet eigenstate with
excited states of the elliptic island.
Let us stress that this is only a very rough estimate of the
switching time,  some numerical
factors (not necessarily close to unity) are not taken into account.

\begin{figure}
\centerline{\psfig{figure=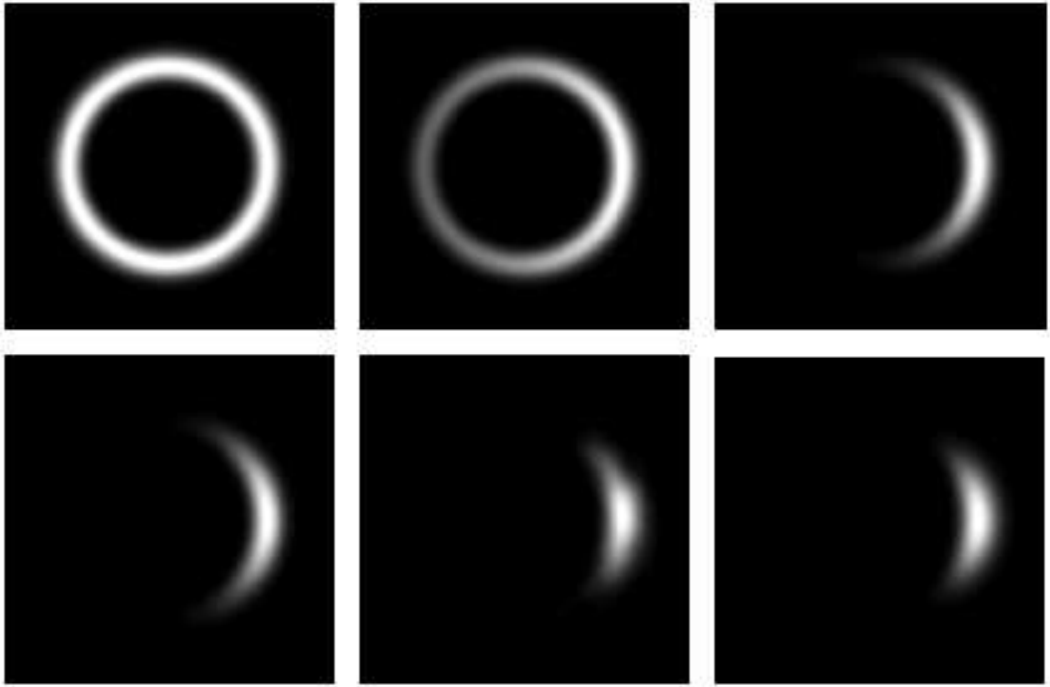,width=14cm}}
\caption{
Snapshots of the electronic density for a two-dimensional
hydrogen atom exposed to a circularly polarized microwave field
with increasing amplitude. The  microwave
field amplitude is switched on
according to eq.~(\protect\ref{abu_ptp_pulse}), with maximum scaled field
$F_{0,\rm max}= 0.03$ and
$T_{\rm switch}=400\times 2\pi/\omega$, where $\omega$ is resonant with
$n_0=60$ (frequency $\omega=1/(60.5)^3$).
The evolution of the initial circular state $n=M=60$ is numerically
computed by solving the time dependent Schr\"odinger equation in a
convenient Sturmian basis.
Top-left -- $t=0$ (initial circular state);
top-middle -- $t=20$ microwave periods;
top-right -- $t=60$ periods, bottom-left -- $t=100$ periods;
bottom-middle -- the final state, $t=400$ periods;
bottom-right -- the non-dispersive wave-packet (exact Floquet
eigenstate): it is almost indistinguishable from the previous wave-function,
what proves that the excitation process efficiently
and almost exclusively populates the state of interest. The box extends
over 10000 Bohr radii in both directions, with the nucleus
at the center. The microwave field is along the horizontal axis,
pointing to the right.}
\label{ptp_fig1}
\end{figure}

The above predictions can be checked, e.g., by a numerical integration of the
time dependent Schr\"odinger equation for a  microwave-driven 
atom, taking into account the  
time-dependent amplitude of the field. An
exemplary calculation on the two-dimensional model atom 
(see sec.~\ref{2d_model}) 
can be found in \cite{kuba97a}, for 
CP driving.
Fig.~\ref{ptp_fig1} shows the evolution
of the electronic density
of the atomic wave-function (initially prepared in the circular Rydberg
state $n=M=60$) during the rising part of the driving
field envelope, modeled by 
\begin{equation}
F(t)=F_{\rm max}\sin^2\left(\frac{\pi t}{2T_{\rm switch}}\right).
\label{abu_ptp_pulse}
\end{equation}

The driving field frequency was chosen according to the resonance condition
with the $n_0=60$ state,
with a maximum scaled amplitude $F_{0,\rm max}=0.03$.
Inspection of 
fig.~\ref{solfig} shows that, for this value of $F_{\rm max}$, the crossing
between the wave-packet eigenstate and the state $|n_+=1,n_-=4\rangle$ 
has to
be passed diabatically after adiabatic
trapping within the principal resonance. By virtue of the above estimations
of the adiabatic and the diabatic time scales, the switching
time (measured in driving field cycles)
has to be chosen such that $n_0<T_{\rm switch}<n_0^3$
(in microwave periods).
Clearly, the pulse 
populates the desired wave-packet 
once the driving field
amplitude 
reaches its maximum value. More quantitatively, 
the overlap of the final state after  
propagation of the time-dependent Schr\"odinger equation 
with the
wave-packet eigenstate of the driven atom in the field
(bottom-right panel) amounts to $94\%$. Since losses of atomic
population due to ionization
are negligible on the time scales considered in the figure, $6\%$ of the
initial atomic population is lost during the switching process.
The same calculation, done for the realistic three-dimensional
atom with $n_0=60$ gives the same result, proving that
the $z$-direction (which is neglected in the 2D model) 
is essentially irrelevant in this problem.
\begin{figure}
\centerline{\psfig{figure=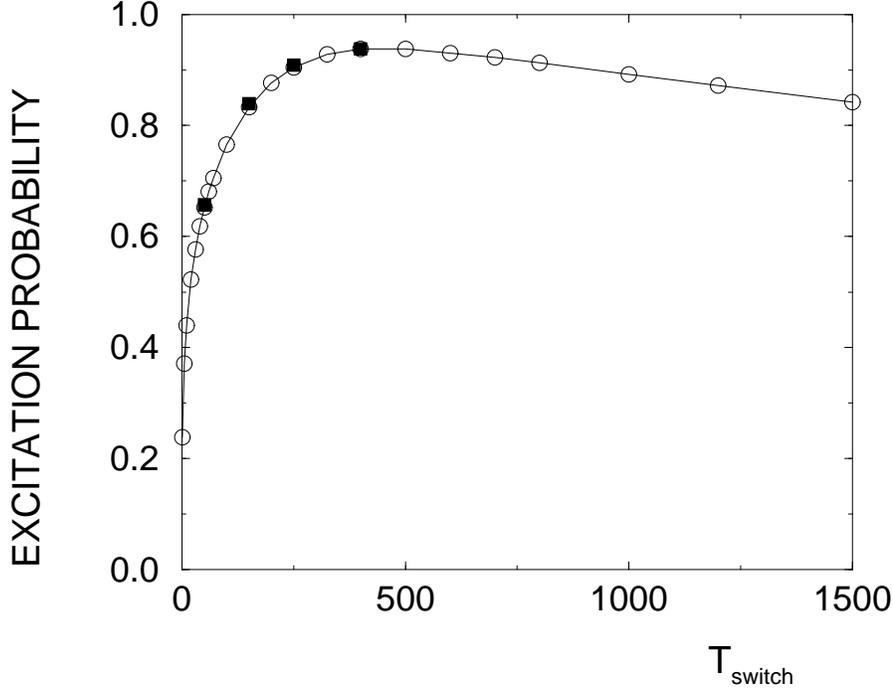,width=12cm,angle=-90}}
\caption{Overlap between the wave-function obtained at the end of
the microwave turn-on and the exact target state representing the
non-dispersive wave-packet, as a function of
the switching time $T_{\rm switch},$ for the two-dimensional
hydrogen atom (circles). The filled squares
indicate the results obtained for a fully
three-dimensional atom. $F_{\rm max}$, $\omega$, and $n_0$  
as in fig.~\protect\ref{ptp_fig1}.}
\label{ptp_fig2}
\end{figure}
Figure~\ref{ptp_fig2}
shows the efficiency of the proposed switching scheme
as a function of the switching time $T_{\rm switch}$, expressed
in units of microwave periods. Observe that too
long switching times tend to be less effective, since the avoided crossings
passed during the switching stage  
are not traversed diabatically. The rough estimate, eq.~(\ref{abu_ptp_tunharm}),
overestimates the maximum switching time by one order of magnitude.
On the other hand, 
too short switching times do not allow the wave-packet 
to localize inside the resonance island. However, a wide range
of switching times remains where good efficiency is achieved.

\begin{figure}
\centerline{\psfig{figure=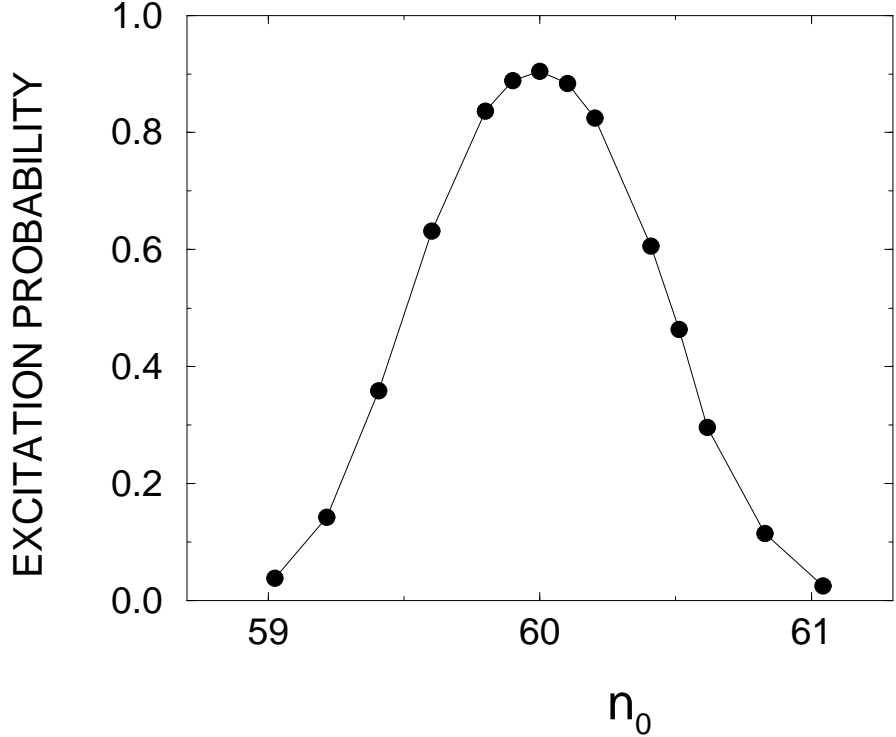,width=12cm,angle=-90}}
\caption{Overlap between the wave-function obtained at the end of
the microwave turn-on and the exact target state representing the
non-dispersive wave-packet as a function of $n_0$,
obtained for the two-dimensional
hydrogen atom (circles). $F_{\rm max}$ is 
as in fig.~\protect\ref{ptp_fig1}, and the switching time 
is $T_{\rm switch}=250$ microwave periods. 
The initial state corresponds to a circular
$n=60$ state of a 2D hydrogen atom.}
\label{ptp_fig3}
\end{figure}

For a given initial state $|\psi_0\rangle$
of the atom, only the
resonance condition defining the driving field frequency is to some extent
restrictive, as depicted in fig.~\ref{ptp_fig3}.
It is crucial that the initially excited field-free state
is adiabatically connected 
(through the Mathieu equation) to the ground state wave-packet. 
The best choice is ``optimal resonance", but the 
adiabaticity is preserved 
if $n_0$ is changed by less than one half, see section~\ref{QD}. 
This corresponds to a relative 
change of $\omega$ of the order of $3/2n_0.$
Given the spectral resolution of presently available
microwave generators, the definition of the
frequency with an accuracy of less than 1\%
is not a limitation. 
The exact numerical calculation displayed in fig.~\ref{ptp_fig3}
fully confirms that efficient excitation is possible as long as 
$n_0=\omega^{-1/3}-1/2$ matches the effective principal 
quantum number of the
 initially excited field-free state within a margin of $\pm 1/2$ 
(in the range [59.5,60.5]).

In conclusion,
the preparation of non-dispersive wave-packets by excitation
of a Rydberg state followed by careful switching of the
microwave field can be considered as an efficient method, provided
a clean experimental preparation of the atomic initial state can be
achieved. Furthermore, the boundaries --
eqs.~(\ref{t_smallF}) and (\ref{abu_ptp_tunharm}) --
imposed on the time scale for the
switching process leave a sufficient flexibility for the experimentalist 
to efficiently 
prepare the wave-packet.
A final word is in place on the homogeneity of the driving field amplitude
experienced by the atoms in the ``flat top region'' of the interaction,
i.e. after the switching from the field free state into
the wave-packet state at $F(t)=F_{\rm max}$. In any laboratory experiment,
a slow drift of the amplitude will be unavoidable over the interaction volume.
Hence, slightly different non--dispersive wave-packets will coexist
at various spatial positions. Since the ionization rate
of nondispersive wave-packets is rather sensitive
with respect to detailed values of the parameters, see
section \ref{ION}, this should manifest itself
by a deviation of the time dependence of the ionization yield from purely
exponential decay.

\subsection{Life time measurements}
\label{LIFE}

Given the above, rather efficient experimental 
schemes for the population of non-dispersive wave-packet eigenstates --
either via direct optical Floquet absorption or through an appropriate
switching procedure --
we still need some means to prove that we
really {\em did} populate the wave-packet. As a matter of fact, to provide 
unambiguous experimental evidence, one has to test various 
characteristic properties of the wave-packet, so as to
exclude accidental coincidences. A natural way is Floquet spectroscopy 
(see sec.~\ref{SPEC}),
i.e. probing the structure of the Floquet quasi-energy levels,
in either
the optical or the microwave regime
(via absorption, stimulated emission, Raman spectroscopy etc.). Another
possibility is to explore unique properties of wave-packet Floquet states.
For example, as discussed in sec.~\ref{ION}, these states exhibit 
extremely small ionization rates. 
Hence, an experimentally accessible quantity to identify
these states is the time-dependence of their survival probability, i.e. 
of the probability not to ionize during an interaction time $t.$
It is
given by \cite{abuth,abu95c}
\begin{equation}
P(t)=\sum_{\epsilon}|c_{\epsilon}|^2\exp(-\Gamma_{\epsilon}t),
\label{abu_lt_pion}
\end{equation}
where
the  $c_{\epsilon}$ denote the expansion coefficients of the initial 
field-free 
state in 
the Floquet basis,
at a given value of microwave amplitude $F$.

If the selective population of the wave-packet is successful, 
only one Floquet state contributes to $P(t)$, and the 
decay of the population to the atomic continuum should manifest in 
its exponential decrease, as opposed to a multiexponential  
decrease in the case of a broad distribution of the 
$c_{\epsilon}$ over the Floquet states 
\cite{abuth,abu95a,abu98b,wimbergerda,wimberger01,abu95b}. 
Of course, the distinction between 
an exponential and an algebraic decay law requires the variability of 
the experimental interaction time over more than one order of magnitude. 
This is a nontrivial task in experiments on atomic Rydberg states of 
hydrogen, since the typical velocities of the atomic beam are of the order
of $1000\ \rm m/s$. That significantly restricts the interval on which
the interaction time may be changed, taking into account the typical
size (in the cm-range) of
the atom-field interaction region 
\cite{koch95b,bayfield85}. 
However, the feasibility of such measurements has already been demonstrated
in microwave experiments on rubidium Rydberg states, where the interaction
time has been scanned from approx. $100$ to approx. $100000$ field cycles,
i.e., over three orders of magnitude \cite{abu95b,arndt91}.
Note that, whereas the dynamics
of the driven Rydberg electron 
along a Kepler ellipse of large eccentricity will 
certainly be affected by the presence of a non-hydrogenic core, 
non-dispersive wave-packets as the ones discussed in 
sections \ref{CP} and \ref{EP} 
can certainly be launched along circular trajectories, since the 
Rydberg electron of the rubidium atom essentially experiences a Coulomb
field on such a circular orbit. 

To use the character of the decay as a means 
to identify the wave-packet,
the microwave field amplitude should be sufficiently large
to guarantee that other Floquet states localized in the chaotic sea (see sec.~\ref{ION}) 
decay rapidly. Otherwise, the observation
of a mono-exponential decay simply suggests that we succeeded in populating
a single Floquet state - not necessarily a wave-packet \cite{abu95b}.
The appropriate choice of the driving 
field amplitude $F$, such that appreciable ionization  
is achieved for the 
longer experimentally accessible 
interaction times,
should therefore allow for the experimental identification of the 
mono-exponential decay from the wave-packet to the atomic continuum, but also
-- by varying $F$ -- of the variations of the decay rate with $F,$
which is predicted to fluctuate wildly over several orders of magnitude,
see section \ref{ION}. 
Note, however, that this requires an excellent homogeneity of the
microwave field (e.g., provided by a high quality microwave cavity), 
as the fluctuations take place over
rather small intervals of $F.$

\section{Conclusions}
\label{CONC}

In this report, we have shown that novel and highly robust eigenstates 
of periodically 
driven quantum systems -- non-dispersive wave-packets -- are born 
out of classically mixed regular-chaotic dynamics. As much as a mixed phase 
space is generic for classical Hamiltonian systems, non-dispersive 
wave-packets are a generic manifestation thereof on the quantum level, given
a sufficiently high density of states (needed to resolve finite-size phase 
space structures). While we described their semiclassical properties and
their experimental preparation, manipulation, and identification during the 
largest part of this report for a specific system -- atomic Rydberg states
driven by a microwave field -- it is clear from our approach 
that such ``quantum particles'' can be anchored to any nonlinear resonance 
between a periodic drive and a periodic trajectory of a Hamiltonian 
system. As an alternative example, we have briefly touched upon the 
atomic realization of the gravitational bouncer, though many other 
realizations in simple quantum optical or atomic and molecular systems can be
thought of. Let us only mention unharmonic traps for ions, atoms, or BEC
condensates, periodically kicked atoms \cite{reinhold01},
as well as molecular dynamics \cite{gerber98,gerber01} 
on adiabatic potential surfaces
(the driven frozen planet briefly discussed in section \ref{HE} may 
be conceived
as opening a perspective in this direction). Nontheless, atomic 
Rydberg states remain arguably the best objects to study the 
fundamental properties of non-dispersive wave-packets as the realization of
Schr\"odinger's dream \cite{schroe26}.
 On one hand, they are 
microscopic realizations of the Keplerian motion and of Bohr's orbitals using
a well understood non-linear dynamical system. 
On the other hand, they possess the essential complication which open 
quantum systems add to bounded Hamiltonian dynamics -- the 
driving-induced, coherent coupling to the atomic continuum of free electronic
states. On top of that, all these features can be controlled in 
real laboratory experiments, and we might actually dream of probing the
characteristic properties of nondispersive wave-packets on single, trapped 
atoms or ions, using novel experimental approaches yet to come. Let us finally 
dare to speculate on the potential use of non-dispersive wave-packets in 
coherent control: given their spectacular robustness -- which we abundantly 
illustrated in this report -- it is clear that they provide a means to store 
and to ``ship'' quantum probability densities in and across phase space,
e.g., under adiabatic changes of the driving field polarization and/or of the 
strength or orientation of additional static fields. Given the recent 
advances in coherent control of molecular reactions employing laser fields
\cite{gerber98} -- which so far do not explore the unique perspectives of 
nonlinear dynamics -- it looks like a promising (and challenging) program to 
systematically study non-dispersive wave-packets in molecular reaction 
dynamics.

\section{Acknowledgments}

It is a pleasure 
to acknowledge a longstanding and fruitful collaboration with Robert
G\c{e}barowski, Beno\^{\i}t Gr\'emaud, 
Klaus Hornberger, Andreas Krug, Romek Marcinek, 
Krzysiek Sacha, Peter Schlagheck, and Sandro
Wimberger on non-dispersive wave-packets and related topics 
over the past five years.

We acknowledge support of bilateral collaborations via
programmes Procope (German-French) and Polonium (Polish-French).
J.Z. acknowledges support by Polish Committee for Scientific Research
under grant 2P03B00915. Laboratoire Kastler Brossel is laboratoire 
de l'Universit{\'e} Pierre et Marie
Curie et de l'Ecole Normale Sup{\'e}rieure, unit{\'e} mixte de
recherche 8552 du CNRS. 
CPU time on various computers has been provided by IDRIS (Orsay) and 
RZG (Garching).


\newpage
\begin{thebibliography}{100}

\bibitem{englert95}
J.~A. Bergou and B.~G. Englert, J. Mod. Opt. {\bf 45},  701  (1998).

\bibitem{schroe26}
E. Schr\"odinger, \natw {\bf 14},  664  (1926).

\bibitem{lichtenberg83}
A.~J. Lichtenberg and M.~A. Lieberman, {\em Regular and Stochastic Motion},
  Vol.~38 of {\em Applied Mathematical Sciences} (Springer, Berlin, 1983).

\bibitem{raman97}
C. Raman, T.~C. Weinacht, and P.~H. Bucksbaum, \pr A {\bf 55},  R3995  (1997).

\bibitem{yeazell90}
J.~A. Yeazell, M. Mallalieu, and J. C.~R.~Stroud, \prL {\bf 64},  2007  (1990).

\bibitem{alber91}
G. Alber and P. Zoller, \phr {\bf 199},  231  (1991).

\bibitem{landau2}
L.~D. Landau and E.~M. Lifschitz, {\em Quantum Mechanics} (Pergamon, Oxford,
  1977).

\bibitem{parker86}
J. Parker and J. C.~R.~Stroud, \prL {\bf 56},  716  (1986).

\bibitem{alber86}
G. Alber, H. Ritsch, and P. Zoller, \pr {\bf 34},  1058  (1986).

\bibitem{averbukh89}
I.~S. Averbukh and N.~F. Perelman, \pla {\bf 139},  449  (1989).

\bibitem{yeazell91}
J.~A. Yeazell and J. C.~R.~Stroud, \pr {\bf 43},  5153  (1991).

\bibitem{hillery84}
M. Hillery, R.~F. O'Connell, M.~O. Scully, and E.~P. Wigner, \phr {\bf 106},
  12  (1984).

\bibitem{moyal49}
J.~E. Moyal, Proc. Camb. Phil. Soc. Math. Phys. Sci. {\bf 45},  99  (1949).

\bibitem{landau1}
L.~D. Landau and E.~M. Lifschitz, {\em Mechanics} (Pergamon, Oxford, 1994).

\bibitem{haake90}
F. Haake, {\em Quantum Signatures of Chaos}, Vol.~54 of {\em Springer Series in
  Synergetics} (Springer, Berlin, 1991).

\bibitem{glauber63}
R.~J. Glauber, \pr {\bf 131},  2766  (1963).

\bibitem{mandel90}
L. Mandel and E. Wolf, {\em Coherence and Quantum Optics} (Cambridge University
  Press, Cambridge, 1995).

\bibitem{cct92}
C. Cohen-Tannoudji, J. Dupont-Roc, and G. Grynberg, {\em Atom-Photon
  Interactions: Basic Processes and Applications} (John Wiley and Sons, New
  York, 1992).

\bibitem{cct73}
C. Cohen-Tannoudji, B. Diu, and F. Lalo\"e, {\em M\'ecanique quantique}
  (Hermann, Paris, 1973).

\bibitem{liboff80}
R.~L. Liboff, {\em Introductory Quantum Mechanics} (Holden-Day, San Francisco,
  1980).

\bibitem{husimi40}
K. Husimi, Proc. Phys. Math. Soc. Japan {\bf 22},  264  (1940).

\bibitem{bethe77}
H.~A. Bethe and E.~E. Salpeter, {\em Quantum Mechanics of One- and Two-Electron
  Atoms} (Plenum Publishing Corp., New York, 1977).

\bibitem{goldberger50}
M.~S. Goldberger and K.~M. Watson, {\em Collision Theory} (Wiley, New York,
  1964).

\bibitem{faisal87}
F.~H.~M. Faisal, {\em Theory of Multiphoton Processes} (Plenum Press, New York,
  1987).

\bibitem{yeazell88}
J.~A. Yeazell and J. C.~R.~Stroud, \prL {\bf 60},  1494  (1988).

\bibitem{marmet94}
L. Marmet {\it et~al.}, \prL {\bf 72},  3779  (1994).

\bibitem{mallalieu94}
M. Mallalieu and J. C.~R.~Stroud, \pr {\bf A49},  2329  (1994).

\bibitem{weinacht99}
T.~C. Weinacht, J. Ahn, and P.~H. Bucksbaum, Nature {\bf 397},  233  (1999).

\bibitem{ottb}
E. Ott, {\em Chaos in Dynamical Systems} (Cambridge University Press,
  Cambridge, 1993).

\bibitem{lee97}
E. Lee, A.~F. Brunello, and D. Farrelly, \pr A {\bf 55},  2203  (1997).

\bibitem{schroe68}
E. Schr\"odinger,  in {\em Sources of Quantum Mechanics}, edited by B.~L.
  van~der Waerden (Dover, New York, 1968).

\bibitem{delande94}
D. Delande and A. Buchleitner, \amop {\bf 35},  85  (1994).

\bibitem{abu95d}
A. Buchleitner and D. Delande, \prL {\bf 75},  1487  (1995).

\bibitem{ibb94}
I. Bia{\l}ynicki-Birula, M. Kalinski, and J.~H. Eberly, \prL {\bf 73},  1777
  (1994).

\bibitem{ahn00}
J. Ahn, T.~C. Weinacht, and P.~H. Bucksbaum, Science {\bf 287},  463  (2000).

\bibitem{meyer00}
D.~A. Meyer, Science {\bf 289},  1431a  (2000).

\bibitem{kwiat00}
P.~G. Kwiat and R.~J. Hughes, Science {\bf 289},  1431a  (2000).

\bibitem{bucksbaum00}
P.~H. Bucksbaum, J. Ahn, and T.~C. Weinacht, Science {\bf 289},  1431a  (2000).

\bibitem{berman77}
G.~P. Berman and G.~M. Zaslavsky, Phys. Lett. {\bf 61A},  295  (1977).

\bibitem{henkel92}
J. Henkel and M. Holthaus, \pr {\bf A45},  1978  (1992).

\bibitem{holthaus94}
M. Holthaus, Prog. Theor. Phys. Supplement {\bf 116},  417  (1994).

\bibitem{holthaus95}
M. Holthaus, \csf {\bf 5},  1143  (1995).

\bibitem{abuth}
A. Buchleitner, Ph.D. thesis, Universit\'e Pierre et Marie Curie, Paris, 1993.

\bibitem{farrelly95a}
D. Farrelly, E. Lee, and T. Uzer, \prL {\bf 75},  972  (1995).

\bibitem{ibb95}
I. Bia{\l}ynicki-Birula, M. Kalinski, and J.~H. Eberly, \prL {\bf 75},  973
  (1995).

\bibitem{farrelly95}
D. Farrelly, E. Lee, and T. Uzer, \pla {\bf 204},  359  (1995).

\bibitem{kalinski95a}
M. Kalinski, J.~H. Eberly, and I. Bia{\l}ynicki-Birula, \pr {\bf A52},  2460
  (1995).

\bibitem{kalinski95b}
M. Kalinski and J.~H. Eberly, \pr {\bf A52},  4285  (1995).

\bibitem{delande95}
D. Delande, J. Zakrzewski, and A. Buchleitner, Europhys. Lett. {\bf 32},  107
  (1995).

\bibitem{kuba95a}
J. Zakrzewski, D. Delande, and A. Buchleitner, Phys. Rev. Lett. {\bf 75},  4015
   (1995).

\bibitem{ibb96}
I. Bia{\l}ynicki-Birula and Z. Bia{\l}ynicka-Birula, Phys. Rev. Lett. {\bf 77},
   4298  (1996).

\bibitem{eberly96}
J.~H. Eberly and M. Kalinski,  in {\em Multiphoton Processes 1996, Proceedings
  of the 7th International Conference on Multiphoton Processes,
  Garmisch-Partenkirchen, Germany, October 1996}, Vol.~154 of {\em Institute of
  Physics Conference Series}, edited by P. Lambropoulos and H. Walther
  (Institute of Physics, Bristol and Philadelphia, 1997), pp.\ 29--36.

\bibitem{kalinski96a}
M. Kalinski and J.~H. Eberly, \prL {\bf 77},  2420  (1996).

\bibitem{brunello96}
A.~F. Brunello, T. Uzer, and D. Farrelly, \prL {\bf 76},  2874  (1996).

\bibitem{kalinski96b}
M. Kalinski and J.~H. Eberly, \pr {\bf A53},  1715  (1996).

\bibitem{kalinski97}
M. Kalinski and J.~H. Eberly, \prL {\bf 79},  3542  (1997).

\bibitem{kalinski98}
M. Kalinski, \pr {\bf 57},  2239  (1998).

\bibitem{ibb97a}
I. Bia{\l}ynicki-Birula and Z. Bia{\l}ynicka-Birula, Phys. Rev. Lett. {\bf 78},
   2539  (1997).

\bibitem{ibb97b}
Z. Bia{\l}ynicka-Birula and I. Bia{\l}ynicki-Birula, \pr {\bf A56},  3629
  (1997).

\bibitem{delande97}
D. Delande, J. Zakrzewski, and A. Buchleitner, \prL {\bf 79},  3541  (1997).

\bibitem{kuba97a}
J. Zakrzewski and D. Delande, \jpb {\bf 30},  L87  (1997).

\bibitem{cerjan97}
C. Cerjan, E. Lee, D. Farrelly, and T. Uzer, \pr {\bf A55},  2222  (1997).

\bibitem{kuba97b}
J. Zakrzewski, A. Buchleitner, and D. Delande, Z. Phys. {\bf B103},  115
  (1997).

\bibitem{abu96}
A. Buchleitner, D. Delande, and J. Zakrzewski,  in {\em Multiphoton Processes
  1996, Proceedings of the 7th International Conference on Multiphoton
  Processes, Garmisch-Partenkirchen, Germany, October 1996}, Vol.~154 of {\em
  Institute of Physics Conference Series}, edited by P. Lambropoulos and H.
  Walther (Institute of Physics, Bristol and Philadelphia, 1997), pp.\ 19--28.

\bibitem{kuba98}
J. Zakrzewski, D. Delande, and A. Buchleitner, Phys. Rev. {\bf E57},  1458
  (1998).

\bibitem{kuba98a}
J. Zakrzewski, D. Delande, and A. Buchleitner, Acta Physica Polon. {\bf A 93},
  179  (1998).

\bibitem{abu98a}
A. Buchleitner, K. Sacha, D. Delande, and J. Zakrzewski, Eur. Phys. J. D {\bf
  5},  145  (1999).

\bibitem{delande98}
D. Delande and J. Zakrzewski, \pr A {\bf 58},  466  (1998).

\bibitem{sachath}
K. Sacha, Ph.D. thesis, Jagellonian University, Krak\'ow, 1998, unpublished.

\bibitem{hornbergerda}
K. Hornberger, Master's thesis, Ludwig-Maximilians-Universit\"at, M\"un\-chen,
  1998.

\bibitem{hornberger98}
K. Hornberger and A. Buchleitner, Europhys. Lett. {\bf 41},  383  (1998).

\bibitem{sacha98a}
K. Sacha, J. Zakrzewski, and D. Delande, Eur. Phys. J. {\bf D1},  231  (1998).

\bibitem{sacha98b}
K. Sacha and J. Zakrzewski, \pr A {\bf 58},  3974  (1998).

\bibitem{sacha99a}
K. Sacha and J. Zakrzewski, \pr A {\bf 59},  1707  (1999).

\bibitem{yeazell00}
D. Farrelly,  in {\em Physics and Chemistry of Wave Packets}, edited by J.~A.
  Yeazell and T. Uzer (John Wiley \& Sons, New York, 2000).

\bibitem{Berry_WKB}
M.~V. Berry and K.~E. Mount, Rep. Prog. Phys. {\bf 35},  315  (1972).

\bibitem{ozorio88}
A.~M.~O. de~Almeida, {\em Hamiltonian Systems: Chaos and Quantization}
  (Cambridge University Press, Cambridge, 1988).

\bibitem{einstein17}
A. Einstein, Verh. d. Dtsch. Phys. Ges.  82  (1917).

\bibitem{bornwolf}
M. Born and E. Wolf, {\em Principles of Optics} (Academic Press, Oxford, 1999).

\bibitem{breuer91}
H.~P. Breuer and M. Holthaus, \anp {\bf 211},  249  (1991).

\bibitem{heller89}
E.~J. Heller,  in {\em Chaos and Quantum Physics}, Vol.~Session LII of {\em Les
  Houches} (North-Holland, Amsterdam, 1991), p.\ 547.

\bibitem{stoeckmann99}
H.~J. St\"ockmann, {\em Quantum Chaos: An Introduction} (Cambridge University
  Press, Cambridge, 1999).

\bibitem{jensen89b}
R.~V. Jensen, M.~M. Sanders, M. Saraceno, and B. Sundaram, \prL {\bf 63},  2771
   (1989).

\bibitem{leopold94}
J.~G. Leopold and D. Richards, \jpb {\bf 27},  2169  (1994).

\bibitem{zaslavsky81}
G.~M. Zaslavsky, \phr {\bf 80},  157  (1981).

\bibitem{ringot00}
J. Ringot, P. Szriftgiser, J.~C. Garreau, and D. Delande, \prL {\bf 85},  2741
  (2000).

\bibitem{abu97}
A. Buchleitner and D. Delande, Phys. Rev. {\bf A55},  R1585  (1997).

\bibitem{chirikov59}
B.~V. Chirikov, Doc. Ac. Sci. USSR {\bf 125},  1015  (1959).

\bibitem{chirikov79}
B.~V. Chirikov, \phr {\bf 52},  263  (1979).

\bibitem{floquet1883}
M.~G. Floquet, Ann. \'Ecole Norm. Sup. {\bf 12},  47  (1883).

\bibitem{mermin76}
N.~W. Ashcroft and N.~D. Mermin, {\em Solid State Physics} (Saunders College
  Publishing, Fort Worth, 1976), college edition.

\bibitem{shirley65}
J.~H. Shirley, \pr {\bf 138},  B979  (1965).

\bibitem{marion}
J.~B. Marion, {\em Classical Dynamics of Particles and Systems}, 2nd ed.
  (Academic Press, New York, 1970).

\bibitem{loudon}
R. Loudon, {\em The Quantum Theory of Light}, 2nd ed. (Clarendon Press, Oxford,
  1986).

\bibitem{abramowitz72}
 in {\em Handbook of Mathematical Functions}, edited by M. Abrammowitz and
  I.~A. Stegun (Dover, New York, 1972).

\bibitem{shakeshaft88}
R. Shakeshaft, Z. Phys. D {\bf 8},  47  (1988).

\bibitem{abu95c}
A. Buchleitner, D. Delande, and J.~C. Gay, \josab {\bf 12},  505  (1995).

\bibitem{cormier96}
E. Cormier and P. Lambropoulos, J. Phys. B {\bf 29},  1667  (1996).

\bibitem{javanainen88}
J. Javanainen, J.~H. Eberly, and Q. Su, \pr {\bf A38},  3430  (1988).

\bibitem{su91}
Q. Su and J.~H. Eberly, \pr {\bf A44},  5997  (1991).

\bibitem{reed91}
V.~C. Reed, P.~L. Knight, and K. Burnett, \prL {\bf 67},  1415  (1991).

\bibitem{burnett92}
K. Burnett, V.~C. Reed, J. Cooper, and P.~L. Knight, \pr {\bf A45},  3347
  (1992).

\bibitem{grobe93}
R. Grobe and J.~H. Eberly, \pr {\bf A47},  R1605  (1993).

\bibitem{grobe94}
R. Grobe, K. Rz\c{a}\.zewski, and J.~H. Eberly, J. Phys. B {\bf 27},  L503
  (1994).

\bibitem{kusta65}
P. Kustaanheimo and E. Stiefel, J. Reine Angew. Math. {\bf 218},  204  (1965).

\bibitem{delande84}
D. Delande and J.-C. Gay, \jpb {\bf 17},  335  (1984).

\bibitem{rath88}
O. Rath and D. Richards, \jpb {\bf 21},  555  (1988).

\bibitem{griffiths92}
J.~A. Griffiths and D. Farrelly, \pr {\bf A45},  R2678  (1992).

\bibitem{gebarowski95}
R. G{\c{e}}barowski and J. Zakrzewski, \pr {\bf A51},  1508  (1995).

\bibitem{schlagheck99}
P. Schlagheck and A. Buchleitner, \phd {\bf 131},  110  (1999).

\bibitem{barut79}
A.~O. Barut, C.~K.~E. Schneider, and R. Wilson, J.\ Math.\ Phys.\ {\bf 20},
  2244  (1979).

\bibitem{chen80}
A.~C. Chen, \pr {\bf A 22},  333  (1980).

\bibitem{chen81}
A.~C. Chen, \pr {\bf A 23},  1655  (1981).

\bibitem{delandeth}
D. Delande, Ph.D. thesis, Universit\'e de Paris, Paris, 1988, th\`ese de
  doctorat d'etat.

\bibitem{lanczos}
C. Lanczos, J. Res. Nat. Bur. Standards, Sect B {\bf 45},  225  (1950).

\bibitem{delande91}
D. Delande, A. Bommier, and J.-C. Gay, \prL {\bf 66},  141  (1991).

\bibitem{ericsson80}
T. Ericsson and A. Ruhe, Math. Comput. {\bf 35},  1251  (1980).

\bibitem{grimes94}
R.~G. Grimes, J.~G. Lewis, and H.~D. Simon, SIAM J. Matrix Anal. Appl. {\bf
  15},  228  (1994).

\bibitem{balslev71}
E. Balslev and J.~M. Combes, Commun.\ math.\ Phys. {\bf 22},  280  (1971).

\bibitem{graffi85}
S. Graffi, V. Grecchi, and H.~J. Silverstone, Ann.\ Inst.\ Henri Poincar\'e
  {\bf 42},  215  (1985).

\bibitem{yajima82}
K. Yajima, Comm.\ Math.\ Phys. {\bf 87},  331  (1982).

\bibitem{nicolaides78}
C.~A. Nicolaides and D.~R. Beck, Int. J. Quant. Chem. {\bf XIV},  457  (1978).

\bibitem{reinhardt83}
B.~R. Johnson and W.~P. Reinhardt, Phys. Rev. A {\bf 28},  1930  (1983).

\bibitem{ho83}
Y.~K. Ho, Phys. Rep. {\bf 99},  1  (1983).

\bibitem{moiseyev98}
N. Moiseyev, Phys. Rep. {\bf 302},  211  (1998).

\bibitem{abu94}
A. Buchleitner, B. Gr\'emaud, and D. Delande, \jpb {\bf 27},  2663  (1994).

\bibitem{englefield72}
M.~J. Englefield, {\em Group Theory and the Coulomb Problem} (Wiley, New-York,
  1972).

\bibitem{goldstein80}
H. Goldstein, {\em Classical Dynamics} (Addison-Wesley, Reading, Ma., 1980),
  p.146.

\bibitem{jensen84}
R.~V. Jensen, \pr {\bf A30},  386  (1984).

\bibitem{meerson82}
B.~I. Meerson, E.~A. Oks, and P.~V. Sasorov, \jpb {\bf 15},  3599  (1982).

\bibitem{casati88}
G. Casati, I. Guarneri, and D.~L. Shepelyansky, IEEE J.\ Quantum Electron. {\bf
  24},  1420  (1988).

\bibitem{bayfield89}
J.~E. Bayfield, G. Casati, I. Guarneri, and D.~W. Sokol, \prL {\bf 63},  364
  (1989).

\bibitem{koch95b}
P.~M. Koch and K.~A.~H. van Leeuwen, \phr {\bf 255},  289  (1995).

\bibitem{bellermann96}
M.~R.~W. Bellermann, P.~M. Koch, D. Mariani, and D. Richards, \prL {\bf 76},
  892  (1996).

\bibitem{bayfield74}
J.~E. Bayfield and P.~M. Koch, \prL {\bf 33},  258  (1974).

\bibitem{bayfield96}
J.~E. Bayfield, S.~Y. Luie, L.~C. Perotti, and M.~P. Skrzypkowski, \pr {\bf
  A53},  R12  (1996).

\bibitem{galvez88}
E.~J. Galvez {\it et~al.}, \prL {\bf 61},  2011  (1988).

\bibitem{fu90}
P. Fu, T.~J. Scholz, J.~M. Hettema, and T.~F. Gallagher, \prL {\bf 64},  511
  (1990).

\bibitem{cheng96}
C.~H. Cheng, C.~Y. Lee, and T.~F. Gallagher, \pr {\bf A54},  3303  (1996).

\bibitem{delande97b}
D. Delande and J. Zakrzewski,  in {\em Classical, Semiclassical and Quantum
  Dynamics in Atoms}, No.~485 in {\em Lecture Notes in Physics}, edited by H.
  Friedrich and B. Eckhardt (Springer, New York, 1997), p.\ 205.

\bibitem{bunkin64}
F.~V. Bunkin and A.~M. Prokhorov, Sov.\ Phys.\ JETP {\bf 19},  739  (1964).

\bibitem{grozdanov92}
T.~P. Grozdanov, M.~J. Rakovi\'c, and E.~A. Solovev, \jpb {\bf 25},  4455
  (1992).

\bibitem{klar89}
H. Klar, \zpd {\bf 11},  45  (1989).

\bibitem{lee95}
E. Lee, A.~F. Brunello, and D. Farrelly, \prL {\bf 75},  3641  (1995).

\bibitem{abu95a}
A. Buchleitner and D. Delande, \csf {\bf 5},  1125  (1995).

\bibitem{heller84}
E.~J. Heller,  in {\em Classical and Quantum Chaos}, edited by A. Voros, M.
  Giannoni, and A. Zinn-Justin (Elsevier, Amsterdam, 1991), Chap.~Scars.

\bibitem{bogomolny88a}
E.~B. Bogomolny, \phd {\bf 31},  169  (1988).

\bibitem{Bth}
A. Brunello, Ph.D. thesis, State University of New York at Stony Brook, Stony
  Brook, 1997, unpublished.

\bibitem{sacha97}
K. Sacha and J. Zakrzewski, \pr {\bf A56},  719  (1997).

\bibitem{sacha98c}
K. Sacha and J. Zakrzewski, \pr A {\bf 58},  488  (1998).

\bibitem{brunello97}
A.~F. Brunello, T. Uzer, and D. Farrelly, \pr A {\bf 55},  3730  (1997).

\bibitem{leopold86}
J.~G. Leopold and D. Richards, \jpb {\bf 19},  1125  (1986).

\bibitem{leopold87}
J.~G. Leopold and D. Richards, \jpb {\bf 20},  2369  (1987).

\bibitem{rakovic98}
M.~J. Rakovi\'c, T. Uzer, and D. Farrelly, \pr A {\bf 57},  2814  (1998).

\bibitem{benvenuto91}
F. Benvenuto, G. Casati, I. Guarneri, and D.~L. Shepelyansky, \zp B {\bf 84},
  159  (1991).

\bibitem{oliveira94}
C.~R. de~Oliveira, G. Casati, and I. Guarneri, \epl {\bf 27},  187  (1994).

\bibitem{flatte96}
M.~E. Flatt\'e and M. Holthaus, \anp {\bf 245},  113  (1996).

\bibitem{steane95}
A. Steane, P. Szriftgiser, P. Desbiolles, and J. Dalibard, \prL {\bf 74},  4972
   (1995).

\bibitem{oberthaler99}
M.~K. Oberthaler {\it et~al.}, \prL  4447  (1999).

\bibitem{bonci98}
L. Bonci, A. Farusi, P. Grigilini, and R. Roncaglia, \pr E {\bf 58},  5689
  (1998).

\bibitem{brodier01}
O. Brodier, P. Schlagheck, and D. Ullmo, \prL {\bf 87},  64101  (2001).

\bibitem{paul92}
W. Paul, Rev. Mod. Phys. {\bf 62},  531  (1992).

\bibitem{benvenuto94}
F. Benvenuto, G. Casati, and D.~L. Shepelyansky, \prL {\bf 72},  1818  (1994).

\bibitem{tanner00}
G. Tanner, K. Richter, and J.~M. Rost, \rmp {\bf 72},  497  (2000).

\bibitem{wintgen93}
D. Wintgen and D. Delande, \jpb {\bf 26},  L399  (1993).

\bibitem{gremaud97}
B. Gr\'emaud and D. Delande, \epl {\bf 40},  363  (1997).

\bibitem{gremaudth}
B. Gr\'emaud, Ph.D. thesis, Universit\'e Paris 6, 1997.

\bibitem{puttner01}
R. P\"uttner {\it et~al.}, \prL {\bf 86},  3747  (2001).

\bibitem{eichmann90}
U. Eichmann, V. Lange, and W. Sandner, \prL {\bf 64},  274  (1990).

\bibitem{richter90}
K. Richter and D. Wintgen, \prL {\bf 65},  1965  (1990).

\bibitem{schlagheck98a}
P. Schlagheck and A. Buchleitner, \jpb {\bf 31},  L489  (1998).

\bibitem{schlagheck99b}
P. Schlagheck and A. Buchleitner, \epl {\bf 46},  24  (1999).

\bibitem{schlagheckth}
P. Schlagheck, Ph.D. thesis, Technische Universit\"at, M\"unchen, 1999,
  published by Herbert Utz Verlag.

\bibitem{hanson95}
L.~G. Hanson and P. Lambropoulos, \prL {\bf 74},  5009  (1995).

\bibitem{zobay96}
O. Zobay and G. Alber, \pr {\bf 54},  5361  (1996).

\bibitem{mecking98}
B. Mecking and P. Lambropoulos, \pr {\bf A57},  2014  (1998).

\bibitem{carpetmen}
M. Berry, I. Marzoli, and W.~P. Schleich, Physics World {\bf 14},    (2001).

\bibitem{stone85}
P.~A. Lee and A.~D. Stone, \prL {\bf 55},  1622  (1985).

\bibitem{LB90}
W.~A. Lin and L.~E. Ballentine, \prL {\bf 65},  2927  (1990).

\bibitem{LB92}
W.~A. Lin and L.~E. Ballentine, \pr {\bf 45},  3637  (1992).

\bibitem{grossmann91}
F. Grossmann, T. Dittrich, P. Jung, and P. H\"anggi, \prL {\bf 67},  516
  (1991).

\bibitem{GDJH91b}
F. Grossmann, T. Dittrich, P. Jung, and P. H\"anggi, Z.\ Phys.\ B\ {\bf 84},
  315  (1991).

\bibitem{GDJH93}
F. Grossmann {\it et~al.}, J.\ Stat.\ Phys.\ {\bf 70},  229  (1993).

\bibitem{plata92}
J. Plata and J.~M.~G. Llorente, \jpa {\bf 25},  L303  (1992).

\bibitem{bohigas93}
O. Bohigas, S. Tomsovic, and D. Ullmo, \phr {\bf 223},  43  (1993).

\bibitem{bohigas93b}
O. Bohigas, D. Boos\'e, R.~E. de~Carvalho, and V. Marvulle, Nucl.\ Phys.\ {\bf
  A560},  197  (1993).

\bibitem{tomsovic94}
S. Tomsovic and D. Ullmo, \pr {\bf E50},  145  (1994).

\bibitem{shudo95}
A. Shudo and K.~S. Ikeda, \prL {\bf 74},  682  (1995).

\bibitem{leyvraz96}
F. Leyvraz and D. Ullmo, \jpa {\bf 29},  2529  (1996).

\bibitem{jackson}
J.~D. Jackson, {\em Classical Electrodynamics} (Wiley, New York, 1975).

\bibitem{haroche92}
S. Haroche,  in {\em Fundamental Systems in Quantum Optics, Le Houches, Session
  LIII,1990}, edited by J. Dalibard, J.~M. Raimond, and J. Zinn-Justin
  (Elsevier, New York, 1992).

\bibitem{lewenstein88}
M. Lewenstein, J. Zakrzewski, T.~W. Mossberg, and J. Mostowski, \jpb {\bf 21},
  L9  (1988).

\bibitem{lewenstein88b}
M. Lewenstein, J. Zakrzewski, and T.~W. Mossberg, \pr {\bf A38},  808  (1988).

\bibitem{abu97b}
D.~D.~R. Buchleitner, D. Delande, and J. Zakrzewski,  in {\em Non-spreading
  Kn\"odel-packets in Microsoft fields, Proceedings of the Third International
  Bavarian Conference on Kn\"odel-Packets}, No.~007 in {\em Food, physics and
  politics}, edited by H. Schalther and W. Weich (Stoiber Comp., Aidling,
  1997), p.\ 294561.

\bibitem{nakamura93}
K. Nakamura, {\em Quantum chaos, a new paradigm of nonlinear dynamics}, {\em
  Cambridge Nonlinear Science Series 3} (Cambridge University Press, Cambridge,
  1993).

\bibitem{casati87b}
G. Casati, I. Guarneri, and D.~L. Shepelyansky, \pr {\bf A36},  3501  (1987).

\bibitem{leopold78}
J.~G. Leopold and I.~C. Percival, \prL {\bf 41},  944  (1978).

\bibitem{bayfield88a}
J.~E. Bayfield and D.~W. Sokol, \prL {\bf 61},  2007  (1988).

\bibitem{bluemel89a}
R. Bl\"umel {\it et~al.}, \prL {\bf 62},  341  (1989).

\bibitem{bluemel91a}
R. Bl\"umel {\it et~al.}, Phys. Rev. {\bf A44},  4521  (1991).

\bibitem{abu91}
A. Buchleitner, L. Sirko, and H. Walther, \epl {\bf 16},  35  (1991).

\bibitem{casati84}
G. Casati, B.~V. Chirikov, and D.~L. Shepelyansky, \prL {\bf 53},  2525
  (1984).

\bibitem{fishman82}
S. Fishman, D.~R. Grempel, and R.~E. Prange, \prL {\bf 49},  509  (1982).

\bibitem{grempel84}
D.~R. Grempel, R.~E. Prange, and S. Fishman, \pr {\bf A29},  1639  (1984).

\bibitem{brenner96}
N. Brenner and S. Fishman, \prL {\bf 77},  3763  (1996).

\bibitem{abu98b}
A. Buchleitner, I. Guarneri, and J. Zakrzewski, Europhys. Lett. {\bf 44},  162
  (1998).

\bibitem{wimbergerda}
S. Wimberger, Master's thesis, Ludwig-Maximilians-Universit\"at M\"unchen,
  2000.

\bibitem{wimberger01}
S. Wimberger and A. Buchleitner, \jpa {\bf 34},  7181  (2001).

\bibitem{sirko96}
L. Sirko, A. Haffmans, M.~R.~W. Bellermann, and P.~M. Koch, \epl {\bf 33},  181
   (1996).

\bibitem{sirko93a}
L. Sirko {\it et~al.}, \prL {\bf 71},  2895  (1993).

\bibitem{sauer92b}
B.~E. Sauer, M.~R.~W. Bellermann, and P.~M. Koch, \prL {\bf 68},  1633  (1992).

\bibitem{leeuwen85}
K.~A.~H. van Leeuwen {\it et~al.}, \prL {\bf 55},  2231  (1985).

\bibitem{koch95a}
P.~M. Koch, Physica D {\bf 83},  178  (1995).

\bibitem{koch92b}
P.~M. Koch,  in {\em Chaos and Quantum Chaos}, Vol.~411 of {\em Lecture Notes
  in Physics}, edited by W.~D. Heiss (Springer, Berlin, 1992).

\bibitem{koch92a}
P.~M. Koch, \ch {\bf 2},  131  (1992).

\bibitem{koch89}
P.~M. Koch {\it et~al.}, \psc {\bf T26},  51  (1989).

\bibitem{noel00}
M.~W. No\"el, M.~W. Griffith, and T.~F. Gallagher, \pr {\bf A62},  063401
  (2000).

\bibitem{bayfield91}
J.~E. Bayfield, \ch {\bf 1},  110  (1991).

\bibitem{kuba95b}
J. Zakrzewski and D. Delande, J. Phys. B {\bf 28},  L667  (1995).

\bibitem{abu95b}
A. Buchleitner {\it et~al.}, Phys. Rev. Lett. {\bf 75},  3818  (1995).

\bibitem{arndt91}
M. Arndt, A. Buchleitner, R.~N. Mantegna, and H. Walther, \prL {\bf 67},  2435
  (1991).

\bibitem{benson95}
O. Benson {\it et~al.}, \pr {\bf A51},  4862  (1995).

\bibitem{noel00b}
M.~W. No\"el, M.~W. Griffith, and T.~F. Gallagher, \prL {\bf 87},  043001
  (2001).

\bibitem{bluemel90a}
R. Bl\"umel and U. Smilansky, \josab {\bf 7},  664  (1990).

\bibitem{krug00}
A. Krug and A. Buchleitner, \epl {\bf 49},  176  (2000).

\bibitem{krug01}
A. Krug and A. Buchleitner, \prL {\bf 86},  3538  (2001).

\bibitem{breuer89b}
H.~P. Breuer and M. Holthaus, \pla {\bf 140},  507  (1989).

\bibitem{bayfield85}
J.~E. Bayfield and L.~A. Pinnaduwage, \prL {\bf 54},  313  (1985).

\bibitem{reinhold01}
C.~O. Reinhold {\it et~al.}, \jpb {\bf 34},  L551  (2001).

\bibitem{gerber98}
A. Assion {\it et~al.}, Science {\bf 282},  919  (1998).

\bibitem{gerber01}
T. Brixner, N.~H. Damrauer, P. Niklaus, and G. Gerber, Nature {\bf 414},  57
  (2001).

\end{thebibliography}
\end{document}